\documentclass[aps,prd,superscriptaddress,preprintnumbers,floatfix,nofootinbib,notitlepage,onecolumn]{revtex4-2}

\usepackage{graphicx}
\usepackage{amsmath,amsfonts,amssymb}
\usepackage{verbatim}
\usepackage[dvipsnames]{xcolor}
\usepackage{aas_macros}

\usepackage{lettrine}
\usepackage{Zallman}

\makeatletter
\renewcommand{\@seccntformat}[1]{%
  \ifcsname prefix@#1\endcsname
    \csname prefix@#1\endcsname
  \else
    \csname the#1\endcsname\quad
  \fi}

\usepackage[breaklinks,colorlinks,urlcolor=MidnightBlue,citecolor=WildStrawberry,linkcolor=Sepia]{hyperref}

\def\lsim{\mathrel{\raise.3ex\hbox{$<$\kern-.75em\lower1ex\hbox{$\sim$}}}}
\def\gsim{\mathrel{\raise.3ex\hbox{$>$\kern-.75em\lower1ex\hbox{$\sim$}}}}

\newcommand{\be}{\begin{equation}}
\newcommand{\ee}{\end{equation}}
\newcommand{\bea}{\begin{equation}\begin{aligned}}
\newcommand{\eea}{\end{aligned}\end{equation}}
\newcommand{\td}{{\rm d}}
\newcommand{\Msun}{M_{\odot}}
\newcommand{\fPBH}{f_{\rm PBH}}
\newcommand{\MPBH}{M_{\rm PBH}}
\newcommand{\Mpl}{M_{\rm Pl}}

\begin{document}

\title{Primordial black holes: constraints, potential evidence and prospects}

\author{Bernard Carr} \email{b.j.carr@qmul.ac.uk}
\affiliation{Astronomy Unit, Queen Mary University of London, Mile End Road, London E1 4NS, UK}

\author{Antonio J. Iovino} \email{a.iovino@nyu.edu}
\affiliation{Center for Astrophysics and Space Science (CASS), New York University Abu Dhabi, PO Box 129188, Abu Dhabi, UAE}

\author{Gabriele Perna} \email{gabriele.perna@kbfi.ee}
\affiliation{Keemilise ja bioloogilise f\"u\"usika instituut, R\"avala pst. 10, 10143 Tallinn, Estonia}

\author{Ville Vaskonen} \email{ville.vaskonen@pd.infn.it}
\affiliation{Keemilise ja bioloogilise f\"u\"usika instituut, R\"avala pst. 10, 10143 Tallinn, Estonia}
\affiliation{Dipartimento di Fisica e Astronomia, Universit\`a degli Studi di Padova, Via Marzolo 8, 35131 Padova, Italy}
\affiliation{Istituto Nazionale di Fisica Nucleare, Sezione di Padova, Via Marzolo 8, 35131 Padova, Italy}

\author{Hardi Veerm\"ae} \email{hardi.veermae@cern.ch}
\affiliation{Keemilise ja bioloogilise f\"u\"usika instituut, R\"avala pst. 10, 10143 Tallinn, Estonia}

\begin{abstract} 
Primordial black holes (PBHs) may have formed in the early Universe and may account for all or part of the dark matter. In this review, we summarize the current observational constraints on PBHs across the full mass range, highlight potential evidence for their existence, and outline the prospects for future searches, particularly with gravitational-wave observatories. We also discuss different PBH formation scenarios, identify the corresponding mass functions, and present the observational constraints in each case.
\end{abstract}


\maketitle

\makeatletter\let\l@subsubsection\@gobbletwo\makeatother
\tableofcontents

\section{Introduction}

\lettrine{P}{rimordial black holes} (PBHs)~\cite{Zeldovich:1967lct,Hawking:1971ei,Carr:1974nx,Carr:1975qj,Chapline:1975ojl} have been widely studied because of their potential cosmological significance. They have been proposed as candidates for dark matter (DM)~\cite{Carr:2016drx,Carr:2020xqk} (for a review of DM see Ref.~\cite{Cirelli:2024ssz}), contributors to observed gravitational wave (GW) events~\cite{Bird:2016dcv}, and plausible progenitors of supermassive black holes (SMBHs) occupying most galactic nuclei~\cite{Carr:2018rid}. If PBHs possess an extended mass function, as many of their formation models predict, they could simultaneously account for several or even all of these phenomena. Indications for PBHs also arise from various other contexts, including microlensing~\cite{Niikura:2019kqi} and $\gamma$-ray bursts~\cite{Cline:1996zg}.

In contrast, numerous constraints on the abundance of PBH have been established on different mass scales, and these have dominated PBH studies so far. The constraints are associated with a wide range of astrophysical processes: quantum evaporation, dynamical effects, lensing, accretion, structure formation, and GW emission. These limits are usually derived under the assumption of a monochromatic mass function. This is, however, unrealistic, since PBHs in most scenarios are expected to have an extended mass distribution. This problem was addressed in~\cite{Carr:2017jsz}. For the resulting constraints, this is a double-edged sword. On the one hand, by spreading the distribution over many decades of mass, one can reduce the density required at any particular mass, thereby avoiding some constraints. On the other hand, PBHs capable of explaining an observation at a particular mass may still contravene the limits at some larger or smaller mass if the mass function is extended.

In recent years, there have been several developments. First, the constraints themselves have been updated, with some becoming stronger and others weaker. Second, PBH formation mechanisms and viable mass functions have been explored in greater detail. Third, as already indicated, in addition to constraints, we now have claims for positive evidence. Regarding DM, the research is mostly focused on asteroid and stellar mass PBHs. In the stellar-mass range, one expects PBHs to form more easily due to QCD effects; however, this range is already subject to several stringent constraints that ostensibly limit PBHs to a small fraction of DM, and avoiding such constraints becomes increasingly difficult. In contrast, the asteroid-mass window is currently free from constraints, but finding positive indications for PBHs there is also more challenging. Nevertheless, current and near-future GW observations offer opportunities to probe both mass ranges.

In this review, we compile the most up-to-date constraints and provide them in digitized tables accompanied by a Mathematica notebook, available at GitHub:~\href{https://github.com/vianvask/PBHconstraints}{PBHconstraints}, to evaluate these limits for extended mass functions. Rather than reiterating previous comprehensive in-depth reviews of PBH constraints (see e.g.,~\cite{Carr:2020gox, Byrnes:2025tji}), we will concentrate on potential positive indications for PBHs (but see also~\cite{Carr:2019kxo,Carr:2023tpt}) and on prospects for probing PBHs with forthcoming experiments.
We will work with units $c=\Mpl=\hbar=k_B=1$.

\section{Formation and mass functions}
\label{Sec::Form_PBH}

For a general population of PBHs, with a range of masses, the abundance and distribution of masses can be characterized by the PBH mass function
\be \label{eq:psidef}
    \psi(\MPBH) 
    \equiv \frac{1}{\Omega_{\rm DM}} \frac{\td \Omega_{\rm PBH}}{\td \ln \MPBH} \,,
\ee
where $\Omega_{\rm DM} \approx 0.12 h^{-1}$ denotes the DM density parameter. Then the fraction of DM in PBHs $\fPBH$ and the mean PBH mass $\langle \MPBH \rangle$ are given by 
\bea
\label{eq::f_PBH_av_M_PBH}
    &\fPBH 
    \equiv \frac{\Omega_{\rm PBH}}{\Omega_{\rm DM}} 
    = \int \td \ln \MPBH \, \psi(\MPBH) \,,
    \\
    &\langle \MPBH \rangle 
    = \fPBH \left[ \int \td \ln \MPBH \, \MPBH^{-1} \psi(\MPBH) \right]^{-1} .
\eea
A generic extended mass function, frequently used as a benchmark case, is the log-normal distribution
\be \label{eq:psiln}
    \psi_{\rm ln}(\MPBH) \propto \frac{1}{\sqrt{2\pi}\sigma} 
    \exp \left[-\frac{\ln^2\left(\MPBH/M_c\right)}{2 \sigma^2}\right] \,,
\ee
characterised by its width $\sigma$ and mode mass $M_c$, the latter being related to the mean mass by $\langle \MPBH\rangle = M_c e^{-\sigma^2/2}$. In the following, we outline the different scenarios for PBH formation and identify the corresponding mass functions appropriate to each case.

Before considering specific cases, a few more generalities can be established about PBH formation. PBHs form in the early universe, which we take to be any moment before recombination. An important quantity relating to the PBH abundance at the time of their formation is the fraction $\beta_k$ of energy density that will be converted into PBHs when the comoving scale $k$ enters the horizon. Given that a PBH with mass $\MPBH$ forms in a horizon-sized volume with probability $\td   P_k(\MPBH|\boldsymbol{\theta})/\td \ln \MPBH$, provided that the conditions parametrised by the vector $\boldsymbol{\theta}$ hold with probability $P(\boldsymbol{\theta})$, the differential collapse fraction per logarithmic mass interval can be expressed as~\footnote{This equation assumes that the PBH forms at $k = 1/aH$. For instance, in critical collapse scenarios, the formation of PBHs takes place after the density perturbation of size $k$ enters the horizon. The delay between the horizon entry of the density perturbation and the formation of the PBH is then accounted for in $P_k(\MPBH|\boldsymbol{\theta})$.}
\be\label{eq:betak_general}
   \beta_k(\MPBH) 
   = \int  \td \boldsymbol{\theta}\, P(\boldsymbol{\theta}) \frac{\MPBH}{M_k}  \frac{\td P_k(\MPBH|\boldsymbol{\theta})}{\td \ln \MPBH} \,,
\ee
where 
\be
    M_k = 12 \Msun \left(\frac{g_{*s}^4/g_{*}^3}{106.75} \right)^{-\frac16} \left( \frac{k}{10^6\,{\rm Mpc}^{-1}} \right)^{-2} \, 
\ee
is the horizon mass at the moment when the mode with wavenumber $k$ re-enters the Hubble horizon, while $g_*$ and $g_{*s}$ are, respectively, the energy and entropy relativistic degrees of freedom~\cite{Borsanyi:2016ksw} evaluated at the temperature  
\be
T_k = 58\,{\rm MeV} \left(\frac{g_*}{106.75}\right)^{-\frac16} \left(\frac{k}{10^6 {\rm Mpc}^{-1}}\right)\, .
\ee
Eq.~\eqref{eq:betak_general} is derived under the assumption that only a single PBH forms within each horizon volume and must be revised if multiple PBHs can be produced per horizon. However, this assumption is justified, since the initial PBH abundance in the early radiation-dominated Universe must be small to avoid PBH overproduction. 

Given the collapse fraction \eqref{eq:betak_general}, the mass function is then obtained as
\be \label{eq:psidef2}
    \psi(\MPBH)  
    = \frac{1}{\Omega_{\rm DM}} \int \td \ln M_k \, \beta_k(\MPBH) \left( \frac{M_{\rm eq}}{M_k} \right)^{1/2} \,,
\ee
$M_{\rm eq} \approx 2.8 \times 10^{17} \Msun$ is the horizon mass at matter–radiation equality. Eq.~\eqref{eq:psidef2} relies on the assumption that PBH masses are constant after formation. Consequently, in scenarios featuring substantial accretion, an extremely high merger rate, or remnants from evaporating sub-asteroid–mass PBHs, Eq.~\eqref{eq:psidef2} must be revised (see, e.g., Eq.~\eqref{eq:fDM_eva}).

\subsection{Critical collapse}

When large-amplitude curvature perturbations re-enter the horizon, regions with sufficient overdensity can gravitationally collapse to form an apparent horizon~\cite{Zeldovich:1967lct, Hawking:1974rv, Chapline:1975ojl}. In the standard picture, a PBH forms once the density contrast $\delta$ exceeds some threshold $\delta_{\rm th}$. Near this threshold, the collapse exhibits critical behaviour. This universal phenomenon was first identified in numerical studies of general relativistic collapse~\cite{Choptuik:1992jv}. The resulting PBH mass follows a critical scaling relation of the form~\cite{Niemeyer:1997mt,Niemeyer:1999ak}
\be\label{eq:criticalscaling}
    m(\delta) = \mathcal{K} M_k \left(\delta-\delta_{\mathrm{th}}\right)^{\gamma} \,,
\ee
where the parameters $\mathcal{K}$ and $\gamma$ and the threshold $\delta_{\rm th}$ depend on the shape of the curvature power spectrum~\cite{Musco:2008hv,Musco:2020jjb,Musco:2023dak,Ianniccari:2024ltb}. The fraction of energy density collapsing into PBHs~\eqref{eq:betak_general} of mass $\MPBH$ then takes the form
\be\label{eq:betak}
   \beta_k(\MPBH) = \int_{\delta_{\rm th}}  \td\delta \, P_k(\delta) \frac{\MPBH}{M_k}  \delta_D\!\left[ \ln\frac{\MPBH}{m(\delta)} \right]\!,
\ee
where $P_k(\delta)$ denotes the probability that a BH will form in the Hubble patch and $\delta_D$ is the Dirac delta function, selecting patches in which the resulting PBH mass is $\MPBH$. Eq.~\eqref{eq:betak} is fully general provided the PBH mass is fixed solely by the density contrast $\delta$ via Eq.~\eqref{eq:criticalscaling}. In particular, it applies in the presence of non-Gaussianity and for computations based on peaks theory or threshold statistics (see, e.g,~\cite{Iovino:2024tyg}). However, if additional variables influence the PBH mass, the dimension of the integration domain must be enlarged accordingly, following Eq.~\eqref{eq:betak_general}.

The critical collapse scenario relies on the presence of large super-horizon curvature perturbations, which can naturally arise during any inflationary epoch. Producing an appreciable PBH abundance requires a substantial enhancement of the curvature perturbation power spectrum, namely $\mathcal{P}_\zeta(k) \sim \mathcal{O}(10^{-3}\!-\!10^{-2})$, on the scales relevant for PBH formation. While the amplitude and shape of the scalar power spectrum are tightly constrained on the largest cosmological scales by direct CMB observations~\cite{Planck:2018vyg,ACT:2025nti}, limits on CMB spectral distortions~\cite{Chluba:2012we,Chluba:2013dna,Sharma:2024img,Byrnes:2024vjt,Pritchard:2025yda} and Lyman-$\alpha$ forest observations~\cite{Bird:2010mp}, much smaller scales remain essentially unconstrained. Consequently, many primordial mechanisms have been proposed that generate a significant enhancement of the scalar spectrum at these scales (e.g.,~\cite{Ivanov:1994pa,Kinney:1997ne,Inoue:2001zt,Kinney:2005vj,Martin:2012pe,Motohashi:2017kbs,Lyth:2001nq,Barnaby:2010vf,Sorbo:2011rz,Pi:2017gih,Wang:2025lti}). Most models capable of generating such an enhancement lead to a power spectrum that can be approximated by one of the following representative ans\"atze:
\begin{align}
    \text{broken power-law:} \qquad & 
       \mathcal{P}_{\zeta}(k)
    =  \frac{ A \left(\alpha+\beta\right)^{\chi}}{\left[ \beta\left(k / k_*\right)^{-\alpha/\chi}+\alpha\left(k / k_*\right)^{\beta/\chi} \right]^{\chi}}\,, \label{eq:BPLPS} \\
    \text{log-normal:} \qquad & 
    \mathcal{P}_{\zeta}(k) = \frac{A}{\sqrt{2\pi}\Delta}\exp\left[-\frac{\ln^2(k/k_{\star})}{2\Delta^2}\right] \,, \label{eq:Narrow}
    \\
    \text{broad:} \qquad & \mathcal{P}_{\zeta}(k) = A\,\Theta(k - k_{\rm min})\,\Theta(k_{\rm max} - k) \,, 
    \label{eq:Broad}
\end{align}
where $\Theta$ denotes the Heaviside step function.
In the broken power-law case, typical of single field models~\cite{Byrnes:2018txb,Frosina:2023nxu,Karam:2022nym}, $\alpha, \beta>0$ describe, respectively, the growth and decay of the spectrum around the peak, while the parameter $\chi$ characterises the flatness of the peak. The log-normal form is typically adopted for a spectrum with a narrow peak arising from spectator field or hybrid models~\cite{Kohri:2012yw,Kawasaki:2012wr,Garcia-Bellido:1996mdl}, characterised by a well-defined wavenumber $k_{\star}$. 
In contrast, in the broad case~\cite{Franciolini:2022pav, Ferrante:2023bgz}, the power spectrum does not exhibit a sharp peak but instead extends over a wide range between $k_{\rm min}$ and $k_{\rm max}\gg k_{\rm min}$. This is the most idealised case, since in realistic scenarios, the transition at the edges of the enhanced region is never sharp. Their spectra could be approximated by a three–power-law template with two breaks, with the middle segment having an almost scale-invariant spectral index. However, as these tails do not significantly impact PBH phenomenology, we will neglect them here in order to limit the number of template parameters.

For relatively narrow power-spectra, the  critical scaling law~\eqref{eq:criticalscaling} determines the shape of the PBH mass function. This is well approximated by a power-law tail at low masses and an exponential cutoff at high masses~\cite{Vaskonen:2020lbd}:
\be \label{eq:psicc0}
    \psi_{\rm cc}(\MPBH) \propto \MPBH^{1+1 / \gamma} \exp \left[-c_1(\MPBH/ \langle \MPBH\rangle)^{c_2}\right]\,,  
\ee
where $\gamma$ is a universal exponent associated with the critical collapse and the mean mass is approximately given by the horizon mass when the largest perturbations re-enter the horizon, $\langle M_{\rm PBH}\rangle \simeq M_{k_*}$. The coefficient $c_1$ is determined by the condition that $\langle M_{\rm PBH} \rangle$ represents the average PBH mass and the coefficient $c_2 \sim 1$ depends on the amplitude and shape of the curvature power spectrum. This ansatz fails to capture the mass functions arising from broad scale-invariant spectra~\cite{Sugiyama:2020roc,DeLuca:2020agl,DeLuca:2020ioi,Ferrante:2023bgz,Crescimbeni:2025ywm}. In such cases, a broken power law provides a more accurate description:
\be \label{eq:psicc}
    \psi_{\rm cc}(\MPBH) \propto \left[\beta\left(\frac{\MPBH}{M_b}\right)^{-\alpha}+\alpha\left(\frac{\MPBH}{M_b}\right)^{\beta} \right]^{-1} \,,
\ee
with typical choices $\alpha=1+1/\gamma$ and $\beta=0.5$, while $M_b$ is a free parameter that determines the location of the maximum of the distribution. We note that $\beta=0.5$ assumes a radiation-dominated expansion history. In non-standard expansion histories, where the equation of state of the dominant fluid is $\rho = w P$, the power law in the UV is $\beta = -2w/(1+w)$~\cite{Carr:1975qj}, so that a non-inflationary universe $w \in (-1/3,1]$ corresponds to $\beta \in [-1,1)$ and radiation-domination corresponds to $\beta = -1/2$ .

When numerically computing the PBH mass functions from a given power spectrum, we adopt the prescription based on the threshold statistics of the compaction function~\cite{Young:2019yug,DeLuca:2019qsy,Ferrante:2022mui,Gow:2022jfb}. An alternative approach uses peak statistics~\cite{Yoo:2018kvb,Yoo:2019pma,Franciolini:2022tfm}. In the absence of strong primordial non-Gaussianity (NG), the two methods differ only by a mild rescaling of the power spectrum amplitude~\cite{Green:2004wb,Young:2014ana,Iovino:2024tyg}, leading to negligible modifications of the resulting mass functions. This justifies our adoption of a single formalism for the analysis presented in this work.

\subsubsection{Impact of the QCD phase transition}

The dynamics of PBH formation is governed by the competition between self-gravity and pressure forces. Within this framework, the equation of state (EoS) of the cosmic fluid plays a pivotal role: PBH formation is enhanced whenever the EoS softens. Prior to big-bang nucleosynthesis, the universe may have experienced exotic phases, which may alter the equation of state~\cite{Allahverdi:2020bys}. In particular, an early matter-dominated epoch can greatly boost PBH formation~\cite{Khlopov:1980mg,Polnarev:1985btg,Harada:2016mhb,Harada:2017fjm,Kokubu:2018fxy,Carr:2017edp}. Such an epoch can occur during reheating, when a heavy oscillating inflaton induces a temporary matter-dominated era~\cite{Green:2000he,Lyth:2005ze,Padilla:2021zgm,Padilla:2024iyr}.

\begin{figure}  
    \includegraphics[width=0.99\textwidth]{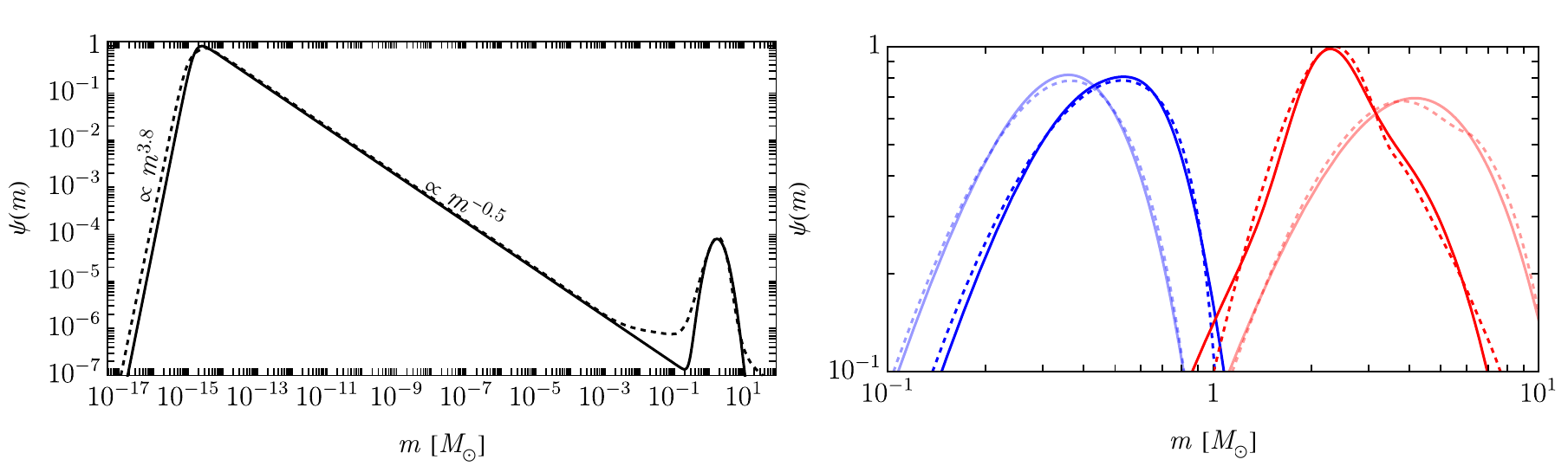}
    \caption{\emph{Left panel:} Broad mass functions obtained from  critical collapse, including the effect of the QCD phase transition. The solid line is given by the ansatz in Eq.~\ref{Eq:QCD}, while the dashed line is computed assuming a broad power spectrum as in Eq.~\ref{eq:Broad}. \emph{Right panel:} Narrow mass functions obtained from critical collapse with the QCD phase transition included (bright lines) and omitted (faded lines). The solid lines are given by Eq.~\ref{Eq:QCD} while dashed lines are numerically computed assuming a log-normal power spectrum with width $\Delta=0.3$ centered at $k_* = 6\cdot10^{6}$ Mpc$^{-1}$ (blue) and $k_* = 2 \times 10^{6}$ Mpc$^{-1}$ (red). In both panels, the mass functions are normalized by $\int \td \ln \MPBH \, \psi(\MPBH) = 1$.}
  \label{fig:mass_functions}
\end{figure}

Within the standard cosmological model, after reheating, the Universe is dominated by relativistic species. As the temperature decreases as a result of cosmic expansion, the thermal plasma undergoes a sequence of transitions where the EoS softens appreciably. Notable examples include the electroweak phase transition at $T \sim 100 \,{\rm GeV}$, epochs of quark annihilation, the QCD phase transition at $T \sim 200 \,{\rm MeV}$, and the stages of electron--positron annihilation and neutrino decoupling around $T \sim 1 \,\text{MeV}$.

Variations in the EoS affect the critical threshold $\delta_{\rm th}$ for PBH formation, a quantity that generally depends on the shape of the primordial power spectrum~\cite{Musco:2018rwt,Germani:2018jgr,Musco:2020jjb}. As a consequence, multiple cosmological transitions can generate multimodal PBH mass distributions, often characterised by a dominant peak accompanied by secondary features. For concreteness, we focus here on the dominant peak, associated with the QCD phase transition, and attempt to parametrise its impact on the PBH mass function\footnote{For additional works studying the effect on the threshold from QCD or other phase transitions see Refs.~\cite{Carr:2019kxo,Jedamzik:1996mr,Jedamzik:1999am,Byrnes:2018clq,Sobrinho:2020cco,Escriva:2022bwe,Escriva:2020tak,Pritchard:2024vix,Pritchard:2025pcn,Blas:2026xws}.}. In particular, when the power spectrum is not nearly scale-invariant, the QCD transition has a significant effect only if the mass function is centred around $\mathcal{O}(0.1\!-\!10)\,\Msun$ as illustrated in Fig.~\ref{fig:mass_functions}. We fit the resulting feature with the log-normal mass function~\eqref{eq:psiln}. Typically, the characteristic mass scale is $M_c = \mathcal{O}(1)\,\Msun$ and the width is $\sigma\sim[0.2,0.5]$, with exact values depending on the shape of the power spectrum and the level of primordial NG~\cite{Ferrante:2022mui, Andres-Carcasona:2024wqk}. For example, for very broad mass spectra without primordial NG, typical values are $M_c=1.7 \,\Msun$, and $\sigma=0.5$. For narrow spectra, these parameters depend sensitively on the location of the main peak. Including the effect of the QCD phase transition, the PBH mass function arising from the critical collapse can be approximated as
\be  \label{Eq:QCD}
    \psi(\MPBH) = A \psi_{\rm cc}(\MPBH) + B \psi_{\rm ln}(\MPBH) \,,
\ee  
where the coefficients $A$ and $B$ characterise the relative contributions arising from the main peak and the QCD part.

In order to numerically determine the impact of the EoS on the thresholds, we follow Ref.~\cite{Musco:2023dak}, in which detailed numerical simulations are presented. Examples of narrow and broad mass functions modified due to QCD phase transitions are shown in Fig.~\ref{fig:mass_functions}. As evident from the right panel, in addition to slightly distorting the overall shape, the QCD phase transition can also shift the average PBH mass $\langle m \rangle$.

\subsubsection{Importance of non-Gaussianity}

\label{sec::PBH_NG}
\begin{figure}
    \includegraphics[width=0.99\textwidth]{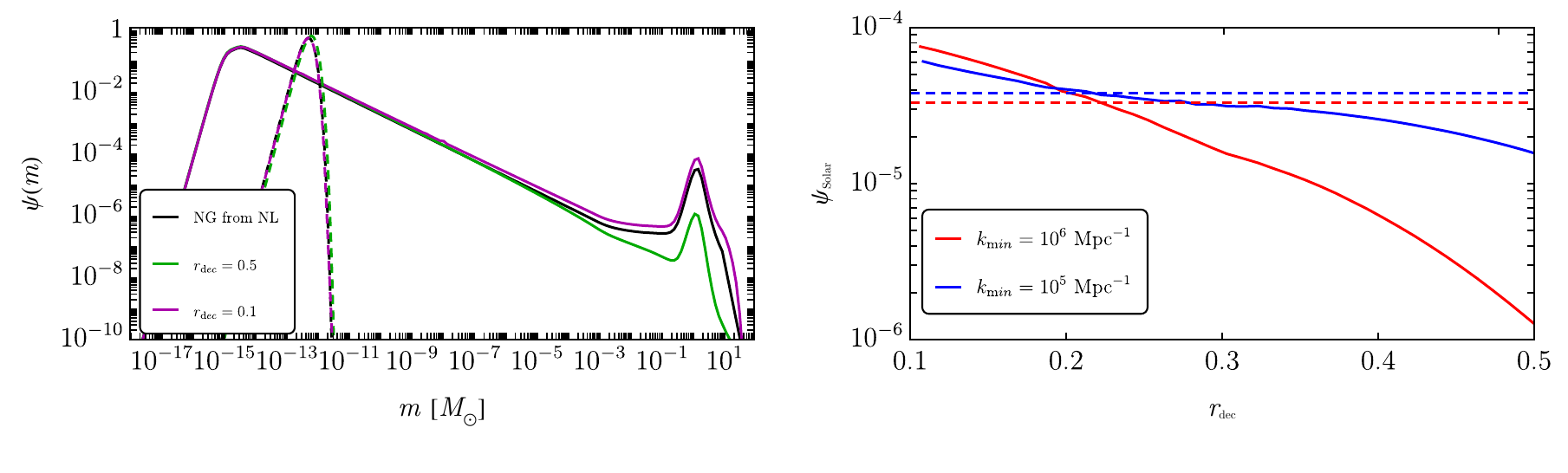}
    \caption{ \emph{Left panel:} Broad (solid) and narrow (dashed) mass functions resulting respectively from a broad power spectrum (with $k_{\rm min}=10^{6}$ $ {\rm Mpc}^{-1}$ and $k_{\rm max} =10^{14.5}$ $ {\rm Mpc}^{-1}$) and a log-normal power spectrum  (with $\Delta=0.5$ and $k_{\rm *}=6 \times 10^{12}$ $ {\rm Mpc}^{-1}$) with different realisations of the NG in the curvaton model, i.e. different $r_{\rm dec}$ (see Ref.~\cite{Ferrante:2023bgz} for more details of the model). The amplitude of the power spectrum has been re-scaled for each $r_{\rm dec}$ such that PBHs comprise the totality of DM.  \emph{Right panel:} Height of the peak at the QCD scale for different $r_{\rm dec}$. We choose the benchmark points $k_{{ \rm min}}=10^{6}$ ${\rm  Mpc}^{-1}$ (red) and $k_{{ \rm min}}=10^{5}$ ${\rm  Mpc}^{-1}$ (blue) respectively. The plot shows a decreasing trend of $\psi_{\rm Solar}\equiv \psi(1\Msun)$ when $r_{\rm dec}$ increases, resulting from the effect of NG reducing the abundance of high masses. The dashed lines represent the height of the peak at the QCD scale, computed by taking into account only non-linearities. The mass functions are normalized in order to get the main peak at $m\simeq 5\times 10^{-14}\Msun$. In both panels, the mass functions are normalized by $\int \td \ln \MPBH \, \psi(\MPBH) = 1$.}
    \label{fig:mass_functionsNG}
\end{figure}

In recent years, it has become increasingly clear that restricting the analysis to the approximation that the probability distribution function (PDF) of density fluctuations $P_k(\delta)$ is Gaussian is not accurate~\cite{Young:2013oia, Bugaev:2013vba,Young:2014ana, Nakama:2016gzw, Byrnes:2012yx, Franciolini:2018vbk, Yoo:2018kvb, Kawasaki:2019mbl,Figueroa:2020jkf, Riccardi:2021rlf, Taoso:2021uvl, Biagetti:2021eep, Kitajima:2021fpq, Hooshangi:2021ubn, Meng:2022ixx, Young:2022phe, Escriva:2022pnz, Hooshangi:2023kss, Ferrante:2022mui, Gow:2022jfb, Ianniccari:2024bkh}. NG of the primordial curvature perturbation is not a marginal technicality but a central element that controls the mapping between an inflationary power spectrum and the resulting PBH mass function. As PBH formation is controlled by the extreme tail of the PDF of the density contrast, even modest deviations from Gaussian statistics produce exponentially large effects on abundances and on the relative weight of different mass ranges, and therefore must be treated explicitly rather than absorbed implicitly into a single rescaling of the power-spectrum amplitude. In practice, NG enters through two conceptually distinct channels that often act together:
\begin{enumerate}
    \item The non-linear relation between curvature perturbations $\zeta$ and the density contrast field $\delta$ induces NG even if $\zeta$ is exactly Gaussian~\cite{DeLuca:2019qsy, Young:2019yug, Germani:2019zez}. This NG is referred to as non-linearity (NL).
    \item In many scenarios that produce large primordial perturbations, including curvaton~\cite{Sasaki:2006kq,Pi:2022ysn,Ferrante:2023bgz} or ultra-slow-roll (USR)~\cite{Atal:2019cdz,Tomberg:2023kli,Iovino:2024sgs} scenarios, the primordial curvature perturbations are already non-Gaussian. This NG is referred to as primordial NG.
\end{enumerate}
When the closed form is unknown, the usual approach is to parametrise the primordial NG with a local expansion
\be\label{eq:FirstExpansion}
    \zeta 
    = \zeta_G + \tfrac{3}{5}f_{\rm NL}\zeta_G^2 + \tfrac{9}{25}g_{\rm NL}\zeta_G^3 + \dots \, ,
\ee
where $\zeta_{\rm G}$ obeys Gaussian statistics, and $f_{\rm NL}$ and $g_{\rm NL}$ encode deviations from the Gaussian limit. We stress that, as shown in Ref.~\cite{Ferrante:2022mui}, this expansion does not capture the correct results for cases with closed relations, as in the curvaton or USR cases. For narrow (peaked) power spectra, the perturbative expansion can converge rapidly, so a truncated series (with convergence checks) may provide reliable estimates of the mass fraction; by contrast, for broad spectra, the perturbative approach typically fails because the region of $\zeta_G$ that dominates the tail lies outside the radius of convergence of the series, and only a non-perturbative treatment of the full function $F(\zeta_G)$ captures the correct result.  This practical dichotomy has immediate consequences for mass functions: in the narrow case, NG mostly rescales and slightly shapes an otherwise localised mass peak, while in the broad case, NG can break naive scale-invariance and introduce an explicit dependence of the PBH fraction $\beta(m,M_k)$ on the horizon mass $M_k$, creating broad redistribution of mass and even secondary features (for example, modifying the prominence of the QCD-related secondary peak). This effect is shown in Fig.~\ref{fig:mass_functionsNG}.

\begin{figure}
  \centering
  \includegraphics[width=0.72\textwidth]{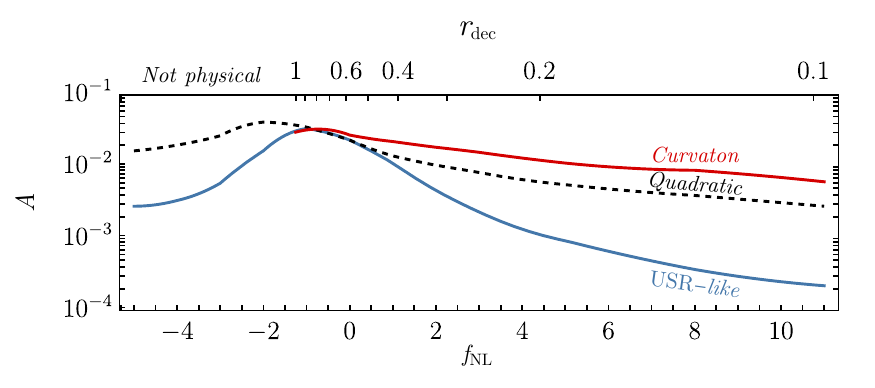}
  \vspace{-5mm}
  \caption{\emph{Left panel}:  Amplitudes for a broken power-law curvature spectrum (Eq.~\eqref{eq:BPLPS}) required to have $\fPBH=1$ for the curvaton model (red solid line) and the USR-like model (blue solid line), using the full NG relation, while the dashed line shows the results using the quadratic approximation. We assume $\alpha=4$, $\beta=3$, $\gamma=1$, $k_*=10^7$ Mpc$^{-1}$ and $\delta_{\rm th}=0.55$.}
  \label{fig:AppFNL}
\end{figure}

The impact of NG on the abundance of PBHs~\footnote{Large negative NG also affects the collapse threshold (see Refs.~\cite{Kehagias:2019eil,Escriva:2022pnz}).} strongly depends on the precise relation between the fully non-Gaussian curvature field $\zeta$ and its Gaussian counterpart $\zeta_G$. Examples of these effects are shown in Fig.~\ref{fig:AppFNL}, where we compare the required amplitude of a given power spectrum shape to achieve $\fPBH=1$ for two cases: quadratic NG, i.e., the first two terms of Eq.~\ref{eq:FirstExpansion} (left panel) and  the curvaton model (right panel).

From Fig.~\ref{fig:AppFNL}, an important observation can be made. One may ask whether the leading-order expansion in $f_{\rm NL}$ can accurately capture the PBH abundance compared to the full curvaton or USR-like cases. For the curvaton, the NG is controlled by the curvaton decay parameter $r_{\rm dec} \in (0,1)$, and the $f_{\rm NL}$ parameter of the curvaton is given by~\cite{Sasaki:2006kq}
\be
    f_{\rm NL} = \frac{5}{3}\left(\frac{3}{4 r_{\rm dec}} - 1 - \frac{r_{\rm dec}}{2}\right)\,,
\ee
so $f_{\rm NL} > -5/4$, as illustrated in the upper $x$-axis of Fig.~\ref{fig:AppFNL}. The USR-like case can be expressed as\footnote{The absolute value follows from Ref.~\cite{Iovino:2024sgs} and mostly serves to guarantee numerical stability for $5f_{\rm NL}\zeta_{G}/6 > 1$. Recent lattice simulations indicate a different continuation beyond that limit~\cite{Caravano:2025diq}. As fluctuations remain perturbative in the examples considered here, the effect of this choice is not significant.}
\be\label{eq:zeta_USR}
    \zeta = - \left(\frac{6f_{\rm NL}}{5}\right)^{-1}\ln\left| 1 - \frac{6f_{\rm NL}}{5}\zeta_{G}\right|\,,
\ee
so that the expansion~(\ref{eq:FirstExpansion}) agrees with it at the quadratic order. Such a relation is responsible for generating exponential tails in the distribution of perturbations of $\zeta$. Although Eq.~\eqref{eq:zeta_USR} was initially derived in the context of inflection point models with a USR to constant-roll transition~\cite{Atal:2018neu,Tomberg:2023kli}, in which case $f_{\rm NL}$ is positive~\cite{Firouzjahi:2023xke}, it has also been observed in the context of first order phase transitions but with a negative $f_{\rm NL}$~\cite{Lewicki:2024ghw}. Thus, we will label~\eqref{eq:zeta_USR} as USR-like and consider it as another generic example of non-Gaussian fluctuations without reference to a specific scenario.

As seen in Fig.~\ref{fig:AppFNL}, for most of the parameter space scanned in both scenarios, the quadratic approximation (dashed line) does not capture the correct  abundance when compared to the full NG relation (solid lines). The quadratic approximation overestimates the PBH abundance for the curvaton model and underestimates it for the USR-like models. In the limit, $r_{\rm dec}\rightarrow1$, the NG relation of the curvaton will take the USR-like form~\eqref{eq:zeta_USR} and, expectedly, both estimates match at that point.

Moreover, one might naively expect that making $f_{\rm NL}$ more negative would monotonically increase the required amplitude of the power spectrum, corresponding to $\fPBH\simeq1$. However, as shown in Fig.~\ref{fig:AppFNL}, the amplitude reaches a maximum around $f_{\rm NL}\simeq-1.5$, with the precise value depending on the model under consideration. The emergence of this peak and the subsequent decrease of $A$ for smaller $f_{\rm NL}$ can be traced back to the fact that the PDF of the density contrast field $\delta$, the fundamental quantity for computing the PBH abundance by Eq.~\eqref{eq:betak}, is not linearly related to that of the curvature perturbation $\zeta$. Consequently, suppressing the PDF of $\zeta$ does not necessarily imply a suppression of the PDF of $\delta$ ~\cite{Franciolini:2023pbf}.

\subsection{Other mechanisms}

In addition to inflationary mechanisms that amplify density perturbations on small scales, PBHs can be produced in several non-inflationary scenarios~\cite{Flores:2024eyy}. These alternative channels exploit processes in the early Universe such as phase transitions or collapse of topological defects, or particle dynamics beyond the Standard Model.

\subsubsection{Phase transitions}

Cosmological first-order phase transitions can lead to formation of PBHs in bubble collisions~\cite{Hawking:1982ga,Jung:2021mku} or from collapsing false vacuum regions~\cite{Sato:1981gv,Kodama:1982sf,Gross:2021qgx,Lewicki:2023ioy}. Collapsing false vacuum regions can also arise from vacuum bubbles nucleated during inflation~\cite{Deng:2017uwc,Kusenko:2020pcg,Maeso:2021xvl,Wang:2025hwc}. While the PBH mass function arising from the collapse of false-vacuum regions during a first-order phase transition is typically very narrow~\cite{Lewicki:2023ioy}, the collapse of vacuum bubbles nucleated during inflation can yield a much broader distribution. In this case, the spectrum can take the form of a cut piecewise power-law,
\be
    \psi_{\rm pl}(\MPBH) \propto 
    \begin{cases}
        (\MPBH/M_b)^{\alpha} \,, & M_{\rm min} \leq \MPBH < M_b \\
        (\MPBH/M_b)^{\beta} \,, & M_b \leq \MPBH \leq M_{\rm max}
    \end{cases} \,.
\ee
The bubble nucleation scenarios of Refs.~\cite{Kusenko:2020pcg,Maeso:2021xvl} predict $\{\alpha,\beta\} = \{0,-1/2\}$ and $\{-1/2,1/2\}$, respectively, with the latter also allowing for $M_b = M_{\rm min}$.

Recent attention has focused on slow and strongly supercooled first-order phase transitions, including a period of thermal inflation during which the Universe remains trapped in a metastable vacuum until well below the critical temperature~\cite{Liu:2021svg,Kawana:2022olo,Gouttenoire:2023naa,Baldes:2023rqv,Lewicki:2024ghw,Lewicki:2024sfw,Franciolini:2025ztf,Kierkla:2025vwp}. In such transitions, the formation of only a few large bubbles per Hubble volume gives rise to considerable stochastic density fluctuations and the PBH formation is based on critical collapse. A subsequent work~\cite{Franciolini:2025ztf} found that, once the coupling between density perturbations and metric perturbations is properly accounted for, PBH formation is highly suppressed. However, accounting for the gradient energy of the bubble walls may change this result~\cite{Gouttenoire:2026xxx}. 

Large density perturbations may also emerge in strongly supercooled transitions that are not of first order~\cite{Dimopoulos:2019wew}. Including both thermal and quantum fluctuations, Ref.~\cite{Bastero-Gil:2023sub} found that the induced curvature perturbations are, however, insufficient to source PBHs. Furthermore, to enhance PBH formation from collapsing false vacuum regions, it has been proposed that particles that are heavy in the true vacuum become trapped inside the false vacuum regions, where they are subsequently compressed into PBHs during collapse~\cite{Kitajima:2020kig,Baker:2021nyl}. However, once the backreaction of these particles on the bubble wall dynamics is taken into account, PBH formation through this mechanism is strongly suppressed~\cite{Lewicki:2023mik} unless the trapped particles possess sufficiently strong attractive self-interactions~\cite{Kawana:2021tde}.

\subsubsection{Topological defects}

The possibility that topological defects, such as cosmic strings, monopoles, and domain walls~\cite{Vilenkin:2000jqa}, could seed PBHs has long been recognized~\cite{Hawking:1987bn}. A cosmic string network rapidly settles into a scaling regime in which loops of sub-Hubble size are continuously generated through string intercommutations. These loops contract under their own tension and can collapse to form PBHs~\cite{Hawking:1987bn,Polnarev:1988dh,Caldwell:1995fu,MacGibbon:1997pu,Hansen:2000jv,Helfer:2018qgv,James-Turner:2019ssu}. Ref.~\cite{James-Turner:2019ssu} found the resulting PBH mass function to have the power-law form, 
\be \label{eq:psipl}
    \psi_{\rm pl}(\MPBH) \propto \MPBH^{\alpha} \,, \quad M_{\rm min} < \MPBH < M_{\rm max}
\ee
with $\alpha = -1/2$ in a radiation-dominated era. The locations of the mass cutoffs and the abundance of PBHs depend on the properties of the string network, most notably the string tension. 

In the case of domain walls, a small bias between vacua is required to prevent the domain walls from dominating the energy-density of the Universe. Although the network eventually annihilates due to this bias, rare higher-energy domains can persist long enough for their collapse to produce PBHs~\cite{Rubin:2000dq,Ferrer:2018uiu,Gelmini:2023ngs,Gouttenoire:2023ftk,Gouttenoire:2023gbn,Ferreira:2024eru,Dunsky:2024zdo}. The PBH mass function in this case is expected to be nearly monochromatic~\cite{Gouttenoire:2023gbn}.

\subsubsection{Other non-inflationary scenarios}

Beyond phase transitions and topological defects, several other non-inflationary mechanisms can generate PBHs. Among them are scenarios including a period dominated by Q-balls, with PBHs forming from the gravitational collapse of regions where the number of Q-balls fluctuates above average due to Poisson statistics~\cite{Cotner:2016cvr,Cotner:2017tir}. Alternatively, in confining gauge theories, PBHs can form because the large amount of energy stored in the color flux tubes connecting quark pairs can induce gravitational collapse~\cite{Dvali:2021byy}. PBHs can also form from large baryon-number fluctuations generated by an inhomogeneous Affleck–Dine mechanism, which later convert into density perturbations at the QCD epoch and collapse gravitationally~\cite{Dolgov:1992pu}.

\section{Evaporation}

BHs lose mass by emitting Hawking radiation~\cite{Hawking:1974rv,Hawking:1975vcx} with an approximately black-body spectrum with  temperature
\be \label{eq:temp}
	T_{\rm BH}(\MPBH) = \frac{1}{8\pi \MPBH}  \approx 10^{4} \left(\frac{\MPBH}{10^9 {\rm g}}\right)^{-1} {\rm GeV} \,.
\ee
The evaporation process decreases both the mass and the angular momentum of the BH. In terms of the dimensionless spin parameter, $a_* \leq 1$, the mass and spin loss rates are~\cite{Page:1976df,Page:1976ki}
\be \label{eq:mlr}
	\frac{\td \MPBH}{\td t} = -\sum_j \alpha_j(\MPBH,a_*) \MPBH^{-2} \,, \qquad \frac{\td a_*}{\td t} = -a_*\sum_j\left(\beta_j(\MPBH,a_*)-2\alpha_j(\MPBH,a_*)\right) \MPBH^{-3} \,,
\ee
where the sums are over all particle species that are lighter than the BH temperature, $m_j<T_{\rm BH}$, and the coefficients $\alpha_j, \beta_j$ depend on the particle species $j$. They are tabulated in Refs.~\cite{Page:1976ki,Taylor:1998dk}. For non-spinning BHs, assuming the Standard Model particle content (including gravitons) and $\MPBH \ll 10^{11}\,$g, which corresponds to $T_{\rm BH} \gg 100\,$GeV, we have $\sum_j \alpha_j \approx 4.4\times 10^{-3}$, and the BHs evaporate entirely on a timescale 
\be \label{eq:tau}
    \tau = 0.4\,{\rm s} \left(\frac{\MPBH}{10^9\,{\rm g}}\right)^3 \,.
\ee
A non-spinning PBH with an initial mass $\MPBH \approx 5 \times 10^{14}$\,g, formed at $10^{-23}$\,s and having the size of a proton, would be completing its evaporation in the present era, while less massive ones would have vanished earlier. For those lighter than $2 \times 10^{14}$\,g, the temperature exceeds the QCD confinement scale, resulting in the emission of quark and gluon jets that subsequently fragment into stable particles. Therefore, the overall radiation comprises both these primary emissions and their secondary products~\cite{MacGibbon:1991vc,MacGibbon:1990zk,Heckler:1997jv}.

While in typical formation scenarios the PBHs have vanishing spins (see e.g.~\cite{DeLuca:2019buf}), those formed during a matter-dominated period can have nearly extremal spins~\cite{Harada:2017fjm}. The evaporation timescale of such BHs is about half that of a non-spinning one (see, e.g.~\cite{Arbey:2019jmj}). Moreover, in scenarios where a population of light PBHs is formed with an extended initial mass function, evaporation reshapes the mass distribution~\cite{Cheek:2022mmy}.

\subsection{Dark matter}

Even when PBHs evaporate, they can contribute to the DM abundance either by partially evaporating into a dark sector or by leaving behind a stable remnant~\cite{Bell:1998jk,Khlopov:2004tn,Fujita:2014hha,Allahverdi:2017sks,Lennon:2017tqq,Rasanen:2018fom,Morrison:2018xla,Salvio:2019llz,Masina:2020xhk,Baldes:2020nuv}. If the DM mass created per PBH is $M_{\rm R}(m_{\rm BH})$, then the contribution to the present DM is given by a slightly modified version of Eqs.~\eqref{eq:psidef2} and~\eqref{eq:betak}:
\be\label{eq:fDM_eva}
    f_{\rm DM} 
    = \frac{1}{\Omega_{\rm DM}} \int \td \ln M_k \left(\frac{M_{\rm eq}}{M_k} \right)^{1/2} \int \td \delta \, P_k(\delta) \frac{M_{\rm R}(m(\delta))}{M_k} \,.
\ee
The semiclassical picture of Hawking radiation may break down as the Hawking temperature approaches the Planck scale. The consequent evolution of the BHs is uncertain, as we do not have a complete quantum theory of gravity. It has been suggested that the evaporation ends, leaving a stable Planck mass relic~\cite{MacGibbon:1987my}. Furthermore, it has been suggested that the evaporation may slow down or even stop much earlier due to the memory burden effect, based on black hole information theory~\cite{Dvali:2020wft,Thoss:2024hsr,Dvali:2024hsb,Dvali:2025ktz,Dondarini:2025ktz}. 

For stable evaporation remnants, the mass produced per PBH is given by the mass of the remnant, $M_{\rm R}(M_{{\rm BH}}) = M_{\rm rem}$ when $M_{{\rm BH}} \geq M_{\rm rem}$. These remnants can overclose the universe~\cite{MacGibbon:1987my} and thus, in addition to being a potential DM candidate, they constrain the initial curvature perturbation. Inflationary models that predict DM in PBH remnants consistent with current CMB observations require an initial PBH mass $\MPBH < 10^6 g$~\cite{Rasanen:2018fom}. 

If PBH remnants do not make up DM, a new stable particle, which we will denote by $\chi$, is needed to explain the DM.
Evaporating PBH will inevitably produce such particles as their temperature increases. If $\chi$ is chemically decoupled from the radiation bath, their comoving number density will be conserved. In this case, given that these particles are cold today, the mass per PBH is 
$M_{\rm R}(M_{\rm PBH}) = m_\chi N_\chi (M_{\rm PBH})$, where $N_\chi$ is the number of $\chi$ particles produced per PBH. $N_\chi$ can be considered constant when the initial temperature of the PBH is higher than $m_\chi$ and will decrease when the initial PBHs are lighter~\cite{Morrison:2018xla}.

\subsection{Gravitational waves and dark radiation}

If produced abundantly, light PBHs can dominate the energy budget of the early Universe before evaporating. Such a period of PBH dominance results in a broad GW background. First, Hawking evaporation itself generates high-frequency gravitons, contributing to the stochastic GW background, whose spectral shape is sensitive to the PBH mass spectrum and the spin distribution \cite{Anantua:2008am,Dolgov:2011cq,Dong:2015yjs,Ireland:2023avg,Gehrman:2023esa}. Second, PBHs form binaries throughout the PBH-dominated period and these binaries then emit GWs~\cite{Anantua:2008am,Dolgov:2011cq,Zagorac:2019ekv,Hooper:2020evu}. Third, the inhomogeneous PBH distribution generates curvature fluctuations that will induce a GW background~\cite{Inomata:2020lmk,Papanikolaou:2020qtd,Domenech:2020ssp,Domenech:2021wkk,Papanikolaou:2022chm,Bhaumik:2022pil}. The transition of the background from matter to radiation domination via evaporation alters the transfer functions connecting primordial curvature perturbations to induced GWs. This affects the locations and magnitudes of the peaks in the GW spectrum when compared to radiation-dominated scenarios.

Finally, because evaporations also inject relativistic degrees of freedom (affecting \(N_{\rm eff}\)) and can produce non-thermal particle populations, joint constraints from BBN, CMB and GW observations become particularly powerful probes of both the initial mass function and its subsequent evaporation history~\cite{Domenech:2021wkk,Bhaumik:2022pil}.

\section{Constraints and potential evidence}

Until recently, most PBH research has focused on obtaining observational limits on the PBH number density for a monochromatic mass function. The limits span an enormous mass range, from $10^{-24}\Msun$ to $10^{20} \Msun$ and can be expressed as upper limits on $\fPBH(\MPBH)$, the fraction of the DM in PBHs with mass $\MPBH$, or $\beta(\MPBH)$, the fraction of the Universe going into such PBHs at formation. We show a compilation of them in Fig.~\ref{fig:mono}. The digitized tables of these constraints and a Mathematica notebook to plot them are publicly available at GitHub:~\href{https://github.com/vianvask/PBHconstraints}{PBHconstraints}. In this section, we briefly list these constraints and provide the relevant references. In addition, we discuss observations accumulated in recent years, which have been claimed as evidence for PBHs. This corresponds to what is termed a ``positivist'' approach to the subject~\cite{Carr:2023tpt}.

\subsection{Evaporation}

\subsubsection{Constraints}

The evaporation of PBHs leads to strong constraints in the mass range $10^{9} - 10^{17}$\,g~\cite{Auffinger:2022khh}. Eq.~\eqref{eq:tau} implies that non-spinning PBHs lighter than about $10^9$\,g evaporate before BBN\footnote{Smaller masses, i.e. $M\lsim10^9$\,g, can be constrained assuming that PBHs leave stable Planck mass relics~\cite{MacGibbon:1987my} but with large theoretical uncertainties~\cite{Lehmann:2019zgt,Bai:2019zcd}}, while the observations of the abundances of light elements generated during the BBN constrain $\beta(\MPBH)$ and $\fPBH(\MPBH)$ in the mass range $10^9 - 10^{13}$\,g~\cite{Carr:2009jm,Carr:2020gox} (for recent similar works see also Refs~\cite{Keith:2020jww,Kohri:1999ex,Boccia:2024nly,Wu:2025ovd}). The abundance of PBHs heavier than $10^{13}$\,g is 
constrained by CMB observations, as they would evaporate during recombination and inject energetic charged particles that damp the CMB temperature anisotropies~\cite{Acharya:2020jbv}. For PBHs that evaporate in the local (low redshift) universe, constraints stronger than those from CMB observations arise from observations of the extragalactic photon background (EGB)~\cite{Carr:2009jm} and observations of the electron-positron spectra with Voyager~\cite{Boudaud:2018hqb}. These constraints extend to PBH masses $m \sim 10^{17}$\,g (see also Refs.~\cite{Clark:2016nst,Boudaud:2018hqb,DeRocco:2019fjq,Dasgupta:2019cae,Laha:2019ssq,Laha:2020ivk,Saha:2021pqf,Mittal:2021egv,Calza:2021czr,Berteaud:2022tws,DelaTorreLuque:2024qms,Saha:2024ies,Su:2024hrp,Khan:2025kag}).

PBH evaporation constraints can be altered in models with large extra dimensions, in which the Planck mass can be substantially lowered~\cite{Dimopoulos:2001hw,Giddings:2001bu}. The evaporation bounds can be further modified by the memory burden effect~\cite{Thoss:2024hsr,Dondarini:2025ktz} or if the collapse ends up in some exotic ultracompact object rather than a BH~\cite{Raidal:2018eoo}.

\subsubsection{Potential evidence}

There are some observations that can be interpreted as signatures of PBH evaporation. In particular, in their final explosive phase, evaporating PBHs produce short-duration $\gamma$-ray bursts with a universal luminosity and spectrum~\cite{Boluna:2023jlo,Baker:2025zxm}. The resulting $\gamma$-ray bursts would be detectable within a few parsecs~\cite{Ukwatta:2010zn}. Studies of the $\gamma$-ray burst catalogues have identified events that could be associated with PBHs~\cite{Cline:1995jf,Cline:1996zg} and, intriguingly, by estimating the distances of $\gamma$-ray bursts using an interplanetary network of detectors, Ref.~\cite{Ukwatta:2015mfb} identified several nearby events within $\mathcal(1-100)$\,AU.

Moreover, the anticipated clustering of evaporating PBHs within our halo should produce a Galactic $\gamma$-ray background, distinguishable from the uniform extragalactic background. Ref.~\cite{Wright:1995bi} claimed that a galactic background had been detected in EGRET observations and attributed this to PBHs. Later analysis of EGRET data, assuming a variety of PBH distributions, reassessed this limit by including a realistic model for the PBH mass spectrum and a more precise relationship between the initial and current PBH mass, leading to constraints on the PBH population~\cite{Lehoucq:2009ge,Carr:2016hva}.

Another observable signature of PBH explosions is the production of extremely energetic particles. Recently, the KM3NeT collaboration reported the detection of the most energetic neutrino to date, with an energy of about $100\,$PeV~\cite{KM3NeT:2025ccp}. One possible interpretation of this event is that it originated from the explosion of a PBH~\cite{Boccia:2025hpm,Baker:2025cff,Aldecoa-Tamayo:2025dqe}.

\begin{figure}
\centering
\includegraphics[width=\textwidth]{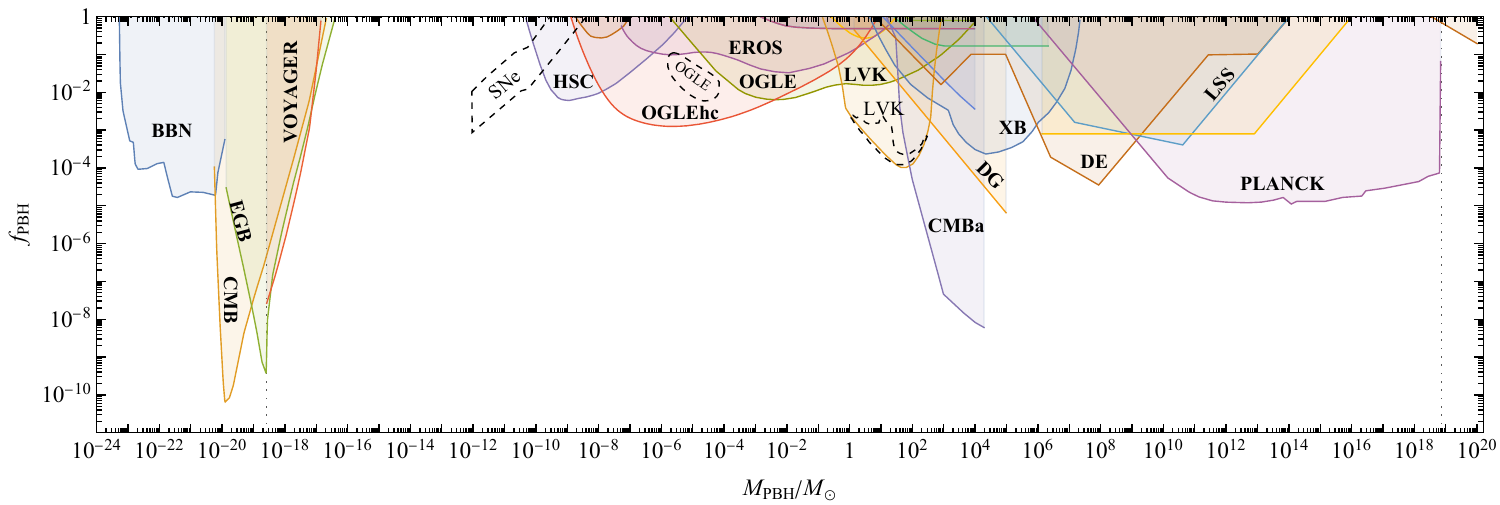}
\caption{Compilation of constraints for monochromatic PBH mass functions (bold) and some relevant potential evidence (dashed) (SNE~\cite{Smirnov:2022zip}, OGLE~\cite{Niikura:2019kqi}, LVK~\cite{Andres-Carcasona:2024wqk}). Only the dominant constraint in each mass range is labelled. The vertical dashed line on the left indicates the PBH mass that would complete its evaporation today, while the vertical dashed line on the right corresponds to the horizon mass at the time of CMB formation.}
\label{fig:mono}
\end{figure}

\subsection{Lensing}

\subsubsection{Constraints}

PBHs can be sought in the present universe through their lensing effects. The most prominent of these is microlensing (ML), which manifests as a temporary brightening of background stars when compact objects traverse their line of sight~\cite{Paczynski:1985jf}. Since the 1990s, several observational campaigns have monitored the brightness of stars in the Magellanic Clouds, the Galactic bulge, and  M31. Currently, the constraints arising from MACHO~\cite{Macho:2000nvd}, EROS~\cite{EROS-2:2006ryy}, Kepler~\cite{Griest:2013aaa}, Subaru/HSC~\cite{Smyth:2019whb} and OGLE~\cite{Mroz:2024mse,Mroz:2024wia,Mroz:2025xbl} ML surveys\footnote{The validity of the recent analysis by the OGLE collaboration has been challenged and is discussed in Refs.~\cite{Hawkins:2025mlo,Mroz:2025aor}.} cover a broad mass range and it has been claimed that this excludes PBHs with $10^{-10} - 10^3\,\Msun$ making up more than $1\%$ of the DM halo mass. The low-mass end of this range is associated with wave-optics and finite-source-size effects that suppress the amplification, while the high-mass end is associated with the long duration of the ML events. These constraints have been extended to higher masses by using lensing of supernovae~\cite{Zumalacarregui:2017qqd} and GW events observed by LVK~\cite{Urrutia:2023mtk}. The former relies on estimates of the supernova luminosities, while the latter is based on the frequency dependence of the lensing effect.

The ML constraints can change if the PBHs are initially clustered~\cite{Petac:2022rio, Gorton:2022fyb}. However, the amount of clustering required to have a significant effect on lensing will affect structure formation and is in conflict with Lyman-$\alpha$ constraints~\cite{DeLuca:2022uvz}. Consequently, ML constraints remain largely unchanged in models with phenomenologically viable initial clustering.

ML constraints depend sensitively on the assumed DM density profile and velocity dispersion in the Milky Way~\cite{Green:2017qoa,Green:2025dut}. It has been claimed that these constraints may be relaxed by up to two orders of magnitude~\cite{Li:2025mqx}, although Ref.~\cite{Mroz:2025xbl} argues that the profiles used here are inconsistent with observations of the Milky Way rotation curve.

\subsubsection{Potential evidence}

Several excess ML events have been observed and might be attributed to a population of PBHs. In particular, the MACHO collaboration detected $13 - 17$ events, which is significantly more than the $2-4$ that could be attributed to known stellar populations~\cite{MACHO:2000qbb}. Moreover, using data from the five-year OGLE survey of microlensing events in the Galactic bulge, Ref.~\cite{Niikura:2019kqi} identified $6$ ultra-short events that can be attributed to planetary-mass objects in the range $10^{-6} - 10^{-4}\Msun$, which is more than expected for free-floating planets. The same group also carried out an observation of Andromeda with the Subaru Hyper Suprime-Camera (HSC) and reported a single ML event by a compact body with mass in the range $10^{-11} - 10^{-5}\Msun$~\cite{Niikura:2017zjd}. 

ML events have also been searched for in the light curves of multiply-lensed quasars. In particular, photometric monitoring of Q0957+561 showed that small variations in brightness in one image were repeated in the second image a year later, which confirmed its identification as a gravitational lens~\cite{1991AJ....101..813S}. Although the ML of quasars might be attributed to ordinary stars, in some cases, the line of sight is too far from the galaxy for this to be viable, and it has been argued in~\cite{Hawkins:2020zie} that stellar mass PBHs are the most plausible candidates for these events. Moreover, based on a large sample of quasar light curves, Ref.~\cite{Hawkins:2022vqo} claims that the observed distribution of their amplitudes implies that stellar mass compact objects constitute an appreciable fraction of the DM.

\subsection{Gravitational waves}
\label{sec::solar}
\subsubsection{Constraints}

PBHs naturally form binary systems in the early Universe, constituting the first gravitationally bound structures to emerge~\cite{Nakamura:1997sm}. The formation mechanism relies on the gravitational decoupling of a PBH pair from cosmic expansion in regions of locally enhanced PBH density (for a review, see~\cite{Raidal:2024bmm}). When $\fPBH \ll 1$, a significant fraction of binaries with coalescence times comparable to the age of the Universe survive until the present~\cite{Ali-Haimoud:2017rtz,Raidal:2018bbj,Vaskonen:2019jpv} and thus contribute to the observed merger rate. 
When $\fPBH = \mathcal{O}(1)$, nearly all these binaries are expected to be perturbed \cite{Raidal:2018bbj,Vaskonen:2019jpv,Delos:2024poq}. However, because binary formation is much more efficient, and the binaries merging presently are mostly formed via 3-body processes in the early universe~\cite{Vaskonen:2019jpv,Raidal:2024bmm}, the $\fPBH = \mathcal{O}(1)$ case results in a much less eccentric PBH binary population than $\fPBH \ll 1$ scenarios. Such a population retains its characteristics, including the merger rates, even if the binaries are frequently disrupted~\cite{Vaskonen:2019jpv,Franciolini:2022ewd}.

The LIGO-Virgo-KAGRA (LVK) collaboration has detected hundreds of binary BH mergers to date~\cite{LIGOScientific:2018mvr,LIGOScientific:2020ibl,KAGRA:2021vkt,LIGOScientific:2025slb}. However, it has been argued that the observed merger rate is inconsistent with scenarios in which stellar-mass PBHs comprise even $0.1 \%$ of the DM. Therefore, the LVK observations place a strong constraint on PBHs~\cite{LIGOScientific:2019kan,Kavanagh:2018ggo,Wong:2020yig,Hutsi:2020sol,DeLuca:2021wjr,Franciolini:2022tfm,Andres-Carcasona:2024wqk,Gasparotto:2025wok}. The constraint shown in Fig.~\ref{fig:mono} is based on Ref.~\cite{Andres-Carcasona:2024wqk}, which incorporates data up to the third LVK observing run. In addition to measurements of the merger rate, unresolved PBH binaries will also induce a stochastic GW background, which implies $\fPBH \lesssim \mathcal{O}(10^-2)$ for stellar-mass PBHs~\cite{Hutsi:2020sol,Boybeyi:2024mhp}. However, this is currently weaker than the constraint obtained from resolvable PBH binary mergers.

Further GW constraints can be obtained from the link between PBH formation and scalar-induced GW (SIGW) production. We do not report those in Fig.~\ref{fig:mono} since they assume PBH formation from critical collapse and depend on the shape of the curvature power spectrum, which are discussed in detail in Sec.~\ref{sec::SIGW}.

\subsubsection{Potential evidence}

The BH population observed by LVK exhibits a mass distribution with two or more peaks~\cite{KAGRA:2021duu,LIGOScientific:2025pvj}. While this could arise from multiple astrophysical binary formation channels, it remains possible that some of the LVK events originate from PBH mergers~\cite{Hutsi:2020sol,DeLuca:2021wjr,Andres-Carcasona:2024wqk}. In particular, LVK has detected events in mass gaps~\cite{KAGRA:2021duu,LIGOScientific:2025pvj} that are difficult to explain with stellar progenitors. In the {\it upper} mass gap between $60 - 120\,\Msun$, progenitor stars are not expected to form BHs, as pair-instability supernovae would completely disrupt them~\cite{1964ApJS....9..201F,Woosley:2016hmi}. In the {\it lower} mass gap between $2 - 5\,\Msun$, compact remnants are disfavoured due to the maximum mass of neutron stars and the minimum mass required for BH formation via stellar collapse~\cite{Bailyn:1997xt,Farr:2010tu}. Although there are astrophysical processes that may fill both mass gaps (see, e.g.,~\cite{LIGOScientific:2025pvj} and references therein), the mass gap events could also point to the existence of a PBH population~\cite{Yuan:2025avq,DeLuca:2025fln}. In addition, a candidate subsolar-mass event with a false alarm
rate of 1 in 5 years was identified during O3 by {\rm GstLAL}~\cite{Cannon:2020qnf}, whose potential primordial origin was considered in Ref.~\cite{Prunier:2023uoo}.

Additionally, the existence of very heavy PBHs, with $\MPBH\geq10^{14}\Msun$, can be tested through a possible detection of the ringdown phase associated with their formation~\cite{DeLuca:2025uov,Yuan:2025bdp}, using future proposed CMB experiments looking for B-modes, such as PIXIE~\cite{Kogut:2011xw}, Super Pixie~\cite{Kogut:2019vqh}, and Voyage2050~\cite{Chluba:2019nxa}.

\subsection{Dynamics}

\subsubsection{Constraints}

There are numerous ways to probe PBHs through their dynamical effects on astrophysical structures. For PBHs larger than a stellar mass, these are associated with dynamical friction effects, heating of galactic discs, 
galactic tidal distortions~\cite{Carr:1997cn}, effects on stellar systems in dwarf galaxies~\cite{Brandt:2016aco,Koushiappas:2017chw}, disruption of wide binary star systems~\cite{Monroy-Rodriguez:2014ula}, and the CMB dipole anisotropy~\cite{Carr:2020erq}. Moreover, the Poisson fluctuations associated with PBH formation influence cosmic structure formation by enhancing the small-scale matter power spectrum. In particular, the Lyman-$\alpha$ forest and the UV luminosity function constrain such an enhancement~\cite{Murgia:2019duy,Gouttenoire:2023nzr,Ellis:2025xju,Ivanov:2025pbu,Gerlach:2025vco}. 
There have also been attempts to constrain the asteroid-mass range from the effects of PBHs on white dwarfs and neutron stars; we discuss this in Sec.~\ref{sec:asteroidmasswindow}.

\subsubsection{Potential evidence}

The main potential dynamical evidence for PBHs is the triggering of explosions of stars. In particular, some recently observed supernov{\ae}, the so-called ``calcium-rich transients'', do not trace the stellar density but are located off-centre from their host galaxies and appear to originate from white dwarfs with masses of around $0.6\Msun$. Ref.~\cite{Smirnov:2022zip} argues that these events could have been triggered by collisions with PBHs in the mass range $10^{-12} - 10^{-10}\Msun$ with $10^{-3} < \fPBH < 0.1$. The wider mass range of $10^{-14} - 10^{-8}\Msun$ has been invoked to explain how some $r$-process elements (i.e.~those generated by fast nuclear reactions) can be produced by the interaction of PBHs with neutron stars and this may also explain the large luminosities and millisecond durations of fast radio bursts (FRBs)~\cite{Fuller:2017uyd,Abramowicz:2017zbp}. Estimates of the rates of such collisions indicate that PBHs can account for the observed FRBs only if the PBH density profile is sufficiently spiky in galactic centers~\cite{Kainulainen:2021rbg,Amaral:2023ekd}.

Other possible dynamical effects include the heating of stars in the Galactic disk~\cite{Cang:2022jyc} and the heating of the cold DM through PBH infall and two-body processes in a scenario where PBHs provide a subdominant DM component~\cite{Boldrini:2019isx}. Interestingly, the latter may lead to the formation of the observed cores in dwarf galaxies. The recent JWST observations have triggered interest in the possible role of PBHs in structure formation~\cite{Cappelluti:2021usg,Hutsi:2022fzw,Colazo:2024jmz,Zhang:2024ytf,Dayal:2024zwq,Dayal:2025aiv,Prole:2025snf,Zhang:2025asq,Matteri:2025klg,Zhang:2025oyl}, for "little red dots"~\cite{Maiolino:2025tih,Hai-LongHuang:2024gtx,Zhang:2025tgm, DeLuca:2025nao} and the SMBHs in galactic nuclei~\cite{Liu:2022bvr,Ziparo:2024nwh,Hai-LongHuang:2024vvz, Zhang:2025grn}.

\subsection{Accretion}

\subsubsection{Constraints}

PBHs with mass $m \gsim 10 \Msun$ can accrete significant amounts of surrounding gas. The radiation emitted in this process could alter the Universe's recombination history~\cite{Carr:2020erq}, generate anisotropies and spectral distortions in the CMB~\cite{Ricotti:2007au,Serpico:2020ehh}, increase the population of X-ray sources~\cite{Inoue:2017csr} and modify the 21-cm signal~\cite{Hektor:2018qqw,Mittal:2021egv,Cang:2021owu, Agius:2025xbj}. These limits rely on uncertain astrophysical inputs, including the accretion mode (disk vs. spherical), and often assume Bondi accretion~\cite{Agius:2024ecw, Facchinetti:2022kbg,Agius:2025xbj}. Moreover, efficient accretion changes the PBH mass function and can substantially weaken the constraints at $m \gsim 10 \Msun$~\cite{DeLuca:2020fpg}. On the other hand, PBH clustering enhances accretion and can strengthen the corresponding constraints~\cite{Hutsi:2019hlw}.

\subsubsection{Potential evidence}

Compelling evidence for stellar-mass PBHs may come from the source-subtracted cosmic infrared and X-ray backgrounds~\cite{Kashlinsky:2016sdv}. The level of the infrared background suggests an overabundance of high-redshift halos that could be explained by the PBH Poisson effect if a significant fraction of the CDM comprises stellar-mass PBHs. In these halos, a few stars form and emit infrared radiation, with this naturally explaining both the amplitude and angular spectrum of the source-subtracted infrared anisotropies. The spatial coherence with the source-subtracted X-ray background may be explained if the X-rays are generated by black hole accretion in these structures~\cite{Kashlinsky:2018mnu}. Although these would not necessarily be primordial, they could be.

\section{Asteroid-mass window}
\label{sec:asteroidmasswindow}

The asteroid-mass window - spanning approximately $10^{-17}\,\Msun$ to $10^{-10}\,\Msun$ - lies between the evaporation and ML constraints and remains the only unconstrained region in the PBH mass spectrum. These PBHs are too heavy to evaporate, and the optical ML searches are insensitive due to wave-optics and finite-source-size effects.

Potential signatures in the asteroid mass range arise from the destruction of stars by PBHs that are swallowed by them. There are two mechanisms by which a PBH can be swallowed. The first is dynamical capture~\cite{Capela:2013yf}, which occurs when a PBH passes through or near a star and loses sufficient kinetic energy through gravitational interactions, allowing it to become gravitationally bound. The orbit of the PBH gradually contracts as it loses energy during subsequent passages. Eventually, the orbit lies entirely within the star, and the PBH settles into the stellar core. The probability for this process is significantly reduced by perturbers affecting the PBH orbit~\cite{Montero-Camacho:2019jte,Caiozzo:2024flz,Holst:2025vgk}. The second mechanism operates during star formation, where the gradual contraction of baryonic matter can adiabatically draw PBHs into the forming star~\cite{Capela:2012jz}. This process effectively captures the PBHs without requiring close encounters. Both processes have potentially catastrophic consequences for the star as the PBH gradually consumes it. The observational consequences of this are a reduction in the stellar population and potential observable transients, such as supernovae. 

Several works have placed constraints on the allowed abundance of PBHs in the asteroid-mass window by analysing the survival of neutron stars in globular clusters~\cite{Capela:2012jz,Capela:2013yf,Capela:2014ita,Holst:2025vgk}. However, these constraints are sensitive to the density of DM in globular clusters, which remains highly uncertain. An alternative approach considers the potential destruction of main-sequence stars in dwarf galaxies, which are known to be DM dominated~\cite{Oncins:2022djq,Esser:2022owk,Esser:2023yut,Esser:2025pnt}. However, although realistic accretion modelling yields consumption times comparable to the simple Bondi time for compact stars~\cite{Baumgarte:2021thx}, recent work indicates that the lifetimes of main-sequence stars are much longer than the Bondi time and remain largely unaffected by asteroid-mass PBHs embedded in their cores~\cite{Bellinger:2023wou}.

Similarly, Ref.~\cite{Graham:2015apa} proposed that a PBH traversing a carbon–oxygen white dwarf (WD) could ignite runaway carbon fusion and thus trigger a thermonuclear explosion by locally heating a narrow cylindrical region through dynamical friction. From the observed survival of WDs, it was inferred that PBHs with masses $10^{19}–10^{20}$\,g cannot make up the bulk of the local DM. Ref.~\cite{Montero-Camacho:2019jte} subsequently revisited this scenario, using hydrodynamic simulations of PBH passages through WDs. The results show that, under realistic conditions, PBHs in the relevant mass range fail to produce self-sustaining ignition. Even for very massive WDs and under optimistic assumptions, the heated region cools or mixes before runaway carbon fusion can develop. This eliminates the constraints proposed in Ref.~\cite{Graham:2015apa}.

Another attempt to constrain the asteroid-mass range was based on femtolensing searches of $\gamma$-ray bursts~\cite{Barnacka:2012bm}. However, closer scrutiny~\cite{Katz:2018zrn} showed that this constraint does not hold because the sizes of most $\gamma$-ray burst sources are too large for femtolensing. 

The asteroid-mass window currently lacks robust observational constraints, making it an interesting target for future searches. X-ray ML has recently emerged as one of the most promising probes of PBHs in this mass range~\cite{Bai:2018bej,Tamta:2024pow,Anchordoqui:2024tdj}. Compact X-ray pulsars overcome the limitations of finite-source-size and wave-optics effects by providing emission regions only tens of kilometres across and photons with sufficiently short wavelengths. As a result, transient-brightening events induced by PBHs passing near the line of sight can be detected with current instruments such as NICER and, more effectively, with next-generation X-ray observatories such as STROBE-X, AstroSat, Athena/Lynx, and eXTP. Depending on instrument sensitivity and exposure time, these searches may decisively test whether PBHs lighter than about $10^{-13}\,\Msun$ can explain DM.
A complementary approach exploits the exquisite precision of Solar System ephemerides to search for transient perturbations in planetary orbits induced by PBH flybys~\cite{Tran:2023jci}. Indeed, If PBHs in the asteroid-mass range provide all the DM, several such events are expected per decade, giving detectable signatures in the Earth–Mars or Earth–Venus distance residuals.

\section{Constraints for extended mass functions}

The PBH constraints are typically given in the literature by assuming a monochromatic mass function. The issue of translating constraints obtained for monochromatic mass functions to extended mass functions was examined in~\cite{Carr:2017jsz}. Importantly, the constraints on $\fPBH$ depend on the assumed mass function. In Fig.~\ref{fig:mono}, for example, $\MPBH$ and $\fPBH$ represent the two parameters of the monochromatic mass function, and the constraint is imposed on the total $\fPBH$, not the PBH fraction in a particular mass range. As a result, one cannot simply compare $\psi(\MPBH) \equiv \td \fPBH/\td \ln \MPBH$ with the constraints on $\fPBH$ in Fig.~\ref{fig:mono} as they represent different quantities. Instead, one must specify the mass function for each scenario at the outset and then constrain whatever parameters describe it.

To extend the constraints to broad mass functions, we will follow the approach of Ref.~\cite{Carr:2017jsz}. For this, one must consider the observable $A$ that imposes the constraint and then examine how the observable is affected by the PBH mass function. In general, any observable must be describable by some functional $A[\psi(\MPBH)]$ of the PBH mass distribution $\psi(\MPBH)$ (defined in Eq.~\eqref{eq:psidef2}) and a bound on that observable can thus be expressed as 
\be \label{eq:constraint}
    A[\psi(\MPBH)] \leq A_{\rm exp} \,.
\ee
where $A_{\rm exp}$ is the observational upper limit on $A$. If each PBH contributes independently to the observables that are constrained, the functional must be linear 
\be \label{eq:A_linear}
    A[\psi(\MPBH)] = A_{0} + \int \td \ln \MPBH \, A_{1}(\MPBH)\psi(\MPBH)\,,
\ee
where $A_{0}$ is the contribution in the absence of PBHs and $A_{1}(\MPBH)$ is an integration kernel that depends on the physics underlying the observable and the details of the experiment. Many observables fall into this pattern. The shape kernel can be explicitly inferred for the lensing and survival of stars~\cite{Belotsky:2014kca}, for evaporation~\cite{Carr:2016drx} and neutron star capture and accretion~\cite{Kuhnel:2017pwq}.

From Eqs.~\eqref{eq:constraint} and \eqref{eq:A_linear}, a constraint for a monochromatic mass function $\psi(\MPBH) \propto \delta(\MPBH - M_c)$ centred around $M_c$ can thus be expressed as
\be
    \fPBH(M_c) 
    \leq \frac{A_{\rm exp} - A_0}{A_{1}(\MPBH)} \equiv f_{\rm max}(M_c)\,.
\ee
where $f_{\rm max}(M_c)$ represents the constraint on the PBH abundance from the observable $A$. In this way, the kernel $A_{1}(\MPBH)$ can be effectively extracted from the monochromatic constraint. Eq.~\eqref{eq:A_linear} now implies that
\be\label{eq:single_observable}
    \int \td \ln \MPBH \,  \frac{\psi(\MPBH)}{f_{{\rm max}}(\MPBH)} \leq 1 \,,
\ee
which gives the upper bound on the total dark matter fraction $\fPBH$ for an extended mass function from  constraint $A$.
Constraints from multiple observables $A_j$, which give an upper bound on the abundance $f_{{\rm max},j}(\MPBH)$ for a monochromatic mass function, can be combined using quadrature (for details see~\cite{Carr:2017jsz}). The combined constraint for extended PBH mass functions from multiple measurements can then be evaluated as
\be\label{eq:master}
    \sum_j \left[ \int \td \ln \MPBH \,  \frac{\psi(\MPBH)}{f_{{\rm max},j}(\MPBH)} \right]^2 \leq 1 \,.
\ee

\begin{figure}
\centering
\includegraphics[height=0.36\textwidth]{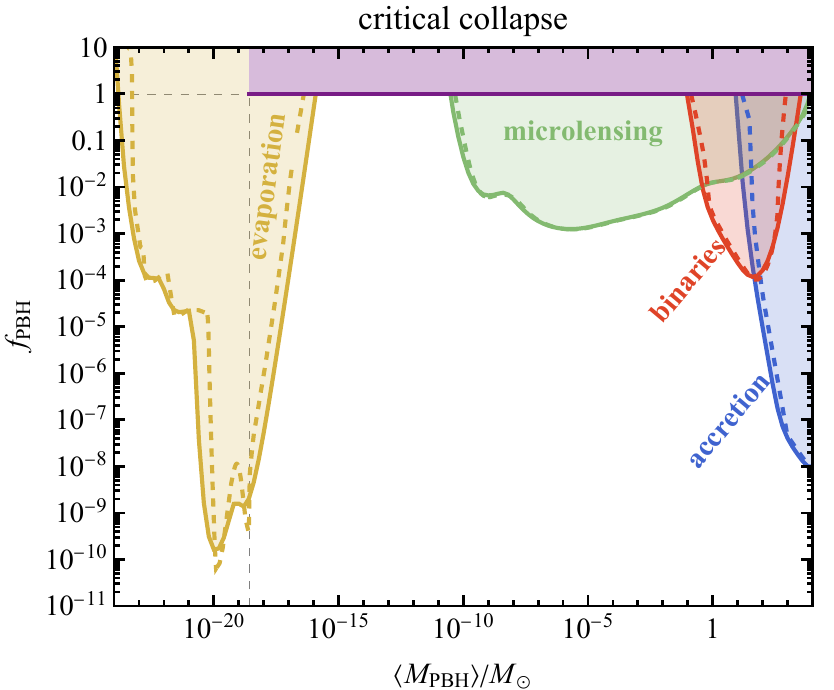} \hspace{4mm}
\includegraphics[height=0.36\textwidth]{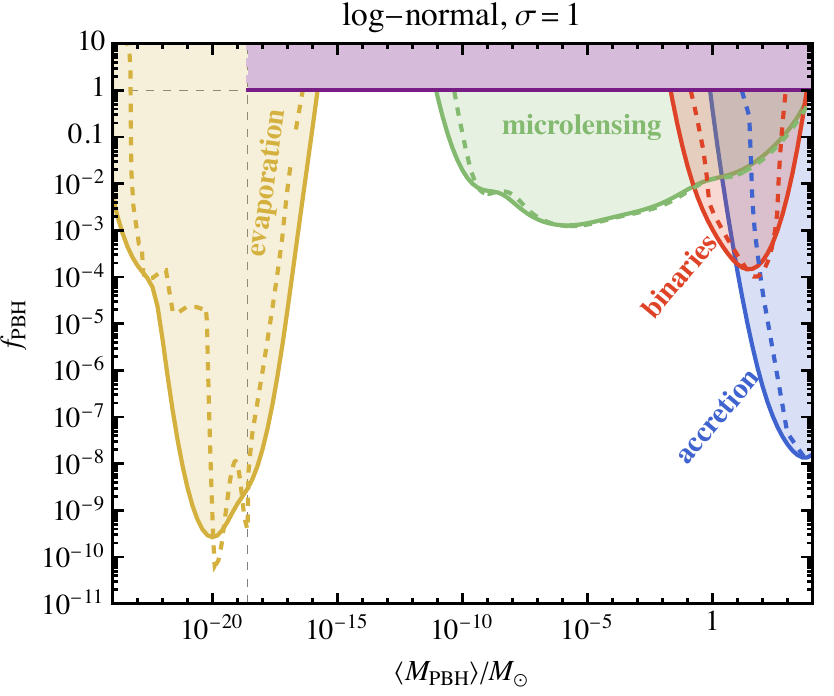} \\ \vspace{2mm}
\includegraphics[height=0.36\textwidth]{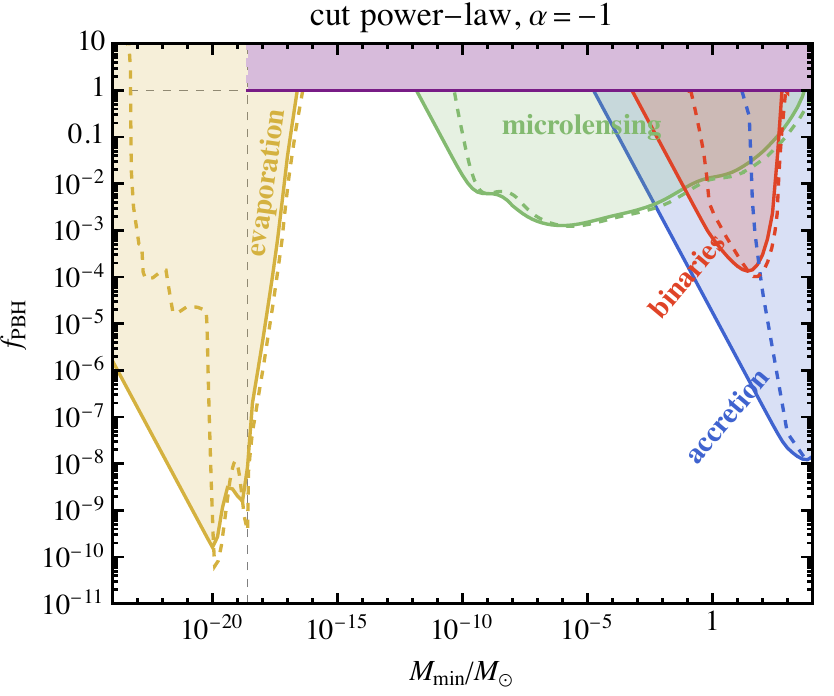} \hspace{4mm}
\includegraphics[height=0.36\textwidth]{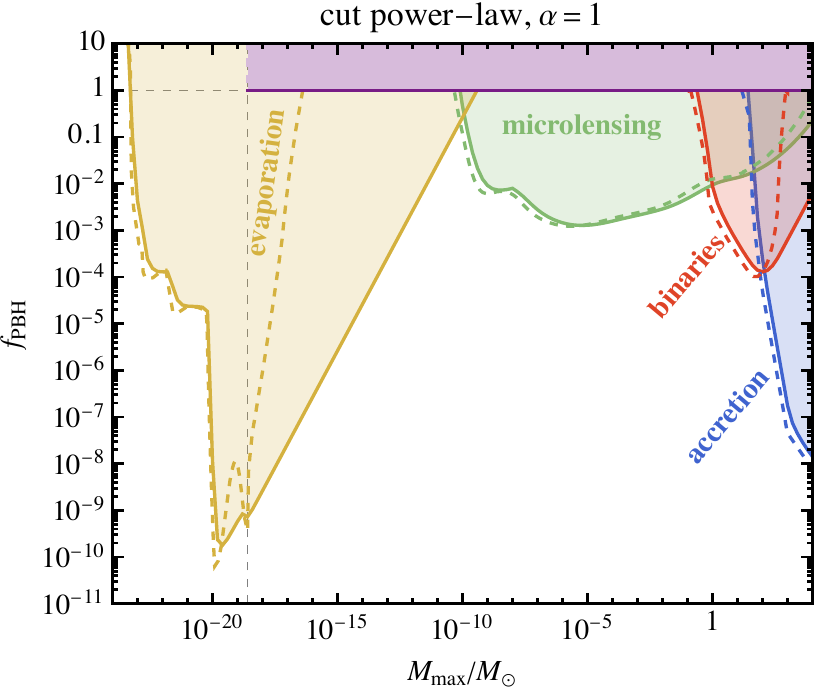}
\caption{The solid curves show the constraints for critical-collapse mass functions~\eqref{eq:psicc0} (upper left), log-normal mass function~\eqref{eq:psiln} with $\sigma=1$ (upper rigtht), and truncated power-law mass function $\psi \propto \MPBH^\alpha$ with $\alpha = -1$ (lower left) and $\alpha = 1$ (lower right). The dashed curves show the constraints for monochromatic  mass functions.}
\label{fig:ext}
\end{figure}

\begin{figure}
\centering
\includegraphics[height=0.34\textwidth]{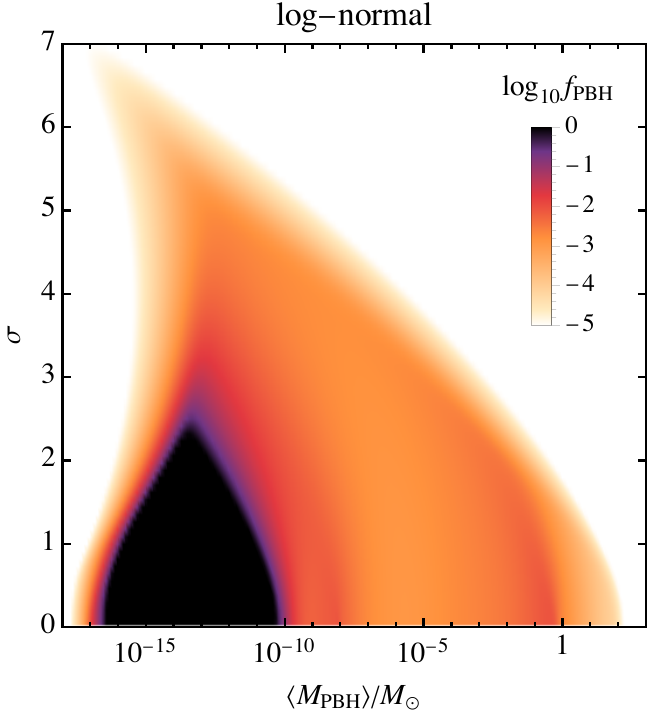}
\includegraphics[height=0.34\textwidth]{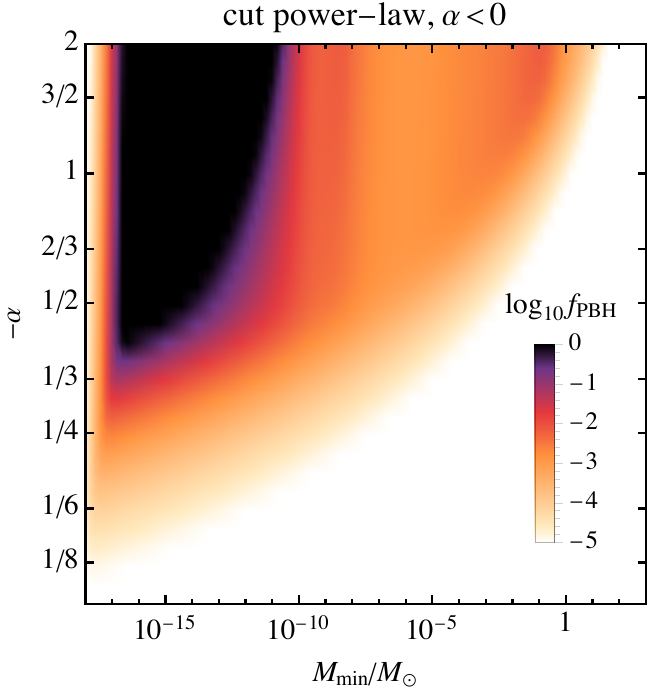}
\includegraphics[height=0.34\textwidth]{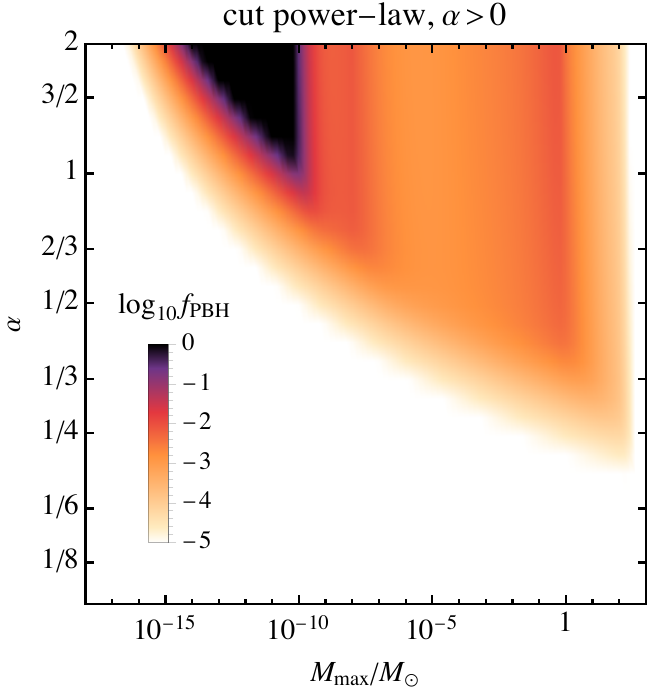}
\caption{Maximal allowed $\fPBH$ for a log-normal PBH mass function~\eqref{eq:psiln} with width $\sigma$ and mean $\langle \MPBH \rangle$ (left panel), a truncated power-law PBH mass function, $\psi \propto \MPBH^\alpha$ with $\alpha < 0$ (middle panel) and $\alpha > 0$ (right panel). The truncated power-law case assumes the separation between the low- and high-mass cut-offs is large, $M_{\rm min} \ll M_{\rm max}$.}
\label{fig:ext2}
\end{figure}

We now discuss some limitations associated with the use of Eq.~\eqref{eq:master}. First, it is derived under the assumption that each PBHs contributes independently to the constrained observable~\eqref{eq:A_linear}, which means that it does not fully capture constraints that depend on phenomena involving the collective effect of multiple PBHs, given that the effect is sensitive to the PBH mass. An example is the constraint from their mergers, as binary formation involves at least two PBHs and binary disruption (see Eq.~\eqref{eq:R2}). It is additionally affected by the Poisson clustering of PBHs, which receive corrections from the width of the mass distribution, especially when $\fPBH = \mathcal{O}(1)$. In general, any constraint that depends on how PBHs cluster and modify structure formation can violate Eq.~\eqref{eq:A_linear} to some extent. Examples of this are Ly-$\alpha$ constraints, dynamical constraints, and to a lesser extent lensing and accretion constraints. Related to that are the effects of mass segregation inside cosmic structures, which cause lighter PBHs to migrate outwards while heavier ones concentrate in the central regions. Such effects introduce corrections to phenomena that depend on assumptions about the environment of PBHs. In particular, this affects constraints from stellar evolution in dwarf galaxies~\cite{Koushiappas:2017chw,Brandt:2016aco}. 

It is also important to consider the possibility that the mass function evolves due to PBH accretion~\cite{DeLuca:2020fpg, DeLuca:2021pls}. This issue can be addressed by interpreting the mass functions introduced in this work not as initial mass functions but rather as effective mass functions that already incorporate these types of corrections. In principle, all the constraints discussed above and those presented in our figures apply to these effective mass functions, which may differ from one constraint to another. However, estimating how these effects modify the initial mass functions would require detailed numerical simulations, which are beyond the scope of the present work.

The constraints for extended mass functions are summarised in Figs.~\ref{fig:ext} and~\ref{fig:ext2}. The left panel of Fig.~\ref{fig:ext} shows the critical-collapse mass function~\eqref{eq:psicc0} with $\gamma = 0.36$ and $c_2 = 1$. Because this mass function is relatively narrow, the change in the constraints compared to the monochromatic case is mild. The right panel of Fig.~\ref{fig:ext} shows the constraints for log-normal mass functions~\eqref{eq:psiln} with different widths $\sigma$ and mean masses $\langle \MPBH\rangle$. The colour-coding indicates the maximal fraction of DM allowed for a given combination of $\langle \MPBH \rangle$ and $\sigma$. In the white region, this fraction is $\fPBH < 10^{-5}$, while in the black region, PBHs can constitute all DM. We see that the constraints become stronger for broader mass functions, and the asteroid-mass window closes for $\sigma \gsim 2.5$. In the same way, Fig.~\ref{fig:ext2} shows the case of the cut power-law mass function~\eqref{eq:psipl}.

\section{Gravitational wave prospects}

\subsection{Scalar-induced gravitational waves}
\label{sec::SIGW}

In the critical-collapse scenario, the same curvature perturbations that lead to PBH formation also generate scalar-induced gravitational waves (SIGWs). More generally, SIGWs are produced whenever curvature perturbations re-enter the horizon, even if their amplitude falls below the threshold required for PBH formation. The shape of the resulting GW spectrum is determined by the properties and statistics of the underlying scalar perturbations~\cite{Tomita:1975kj, Matarrese:1992rp,Matarrese:1993zf,Matarrese:1997ay, Acquaviva:2002ud, Mollerach:2003nq,Carbone:2004iv, Ananda:2006af, Baumann:2007zm, Domenech:2021ztg}. On CMB scales, the scalar spectrum is tightly constrained, and SIGWs have undetectably small amplitudes. However, PBH production requires a boost in the power spectrum at smaller scales, and this would inevitably enhance the SIGW signal. In this section, we focus on the interplay between SIGWs and PBHs, exploiting their common origin and relating current and future GW data to PBH production (see also~\cite{Saito:2008jc, ZhengRuiFeng:2021zoz, Zhang:2021vak, Yuan:2021qgz, Yi:2022ymw, Zhao:2023xnh,Gouttenoire:2025jxe}).

\subsubsection{Spectrum}

SIGWs were first theorized in~\cite{Tomita:1975kj, Matarrese:1992rp,Matarrese:1993zf,Matarrese:1997ay}, where it was noticed that scalar perturbations could generate GWs at second order in perturbation theory, and this possibility was fully analysed in~\cite{Ananda:2006af,Baumann:2007zm}. At first order in perturbation theory, tensors, vectors and scalars are independent, but at second order a coupling between scalars and tensors arises. Neglecting first-order tensor and vector perturbations (for more details on tensor-induced GWs see~\cite{Bari:2023rcw, Picard:2023sbz}) and anisotropic stress, the Einstein equations give~\cite{Acquaviva:2002ud, Mollerach:2003nq, Ananda:2006af, Baumann:2007zm}
\be
    h_{ij}''(\textbf{x},\eta) + 2\mathcal{H}h_{ij}'(\textbf{x},\eta) - \nabla^2h_{ij}(\textbf{x},\eta) = -4\hat{\mathbb{T}}_{ij}^{lm}\mathcal{S}_{lm}(\textbf{x},\eta)\,,
\ee
where $\hat{\mathbb{T}}_{ij}^{lm}$ is the projector in the transverse-traceless (TT) gauge, needed to extract the propagating degrees of freedom, and $\eta$ is the conformal time. The source term $\mathcal{S}_{ij}(\textbf{x},\eta)$ is given by
\bea \label{source1}
    \mathcal{S}_{ij} \equiv&\hspace{0.1cm} 4\Phi\partial_i\partial_j\Phi + \frac{2(1+3w)}{3(1+w)}\partial_i\Phi\partial_j\Phi - \frac{4}{3(1+w)\mathcal{H}^2}\left[\partial_i\Phi'\partial_j\Phi' + \mathcal{H}\partial_i\Phi\partial_j\Phi' + \mathcal{H}\partial_i\Phi'\partial_j\Phi\right]\,,
\eea
where $\Phi$ is the first-order gravitational potential and the dependence on $\textbf{x}$ and $\eta$ is implicit. The last equation shows explicitly that second-order tensors are sourced by scalars (see~\cite{Iovino:2025xkq} for a recent physical explanation of the origin of SIGW). In Fourier space, $\Phi(\mathbf{k},\eta) = {T}_\Phi(\mathbf{k},\eta)\zeta(\mathbf{k})$ where ${T}_{\Phi}(\mathbf{k},\eta)$ denotes the transfer function and $\zeta(\mathbf{k})$ the comoving curvature perturbation, so the equation of motion can be solved with the Green's method, leading to the following SIGW spectrum~\cite{Baumann:2007zm,Espinosa:2018eve,Kohri:2018awv,Domenech:2021ztg}: 
\bea \label{eq::SIGW_spectrum}
    &\Omega_{\rm SIGW,0}(k) = \Omega_{\rm rad,0} \left(\frac{g_*(\eta)}{g_*^0}\right) \left(\frac{g_{*s}^0}{g_{*s}(\eta)}\right)^{4/3} \left(\frac{k}{\mathcal{H}(\eta)}\right) \frac{\pi k^3}{(2\pi)^6} \\
    & \quad \times \sum_\lambda \int\frac{d^3\textbf{q}_1}{(2\pi)^{3}}\frac{d^3\textbf{q}_2}{(2\pi)^{3}}Q_{\lambda}(\textbf{k},\textbf{q}_1)Q_{\lambda}(-\textbf{k},\textbf{q}_2)  I(|\textbf{k}-\textbf{q}_1|,q_1,\eta) I(|-\textbf{k}-\textbf{q}_2|,q_2,\eta)\mathcal{T}_{\zeta}( \textbf{q}_1,{\textbf{k}-\textbf{q}_1},{\textbf{q}_2},{-\textbf{k}-\textbf{q}_2}) \,,
\eea
where $\Omega_{\rm rad,0}$ is the current radiation density parameter~\cite{Planck:2018vyg}. The kernel $I(|\textbf{k}-\textbf{q}_1|,q_1,\eta)$ contains all the information about the evolution of scalar and tensor perturbations at any epoch. The $Q_{\lambda}(\textbf{k},\textbf{q}_1)$ functions project the internal momenta to the GW ones by using the corresponding polarization tensors; their explicit form can be found in~\cite{Kohri:2018awv, Domenech:2021ztg,Adshead:2021hnm, Perna:2024ehx, LISACosmologyWorkingGroup:2025vdz}. Finally, $\mathcal{T}_\zeta$ is the trispectrum of scalar perturbations, which we will briefly discuss in the next section.

\subsubsection{Imprints of non-Gaussianity}
\label{sec::NG_SIGW}

A noteworthy aspect is that SIGWs are not only sensitive to the expansion history of the Universe, but also to primordial NG, as is evident from Eq.~\eqref{eq::SIGW_spectrum}, where the tensor spectrum is dependent on the trispectrum of scalar perturbations~\cite{Gangui:1993tt, Matarrese:2000iz, Bartolo:2001cw, Maldacena:2002vr, Bartolo:2004if,Chen:2010xka,Byrnes:2010em,Wands:2010af,Renaux-Petel:2015bja,Achucarro:2022qrl}. We argued in Sec.~\ref{sec::PBH_NG} that NG can have an enormous impact on the PBH abundances. By contrast, the impact of NG on the SIGW is not fully understood~\cite{Iovino:2024sgs}. Regardless of how one includes NG in the computation, it is expected to modify the GW spectrum, not only enhancing or suppressing it but also leaving specific spectral features. Explicitly, this can be understood by decomposing the trispectrum into a connected and a disconnected component $\mathcal{T}( \textbf{k}_1,\textbf{k}_2,{\textbf{k}_3},{\textbf{k}_4})=\mathcal{T}( \textbf{k}_1,\textbf{k}_2,{\textbf{k}_3},{\textbf{k}_4})|_c+\mathcal{T}( \textbf{k}_1,\textbf{k}_2,{\textbf{k}_3},{\textbf{k}_4})|_d$. In the absence of NG, the connected contribution vanishes and only the disconnected contribution survives, resulting in a term proportional to products of curvature power spectra $\mathcal{P}_\zeta(k_1)\mathcal{P}_\zeta(k_2)$. A simple way to model primordial NG is to adopt the local ansatz of Eq.~\eqref{eq:FirstExpansion}. As in the PBH case, the use of this ansatz has many limitations~\cite{Iovino:2024sgs}, but it nevertheless reveals the effects of NG on the GW spectrum~\cite{Cai:2018dig,Unal:2018yaa,Adshead:2021hnm,Garcia-Saenz:2022tzu,Perna:2024ehx}. We stress that this affects the GW spectrum quite differently from the PBH abundance. In the latter case, it substantially changes the tail of the distribution, inducing a variation in the abundance of many orders of magnitude. For SIGWs, the main effect is the presence of additional features in the GW spectrum with a relatively small variation in the amplitude.

\subsubsection{Interplay between SIGW and PBHs}

SIGWs are a guaranteed background that is potentially observable by future and current GW detectors~\cite{Perna:2024ehx,LISACosmologyWorkingGroup:2025vdz,ET:2025xjr}. Here we explore their interconnection with PBHs. The detection of a SIGW background can be mapped into constraints on the amplitude of the scalar curvature power spectrum, $A$, at the corresponding pivot frequency $f_\star$ (or $k_\star= 2 \pi f_\star$). The implications for the PBH abundance are straightforward since the pair $(A,k_\star)$ will, in turn, imply some abundance and mean mass of PBHs $(\fPBH, \langle \MPBH \rangle)$ (see Sec.~\ref{Sec::Form_PBH}). So bounds on the amplitude of scalar-curvature perturbations will also constrain the PBH abundance in critical-collapse scenarios. 

The most interesting mass ranges are in the otherwise unconstrained asteroid-mass window (see Sec.~\ref{sec:asteroidmasswindow}), where DM may consist entirely of PBHs, and the solar/sub-solar mass ranges (see Sec.~\ref{sec::solar}), in which the observed BH binary mergers have generated substantial interest in PBHs.  In addition, recent PTA measurements~\cite{NANOGrav:2023gor,EPTA:2023fyk,Reardon:2023gzh,Xu:2023wog} provide evidence of a stochastic GW background in the nano-Hz band, while the Laser Interferometer Space Antenna (LISA)~\cite{LISA:2017pwj}, expected to operate within the next decade, will survey the milli-Hz region. Coincidentally, those ranges correspond exactly to the formation of solar/subsolar and asteroid-mass PBHs. Either a detection or non-detection of a SIGW background in these experiments would thus have decisive consequences for PBHs.

\begin{figure}
    \centering
    \includegraphics[width=0.84\linewidth]{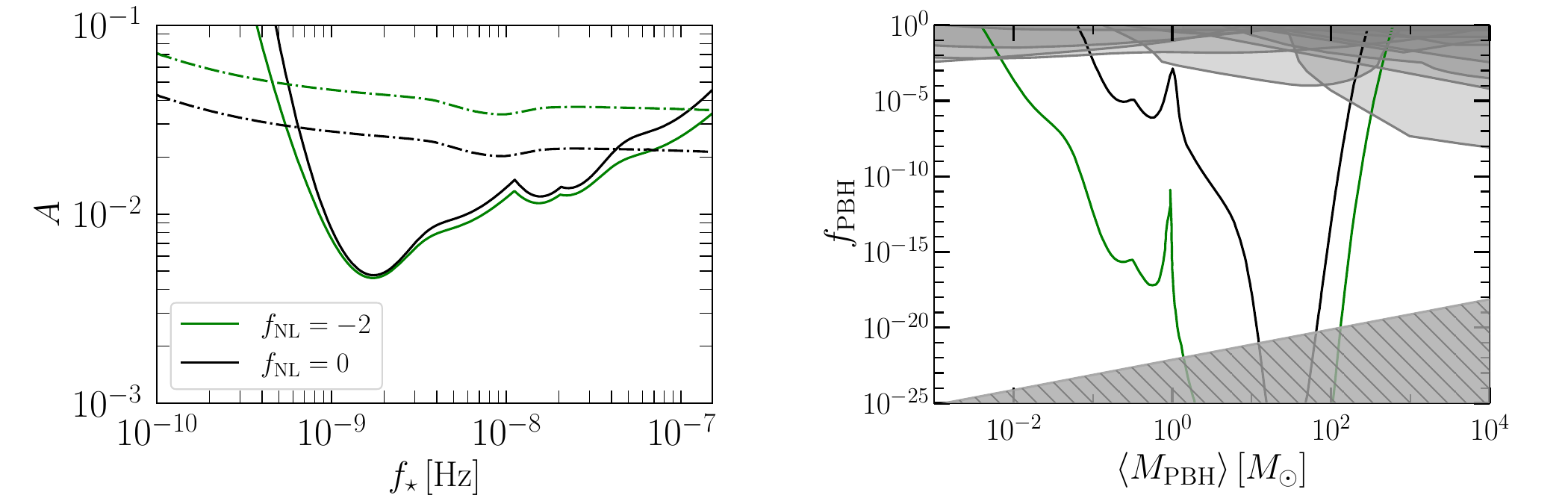}
    \caption{\emph{Left panel:} Constraints on the amplitude of the scalar power spectrum, obtained using the latest NANOGrav results~\cite{NANOGrav:2023gor}, for a log-normal curvature power spectrum with $\Delta=0.5$. The Gaussian case is shown in black, while the green curves show a benchmark non-Gaussian case with $|f_{\rm NL}|=2$. The dot-dashed lines correspond to $\fPBH=1$. In the non-Gaussian case we show the line corresponding to $f_{\rm NL}=-2$, which gives the most stringent estimate. \emph{Right panel:} Corresponding constraints on the PBH abundance. The shaded region on the bottom of the plot excludes regions with less than a single PBH in the current Hubble volume.}
    \label{Fig::GW_Solar}
\end{figure}

\begin{figure}
    \centering
    \includegraphics[width=0.84\linewidth]{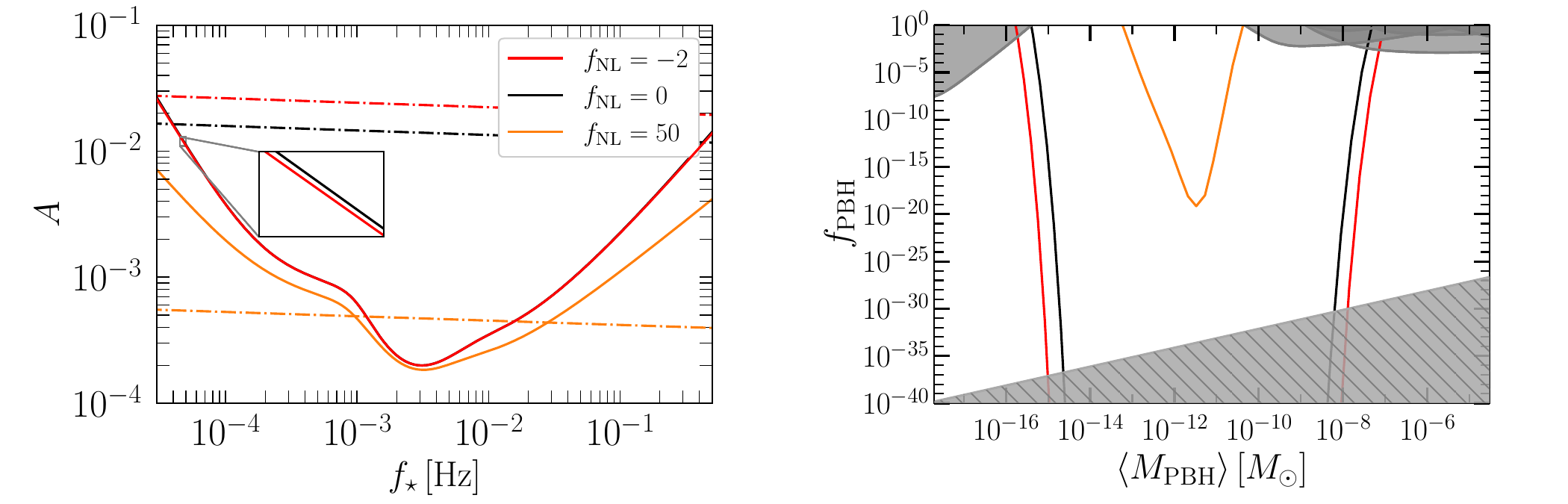}
    \caption{\emph{Left panel:} Constraints on the amplitude of the scalar power spectrum, assuming a null-detection by LISA~\cite{LISA:2017pwj}, for a log-normal curvature power spectrum with $\Delta=0.5$. The Gaussian case is shown in black, while the red and orange curves show a benchmark non-Gaussian case with $|f_{\rm NL}|=2$ and $|f_{\rm NL}|=50$, respectively. The dot-dashed lines correspond to $\fPBH=1$. In the non-Gaussian case we show the line corresponding to $f_{\rm NL}=-2$, which returns the most stringent estimate, and $f_{\rm NL}=50$. \emph{Right panel:} Corresponding constraints on the PBH abundance. The shaded region at the bottom of the plot excludes regions with less than a single PBH in the current Hubble volume.}
    \label{Fig::GW_Asteroidal}
\end{figure}

In the following, we will discuss what existing NANOGrav and prospective LISA observations can say about PBHs and discuss the impact of NG. Fig.~\ref{Fig::GW_Solar} shows the constraints on $\fPBH$ in the solar-mass range~\cite{Iovino:2024tyg} implied by PTA observations~\cite{NANOGrav:2023gor}. The left panel shows the constraints on the amplitude of the scalar power spectrum for $f_{\rm NL} = -2$ and $f_{\rm NL} = 0$ if one has a log-normal scalar power spectrum~\eqref{eq:Narrow} with $\Delta=0.5$. The amplitude corresponding to $\fPBH=1$ is indicated by the dot-dashed lines. In the right panel, the implied constraints on the abundance of PBHs are shown. In the Gaussian ($f_{\rm NL} = 0$) case, current GW data can exclude PBHs in the 10–100 $\Msun$ range and place stronger constraints in the surrounding mass range than those indicated in Fig.~\ref{fig:mono}. The presence of NG can significantly alter the constraints. While the SIGW deviations are not remarkable\footnote{In this case, the NG correction is proportional to $f_{\rm NL}^2$, so it does not distinguish between positive and negative $f_{\rm NL}$.} the PBH abundance is strongly affected, with differences of many orders of magnitude with respect to the Gaussian case. For $f_{\rm NL} = -2$, the formation of PBHs in the $1$--$100\,\Msun$ range is entirely excluded, while their abundance is strongly suppressed, down to $\fPBH \sim 10^{-15}$, for $\langle \MPBH \rangle \sim 0.1\,\Msun$. As illustrated by Fig.~\ref{fig:AppFNL}, the strongest non-Gaussian suppression of PBH formation, and thus the largest amplitudes permitting $\fPBH=1$, are obtained when $f_{\rm NL} \approx -2$ \footnote{The Figure holds for a broken power law power spectrum, but similar results are obtained for a log-normal.}. As a result, the $f_{\rm NL} = -2$ case gives the weakest (i.e., most conservative) constraint, indicating that PTA data excludes $\fPBH = 1$ in the range $0.1 - 100 \Msun$ in critical-collapse scenarios.

In a similar way, LISA can probe PBHs in the asteroid mass range~\cite{Bartolo:2018evs,Iovino:2025cdy}. Fig.~\ref{Fig::GW_Asteroidal} shows the sensitivity to $\fPBH$, including the impact of NG in the range $f_{\rm NL} \in [-2, 50]$~\cite{Iovino:2025cdy}. As seen in the left panel, SIGWs can probe the amplitude of the power spectrum down to $\mathcal{O}(10^{-4})$. The right panel shows that LISA could, in principle, probe any conceivable PBH abundance in the asteroid mass range, except in the case of strong NG ($f_{\rm NL} = 50$). As the amplitude required to produce $\fPBH=1$ is $\mathcal{O}(10^{-3})$ and therefore much smaller than in the Gaussian case, such large NG can still be in the perturbative regime (see~\cite{Iovino:2025cdy} for more details). We add that since weaker signals lead to larger uncertainties in estimating $f_{\rm NL}$, distinguishing whether SIGWs arise from Gaussian or non-Gaussian fluctuations becomes challenging, thereby limiting LISA’s ability to probe PBHs. Therefore, the low-mass corner of the asteroid mass window may be left untested by LISA, motivating GW detectors such as AEDGE~\cite{AEDGE:2019nxb} that cover the frequency range between LISA and ET.

\subsection{Gravitational waves from binaries}

\begin{figure}
\centering
\includegraphics[height=0.29\textwidth]{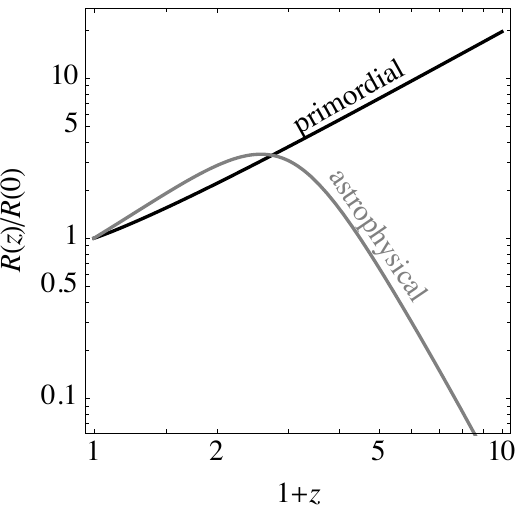}
\hspace{2mm}
\includegraphics[height=0.31\textwidth]{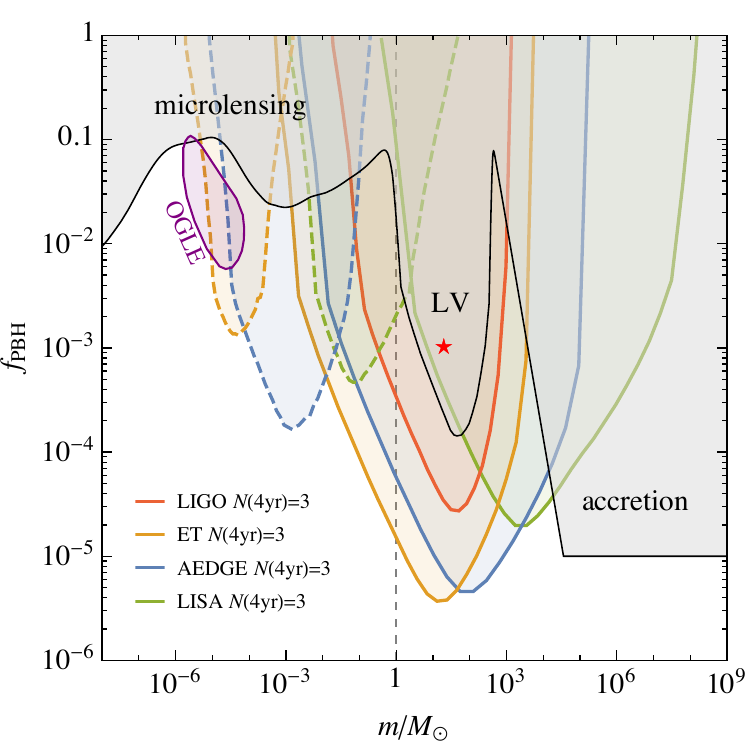}
\hspace{2mm}
\includegraphics[height=0.31\textwidth]{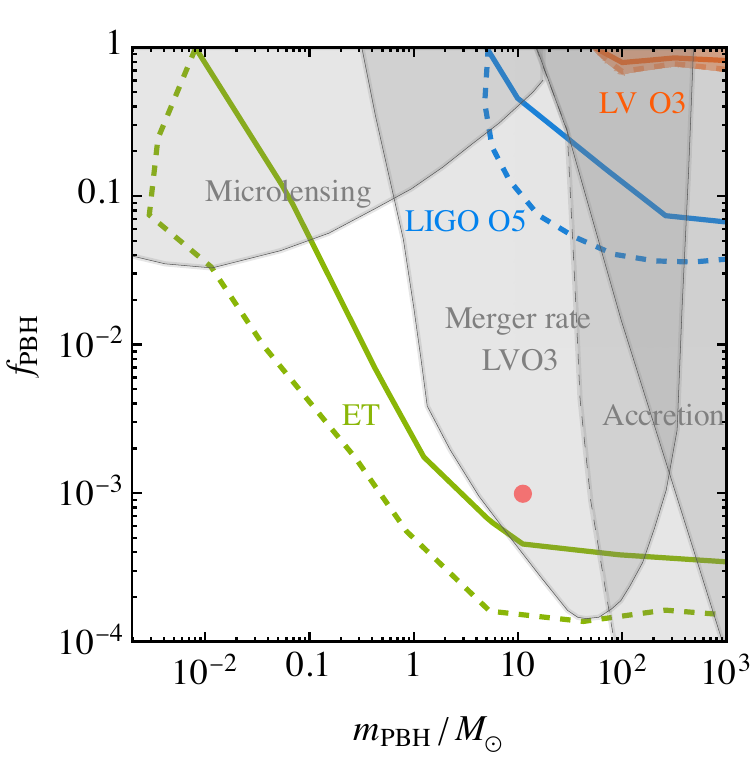}
\caption{\emph{Left panel:} The redshift dependence of the PBH merger rate (black) and the merger rate of astrophysical BHs (gray) (from~\cite{Hutsi:2020sol}). \emph{Middle panel:} The sensitivities of future observatories to the DM fraction in PBHs through GW observations of individual binaries (from~\cite{Pujolas:2021yaw}). \emph{Right panel:} The sensitivities of future observatories on the DM fraction in PBHs through observations of lensing effects in GW signals (from~\cite{Urrutia:2023mtk}).}
\label{fig:mergers}
\end{figure}

The PBH merger rate is dominated by binaries formed through the early two-body channel. The merger rate is given by~\cite{Raidal:2018bbj,Vaskonen:2019jpv}
\be\label{eq:R2}
    \frac{\td R_{\rm PBH}}{\td m_1 \td m_2} = \frac{1.6\times 10^6}{\text{Gpc}^3\text{yr}^1} \fPBH^{-\frac{21}{37}}
    \left[ \frac{t}{t_0}\right]^{-\frac{34}{37}}
    \left[ \frac{M}{\Msun}\right]^{-\frac{32}{37}} \eta^{-\frac{34}{37}} S[\psi(\MPBH),t] \,\frac{\psi(m_1)\psi(m_2)}{m_1 m_2} \,,
\ee
where $M \equiv m_1+m_2$, $\eta \equiv m_1m_2/M^2$, and $S$ is a suppression factor that accounts for the disruption of binaries after their formation. Neglecting the mild $z$-dependence in $S$, this merger rate grows as $R_{\rm PBH} \propto (1+z)^{51/37}$ in a matter-dominated background. As shown in the left panel of Fig.~\ref{fig:mergers}, this strongly contrasts with the merger rate of astrophysical BHs, which decreases at $z \gsim 3$~\cite{Madau:2016jbv}. Therefore, observing a substantial population of BHs at high redshift would provide compelling evidence for PBHs~\cite{Ng:2022agi}. Such observations will be possible with the next-generation GW observatories (e.g. ET~\cite{ET:2025xjr}, AION/AEDGE~\cite{Badurina:2019hst,AEDGE:2019nxb}, LISA~\cite{LISA:2017pwj}). Another strong indicator of PBHs would come from GW detections of subsolar mass binaries, since such systems are not expected to form through astrophysical channels. The middle panel of Fig.~\ref{fig:mergers} indicates the parameter ranges accessible to upcoming GW observatories. The solid curves denote the sensitivity to extragalactic binaries, while the dashed curves show the reach for galactic sources. The latter extend the prospects for detecting PBH binaries down to masses of order $10^{-5}\,\Msun$. A comparable mass range can also be accessed through searches for the stochastic GW background from PBH binaries~\cite{Pujolas:2021yaw}.

\subsection{Lensing of gravitational waves}

In the solar-mass range, optical ML searches lose sensitivity because the potential events become longer than the monitoring time. However, for GW signals, the coherent nature of the waveform allows the detection of lensing effects even when the lens motion is negligible. In this regime, lensing produces a frequency-dependent magnification that imprints a characteristic frequency dependence on the signal's amplitude~\cite{Takahashi:2003ix}.  In fact, the event GW231123~\cite{LIGOScientific:2025rsn} has been recently claimed to be a potential candidate for the first lensed GW event~\cite{LIGOScientific:2025cwb,Chan:2025kyu,Goyal:2025eqo,Chakraborty:2025pxt}. 

A lower cutoff on the lens mass to which GW lensing searches are sensitive is set by the onset of the wave-optics regime, where diffraction suppresses the observable modulations. With the current LVK sensitivity, this restricts the constraining power to PBH masses above $100,\Msun$~\cite{Jung:2017flg,Diego:2019rzc,Liao:2020hnx,Urrutia:2021qak,Basak:2021ten,Urrutia:2023mtk}, as shown by the red shaded region in the right panel of Fig.~\ref{fig:mergers}. As shown by the green curve, the next generation GW detectors, such as ET, are expected to reach sensitivities which probe PBH abundances as low as $\fPBH \sim 10^{-4}$ and masses down to $\MPBH \sim 10^{-2}\,\Msun$ through GW lensing~\cite{Liao:2020hnx,Urrutia:2021qak,Urrutia:2023mtk,Zumalacarregui:2024ocb,Kim:2025njb}. The sensitivity range is further enhanced for $\fPBH < 1$ when a particle DM minihalo surrounding the PBHs is taken into account~\cite{Urrutia:2023mtk}, as indicated by the dashed curves.

\section{Summary and outlook}

Over the past decades, PBHs have been extensively studied as potential DM candidates. In this work, we have reviewed the main formation mechanisms, including those associated with inflationary perturbations, phase transitions, and topological defects, and discussed the mass functions predicted in these scenarios. We have also summarised the key observational constraints, from quantum evaporation, dynamical effects, lensing, accretion, and GW emission, highlighting the mass windows where PBHs could still constitute a significant fraction of DM. The digitised tables of the constraints, together with a Mathematica notebook that displays them for extended mass functions, are publicly available at GitHub:~\href{https://github.com/vianvask/PBHconstraints}{PBHconstraints}. In addition to constraints, we have reviewed the hints for PBHs from various observations, such as ML and GW events.

We have also discussed the prospects for future searches. In particular, GW observations offer a powerful avenue to probe PBHs. Current and upcoming detectors can potentially observe mergers across a wide range of masses and redshifts, including the sub-solar and high-redshift regimes, where astrophysical black holes are not expected. In addition, PBHs can be associated to a stochastic background of SIGWs sourced by primordial scalar perturbations. The detection of such a background would provide a complementary and independent probe of PBHs, especially in the asteroid-mass range, which is hard to access with other observations.

PBHs remain one of the most intriguing DM candidates. Theoretical work continues to refine our understanding of their formation and is exploring how different formation mechanisms influence their mass and abundance. At the same time, ongoing and upcoming observational programs, ranging from GW detectors to high-precision ML surveys and X-ray observations, will probe both the asteroid and solar mass ranges, testing scenarios that were previously inaccessible and searching for direct or indirect signatures of PBH dark matter.

\vspace{4mm}
\noindent
\emph{Acknowledgments --}This work was supported by the Estonian Research Council grants PSG869, TARISTU24-TK3, TARISTU24-TK10, KOHTO34 and the Center of Excellence program TK202. The work of V.V. was partially funded by the European Union's Horizon Europe program under the Marie Sk\l{}odowska-Curie grant agreement No. 101065736.

\bibliography{PBH}

\begin{thebibliography}{430}%
\makeatletter
\providecommand \@ifxundefined [1]{%
 \@ifx{#1\undefined}
}%
\providecommand \@ifnum [1]{%
 \ifnum #1\expandafter \@firstoftwo
 \else \expandafter \@secondoftwo
 \fi
}%
\providecommand \@ifx [1]{%
 \ifx #1\expandafter \@firstoftwo
 \else \expandafter \@secondoftwo
 \fi
}%
\providecommand \natexlab [1]{#1}%
\providecommand \enquote  [1]{``#1''}%
\providecommand \bibnamefont  [1]{#1}%
\providecommand \bibfnamefont [1]{#1}%
\providecommand \citenamefont [1]{#1}%
\providecommand \href@noop [0]{\@secondoftwo}%
\providecommand \href [0]{\begingroup \@sanitize@url \@href}%
\providecommand \@href[1]{\@@startlink{#1}\@@href}%
\providecommand \@@href[1]{\endgroup#1\@@endlink}%
\providecommand \@sanitize@url [0]{\catcode `\\12\catcode `\$12\catcode
  `\&12\catcode `\#12\catcode `\^12\catcode `\_12\catcode `\%12\relax}%
\providecommand \@@startlink[1]{}%
\providecommand \@@endlink[0]{}%
\providecommand \url  [0]{\begingroup\@sanitize@url \@url }%
\providecommand \@url [1]{\endgroup\@href {#1}{\urlprefix }}%
\providecommand \urlprefix  [0]{URL }%
\providecommand \Eprint [0]{\href }%
\providecommand \doibase [0]{https://doi.org/}%
\providecommand \selectlanguage [0]{\@gobble}%
\providecommand \bibinfo  [0]{\@secondoftwo}%
\providecommand \bibfield  [0]{\@secondoftwo}%
\providecommand \translation [1]{[#1]}%
\providecommand \BibitemOpen [0]{}%
\providecommand \bibitemStop [0]{}%
\providecommand \bibitemNoStop [0]{.\EOS\space}%
\providecommand \EOS [0]{\spacefactor3000\relax}%
\providecommand \BibitemShut  [1]{\csname bibitem#1\endcsname}%
\let\auto@bib@innerbib\@empty
\bibitem [{\citenamefont {Zel'dovich}\ and\ \citenamefont
  {Novikov}(1967)}]{Zeldovich:1967lct}%
  \BibitemOpen
  \bibfield  {author} {\bibinfo {author} {\bibfnamefont {Y.~B.}\ \bibnamefont
  {Zel'dovich}}\ and\ \bibinfo {author} {\bibfnamefont {I.~D.}\ \bibnamefont
  {Novikov}},\ }\bibfield  {title} {\bibinfo {title} {{The Hypothesis of Cores
  Retarded during Expansion and the Hot Cosmological Model}},\ }\href@noop {}
  {\bibfield  {journal} {\bibinfo  {journal} {Sov. Astron.}\ }\textbf {\bibinfo
  {volume} {10}},\ \bibinfo {pages} {602} (\bibinfo {year} {1967})}\BibitemShut
  {NoStop}%
\bibitem [{\citenamefont {Hawking}(1971)}]{Hawking:1971ei}%
  \BibitemOpen
  \bibfield  {author} {\bibinfo {author} {\bibfnamefont {S.}~\bibnamefont
  {Hawking}},\ }\bibfield  {title} {\bibinfo {title} {{Gravitationally
  collapsed objects of very low mass}},\ }\href
  {https://doi.org/10.1093/mnras/152.1.75} {\bibfield  {journal} {\bibinfo
  {journal} {Mon. Not. Roy. Astron. Soc.}\ }\textbf {\bibinfo {volume} {152}},\
  \bibinfo {pages} {75} (\bibinfo {year} {1971})}\BibitemShut {NoStop}%
\bibitem [{\citenamefont {Carr}\ and\ \citenamefont
  {Hawking}(1974)}]{Carr:1974nx}%
  \BibitemOpen
  \bibfield  {author} {\bibinfo {author} {\bibfnamefont {B.~J.}\ \bibnamefont
  {Carr}}\ and\ \bibinfo {author} {\bibfnamefont {S.~W.}\ \bibnamefont
  {Hawking}},\ }\bibfield  {title} {\bibinfo {title} {{Black holes in the early
  Universe}},\ }\href {https://doi.org/10.1093/mnras/168.2.399} {\bibfield
  {journal} {\bibinfo  {journal} {Mon. Not. Roy. Astron. Soc.}\ }\textbf
  {\bibinfo {volume} {168}},\ \bibinfo {pages} {399} (\bibinfo {year}
  {1974})}\BibitemShut {NoStop}%
\bibitem [{\citenamefont {Carr}(1975)}]{Carr:1975qj}%
  \BibitemOpen
  \bibfield  {author} {\bibinfo {author} {\bibfnamefont {B.~J.}\ \bibnamefont
  {Carr}},\ }\bibfield  {title} {\bibinfo {title} {{The Primordial black hole
  mass spectrum}},\ }\href {https://doi.org/10.1086/153853} {\bibfield
  {journal} {\bibinfo  {journal} {Astrophys. J.}\ }\textbf {\bibinfo {volume}
  {201}},\ \bibinfo {pages} {1} (\bibinfo {year} {1975})}\BibitemShut {NoStop}%
\bibitem [{\citenamefont {Chapline}(1975)}]{Chapline:1975ojl}%
  \BibitemOpen
  \bibfield  {author} {\bibinfo {author} {\bibfnamefont {G.~F.}\ \bibnamefont
  {Chapline}},\ }\bibfield  {title} {\bibinfo {title} {{Cosmological effects of
  primordial black holes}},\ }\href {https://doi.org/10.1038/253251a0}
  {\bibfield  {journal} {\bibinfo  {journal} {Nature}\ }\textbf {\bibinfo
  {volume} {253}},\ \bibinfo {pages} {251} (\bibinfo {year}
  {1975})}\BibitemShut {NoStop}%
\bibitem [{\citenamefont {Carr}\ \emph
  {et~al.}(2016{\natexlab{a}})\citenamefont {Carr}, \citenamefont {Kuhnel},\
  and\ \citenamefont {Sandstad}}]{Carr:2016drx}%
  \BibitemOpen
  \bibfield  {author} {\bibinfo {author} {\bibfnamefont {B.}~\bibnamefont
  {Carr}}, \bibinfo {author} {\bibfnamefont {F.}~\bibnamefont {Kuhnel}},\ and\
  \bibinfo {author} {\bibfnamefont {M.}~\bibnamefont {Sandstad}},\ }\bibfield
  {title} {\bibinfo {title} {{Primordial Black Holes as Dark Matter}},\ }\href
  {https://doi.org/10.1103/PhysRevD.94.083504} {\bibfield  {journal} {\bibinfo
  {journal} {Phys. Rev. D}\ }\textbf {\bibinfo {volume} {94}},\ \bibinfo
  {pages} {083504} (\bibinfo {year} {2016}{\natexlab{a}})},\ \Eprint
  {https://arxiv.org/abs/1607.06077} {arXiv:1607.06077 [astro-ph.CO]}
  \BibitemShut {NoStop}%
\bibitem [{\citenamefont {Carr}\ and\ \citenamefont
  {Kuhnel}(2020)}]{Carr:2020xqk}%
  \BibitemOpen
  \bibfield  {author} {\bibinfo {author} {\bibfnamefont {B.}~\bibnamefont
  {Carr}}\ and\ \bibinfo {author} {\bibfnamefont {F.}~\bibnamefont {Kuhnel}},\
  }\bibfield  {title} {\bibinfo {title} {{Primordial Black Holes as Dark
  Matter: Recent Developments}},\ }\href
  {https://doi.org/10.1146/annurev-nucl-050520-125911} {\bibfield  {journal}
  {\bibinfo  {journal} {Ann. Rev. Nucl. Part. Sci.}\ }\textbf {\bibinfo
  {volume} {70}},\ \bibinfo {pages} {355} (\bibinfo {year} {2020})},\ \Eprint
  {https://arxiv.org/abs/2006.02838} {arXiv:2006.02838 [astro-ph.CO]}
  \BibitemShut {NoStop}%
\bibitem [{\citenamefont {Cirelli}\ \emph {et~al.}(2024)\citenamefont
  {Cirelli}, \citenamefont {Strumia},\ and\ \citenamefont
  {Zupan}}]{Cirelli:2024ssz}%
  \BibitemOpen
  \bibfield  {author} {\bibinfo {author} {\bibfnamefont {M.}~\bibnamefont
  {Cirelli}}, \bibinfo {author} {\bibfnamefont {A.}~\bibnamefont {Strumia}},\
  and\ \bibinfo {author} {\bibfnamefont {J.}~\bibnamefont {Zupan}},\ }\bibfield
   {title} {\bibinfo {title} {{Dark Matter}},\ }\href@noop {} {\  (\bibinfo
  {year} {2024})},\ \Eprint {https://arxiv.org/abs/2406.01705}
  {arXiv:2406.01705 [hep-ph]} \BibitemShut {NoStop}%
\bibitem [{\citenamefont {Bird}\ \emph {et~al.}(2016)\citenamefont {Bird},
  \citenamefont {Cholis}, \citenamefont {Mu{\~n}oz}, \citenamefont
  {Ali-Ha{\"\i}moud}, \citenamefont {Kamionkowski}, \citenamefont {Kovetz},
  \citenamefont {Raccanelli},\ and\ \citenamefont {Riess}}]{Bird:2016dcv}%
  \BibitemOpen
  \bibfield  {author} {\bibinfo {author} {\bibfnamefont {S.}~\bibnamefont
  {Bird}}, \bibinfo {author} {\bibfnamefont {I.}~\bibnamefont {Cholis}},
  \bibinfo {author} {\bibfnamefont {J.~B.}\ \bibnamefont {Mu{\~n}oz}}, \bibinfo
  {author} {\bibfnamefont {Y.}~\bibnamefont {Ali-Ha{\"\i}moud}}, \bibinfo
  {author} {\bibfnamefont {M.}~\bibnamefont {Kamionkowski}}, \bibinfo {author}
  {\bibfnamefont {E.~D.}\ \bibnamefont {Kovetz}}, \bibinfo {author}
  {\bibfnamefont {A.}~\bibnamefont {Raccanelli}},\ and\ \bibinfo {author}
  {\bibfnamefont {A.~G.}\ \bibnamefont {Riess}},\ }\bibfield  {title} {\bibinfo
  {title} {{Did LIGO detect dark matter?}},\ }\href
  {https://doi.org/10.1103/PhysRevLett.116.201301} {\bibfield  {journal}
  {\bibinfo  {journal} {Phys. Rev. Lett.}\ }\textbf {\bibinfo {volume} {116}},\
  \bibinfo {pages} {201301} (\bibinfo {year} {2016})},\ \Eprint
  {https://arxiv.org/abs/1603.00464} {arXiv:1603.00464 [astro-ph.CO]}
  \BibitemShut {NoStop}%
\bibitem [{\citenamefont {Carr}\ and\ \citenamefont
  {Silk}(2018)}]{Carr:2018rid}%
  \BibitemOpen
  \bibfield  {author} {\bibinfo {author} {\bibfnamefont {B.}~\bibnamefont
  {Carr}}\ and\ \bibinfo {author} {\bibfnamefont {J.}~\bibnamefont {Silk}},\
  }\bibfield  {title} {\bibinfo {title} {{Primordial Black Holes as Generators
  of Cosmic Structures}},\ }\href {https://doi.org/10.1093/mnras/sty1204}
  {\bibfield  {journal} {\bibinfo  {journal} {Mon. Not. Roy. Astron. Soc.}\
  }\textbf {\bibinfo {volume} {478}},\ \bibinfo {pages} {3756} (\bibinfo {year}
  {2018})},\ \Eprint {https://arxiv.org/abs/1801.00672} {arXiv:1801.00672
  [astro-ph.CO]} \BibitemShut {NoStop}%
\bibitem [{\citenamefont {Niikura}\ \emph
  {et~al.}(2019{\natexlab{a}})\citenamefont {Niikura}, \citenamefont {Takada},
  \citenamefont {Yokoyama}, \citenamefont {Sumi},\ and\ \citenamefont
  {Masaki}}]{Niikura:2019kqi}%
  \BibitemOpen
  \bibfield  {author} {\bibinfo {author} {\bibfnamefont {H.}~\bibnamefont
  {Niikura}}, \bibinfo {author} {\bibfnamefont {M.}~\bibnamefont {Takada}},
  \bibinfo {author} {\bibfnamefont {S.}~\bibnamefont {Yokoyama}}, \bibinfo
  {author} {\bibfnamefont {T.}~\bibnamefont {Sumi}},\ and\ \bibinfo {author}
  {\bibfnamefont {S.}~\bibnamefont {Masaki}},\ }\bibfield  {title} {\bibinfo
  {title} {{Constraints on Earth-mass primordial black holes from OGLE 5-year
  microlensing events}},\ }\href {https://doi.org/10.1103/PhysRevD.99.083503}
  {\bibfield  {journal} {\bibinfo  {journal} {Phys. Rev. D}\ }\textbf {\bibinfo
  {volume} {99}},\ \bibinfo {pages} {083503} (\bibinfo {year}
  {2019}{\natexlab{a}})},\ \Eprint {https://arxiv.org/abs/1901.07120}
  {arXiv:1901.07120 [astro-ph.CO]} \BibitemShut {NoStop}%
\bibitem [{\citenamefont {Cline}\ \emph {et~al.}(1997)\citenamefont {Cline},
  \citenamefont {Sanders},\ and\ \citenamefont {Hong}}]{Cline:1996zg}%
  \BibitemOpen
  \bibfield  {author} {\bibinfo {author} {\bibfnamefont {D.~B.}\ \bibnamefont
  {Cline}}, \bibinfo {author} {\bibfnamefont {D.~A.}\ \bibnamefont {Sanders}},\
  and\ \bibinfo {author} {\bibfnamefont {W.}~\bibnamefont {Hong}},\ }\bibfield
  {title} {\bibinfo {title} {{Further evidence for gamma-ray bursts consistent
  with primordial black hole evaporation}},\ }\href
  {https://doi.org/10.1086/304480} {\bibfield  {journal} {\bibinfo  {journal}
  {Astrophys. J.}\ }\textbf {\bibinfo {volume} {486}},\ \bibinfo {pages} {169}
  (\bibinfo {year} {1997})}\BibitemShut {NoStop}%
\bibitem [{\citenamefont {Carr}\ \emph
  {et~al.}(2017{\natexlab{a}})\citenamefont {Carr}, \citenamefont {Raidal},
  \citenamefont {Tenkanen}, \citenamefont {Vaskonen},\ and\ \citenamefont
  {Veerm\"ae}}]{Carr:2017jsz}%
  \BibitemOpen
  \bibfield  {author} {\bibinfo {author} {\bibfnamefont {B.}~\bibnamefont
  {Carr}}, \bibinfo {author} {\bibfnamefont {M.}~\bibnamefont {Raidal}},
  \bibinfo {author} {\bibfnamefont {T.}~\bibnamefont {Tenkanen}}, \bibinfo
  {author} {\bibfnamefont {V.}~\bibnamefont {Vaskonen}},\ and\ \bibinfo
  {author} {\bibfnamefont {H.}~\bibnamefont {Veerm\"ae}},\ }\bibfield  {title}
  {\bibinfo {title} {{Primordial black hole constraints for extended mass
  functions}},\ }\href {https://doi.org/10.1103/PhysRevD.96.023514} {\bibfield
  {journal} {\bibinfo  {journal} {Phys. Rev. D}\ }\textbf {\bibinfo {volume}
  {96}},\ \bibinfo {pages} {023514} (\bibinfo {year} {2017}{\natexlab{a}})},\
  \Eprint {https://arxiv.org/abs/1705.05567} {arXiv:1705.05567 [astro-ph.CO]}
  \BibitemShut {NoStop}%
\bibitem [{\citenamefont {Carr}\ \emph
  {et~al.}(2021{\natexlab{a}})\citenamefont {Carr}, \citenamefont {Kohri},
  \citenamefont {Sendouda},\ and\ \citenamefont {Yokoyama}}]{Carr:2020gox}%
  \BibitemOpen
  \bibfield  {author} {\bibinfo {author} {\bibfnamefont {B.}~\bibnamefont
  {Carr}}, \bibinfo {author} {\bibfnamefont {K.}~\bibnamefont {Kohri}},
  \bibinfo {author} {\bibfnamefont {Y.}~\bibnamefont {Sendouda}},\ and\
  \bibinfo {author} {\bibfnamefont {J.}~\bibnamefont {Yokoyama}},\ }\bibfield
  {title} {\bibinfo {title} {{Constraints on primordial black holes}},\ }\href
  {https://doi.org/10.1088/1361-6633/ac1e31} {\bibfield  {journal} {\bibinfo
  {journal} {Rept. Prog. Phys.}\ }\textbf {\bibinfo {volume} {84}},\ \bibinfo
  {pages} {116902} (\bibinfo {year} {2021}{\natexlab{a}})},\ \Eprint
  {https://arxiv.org/abs/2002.12778} {arXiv:2002.12778 [astro-ph.CO]}
  \BibitemShut {NoStop}%
\bibitem [{\citenamefont {Byrnes}\ \emph {et~al.}(2025)\citenamefont {Byrnes},
  \citenamefont {Franciolini}, \citenamefont {Harada}, \citenamefont {Pani},\
  and\ \citenamefont {Sasaki}}]{Byrnes:2025tji}%
  \BibitemOpen
  \bibinfo {editor} {\bibfnamefont {C.}~\bibnamefont {Byrnes}}, \bibinfo
  {editor} {\bibfnamefont {G.}~\bibnamefont {Franciolini}}, \bibinfo {editor}
  {\bibfnamefont {T.}~\bibnamefont {Harada}}, \bibinfo {editor} {\bibfnamefont
  {P.}~\bibnamefont {Pani}},\ and\ \bibinfo {editor} {\bibfnamefont
  {M.}~\bibnamefont {Sasaki}},\ eds.,\ \href
  {https://doi.org/10.1007/978-981-97-8887-3} {\emph {\bibinfo {title}
  {{Primordial Black Holes}}}},\ Springer Series in Astrophysics and Cosmology\
  (\bibinfo  {publisher} {Springer},\ \bibinfo {year} {2025})\BibitemShut
  {NoStop}%
\bibitem [{\citenamefont {Carr}\ \emph
  {et~al.}(2021{\natexlab{b}})\citenamefont {Carr}, \citenamefont {Clesse},
  \citenamefont {Garc{\'\i}a-Bellido},\ and\ \citenamefont
  {K{\"u}hnel}}]{Carr:2019kxo}%
  \BibitemOpen
  \bibfield  {author} {\bibinfo {author} {\bibfnamefont {B.}~\bibnamefont
  {Carr}}, \bibinfo {author} {\bibfnamefont {S.}~\bibnamefont {Clesse}},
  \bibinfo {author} {\bibfnamefont {J.}~\bibnamefont {Garc{\'\i}a-Bellido}},\
  and\ \bibinfo {author} {\bibfnamefont {F.}~\bibnamefont {K{\"u}hnel}},\
  }\bibfield  {title} {\bibinfo {title} {{Cosmic conundra explained by thermal
  history and primordial black holes}},\ }\href
  {https://doi.org/10.1016/j.dark.2020.100755} {\bibfield  {journal} {\bibinfo
  {journal} {Phys. Dark Univ.}\ }\textbf {\bibinfo {volume} {31}},\ \bibinfo
  {pages} {100755} (\bibinfo {year} {2021}{\natexlab{b}})},\ \Eprint
  {https://arxiv.org/abs/1906.08217} {arXiv:1906.08217 [astro-ph.CO]}
  \BibitemShut {NoStop}%
\bibitem [{\citenamefont {Carr}\ \emph {et~al.}(2024)\citenamefont {Carr},
  \citenamefont {Clesse}, \citenamefont {Garcia-Bellido}, \citenamefont
  {Hawkins},\ and\ \citenamefont {Kuhnel}}]{Carr:2023tpt}%
  \BibitemOpen
  \bibfield  {author} {\bibinfo {author} {\bibfnamefont {B.}~\bibnamefont
  {Carr}}, \bibinfo {author} {\bibfnamefont {S.}~\bibnamefont {Clesse}},
  \bibinfo {author} {\bibfnamefont {J.}~\bibnamefont {Garcia-Bellido}},
  \bibinfo {author} {\bibfnamefont {M.}~\bibnamefont {Hawkins}},\ and\ \bibinfo
  {author} {\bibfnamefont {F.}~\bibnamefont {Kuhnel}},\ }\bibfield  {title}
  {\bibinfo {title} {{Observational evidence for primordial black holes: A
  positivist perspective}},\ }\href
  {https://doi.org/10.1016/j.physrep.2023.11.005} {\bibfield  {journal}
  {\bibinfo  {journal} {Phys. Rept.}\ }\textbf {\bibinfo {volume} {1054}},\
  \bibinfo {pages} {1} (\bibinfo {year} {2024})},\ \Eprint
  {https://arxiv.org/abs/2306.03903} {arXiv:2306.03903 [astro-ph.CO]}
  \BibitemShut {NoStop}%
\bibitem [{\citenamefont {Borsanyi}\ \emph {et~al.}(2016)\citenamefont
  {Borsanyi} \emph {et~al.}}]{Borsanyi:2016ksw}%
  \BibitemOpen
  \bibfield  {author} {\bibinfo {author} {\bibfnamefont {S.}~\bibnamefont
  {Borsanyi}} \emph {et~al.},\ }\bibfield  {title} {\bibinfo {title}
  {{Calculation of the axion mass based on high-temperature lattice quantum
  chromodynamics}},\ }\href {https://doi.org/10.1038/nature20115} {\bibfield
  {journal} {\bibinfo  {journal} {Nature}\ }\textbf {\bibinfo {volume} {539}},\
  \bibinfo {pages} {69} (\bibinfo {year} {2016})},\ \Eprint
  {https://arxiv.org/abs/1606.07494} {arXiv:1606.07494 [hep-lat]} \BibitemShut
  {NoStop}%
\bibitem [{\citenamefont {Hawking}(1974)}]{Hawking:1974rv}%
  \BibitemOpen
  \bibfield  {author} {\bibinfo {author} {\bibfnamefont {S.~W.}\ \bibnamefont
  {Hawking}},\ }\bibfield  {title} {\bibinfo {title} {{Black hole
  explosions}},\ }\href {https://doi.org/10.1038/248030a0} {\bibfield
  {journal} {\bibinfo  {journal} {Nature}\ }\textbf {\bibinfo {volume} {248}},\
  \bibinfo {pages} {30} (\bibinfo {year} {1974})}\BibitemShut {NoStop}%
\bibitem [{\citenamefont {Choptuik}(1993)}]{Choptuik:1992jv}%
  \BibitemOpen
  \bibfield  {author} {\bibinfo {author} {\bibfnamefont {M.~W.}\ \bibnamefont
  {Choptuik}},\ }\bibfield  {title} {\bibinfo {title} {{Universality and
  scaling in gravitational collapse of a massless scalar field}},\ }\href
  {https://doi.org/10.1103/PhysRevLett.70.9} {\bibfield  {journal} {\bibinfo
  {journal} {Phys. Rev. Lett.}\ }\textbf {\bibinfo {volume} {70}},\ \bibinfo
  {pages} {9} (\bibinfo {year} {1993})}\BibitemShut {NoStop}%
\bibitem [{\citenamefont {Niemeyer}\ and\ \citenamefont
  {Jedamzik}(1998)}]{Niemeyer:1997mt}%
  \BibitemOpen
  \bibfield  {author} {\bibinfo {author} {\bibfnamefont {J.~C.}\ \bibnamefont
  {Niemeyer}}\ and\ \bibinfo {author} {\bibfnamefont {K.}~\bibnamefont
  {Jedamzik}},\ }\bibfield  {title} {\bibinfo {title} {{Near-critical
  gravitational collapse and the initial mass function of primordial black
  holes}},\ }\href {https://doi.org/10.1103/PhysRevLett.80.5481} {\bibfield
  {journal} {\bibinfo  {journal} {Phys. Rev. Lett.}\ }\textbf {\bibinfo
  {volume} {80}},\ \bibinfo {pages} {5481} (\bibinfo {year} {1998})},\ \Eprint
  {https://arxiv.org/abs/astro-ph/9709072} {arXiv:astro-ph/9709072}
  \BibitemShut {NoStop}%
\bibitem [{\citenamefont {Niemeyer}\ and\ \citenamefont
  {Jedamzik}(1999)}]{Niemeyer:1999ak}%
  \BibitemOpen
  \bibfield  {author} {\bibinfo {author} {\bibfnamefont {J.~C.}\ \bibnamefont
  {Niemeyer}}\ and\ \bibinfo {author} {\bibfnamefont {K.}~\bibnamefont
  {Jedamzik}},\ }\bibfield  {title} {\bibinfo {title} {{Dynamics of primordial
  black hole formation}},\ }\href {https://doi.org/10.1103/PhysRevD.59.124013}
  {\bibfield  {journal} {\bibinfo  {journal} {Phys. Rev. D}\ }\textbf {\bibinfo
  {volume} {59}},\ \bibinfo {pages} {124013} (\bibinfo {year} {1999})},\
  \Eprint {https://arxiv.org/abs/astro-ph/9901292} {arXiv:astro-ph/9901292}
  \BibitemShut {NoStop}%
\bibitem [{\citenamefont {Musco}\ \emph {et~al.}(2009)\citenamefont {Musco},
  \citenamefont {Miller},\ and\ \citenamefont {Polnarev}}]{Musco:2008hv}%
  \BibitemOpen
  \bibfield  {author} {\bibinfo {author} {\bibfnamefont {I.}~\bibnamefont
  {Musco}}, \bibinfo {author} {\bibfnamefont {J.~C.}\ \bibnamefont {Miller}},\
  and\ \bibinfo {author} {\bibfnamefont {A.~G.}\ \bibnamefont {Polnarev}},\
  }\bibfield  {title} {\bibinfo {title} {{Primordial black hole formation in
  the radiative era: Investigation of the critical nature of the collapse}},\
  }\href {https://doi.org/10.1088/0264-9381/26/23/235001} {\bibfield  {journal}
  {\bibinfo  {journal} {Class. Quant. Grav.}\ }\textbf {\bibinfo {volume}
  {26}},\ \bibinfo {pages} {235001} (\bibinfo {year} {2009})},\ \Eprint
  {https://arxiv.org/abs/0811.1452} {arXiv:0811.1452 [gr-qc]} \BibitemShut
  {NoStop}%
\bibitem [{\citenamefont {Musco}\ \emph {et~al.}(2021)\citenamefont {Musco},
  \citenamefont {De~Luca}, \citenamefont {Franciolini},\ and\ \citenamefont
  {Riotto}}]{Musco:2020jjb}%
  \BibitemOpen
  \bibfield  {author} {\bibinfo {author} {\bibfnamefont {I.}~\bibnamefont
  {Musco}}, \bibinfo {author} {\bibfnamefont {V.}~\bibnamefont {De~Luca}},
  \bibinfo {author} {\bibfnamefont {G.}~\bibnamefont {Franciolini}},\ and\
  \bibinfo {author} {\bibfnamefont {A.}~\bibnamefont {Riotto}},\ }\bibfield
  {title} {\bibinfo {title} {{Threshold for primordial black holes. II. A
  simple analytic prescription}},\ }\href
  {https://doi.org/10.1103/PhysRevD.103.063538} {\bibfield  {journal} {\bibinfo
   {journal} {Phys. Rev. D}\ }\textbf {\bibinfo {volume} {103}},\ \bibinfo
  {pages} {063538} (\bibinfo {year} {2021})},\ \Eprint
  {https://arxiv.org/abs/2011.03014} {arXiv:2011.03014 [astro-ph.CO]}
  \BibitemShut {NoStop}%
\bibitem [{\citenamefont {Musco}\ \emph {et~al.}(2024)\citenamefont {Musco},
  \citenamefont {Jedamzik},\ and\ \citenamefont {Young}}]{Musco:2023dak}%
  \BibitemOpen
  \bibfield  {author} {\bibinfo {author} {\bibfnamefont {I.}~\bibnamefont
  {Musco}}, \bibinfo {author} {\bibfnamefont {K.}~\bibnamefont {Jedamzik}},\
  and\ \bibinfo {author} {\bibfnamefont {S.}~\bibnamefont {Young}},\ }\bibfield
   {title} {\bibinfo {title} {{Primordial black hole formation during the QCD
  phase transition: Threshold, mass distribution, and abundance}},\ }\href
  {https://doi.org/10.1103/PhysRevD.109.083506} {\bibfield  {journal} {\bibinfo
   {journal} {Phys. Rev. D}\ }\textbf {\bibinfo {volume} {109}},\ \bibinfo
  {pages} {083506} (\bibinfo {year} {2024})},\ \Eprint
  {https://arxiv.org/abs/2303.07980} {arXiv:2303.07980 [astro-ph.CO]}
  \BibitemShut {NoStop}%
\bibitem [{\citenamefont {Ianniccari}\ \emph
  {et~al.}(2024{\natexlab{a}})\citenamefont {Ianniccari}, \citenamefont
  {Iovino}, \citenamefont {Kehagias}, \citenamefont {Perrone},\ and\
  \citenamefont {Riotto}}]{Ianniccari:2024ltb}%
  \BibitemOpen
  \bibfield  {author} {\bibinfo {author} {\bibfnamefont {A.}~\bibnamefont
  {Ianniccari}}, \bibinfo {author} {\bibfnamefont {A.~J.}\ \bibnamefont
  {Iovino}}, \bibinfo {author} {\bibfnamefont {A.}~\bibnamefont {Kehagias}},
  \bibinfo {author} {\bibfnamefont {D.}~\bibnamefont {Perrone}},\ and\ \bibinfo
  {author} {\bibfnamefont {A.}~\bibnamefont {Riotto}},\ }\bibfield  {title}
  {\bibinfo {title} {{The Black Hole Formation -- Null Geodesic
  Correspondence}},\ }\href {https://doi.org/10.1103/PhysRevLett.133.081401}
  {\bibfield  {journal} {\bibinfo  {journal} {Phys. Rev. Lett.}\ }\textbf
  {\bibinfo {volume} {133}},\ \bibinfo {pages} {081401} (\bibinfo {year}
  {2024}{\natexlab{a}})},\ \Eprint {https://arxiv.org/abs/2404.02801}
  {arXiv:2404.02801 [astro-ph.CO]} \BibitemShut {NoStop}%
\bibitem [{\citenamefont {Iovino}\ \emph {et~al.}(2024)\citenamefont {Iovino},
  \citenamefont {Perna}, \citenamefont {Riotto},\ and\ \citenamefont
  {Veerm{\"a}e}}]{Iovino:2024tyg}%
  \BibitemOpen
  \bibfield  {author} {\bibinfo {author} {\bibfnamefont {A.~J.}\ \bibnamefont
  {Iovino}}, \bibinfo {author} {\bibfnamefont {G.}~\bibnamefont {Perna}},
  \bibinfo {author} {\bibfnamefont {A.}~\bibnamefont {Riotto}},\ and\ \bibinfo
  {author} {\bibfnamefont {H.}~\bibnamefont {Veerm{\"a}e}},\ }\bibfield
  {title} {\bibinfo {title} {{Curbing PBHs with PTAs}},\ }\href
  {https://doi.org/10.1088/1475-7516/2024/10/050} {\bibfield  {journal}
  {\bibinfo  {journal} {JCAP}\ }\textbf {\bibinfo {volume} {10}},\ \bibinfo
  {pages} {050}},\ \Eprint {https://arxiv.org/abs/2406.20089} {arXiv:2406.20089
  [astro-ph.CO]} \BibitemShut {NoStop}%
\bibitem [{\citenamefont {Aghanim}\ \emph {et~al.}(2020)\citenamefont {Aghanim}
  \emph {et~al.}}]{Planck:2018vyg}%
  \BibitemOpen
  \bibfield  {author} {\bibinfo {author} {\bibfnamefont {N.}~\bibnamefont
  {Aghanim}} \emph {et~al.} (\bibinfo {collaboration} {Planck}),\ }\bibfield
  {title} {\bibinfo {title} {{Planck 2018 results. VI. Cosmological
  parameters}},\ }\href {https://doi.org/10.1051/0004-6361/201833910}
  {\bibfield  {journal} {\bibinfo  {journal} {Astron. Astrophys.}\ }\textbf
  {\bibinfo {volume} {641}},\ \bibinfo {pages} {A6} (\bibinfo {year} {2020})},\
  \bibinfo {note} {[Erratum: Astron.Astrophys. 652, C4 (2021)]},\ \Eprint
  {https://arxiv.org/abs/1807.06209} {arXiv:1807.06209 [astro-ph.CO]}
  \BibitemShut {NoStop}%
\bibitem [{\citenamefont {Calabrese}\ \emph {et~al.}(2025)\citenamefont
  {Calabrese} \emph {et~al.}}]{ACT:2025nti}%
  \BibitemOpen
  \bibfield  {author} {\bibinfo {author} {\bibfnamefont {E.}~\bibnamefont
  {Calabrese}} \emph {et~al.} (\bibinfo {collaboration} {Atacama Cosmology
  Telescope}),\ }\bibfield  {title} {\bibinfo {title} {{The Atacama Cosmology
  Telescope: DR6 constraints on extended cosmological models}},\ }\href
  {https://doi.org/10.1088/1475-7516/2025/11/063} {\bibfield  {journal}
  {\bibinfo  {journal} {JCAP}\ }\textbf {\bibinfo {volume} {11}},\ \bibinfo
  {pages} {063}},\ \Eprint {https://arxiv.org/abs/2503.14454} {arXiv:2503.14454
  [astro-ph.CO]} \BibitemShut {NoStop}%
\bibitem [{\citenamefont {Chluba}\ \emph {et~al.}(2012)\citenamefont {Chluba},
  \citenamefont {Erickcek},\ and\ \citenamefont {Ben-Dayan}}]{Chluba:2012we}%
  \BibitemOpen
  \bibfield  {author} {\bibinfo {author} {\bibfnamefont {J.}~\bibnamefont
  {Chluba}}, \bibinfo {author} {\bibfnamefont {A.~L.}\ \bibnamefont
  {Erickcek}},\ and\ \bibinfo {author} {\bibfnamefont {I.}~\bibnamefont
  {Ben-Dayan}},\ }\bibfield  {title} {\bibinfo {title} {{Probing the inflaton:
  Small-scale power spectrum constraints from measurements of the CMB energy
  spectrum}},\ }\href {https://doi.org/10.1088/0004-637X/758/2/76} {\bibfield
  {journal} {\bibinfo  {journal} {Astrophys. J.}\ }\textbf {\bibinfo {volume}
  {758}},\ \bibinfo {pages} {76} (\bibinfo {year} {2012})},\ \Eprint
  {https://arxiv.org/abs/1203.2681} {arXiv:1203.2681 [astro-ph.CO]}
  \BibitemShut {NoStop}%
\bibitem [{\citenamefont {Chluba}\ and\ \citenamefont
  {Grin}(2013)}]{Chluba:2013dna}%
  \BibitemOpen
  \bibfield  {author} {\bibinfo {author} {\bibfnamefont {J.}~\bibnamefont
  {Chluba}}\ and\ \bibinfo {author} {\bibfnamefont {D.}~\bibnamefont {Grin}},\
  }\bibfield  {title} {\bibinfo {title} {{CMB spectral distortions from
  small-scale isocurvature fluctuations}},\ }\href
  {https://doi.org/10.1093/mnras/stt1129} {\bibfield  {journal} {\bibinfo
  {journal} {Mon. Not. Roy. Astron. Soc.}\ }\textbf {\bibinfo {volume} {434}},\
  \bibinfo {pages} {1619} (\bibinfo {year} {2013})},\ \Eprint
  {https://arxiv.org/abs/1304.4596} {arXiv:1304.4596 [astro-ph.CO]}
  \BibitemShut {NoStop}%
\bibitem [{\citenamefont {Sharma}\ \emph {et~al.}(2024)\citenamefont {Sharma},
  \citenamefont {Lesgourgues},\ and\ \citenamefont {Byrnes}}]{Sharma:2024img}%
  \BibitemOpen
  \bibfield  {author} {\bibinfo {author} {\bibfnamefont {D.}~\bibnamefont
  {Sharma}}, \bibinfo {author} {\bibfnamefont {J.}~\bibnamefont
  {Lesgourgues}},\ and\ \bibinfo {author} {\bibfnamefont {C.~T.}\ \bibnamefont
  {Byrnes}},\ }\bibfield  {title} {\bibinfo {title} {{Spectral distortions from
  acoustic dissipation with non-Gaussian (or not) perturbations}},\ }\href
  {https://doi.org/10.1088/1475-7516/2024/07/090} {\bibfield  {journal}
  {\bibinfo  {journal} {JCAP}\ }\textbf {\bibinfo {volume} {07}},\ \bibinfo
  {pages} {090}},\ \Eprint {https://arxiv.org/abs/2404.18474} {arXiv:2404.18474
  [astro-ph.CO]} \BibitemShut {NoStop}%
\bibitem [{\citenamefont {Byrnes}\ \emph {et~al.}(2024)\citenamefont {Byrnes},
  \citenamefont {Lesgourgues},\ and\ \citenamefont {Sharma}}]{Byrnes:2024vjt}%
  \BibitemOpen
  \bibfield  {author} {\bibinfo {author} {\bibfnamefont {C.~T.}\ \bibnamefont
  {Byrnes}}, \bibinfo {author} {\bibfnamefont {J.}~\bibnamefont
  {Lesgourgues}},\ and\ \bibinfo {author} {\bibfnamefont {D.}~\bibnamefont
  {Sharma}},\ }\bibfield  {title} {\bibinfo {title} {{Robust
  {\ensuremath{\mu}}-distortion constraints on primordial supermassive black
  holes from non-Gaussian perturbations}},\ }\href
  {https://doi.org/10.1088/1475-7516/2024/09/012} {\bibfield  {journal}
  {\bibinfo  {journal} {JCAP}\ }\textbf {\bibinfo {volume} {09}},\ \bibinfo
  {pages} {012}},\ \Eprint {https://arxiv.org/abs/2404.18475} {arXiv:2404.18475
  [astro-ph.CO]} \BibitemShut {NoStop}%
\bibitem [{\citenamefont {Pritchard}\ \emph
  {et~al.}(2025{\natexlab{a}})\citenamefont {Pritchard}, \citenamefont
  {Byrnes}, \citenamefont {Lesgourgues},\ and\ \citenamefont
  {Sharma}}]{Pritchard:2025yda}%
  \BibitemOpen
  \bibfield  {author} {\bibinfo {author} {\bibfnamefont {X.}~\bibnamefont
  {Pritchard}}, \bibinfo {author} {\bibfnamefont {C.~T.}\ \bibnamefont
  {Byrnes}}, \bibinfo {author} {\bibfnamefont {J.}~\bibnamefont
  {Lesgourgues}},\ and\ \bibinfo {author} {\bibfnamefont {D.}~\bibnamefont
  {Sharma}},\ }\bibfield  {title} {\bibinfo {title} {{Robust
  {\ensuremath{\mu}}-distortion constraints on primordial supermassive black
  holes from cubic (gNL) non-Gaussian perturbations}},\ }\href
  {https://doi.org/10.1088/1475-7516/2025/07/079} {\bibfield  {journal}
  {\bibinfo  {journal} {JCAP}\ }\textbf {\bibinfo {volume} {07}},\ \bibinfo
  {pages} {079}},\ \Eprint {https://arxiv.org/abs/2505.08442} {arXiv:2505.08442
  [astro-ph.CO]} \BibitemShut {NoStop}%
\bibitem [{\citenamefont {Bird}\ \emph {et~al.}(2011)\citenamefont {Bird},
  \citenamefont {Peiris}, \citenamefont {Viel},\ and\ \citenamefont
  {Verde}}]{Bird:2010mp}%
  \BibitemOpen
  \bibfield  {author} {\bibinfo {author} {\bibfnamefont {S.}~\bibnamefont
  {Bird}}, \bibinfo {author} {\bibfnamefont {H.~V.}\ \bibnamefont {Peiris}},
  \bibinfo {author} {\bibfnamefont {M.}~\bibnamefont {Viel}},\ and\ \bibinfo
  {author} {\bibfnamefont {L.}~\bibnamefont {Verde}},\ }\bibfield  {title}
  {\bibinfo {title} {{Minimally Parametric Power Spectrum Reconstruction from
  the Lyman-alpha Forest}},\ }\href
  {https://doi.org/10.1111/j.1365-2966.2011.18245.x} {\bibfield  {journal}
  {\bibinfo  {journal} {Mon. Not. Roy. Astron. Soc.}\ }\textbf {\bibinfo
  {volume} {413}},\ \bibinfo {pages} {1717} (\bibinfo {year} {2011})},\ \Eprint
  {https://arxiv.org/abs/1010.1519} {arXiv:1010.1519 [astro-ph.CO]}
  \BibitemShut {NoStop}%
\bibitem [{\citenamefont {Ivanov}\ \emph {et~al.}(1994)\citenamefont {Ivanov},
  \citenamefont {Naselsky},\ and\ \citenamefont {Novikov}}]{Ivanov:1994pa}%
  \BibitemOpen
  \bibfield  {author} {\bibinfo {author} {\bibfnamefont {P.}~\bibnamefont
  {Ivanov}}, \bibinfo {author} {\bibfnamefont {P.}~\bibnamefont {Naselsky}},\
  and\ \bibinfo {author} {\bibfnamefont {I.}~\bibnamefont {Novikov}},\
  }\bibfield  {title} {\bibinfo {title} {{Inflation and primordial black holes
  as dark matter}},\ }\href {https://doi.org/10.1103/PhysRevD.50.7173}
  {\bibfield  {journal} {\bibinfo  {journal} {Phys. Rev. D}\ }\textbf {\bibinfo
  {volume} {50}},\ \bibinfo {pages} {7173} (\bibinfo {year}
  {1994})}\BibitemShut {NoStop}%
\bibitem [{\citenamefont {Kinney}(1997)}]{Kinney:1997ne}%
  \BibitemOpen
  \bibfield  {author} {\bibinfo {author} {\bibfnamefont {W.~H.}\ \bibnamefont
  {Kinney}},\ }\bibfield  {title} {\bibinfo {title} {{A Hamilton-Jacobi
  approach to nonslow roll inflation}},\ }\href
  {https://doi.org/10.1103/PhysRevD.56.2002} {\bibfield  {journal} {\bibinfo
  {journal} {Phys. Rev. D}\ }\textbf {\bibinfo {volume} {56}},\ \bibinfo
  {pages} {2002} (\bibinfo {year} {1997})},\ \Eprint
  {https://arxiv.org/abs/hep-ph/9702427} {arXiv:hep-ph/9702427} \BibitemShut
  {NoStop}%
\bibitem [{\citenamefont {Inoue}\ and\ \citenamefont
  {Yokoyama}(2002)}]{Inoue:2001zt}%
  \BibitemOpen
  \bibfield  {author} {\bibinfo {author} {\bibfnamefont {S.}~\bibnamefont
  {Inoue}}\ and\ \bibinfo {author} {\bibfnamefont {J.}~\bibnamefont
  {Yokoyama}},\ }\bibfield  {title} {\bibinfo {title} {{Curvature perturbation
  at the local extremum of the inflaton's potential}},\ }\href
  {https://doi.org/10.1016/S0370-2693(01)01369-7} {\bibfield  {journal}
  {\bibinfo  {journal} {Phys. Lett. B}\ }\textbf {\bibinfo {volume} {524}},\
  \bibinfo {pages} {15} (\bibinfo {year} {2002})},\ \Eprint
  {https://arxiv.org/abs/hep-ph/0104083} {arXiv:hep-ph/0104083} \BibitemShut
  {NoStop}%
\bibitem [{\citenamefont {Kinney}(2005)}]{Kinney:2005vj}%
  \BibitemOpen
  \bibfield  {author} {\bibinfo {author} {\bibfnamefont {W.~H.}\ \bibnamefont
  {Kinney}},\ }\bibfield  {title} {\bibinfo {title} {{Horizon crossing and
  inflation with large eta}},\ }\href
  {https://doi.org/10.1103/PhysRevD.72.023515} {\bibfield  {journal} {\bibinfo
  {journal} {Phys. Rev. D}\ }\textbf {\bibinfo {volume} {72}},\ \bibinfo
  {pages} {023515} (\bibinfo {year} {2005})},\ \Eprint
  {https://arxiv.org/abs/gr-qc/0503017} {arXiv:gr-qc/0503017} \BibitemShut
  {NoStop}%
\bibitem [{\citenamefont {Martin}\ \emph {et~al.}(2013)\citenamefont {Martin},
  \citenamefont {Motohashi},\ and\ \citenamefont {Suyama}}]{Martin:2012pe}%
  \BibitemOpen
  \bibfield  {author} {\bibinfo {author} {\bibfnamefont {J.}~\bibnamefont
  {Martin}}, \bibinfo {author} {\bibfnamefont {H.}~\bibnamefont {Motohashi}},\
  and\ \bibinfo {author} {\bibfnamefont {T.}~\bibnamefont {Suyama}},\
  }\bibfield  {title} {\bibinfo {title} {{Ultra Slow-Roll Inflation and the
  non-Gaussianity Consistency Relation}},\ }\href
  {https://doi.org/10.1103/PhysRevD.87.023514} {\bibfield  {journal} {\bibinfo
  {journal} {Phys. Rev. D}\ }\textbf {\bibinfo {volume} {87}},\ \bibinfo
  {pages} {023514} (\bibinfo {year} {2013})},\ \Eprint
  {https://arxiv.org/abs/1211.0083} {arXiv:1211.0083 [astro-ph.CO]}
  \BibitemShut {NoStop}%
\bibitem [{\citenamefont {Motohashi}\ and\ \citenamefont
  {Hu}(2017)}]{Motohashi:2017kbs}%
  \BibitemOpen
  \bibfield  {author} {\bibinfo {author} {\bibfnamefont {H.}~\bibnamefont
  {Motohashi}}\ and\ \bibinfo {author} {\bibfnamefont {W.}~\bibnamefont {Hu}},\
  }\bibfield  {title} {\bibinfo {title} {{Primordial Black Holes and Slow-Roll
  Violation}},\ }\href {https://doi.org/10.1103/PhysRevD.96.063503} {\bibfield
  {journal} {\bibinfo  {journal} {Phys. Rev. D}\ }\textbf {\bibinfo {volume}
  {96}},\ \bibinfo {pages} {063503} (\bibinfo {year} {2017})},\ \Eprint
  {https://arxiv.org/abs/1706.06784} {arXiv:1706.06784 [astro-ph.CO]}
  \BibitemShut {NoStop}%
\bibitem [{\citenamefont {Lyth}\ and\ \citenamefont
  {Wands}(2002)}]{Lyth:2001nq}%
  \BibitemOpen
  \bibfield  {author} {\bibinfo {author} {\bibfnamefont {D.~H.}\ \bibnamefont
  {Lyth}}\ and\ \bibinfo {author} {\bibfnamefont {D.}~\bibnamefont {Wands}},\
  }\bibfield  {title} {\bibinfo {title} {{Generating the curvature perturbation
  without an inflaton}},\ }\href
  {https://doi.org/10.1016/S0370-2693(01)01366-1} {\bibfield  {journal}
  {\bibinfo  {journal} {Phys. Lett. B}\ }\textbf {\bibinfo {volume} {524}},\
  \bibinfo {pages} {5} (\bibinfo {year} {2002})},\ \Eprint
  {https://arxiv.org/abs/hep-ph/0110002} {arXiv:hep-ph/0110002} \BibitemShut
  {NoStop}%
\bibitem [{\citenamefont {Barnaby}\ and\ \citenamefont
  {Peloso}(2011)}]{Barnaby:2010vf}%
  \BibitemOpen
  \bibfield  {author} {\bibinfo {author} {\bibfnamefont {N.}~\bibnamefont
  {Barnaby}}\ and\ \bibinfo {author} {\bibfnamefont {M.}~\bibnamefont
  {Peloso}},\ }\bibfield  {title} {\bibinfo {title} {{Large Nongaussianity in
  Axion Inflation}},\ }\href {https://doi.org/10.1103/PhysRevLett.106.181301}
  {\bibfield  {journal} {\bibinfo  {journal} {Phys. Rev. Lett.}\ }\textbf
  {\bibinfo {volume} {106}},\ \bibinfo {pages} {181301} (\bibinfo {year}
  {2011})},\ \Eprint {https://arxiv.org/abs/1011.1500} {arXiv:1011.1500
  [hep-ph]} \BibitemShut {NoStop}%
\bibitem [{\citenamefont {Sorbo}(2011)}]{Sorbo:2011rz}%
  \BibitemOpen
  \bibfield  {author} {\bibinfo {author} {\bibfnamefont {L.}~\bibnamefont
  {Sorbo}},\ }\bibfield  {title} {\bibinfo {title} {{Parity violation in the
  Cosmic Microwave Background from a pseudoscalar inflaton}},\ }\href
  {https://doi.org/10.1088/1475-7516/2011/06/003} {\bibfield  {journal}
  {\bibinfo  {journal} {JCAP}\ }\textbf {\bibinfo {volume} {06}},\ \bibinfo
  {pages} {003}},\ \Eprint {https://arxiv.org/abs/1101.1525} {arXiv:1101.1525
  [astro-ph.CO]} \BibitemShut {NoStop}%
\bibitem [{\citenamefont {Pi}\ \emph {et~al.}(2018)\citenamefont {Pi},
  \citenamefont {Zhang}, \citenamefont {Huang},\ and\ \citenamefont
  {Sasaki}}]{Pi:2017gih}%
  \BibitemOpen
  \bibfield  {author} {\bibinfo {author} {\bibfnamefont {S.}~\bibnamefont
  {Pi}}, \bibinfo {author} {\bibfnamefont {Y.-l.}\ \bibnamefont {Zhang}},
  \bibinfo {author} {\bibfnamefont {Q.-G.}\ \bibnamefont {Huang}},\ and\
  \bibinfo {author} {\bibfnamefont {M.}~\bibnamefont {Sasaki}},\ }\bibfield
  {title} {\bibinfo {title} {{Scalaron from $R^2$-gravity as a heavy field}},\
  }\href {https://doi.org/10.1088/1475-7516/2018/05/042} {\bibfield  {journal}
  {\bibinfo  {journal} {JCAP}\ }\textbf {\bibinfo {volume} {05}},\ \bibinfo
  {pages} {042}},\ \Eprint {https://arxiv.org/abs/1712.09896} {arXiv:1712.09896
  [astro-ph.CO]} \BibitemShut {NoStop}%
\bibitem [{\citenamefont {Wang}\ \emph
  {et~al.}(2025{\natexlab{a}})\citenamefont {Wang}, \citenamefont {Sasaki},\
  and\ \citenamefont {Zhang}}]{Wang:2025lti}%
  \BibitemOpen
  \bibfield  {author} {\bibinfo {author} {\bibfnamefont {X.}~\bibnamefont
  {Wang}}, \bibinfo {author} {\bibfnamefont {M.}~\bibnamefont {Sasaki}},\ and\
  \bibinfo {author} {\bibfnamefont {Y.-l.}\ \bibnamefont {Zhang}},\ }\bibfield
  {title} {\bibinfo {title} {{The Dual Primordial Black Hole Formation
  Scenario}},\ }\href@noop {} {\  (\bibinfo {year} {2025}{\natexlab{a}})},\
  \Eprint {https://arxiv.org/abs/2505.09337} {arXiv:2505.09337 [astro-ph.CO]}
  \BibitemShut {NoStop}%
\bibitem [{\citenamefont {Byrnes}\ \emph {et~al.}(2019)\citenamefont {Byrnes},
  \citenamefont {Cole},\ and\ \citenamefont {Patil}}]{Byrnes:2018txb}%
  \BibitemOpen
  \bibfield  {author} {\bibinfo {author} {\bibfnamefont {C.~T.}\ \bibnamefont
  {Byrnes}}, \bibinfo {author} {\bibfnamefont {P.~S.}\ \bibnamefont {Cole}},\
  and\ \bibinfo {author} {\bibfnamefont {S.~P.}\ \bibnamefont {Patil}},\
  }\bibfield  {title} {\bibinfo {title} {{Steepest growth of the power spectrum
  and primordial black holes}},\ }\href
  {https://doi.org/10.1088/1475-7516/2019/06/028} {\bibfield  {journal}
  {\bibinfo  {journal} {JCAP}\ }\textbf {\bibinfo {volume} {06}},\ \bibinfo
  {pages} {028}},\ \Eprint {https://arxiv.org/abs/1811.11158} {arXiv:1811.11158
  [astro-ph.CO]} \BibitemShut {NoStop}%
\bibitem [{\citenamefont {Frosina}\ and\ \citenamefont
  {Urbano}(2023)}]{Frosina:2023nxu}%
  \BibitemOpen
  \bibfield  {author} {\bibinfo {author} {\bibfnamefont {L.}~\bibnamefont
  {Frosina}}\ and\ \bibinfo {author} {\bibfnamefont {A.}~\bibnamefont
  {Urbano}},\ }\bibfield  {title} {\bibinfo {title} {{Inflationary
  interpretation of the nHz gravitational-wave background}},\ }\href
  {https://doi.org/10.1103/PhysRevD.108.103544} {\bibfield  {journal} {\bibinfo
   {journal} {Phys. Rev. D}\ }\textbf {\bibinfo {volume} {108}},\ \bibinfo
  {pages} {103544} (\bibinfo {year} {2023})},\ \Eprint
  {https://arxiv.org/abs/2308.06915} {arXiv:2308.06915 [astro-ph.CO]}
  \BibitemShut {NoStop}%
\bibitem [{\citenamefont {Karam}\ \emph {et~al.}(2023)\citenamefont {Karam},
  \citenamefont {Koivunen}, \citenamefont {Tomberg}, \citenamefont {Vaskonen},\
  and\ \citenamefont {Veerm{\"a}e}}]{Karam:2022nym}%
  \BibitemOpen
  \bibfield  {author} {\bibinfo {author} {\bibfnamefont {A.}~\bibnamefont
  {Karam}}, \bibinfo {author} {\bibfnamefont {N.}~\bibnamefont {Koivunen}},
  \bibinfo {author} {\bibfnamefont {E.}~\bibnamefont {Tomberg}}, \bibinfo
  {author} {\bibfnamefont {V.}~\bibnamefont {Vaskonen}},\ and\ \bibinfo
  {author} {\bibfnamefont {H.}~\bibnamefont {Veerm{\"a}e}},\ }\bibfield
  {title} {\bibinfo {title} {{Anatomy of single-field inflationary models for
  primordial black holes}},\ }\href
  {https://doi.org/10.1088/1475-7516/2023/03/013} {\bibfield  {journal}
  {\bibinfo  {journal} {JCAP}\ }\textbf {\bibinfo {volume} {03}},\ \bibinfo
  {pages} {013}},\ \Eprint {https://arxiv.org/abs/2205.13540} {arXiv:2205.13540
  [astro-ph.CO]} \BibitemShut {NoStop}%
\bibitem [{\citenamefont {Kohri}\ \emph {et~al.}(2013)\citenamefont {Kohri},
  \citenamefont {Lin},\ and\ \citenamefont {Matsuda}}]{Kohri:2012yw}%
  \BibitemOpen
  \bibfield  {author} {\bibinfo {author} {\bibfnamefont {K.}~\bibnamefont
  {Kohri}}, \bibinfo {author} {\bibfnamefont {C.-M.}\ \bibnamefont {Lin}},\
  and\ \bibinfo {author} {\bibfnamefont {T.}~\bibnamefont {Matsuda}},\
  }\bibfield  {title} {\bibinfo {title} {{Primordial black holes from the
  inflating curvaton}},\ }\href {https://doi.org/10.1103/PhysRevD.87.103527}
  {\bibfield  {journal} {\bibinfo  {journal} {Phys. Rev. D}\ }\textbf {\bibinfo
  {volume} {87}},\ \bibinfo {pages} {103527} (\bibinfo {year} {2013})},\
  \Eprint {https://arxiv.org/abs/1211.2371} {arXiv:1211.2371 [hep-ph]}
  \BibitemShut {NoStop}%
\bibitem [{\citenamefont {Kawasaki}\ \emph {et~al.}(2013)\citenamefont
  {Kawasaki}, \citenamefont {Kitajima},\ and\ \citenamefont
  {Yanagida}}]{Kawasaki:2012wr}%
  \BibitemOpen
  \bibfield  {author} {\bibinfo {author} {\bibfnamefont {M.}~\bibnamefont
  {Kawasaki}}, \bibinfo {author} {\bibfnamefont {N.}~\bibnamefont {Kitajima}},\
  and\ \bibinfo {author} {\bibfnamefont {T.~T.}\ \bibnamefont {Yanagida}},\
  }\bibfield  {title} {\bibinfo {title} {{Primordial black hole formation from
  an axionlike curvaton model}},\ }\href
  {https://doi.org/10.1103/PhysRevD.87.063519} {\bibfield  {journal} {\bibinfo
  {journal} {Phys. Rev. D}\ }\textbf {\bibinfo {volume} {87}},\ \bibinfo
  {pages} {063519} (\bibinfo {year} {2013})},\ \Eprint
  {https://arxiv.org/abs/1207.2550} {arXiv:1207.2550 [hep-ph]} \BibitemShut
  {NoStop}%
\bibitem [{\citenamefont {Garcia-Bellido}\ \emph {et~al.}(1996)\citenamefont
  {Garcia-Bellido}, \citenamefont {Linde},\ and\ \citenamefont
  {Wands}}]{Garcia-Bellido:1996mdl}%
  \BibitemOpen
  \bibfield  {author} {\bibinfo {author} {\bibfnamefont {J.}~\bibnamefont
  {Garcia-Bellido}}, \bibinfo {author} {\bibfnamefont {A.~D.}\ \bibnamefont
  {Linde}},\ and\ \bibinfo {author} {\bibfnamefont {D.}~\bibnamefont {Wands}},\
  }\bibfield  {title} {\bibinfo {title} {{Density perturbations and black hole
  formation in hybrid inflation}},\ }\href
  {https://doi.org/10.1103/PhysRevD.54.6040} {\bibfield  {journal} {\bibinfo
  {journal} {Phys. Rev. D}\ }\textbf {\bibinfo {volume} {54}},\ \bibinfo
  {pages} {6040} (\bibinfo {year} {1996})},\ \Eprint
  {https://arxiv.org/abs/astro-ph/9605094} {arXiv:astro-ph/9605094}
  \BibitemShut {NoStop}%
\bibitem [{\citenamefont {Franciolini}\ and\ \citenamefont
  {Urbano}(2022)}]{Franciolini:2022pav}%
  \BibitemOpen
  \bibfield  {author} {\bibinfo {author} {\bibfnamefont {G.}~\bibnamefont
  {Franciolini}}\ and\ \bibinfo {author} {\bibfnamefont {A.}~\bibnamefont
  {Urbano}},\ }\bibfield  {title} {\bibinfo {title} {{Primordial black hole
  dark matter from inflation: The reverse engineering approach}},\ }\href
  {https://doi.org/10.1103/PhysRevD.106.123519} {\bibfield  {journal} {\bibinfo
   {journal} {Phys. Rev. D}\ }\textbf {\bibinfo {volume} {106}},\ \bibinfo
  {pages} {123519} (\bibinfo {year} {2022})},\ \Eprint
  {https://arxiv.org/abs/2207.10056} {arXiv:2207.10056 [astro-ph.CO]}
  \BibitemShut {NoStop}%
\bibitem [{\citenamefont {Ferrante}\ \emph
  {et~al.}(2023{\natexlab{a}})\citenamefont {Ferrante}, \citenamefont
  {Franciolini}, \citenamefont {Iovino},\ and\ \citenamefont
  {Urbano}}]{Ferrante:2023bgz}%
  \BibitemOpen
  \bibfield  {author} {\bibinfo {author} {\bibfnamefont {G.}~\bibnamefont
  {Ferrante}}, \bibinfo {author} {\bibfnamefont {G.}~\bibnamefont
  {Franciolini}}, \bibinfo {author} {\bibfnamefont {A.}~\bibnamefont {Iovino},
  \bibfnamefont {Junior.}},\ and\ \bibinfo {author} {\bibfnamefont
  {A.}~\bibnamefont {Urbano}},\ }\bibfield  {title} {\bibinfo {title}
  {{Primordial black holes in the curvaton model: possible connections to
  pulsar timing arrays and dark matter}},\ }\href
  {https://doi.org/10.1088/1475-7516/2023/06/057} {\bibfield  {journal}
  {\bibinfo  {journal} {JCAP}\ }\textbf {\bibinfo {volume} {06}},\ \bibinfo
  {pages} {057}},\ \Eprint {https://arxiv.org/abs/2305.13382} {arXiv:2305.13382
  [astro-ph.CO]} \BibitemShut {NoStop}%
\bibitem [{\citenamefont {Vaskonen}\ and\ \citenamefont
  {Veerm{\"a}e}(2021)}]{Vaskonen:2020lbd}%
  \BibitemOpen
  \bibfield  {author} {\bibinfo {author} {\bibfnamefont {V.}~\bibnamefont
  {Vaskonen}}\ and\ \bibinfo {author} {\bibfnamefont {H.}~\bibnamefont
  {Veerm{\"a}e}},\ }\bibfield  {title} {\bibinfo {title} {{Did NANOGrav see a
  signal from primordial black hole formation?}},\ }\href
  {https://doi.org/10.1103/PhysRevLett.126.051303} {\bibfield  {journal}
  {\bibinfo  {journal} {Phys. Rev. Lett.}\ }\textbf {\bibinfo {volume} {126}},\
  \bibinfo {pages} {051303} (\bibinfo {year} {2021})},\ \Eprint
  {https://arxiv.org/abs/2009.07832} {arXiv:2009.07832 [astro-ph.CO]}
  \BibitemShut {NoStop}%
\bibitem [{\citenamefont {Sugiyama}\ \emph {et~al.}(2021)\citenamefont
  {Sugiyama}, \citenamefont {Takhistov}, \citenamefont {Vitagliano},
  \citenamefont {Kusenko}, \citenamefont {Sasaki},\ and\ \citenamefont
  {Takada}}]{Sugiyama:2020roc}%
  \BibitemOpen
  \bibfield  {author} {\bibinfo {author} {\bibfnamefont {S.}~\bibnamefont
  {Sugiyama}}, \bibinfo {author} {\bibfnamefont {V.}~\bibnamefont {Takhistov}},
  \bibinfo {author} {\bibfnamefont {E.}~\bibnamefont {Vitagliano}}, \bibinfo
  {author} {\bibfnamefont {A.}~\bibnamefont {Kusenko}}, \bibinfo {author}
  {\bibfnamefont {M.}~\bibnamefont {Sasaki}},\ and\ \bibinfo {author}
  {\bibfnamefont {M.}~\bibnamefont {Takada}},\ }\bibfield  {title} {\bibinfo
  {title} {{Testing Stochastic Gravitational Wave Signals from Primordial Black
  Holes with Optical Telescopes}},\ }\href
  {https://doi.org/10.1016/j.physletb.2021.136097} {\bibfield  {journal}
  {\bibinfo  {journal} {Phys. Lett. B}\ }\textbf {\bibinfo {volume} {814}},\
  \bibinfo {pages} {136097} (\bibinfo {year} {2021})},\ \Eprint
  {https://arxiv.org/abs/2010.02189} {arXiv:2010.02189 [astro-ph.CO]}
  \BibitemShut {NoStop}%
\bibitem [{\citenamefont {De~Luca}\ \emph
  {et~al.}(2021{\natexlab{a}})\citenamefont {De~Luca}, \citenamefont
  {Franciolini},\ and\ \citenamefont {Riotto}}]{DeLuca:2020agl}%
  \BibitemOpen
  \bibfield  {author} {\bibinfo {author} {\bibfnamefont {V.}~\bibnamefont
  {De~Luca}}, \bibinfo {author} {\bibfnamefont {G.}~\bibnamefont
  {Franciolini}},\ and\ \bibinfo {author} {\bibfnamefont {A.}~\bibnamefont
  {Riotto}},\ }\bibfield  {title} {\bibinfo {title} {{NANOGrav Data Hints at
  Primordial Black Holes as Dark Matter}},\ }\href
  {https://doi.org/10.1103/PhysRevLett.126.041303} {\bibfield  {journal}
  {\bibinfo  {journal} {Phys. Rev. Lett.}\ }\textbf {\bibinfo {volume} {126}},\
  \bibinfo {pages} {041303} (\bibinfo {year} {2021}{\natexlab{a}})},\ \Eprint
  {https://arxiv.org/abs/2009.08268} {arXiv:2009.08268 [astro-ph.CO]}
  \BibitemShut {NoStop}%
\bibitem [{\citenamefont {De~Luca}\ \emph
  {et~al.}(2020{\natexlab{a}})\citenamefont {De~Luca}, \citenamefont
  {Franciolini},\ and\ \citenamefont {Riotto}}]{DeLuca:2020ioi}%
  \BibitemOpen
  \bibfield  {author} {\bibinfo {author} {\bibfnamefont {V.}~\bibnamefont
  {De~Luca}}, \bibinfo {author} {\bibfnamefont {G.}~\bibnamefont
  {Franciolini}},\ and\ \bibinfo {author} {\bibfnamefont {A.}~\bibnamefont
  {Riotto}},\ }\bibfield  {title} {\bibinfo {title} {{On the primordial black
  hole mass function for broad spectra}},\ }\href
  {https://doi.org/10.1016/j.physletb.2020.135550} {\bibfield  {journal}
  {\bibinfo  {journal} {Phys. Lett. B}\ }\textbf {\bibinfo {volume} {807}},\
  \bibinfo {pages} {135550} (\bibinfo {year} {2020}{\natexlab{a}})},\ \Eprint
  {https://arxiv.org/abs/2001.04371} {arXiv:2001.04371 [astro-ph.CO]}
  \BibitemShut {NoStop}%
\bibitem [{\citenamefont {Crescimbeni}\ \emph {et~al.}(2025)\citenamefont
  {Crescimbeni}, \citenamefont {Desjacques}, \citenamefont {Franciolini},
  \citenamefont {Ianniccari}, \citenamefont {Iovino}, \citenamefont {Perna},
  \citenamefont {Perrone}, \citenamefont {Riotto},\ and\ \citenamefont
  {Veerm{\"a}e}}]{Crescimbeni:2025ywm}%
  \BibitemOpen
  \bibfield  {author} {\bibinfo {author} {\bibfnamefont {F.}~\bibnamefont
  {Crescimbeni}}, \bibinfo {author} {\bibfnamefont {V.}~\bibnamefont
  {Desjacques}}, \bibinfo {author} {\bibfnamefont {G.}~\bibnamefont
  {Franciolini}}, \bibinfo {author} {\bibfnamefont {A.}~\bibnamefont
  {Ianniccari}}, \bibinfo {author} {\bibfnamefont {A.~J.}\ \bibnamefont
  {Iovino}}, \bibinfo {author} {\bibfnamefont {G.}~\bibnamefont {Perna}},
  \bibinfo {author} {\bibfnamefont {D.}~\bibnamefont {Perrone}}, \bibinfo
  {author} {\bibfnamefont {A.}~\bibnamefont {Riotto}},\ and\ \bibinfo {author}
  {\bibfnamefont {H.}~\bibnamefont {Veerm{\"a}e}},\ }\bibfield  {title}
  {\bibinfo {title} {{The irrelevance of primordial black hole clustering in
  the LVK mass range}},\ }\href {https://doi.org/10.1088/1475-7516/2025/05/001}
  {\bibfield  {journal} {\bibinfo  {journal} {JCAP}\ }\textbf {\bibinfo
  {volume} {05}},\ \bibinfo {pages} {001}},\ \Eprint
  {https://arxiv.org/abs/2502.01617} {arXiv:2502.01617 [astro-ph.CO]}
  \BibitemShut {NoStop}%
\bibitem [{\citenamefont {Young}\ \emph {et~al.}(2019)\citenamefont {Young},
  \citenamefont {Musco},\ and\ \citenamefont {Byrnes}}]{Young:2019yug}%
  \BibitemOpen
  \bibfield  {author} {\bibinfo {author} {\bibfnamefont {S.}~\bibnamefont
  {Young}}, \bibinfo {author} {\bibfnamefont {I.}~\bibnamefont {Musco}},\ and\
  \bibinfo {author} {\bibfnamefont {C.~T.}\ \bibnamefont {Byrnes}},\ }\bibfield
   {title} {\bibinfo {title} {{Primordial black hole formation and abundance:
  contribution from the non-linear relation between the density and curvature
  perturbation}},\ }\href {https://doi.org/10.1088/1475-7516/2019/11/012}
  {\bibfield  {journal} {\bibinfo  {journal} {JCAP}\ }\textbf {\bibinfo
  {volume} {11}},\ \bibinfo {pages} {012}},\ \Eprint
  {https://arxiv.org/abs/1904.00984} {arXiv:1904.00984 [astro-ph.CO]}
  \BibitemShut {NoStop}%
\bibitem [{\citenamefont {De~Luca}\ \emph
  {et~al.}(2019{\natexlab{a}})\citenamefont {De~Luca}, \citenamefont
  {Franciolini}, \citenamefont {Kehagias}, \citenamefont {Peloso},
  \citenamefont {Riotto},\ and\ \citenamefont {{\"U}nal}}]{DeLuca:2019qsy}%
  \BibitemOpen
  \bibfield  {author} {\bibinfo {author} {\bibfnamefont {V.}~\bibnamefont
  {De~Luca}}, \bibinfo {author} {\bibfnamefont {G.}~\bibnamefont
  {Franciolini}}, \bibinfo {author} {\bibfnamefont {A.}~\bibnamefont
  {Kehagias}}, \bibinfo {author} {\bibfnamefont {M.}~\bibnamefont {Peloso}},
  \bibinfo {author} {\bibfnamefont {A.}~\bibnamefont {Riotto}},\ and\ \bibinfo
  {author} {\bibfnamefont {C.}~\bibnamefont {{\"U}nal}},\ }\bibfield  {title}
  {\bibinfo {title} {{The Ineludible non-Gaussianity of the Primordial Black
  Hole Abundance}},\ }\href {https://doi.org/10.1088/1475-7516/2019/07/048}
  {\bibfield  {journal} {\bibinfo  {journal} {JCAP}\ }\textbf {\bibinfo
  {volume} {07}},\ \bibinfo {pages} {048}},\ \Eprint
  {https://arxiv.org/abs/1904.00970} {arXiv:1904.00970 [astro-ph.CO]}
  \BibitemShut {NoStop}%
\bibitem [{\citenamefont {Ferrante}\ \emph
  {et~al.}(2023{\natexlab{b}})\citenamefont {Ferrante}, \citenamefont
  {Franciolini}, \citenamefont {Iovino},\ and\ \citenamefont
  {Urbano}}]{Ferrante:2022mui}%
  \BibitemOpen
  \bibfield  {author} {\bibinfo {author} {\bibfnamefont {G.}~\bibnamefont
  {Ferrante}}, \bibinfo {author} {\bibfnamefont {G.}~\bibnamefont
  {Franciolini}}, \bibinfo {author} {\bibfnamefont {A.}~\bibnamefont {Iovino},
  \bibfnamefont {Junior.}},\ and\ \bibinfo {author} {\bibfnamefont
  {A.}~\bibnamefont {Urbano}},\ }\bibfield  {title} {\bibinfo {title}
  {{Primordial non-Gaussianity up to all orders: Theoretical aspects and
  implications for primordial black hole models}},\ }\href
  {https://doi.org/10.1103/PhysRevD.107.043520} {\bibfield  {journal} {\bibinfo
   {journal} {Phys. Rev. D}\ }\textbf {\bibinfo {volume} {107}},\ \bibinfo
  {pages} {043520} (\bibinfo {year} {2023}{\natexlab{b}})},\ \Eprint
  {https://arxiv.org/abs/2211.01728} {arXiv:2211.01728 [astro-ph.CO]}
  \BibitemShut {NoStop}%
\bibitem [{\citenamefont {Gow}\ \emph {et~al.}(2023)\citenamefont {Gow},
  \citenamefont {Assadullahi}, \citenamefont {Jackson}, \citenamefont {Koyama},
  \citenamefont {Vennin},\ and\ \citenamefont {Wands}}]{Gow:2022jfb}%
  \BibitemOpen
  \bibfield  {author} {\bibinfo {author} {\bibfnamefont {A.~D.}\ \bibnamefont
  {Gow}}, \bibinfo {author} {\bibfnamefont {H.}~\bibnamefont {Assadullahi}},
  \bibinfo {author} {\bibfnamefont {J.~H.~P.}\ \bibnamefont {Jackson}},
  \bibinfo {author} {\bibfnamefont {K.}~\bibnamefont {Koyama}}, \bibinfo
  {author} {\bibfnamefont {V.}~\bibnamefont {Vennin}},\ and\ \bibinfo {author}
  {\bibfnamefont {D.}~\bibnamefont {Wands}},\ }\bibfield  {title} {\bibinfo
  {title} {{Non-perturbative non-Gaussianity and primordial black holes}},\
  }\href {https://doi.org/10.1209/0295-5075/acd417} {\bibfield  {journal}
  {\bibinfo  {journal} {EPL}\ }\textbf {\bibinfo {volume} {142}},\ \bibinfo
  {pages} {49001} (\bibinfo {year} {2023})},\ \Eprint
  {https://arxiv.org/abs/2211.08348} {arXiv:2211.08348 [astro-ph.CO]}
  \BibitemShut {NoStop}%
\bibitem [{\citenamefont {Yoo}\ \emph {et~al.}(2018)\citenamefont {Yoo},
  \citenamefont {Harada}, \citenamefont {Garriga},\ and\ \citenamefont
  {Kohri}}]{Yoo:2018kvb}%
  \BibitemOpen
  \bibfield  {author} {\bibinfo {author} {\bibfnamefont {C.-M.}\ \bibnamefont
  {Yoo}}, \bibinfo {author} {\bibfnamefont {T.}~\bibnamefont {Harada}},
  \bibinfo {author} {\bibfnamefont {J.}~\bibnamefont {Garriga}},\ and\ \bibinfo
  {author} {\bibfnamefont {K.}~\bibnamefont {Kohri}},\ }\bibfield  {title}
  {\bibinfo {title} {{Primordial black hole abundance from random Gaussian
  curvature perturbations and a local density threshold}},\ }\href
  {https://doi.org/10.1093/ptep/pty120} {\bibfield  {journal} {\bibinfo
  {journal} {PTEP}\ }\textbf {\bibinfo {volume} {2018}},\ \bibinfo {pages}
  {123E01} (\bibinfo {year} {2018})},\ \bibinfo {note} {[Erratum: PTEP 2024,
  049202 (2024)]},\ \Eprint {https://arxiv.org/abs/1805.03946}
  {arXiv:1805.03946 [astro-ph.CO]} \BibitemShut {NoStop}%
\bibitem [{\citenamefont {Yoo}\ \emph {et~al.}(2019)\citenamefont {Yoo},
  \citenamefont {Gong},\ and\ \citenamefont {Yokoyama}}]{Yoo:2019pma}%
  \BibitemOpen
  \bibfield  {author} {\bibinfo {author} {\bibfnamefont {C.-M.}\ \bibnamefont
  {Yoo}}, \bibinfo {author} {\bibfnamefont {J.-O.}\ \bibnamefont {Gong}},\ and\
  \bibinfo {author} {\bibfnamefont {S.}~\bibnamefont {Yokoyama}},\ }\bibfield
  {title} {\bibinfo {title} {{Abundance of primordial black holes with local
  non-Gaussianity in peak theory}},\ }\href
  {https://doi.org/10.1088/1475-7516/2019/09/033} {\bibfield  {journal}
  {\bibinfo  {journal} {JCAP}\ }\textbf {\bibinfo {volume} {09}},\ \bibinfo
  {pages} {033}},\ \Eprint {https://arxiv.org/abs/1906.06790} {arXiv:1906.06790
  [astro-ph.CO]} \BibitemShut {NoStop}%
\bibitem [{\citenamefont {Franciolini}\ \emph
  {et~al.}(2022{\natexlab{a}})\citenamefont {Franciolini}, \citenamefont
  {Musco}, \citenamefont {Pani},\ and\ \citenamefont
  {Urbano}}]{Franciolini:2022tfm}%
  \BibitemOpen
  \bibfield  {author} {\bibinfo {author} {\bibfnamefont {G.}~\bibnamefont
  {Franciolini}}, \bibinfo {author} {\bibfnamefont {I.}~\bibnamefont {Musco}},
  \bibinfo {author} {\bibfnamefont {P.}~\bibnamefont {Pani}},\ and\ \bibinfo
  {author} {\bibfnamefont {A.}~\bibnamefont {Urbano}},\ }\bibfield  {title}
  {\bibinfo {title} {{From inflation to black hole mergers and back again:
  Gravitational-wave data-driven constraints on inflationary scenarios with a
  first-principle model of primordial black holes across the QCD epoch}},\
  }\href {https://doi.org/10.1103/PhysRevD.106.123526} {\bibfield  {journal}
  {\bibinfo  {journal} {Phys. Rev. D}\ }\textbf {\bibinfo {volume} {106}},\
  \bibinfo {pages} {123526} (\bibinfo {year} {2022}{\natexlab{a}})},\ \Eprint
  {https://arxiv.org/abs/2209.05959} {arXiv:2209.05959 [astro-ph.CO]}
  \BibitemShut {NoStop}%
\bibitem [{\citenamefont {Green}\ \emph {et~al.}(2004)\citenamefont {Green},
  \citenamefont {Liddle}, \citenamefont {Malik},\ and\ \citenamefont
  {Sasaki}}]{Green:2004wb}%
  \BibitemOpen
  \bibfield  {author} {\bibinfo {author} {\bibfnamefont {A.~M.}\ \bibnamefont
  {Green}}, \bibinfo {author} {\bibfnamefont {A.~R.}\ \bibnamefont {Liddle}},
  \bibinfo {author} {\bibfnamefont {K.~A.}\ \bibnamefont {Malik}},\ and\
  \bibinfo {author} {\bibfnamefont {M.}~\bibnamefont {Sasaki}},\ }\bibfield
  {title} {\bibinfo {title} {{A New calculation of the mass fraction of
  primordial black holes}},\ }\href
  {https://doi.org/10.1103/PhysRevD.70.041502} {\bibfield  {journal} {\bibinfo
  {journal} {Phys. Rev. D}\ }\textbf {\bibinfo {volume} {70}},\ \bibinfo
  {pages} {041502} (\bibinfo {year} {2004})},\ \Eprint
  {https://arxiv.org/abs/astro-ph/0403181} {arXiv:astro-ph/0403181}
  \BibitemShut {NoStop}%
\bibitem [{\citenamefont {Young}\ \emph {et~al.}(2014)\citenamefont {Young},
  \citenamefont {Byrnes},\ and\ \citenamefont {Sasaki}}]{Young:2014ana}%
  \BibitemOpen
  \bibfield  {author} {\bibinfo {author} {\bibfnamefont {S.}~\bibnamefont
  {Young}}, \bibinfo {author} {\bibfnamefont {C.~T.}\ \bibnamefont {Byrnes}},\
  and\ \bibinfo {author} {\bibfnamefont {M.}~\bibnamefont {Sasaki}},\
  }\bibfield  {title} {\bibinfo {title} {{Calculating the mass fraction of
  primordial black holes}},\ }\href
  {https://doi.org/10.1088/1475-7516/2014/07/045} {\bibfield  {journal}
  {\bibinfo  {journal} {JCAP}\ }\textbf {\bibinfo {volume} {07}},\ \bibinfo
  {pages} {045}},\ \Eprint {https://arxiv.org/abs/1405.7023} {arXiv:1405.7023
  [gr-qc]} \BibitemShut {NoStop}%
\bibitem [{\citenamefont {Allahverdi}\ \emph {et~al.}(2021)\citenamefont
  {Allahverdi} \emph {et~al.}}]{Allahverdi:2020bys}%
  \BibitemOpen
  \bibfield  {author} {\bibinfo {author} {\bibfnamefont {R.}~\bibnamefont
  {Allahverdi}} \emph {et~al.},\ }\bibfield  {title} {\bibinfo {title} {{The
  First Three Seconds: a Review of Possible Expansion Histories of the Early
  Universe}},\ }\href {https://doi.org/10.21105/astro.2006.16182} {\bibfield
  {journal} {\bibinfo  {journal} {Open J. Astrophys.}\ }\textbf {\bibinfo
  {volume} {4}},\ \bibinfo {pages} {astro.2006.16182} (\bibinfo {year}
  {2021})},\ \Eprint {https://arxiv.org/abs/2006.16182} {arXiv:2006.16182
  [astro-ph.CO]} \BibitemShut {NoStop}%
\bibitem [{\citenamefont {Khlopov}\ and\ \citenamefont
  {Polnarev}(1980)}]{Khlopov:1980mg}%
  \BibitemOpen
  \bibfield  {author} {\bibinfo {author} {\bibfnamefont {M.~Y.}\ \bibnamefont
  {Khlopov}}\ and\ \bibinfo {author} {\bibfnamefont {A.~G.}\ \bibnamefont
  {Polnarev}},\ }\bibfield  {title} {\bibinfo {title} {{PRIMORDIAL BLACK HOLES
  AS A COSMOLOGICAL TEST OF GRAND UNIFICATION}},\ }\href
  {https://doi.org/10.1016/0370-2693(80)90624-3} {\bibfield  {journal}
  {\bibinfo  {journal} {Phys. Lett. B}\ }\textbf {\bibinfo {volume} {97}},\
  \bibinfo {pages} {383} (\bibinfo {year} {1980})}\BibitemShut {NoStop}%
\bibitem [{\citenamefont {Polnarev}\ and\ \citenamefont
  {Khlopov}(1985)}]{Polnarev:1985btg}%
  \BibitemOpen
  \bibfield  {author} {\bibinfo {author} {\bibfnamefont {A.~G.}\ \bibnamefont
  {Polnarev}}\ and\ \bibinfo {author} {\bibfnamefont {M.~Y.}\ \bibnamefont
  {Khlopov}},\ }\bibfield  {title} {\bibinfo {title} {{COSMOLOGY, PRIMORDIAL
  BLACK HOLES, AND SUPERMASSIVE PARTICLES}},\ }\href
  {https://doi.org/10.1070/PU1985v028n03ABEH003858} {\bibfield  {journal}
  {\bibinfo  {journal} {Sov. Phys. Usp.}\ }\textbf {\bibinfo {volume} {28}},\
  \bibinfo {pages} {213} (\bibinfo {year} {1985})}\BibitemShut {NoStop}%
\bibitem [{\citenamefont {Harada}\ \emph {et~al.}(2016)\citenamefont {Harada},
  \citenamefont {Yoo}, \citenamefont {Kohri}, \citenamefont {Nakao},\ and\
  \citenamefont {Jhingan}}]{Harada:2016mhb}%
  \BibitemOpen
  \bibfield  {author} {\bibinfo {author} {\bibfnamefont {T.}~\bibnamefont
  {Harada}}, \bibinfo {author} {\bibfnamefont {C.-M.}\ \bibnamefont {Yoo}},
  \bibinfo {author} {\bibfnamefont {K.}~\bibnamefont {Kohri}}, \bibinfo
  {author} {\bibfnamefont {K.-i.}\ \bibnamefont {Nakao}},\ and\ \bibinfo
  {author} {\bibfnamefont {S.}~\bibnamefont {Jhingan}},\ }\bibfield  {title}
  {\bibinfo {title} {{Primordial black hole formation in the matter-dominated
  phase of the Universe}},\ }\href {https://doi.org/10.3847/1538-4357/833/1/61}
  {\bibfield  {journal} {\bibinfo  {journal} {Astrophys. J.}\ }\textbf
  {\bibinfo {volume} {833}},\ \bibinfo {pages} {61} (\bibinfo {year} {2016})},\
  \Eprint {https://arxiv.org/abs/1609.01588} {arXiv:1609.01588 [astro-ph.CO]}
  \BibitemShut {NoStop}%
\bibitem [{\citenamefont {Harada}\ \emph {et~al.}(2017)\citenamefont {Harada},
  \citenamefont {Yoo}, \citenamefont {Kohri},\ and\ \citenamefont
  {Nakao}}]{Harada:2017fjm}%
  \BibitemOpen
  \bibfield  {author} {\bibinfo {author} {\bibfnamefont {T.}~\bibnamefont
  {Harada}}, \bibinfo {author} {\bibfnamefont {C.-M.}\ \bibnamefont {Yoo}},
  \bibinfo {author} {\bibfnamefont {K.}~\bibnamefont {Kohri}},\ and\ \bibinfo
  {author} {\bibfnamefont {K.-I.}\ \bibnamefont {Nakao}},\ }\bibfield  {title}
  {\bibinfo {title} {{Spins of primordial black holes formed in the
  matter-dominated phase of the Universe}},\ }\href
  {https://doi.org/10.1103/PhysRevD.96.083517} {\bibfield  {journal} {\bibinfo
  {journal} {Phys. Rev. D}\ }\textbf {\bibinfo {volume} {96}},\ \bibinfo
  {pages} {083517} (\bibinfo {year} {2017})},\ \bibinfo {note} {[Erratum:
  Phys.Rev.D 99, 069904 (2019)]},\ \Eprint {https://arxiv.org/abs/1707.03595}
  {arXiv:1707.03595 [gr-qc]} \BibitemShut {NoStop}%
\bibitem [{\citenamefont {Kokubu}\ \emph {et~al.}(2018)\citenamefont {Kokubu},
  \citenamefont {Kyutoku}, \citenamefont {Kohri},\ and\ \citenamefont
  {Harada}}]{Kokubu:2018fxy}%
  \BibitemOpen
  \bibfield  {author} {\bibinfo {author} {\bibfnamefont {T.}~\bibnamefont
  {Kokubu}}, \bibinfo {author} {\bibfnamefont {K.}~\bibnamefont {Kyutoku}},
  \bibinfo {author} {\bibfnamefont {K.}~\bibnamefont {Kohri}},\ and\ \bibinfo
  {author} {\bibfnamefont {T.}~\bibnamefont {Harada}},\ }\bibfield  {title}
  {\bibinfo {title} {{Effect of Inhomogeneity on Primordial Black Hole
  Formation in the Matter Dominated Era}},\ }\href
  {https://doi.org/10.1103/PhysRevD.98.123024} {\bibfield  {journal} {\bibinfo
  {journal} {Phys. Rev. D}\ }\textbf {\bibinfo {volume} {98}},\ \bibinfo
  {pages} {123024} (\bibinfo {year} {2018})},\ \Eprint
  {https://arxiv.org/abs/1810.03490} {arXiv:1810.03490 [astro-ph.CO]}
  \BibitemShut {NoStop}%
\bibitem [{\citenamefont {Carr}\ \emph
  {et~al.}(2017{\natexlab{b}})\citenamefont {Carr}, \citenamefont {Tenkanen},\
  and\ \citenamefont {Vaskonen}}]{Carr:2017edp}%
  \BibitemOpen
  \bibfield  {author} {\bibinfo {author} {\bibfnamefont {B.}~\bibnamefont
  {Carr}}, \bibinfo {author} {\bibfnamefont {T.}~\bibnamefont {Tenkanen}},\
  and\ \bibinfo {author} {\bibfnamefont {V.}~\bibnamefont {Vaskonen}},\
  }\bibfield  {title} {\bibinfo {title} {{Primordial black holes from inflaton
  and spectator field perturbations in a matter-dominated era}},\ }\href
  {https://doi.org/10.1103/PhysRevD.96.063507} {\bibfield  {journal} {\bibinfo
  {journal} {Phys. Rev. D}\ }\textbf {\bibinfo {volume} {96}},\ \bibinfo
  {pages} {063507} (\bibinfo {year} {2017}{\natexlab{b}})},\ \Eprint
  {https://arxiv.org/abs/1706.03746} {arXiv:1706.03746 [astro-ph.CO]}
  \BibitemShut {NoStop}%
\bibitem [{\citenamefont {Green}\ and\ \citenamefont
  {Malik}(2001)}]{Green:2000he}%
  \BibitemOpen
  \bibfield  {author} {\bibinfo {author} {\bibfnamefont {A.~M.}\ \bibnamefont
  {Green}}\ and\ \bibinfo {author} {\bibfnamefont {K.~A.}\ \bibnamefont
  {Malik}},\ }\bibfield  {title} {\bibinfo {title} {{Primordial black hole
  production due to preheating}},\ }\href
  {https://doi.org/10.1103/PhysRevD.64.021301} {\bibfield  {journal} {\bibinfo
  {journal} {Phys. Rev. D}\ }\textbf {\bibinfo {volume} {64}},\ \bibinfo
  {pages} {021301} (\bibinfo {year} {2001})},\ \Eprint
  {https://arxiv.org/abs/hep-ph/0008113} {arXiv:hep-ph/0008113} \BibitemShut
  {NoStop}%
\bibitem [{\citenamefont {Lyth}\ \emph {et~al.}(2006)\citenamefont {Lyth},
  \citenamefont {Malik}, \citenamefont {Sasaki},\ and\ \citenamefont
  {Zaballa}}]{Lyth:2005ze}%
  \BibitemOpen
  \bibfield  {author} {\bibinfo {author} {\bibfnamefont {D.~H.}\ \bibnamefont
  {Lyth}}, \bibinfo {author} {\bibfnamefont {K.~A.}\ \bibnamefont {Malik}},
  \bibinfo {author} {\bibfnamefont {M.}~\bibnamefont {Sasaki}},\ and\ \bibinfo
  {author} {\bibfnamefont {I.}~\bibnamefont {Zaballa}},\ }\bibfield  {title}
  {\bibinfo {title} {{Forming sub-horizon black holes at the end of
  inflation}},\ }\href {https://doi.org/10.1088/1475-7516/2006/01/011}
  {\bibfield  {journal} {\bibinfo  {journal} {JCAP}\ }\textbf {\bibinfo
  {volume} {01}},\ \bibinfo {pages} {011}},\ \Eprint
  {https://arxiv.org/abs/astro-ph/0510647} {arXiv:astro-ph/0510647}
  \BibitemShut {NoStop}%
\bibitem [{\citenamefont {Padilla}\ \emph {et~al.}(2022)\citenamefont
  {Padilla}, \citenamefont {Hidalgo},\ and\ \citenamefont
  {Malik}}]{Padilla:2021zgm}%
  \BibitemOpen
  \bibfield  {author} {\bibinfo {author} {\bibfnamefont {L.~E.}\ \bibnamefont
  {Padilla}}, \bibinfo {author} {\bibfnamefont {J.~C.}\ \bibnamefont
  {Hidalgo}},\ and\ \bibinfo {author} {\bibfnamefont {K.~A.}\ \bibnamefont
  {Malik}},\ }\bibfield  {title} {\bibinfo {title} {{New mechanism for
  primordial black hole formation during reheating}},\ }\href
  {https://doi.org/10.1103/PhysRevD.106.023519} {\bibfield  {journal} {\bibinfo
   {journal} {Phys. Rev. D}\ }\textbf {\bibinfo {volume} {106}},\ \bibinfo
  {pages} {023519} (\bibinfo {year} {2022})},\ \Eprint
  {https://arxiv.org/abs/2110.14584} {arXiv:2110.14584 [astro-ph.CO]}
  \BibitemShut {NoStop}%
\bibitem [{\citenamefont {Padilla}\ \emph {et~al.}(2024)\citenamefont
  {Padilla}, \citenamefont {Hidalgo}, \citenamefont {Gomez-Aguilar},
  \citenamefont {Malik},\ and\ \citenamefont {German}}]{Padilla:2024iyr}%
  \BibitemOpen
  \bibfield  {author} {\bibinfo {author} {\bibfnamefont {L.~E.}\ \bibnamefont
  {Padilla}}, \bibinfo {author} {\bibfnamefont {J.~C.}\ \bibnamefont
  {Hidalgo}}, \bibinfo {author} {\bibfnamefont {T.~D.}\ \bibnamefont
  {Gomez-Aguilar}}, \bibinfo {author} {\bibfnamefont {K.~A.}\ \bibnamefont
  {Malik}},\ and\ \bibinfo {author} {\bibfnamefont {G.}~\bibnamefont
  {German}},\ }\bibfield  {title} {\bibinfo {title} {{Primordial black hole
  formation during slow-reheating: a review}},\ }\href
  {https://doi.org/10.3389/fspas.2024.1361399} {\bibfield  {journal} {\bibinfo
  {journal} {Front. Astron. Space Sci.}\ }\textbf {\bibinfo {volume} {11}},\
  \bibinfo {pages} {1361399} (\bibinfo {year} {2024})},\ \Eprint
  {https://arxiv.org/abs/2402.03542} {arXiv:2402.03542 [astro-ph.CO]}
  \BibitemShut {NoStop}%
\bibitem [{\citenamefont {Musco}(2019)}]{Musco:2018rwt}%
  \BibitemOpen
  \bibfield  {author} {\bibinfo {author} {\bibfnamefont {I.}~\bibnamefont
  {Musco}},\ }\bibfield  {title} {\bibinfo {title} {{Threshold for primordial
  black holes: Dependence on the shape of the cosmological perturbations}},\
  }\href {https://doi.org/10.1103/PhysRevD.100.123524} {\bibfield  {journal}
  {\bibinfo  {journal} {Phys. Rev. D}\ }\textbf {\bibinfo {volume} {100}},\
  \bibinfo {pages} {123524} (\bibinfo {year} {2019})},\ \Eprint
  {https://arxiv.org/abs/1809.02127} {arXiv:1809.02127 [gr-qc]} \BibitemShut
  {NoStop}%
\bibitem [{\citenamefont {Germani}\ and\ \citenamefont
  {Musco}(2019)}]{Germani:2018jgr}%
  \BibitemOpen
  \bibfield  {author} {\bibinfo {author} {\bibfnamefont {C.}~\bibnamefont
  {Germani}}\ and\ \bibinfo {author} {\bibfnamefont {I.}~\bibnamefont
  {Musco}},\ }\bibfield  {title} {\bibinfo {title} {{Abundance of Primordial
  Black Holes Depends on the Shape of the Inflationary Power Spectrum}},\
  }\href {https://doi.org/10.1103/PhysRevLett.122.141302} {\bibfield  {journal}
  {\bibinfo  {journal} {Phys. Rev. Lett.}\ }\textbf {\bibinfo {volume} {122}},\
  \bibinfo {pages} {141302} (\bibinfo {year} {2019})},\ \Eprint
  {https://arxiv.org/abs/1805.04087} {arXiv:1805.04087 [astro-ph.CO]}
  \BibitemShut {NoStop}%
\bibitem [{\citenamefont {Jedamzik}(1997)}]{Jedamzik:1996mr}%
  \BibitemOpen
  \bibfield  {author} {\bibinfo {author} {\bibfnamefont {K.}~\bibnamefont
  {Jedamzik}},\ }\bibfield  {title} {\bibinfo {title} {{Primordial black hole
  formation during the QCD epoch}},\ }\href
  {https://doi.org/10.1103/PhysRevD.55.R5871} {\bibfield  {journal} {\bibinfo
  {journal} {Phys. Rev. D}\ }\textbf {\bibinfo {volume} {55}},\ \bibinfo
  {pages} {5871} (\bibinfo {year} {1997})},\ \Eprint
  {https://arxiv.org/abs/astro-ph/9605152} {arXiv:astro-ph/9605152}
  \BibitemShut {NoStop}%
\bibitem [{\citenamefont {Jedamzik}\ and\ \citenamefont
  {Niemeyer}(1999)}]{Jedamzik:1999am}%
  \BibitemOpen
  \bibfield  {author} {\bibinfo {author} {\bibfnamefont {K.}~\bibnamefont
  {Jedamzik}}\ and\ \bibinfo {author} {\bibfnamefont {J.~C.}\ \bibnamefont
  {Niemeyer}},\ }\bibfield  {title} {\bibinfo {title} {{Primordial black hole
  formation during first order phase transitions}},\ }\href
  {https://doi.org/10.1103/PhysRevD.59.124014} {\bibfield  {journal} {\bibinfo
  {journal} {Phys. Rev. D}\ }\textbf {\bibinfo {volume} {59}},\ \bibinfo
  {pages} {124014} (\bibinfo {year} {1999})},\ \Eprint
  {https://arxiv.org/abs/astro-ph/9901293} {arXiv:astro-ph/9901293}
  \BibitemShut {NoStop}%
\bibitem [{\citenamefont {Byrnes}\ \emph {et~al.}(2018)\citenamefont {Byrnes},
  \citenamefont {Hindmarsh}, \citenamefont {Young},\ and\ \citenamefont
  {Hawkins}}]{Byrnes:2018clq}%
  \BibitemOpen
  \bibfield  {author} {\bibinfo {author} {\bibfnamefont {C.~T.}\ \bibnamefont
  {Byrnes}}, \bibinfo {author} {\bibfnamefont {M.}~\bibnamefont {Hindmarsh}},
  \bibinfo {author} {\bibfnamefont {S.}~\bibnamefont {Young}},\ and\ \bibinfo
  {author} {\bibfnamefont {M.~R.~S.}\ \bibnamefont {Hawkins}},\ }\bibfield
  {title} {\bibinfo {title} {{Primordial black holes with an accurate QCD
  equation of state}},\ }\href {https://doi.org/10.1088/1475-7516/2018/08/041}
  {\bibfield  {journal} {\bibinfo  {journal} {JCAP}\ }\textbf {\bibinfo
  {volume} {08}},\ \bibinfo {pages} {041}},\ \Eprint
  {https://arxiv.org/abs/1801.06138} {arXiv:1801.06138 [astro-ph.CO]}
  \BibitemShut {NoStop}%
\bibitem [{\citenamefont {Sobrinho}\ and\ \citenamefont
  {Augusto}(2020)}]{Sobrinho:2020cco}%
  \BibitemOpen
  \bibfield  {author} {\bibinfo {author} {\bibfnamefont {J.~L.~G.}\
  \bibnamefont {Sobrinho}}\ and\ \bibinfo {author} {\bibfnamefont
  {P.}~\bibnamefont {Augusto}},\ }\bibfield  {title} {\bibinfo {title}
  {{Stellar mass Primordial Black Holes as Cold Dark Matter}},\ }\href
  {https://doi.org/10.1093/mnras/staa1437} {\bibfield  {journal} {\bibinfo
  {journal} {Mon. Not. Roy. Astron. Soc.}\ }\textbf {\bibinfo {volume} {496}},\
  \bibinfo {pages} {60} (\bibinfo {year} {2020})},\ \Eprint
  {https://arxiv.org/abs/2005.10037} {arXiv:2005.10037 [astro-ph.CO]}
  \BibitemShut {NoStop}%
\bibitem [{\citenamefont {Escriv{\`a}}\ \emph {et~al.}(2023)\citenamefont
  {Escriv{\`a}}, \citenamefont {Bagui},\ and\ \citenamefont
  {Clesse}}]{Escriva:2022bwe}%
  \BibitemOpen
  \bibfield  {author} {\bibinfo {author} {\bibfnamefont {A.}~\bibnamefont
  {Escriv{\`a}}}, \bibinfo {author} {\bibfnamefont {E.}~\bibnamefont {Bagui}},\
  and\ \bibinfo {author} {\bibfnamefont {S.}~\bibnamefont {Clesse}},\
  }\bibfield  {title} {\bibinfo {title} {{Simulations of PBH formation at the
  QCD epoch and comparison with the GWTC-3 catalog}},\ }\href
  {https://doi.org/10.1088/1475-7516/2023/05/004} {\bibfield  {journal}
  {\bibinfo  {journal} {JCAP}\ }\textbf {\bibinfo {volume} {05}},\ \bibinfo
  {pages} {004}},\ \Eprint {https://arxiv.org/abs/2209.06196} {arXiv:2209.06196
  [astro-ph.CO]} \BibitemShut {NoStop}%
\bibitem [{\citenamefont {Escriv{\`a}}\ \emph {et~al.}(2021)\citenamefont
  {Escriv{\`a}}, \citenamefont {Germani},\ and\ \citenamefont
  {Sheth}}]{Escriva:2020tak}%
  \BibitemOpen
  \bibfield  {author} {\bibinfo {author} {\bibfnamefont {A.}~\bibnamefont
  {Escriv{\`a}}}, \bibinfo {author} {\bibfnamefont {C.}~\bibnamefont
  {Germani}},\ and\ \bibinfo {author} {\bibfnamefont {R.~K.}\ \bibnamefont
  {Sheth}},\ }\bibfield  {title} {\bibinfo {title} {{Analytical thresholds for
  black hole formation in general cosmological backgrounds}},\ }\href
  {https://doi.org/10.1088/1475-7516/2021/01/030} {\bibfield  {journal}
  {\bibinfo  {journal} {JCAP}\ }\textbf {\bibinfo {volume} {01}},\ \bibinfo
  {pages} {030}},\ \Eprint {https://arxiv.org/abs/2007.05564} {arXiv:2007.05564
  [gr-qc]} \BibitemShut {NoStop}%
\bibitem [{\citenamefont {Pritchard}\ and\ \citenamefont
  {Byrnes}(2025)}]{Pritchard:2024vix}%
  \BibitemOpen
  \bibfield  {author} {\bibinfo {author} {\bibfnamefont {X.}~\bibnamefont
  {Pritchard}}\ and\ \bibinfo {author} {\bibfnamefont {C.~T.}\ \bibnamefont
  {Byrnes}},\ }\bibfield  {title} {\bibinfo {title} {{Constraining the impact
  of standard model phase transitions on primordial black holes}},\ }\href
  {https://doi.org/10.1088/1475-7516/2025/01/076} {\bibfield  {journal}
  {\bibinfo  {journal} {JCAP}\ }\textbf {\bibinfo {volume} {01}},\ \bibinfo
  {pages} {076}},\ \Eprint {https://arxiv.org/abs/2407.16563} {arXiv:2407.16563
  [astro-ph.CO]} \BibitemShut {NoStop}%
\bibitem [{\citenamefont {Pritchard}\ \emph
  {et~al.}(2025{\natexlab{b}})\citenamefont {Pritchard}, \citenamefont
  {Starbuck},\ and\ \citenamefont {Leung}}]{Pritchard:2025pcn}%
  \BibitemOpen
  \bibfield  {author} {\bibinfo {author} {\bibfnamefont {X.}~\bibnamefont
  {Pritchard}}, \bibinfo {author} {\bibfnamefont {M.}~\bibnamefont
  {Starbuck}},\ and\ \bibinfo {author} {\bibfnamefont {W.}~\bibnamefont
  {Leung}},\ }\bibfield  {title} {\bibinfo {title} {{Beyond Standard Model
  equation of state and primordial black holes}},\ }\href@noop {} {\  (\bibinfo
  {year} {2025}{\natexlab{b}})},\ \Eprint {https://arxiv.org/abs/2510.19629}
  {arXiv:2510.19629 [astro-ph.CO]} \BibitemShut {NoStop}%
\bibitem [{\citenamefont {Blas}\ \emph {et~al.}(2026)\citenamefont {Blas},
  \citenamefont {Foster}, \citenamefont {Gouttenoire}, \citenamefont {Iovino},
  \citenamefont {Musco}, \citenamefont {Trifinopoulos},\ and\ \citenamefont
  {Vanvlasselaer}}]{Blas:2026xws}%
  \BibitemOpen
  \bibfield  {author} {\bibinfo {author} {\bibfnamefont {D.}~\bibnamefont
  {Blas}}, \bibinfo {author} {\bibfnamefont {J.~W.}\ \bibnamefont {Foster}},
  \bibinfo {author} {\bibfnamefont {Y.}~\bibnamefont {Gouttenoire}}, \bibinfo
  {author} {\bibfnamefont {A.~J.}\ \bibnamefont {Iovino}}, \bibinfo {author}
  {\bibfnamefont {I.}~\bibnamefont {Musco}}, \bibinfo {author} {\bibfnamefont
  {S.}~\bibnamefont {Trifinopoulos}},\ and\ \bibinfo {author} {\bibfnamefont
  {M.}~\bibnamefont {Vanvlasselaer}},\ }\bibfield  {title} {\bibinfo {title}
  {{The Dark Side of the Moon: Listening to Scalar-Induced Gravitational
  Waves}},\ }\href@noop {} {\  (\bibinfo {year} {2026})},\ \Eprint
  {https://arxiv.org/abs/2602.12252} {arXiv:2602.12252 [astro-ph.CO]}
  \BibitemShut {NoStop}%
\bibitem [{\citenamefont {Andr\'es-Carcasona}\ \emph
  {et~al.}(2024)\citenamefont {Andr\'es-Carcasona}, \citenamefont {Iovino},
  \citenamefont {Vaskonen}, \citenamefont {Veerm\"ae}, \citenamefont
  {Mart\'\i{}nez}, \citenamefont {Pujol\`as},\ and\ \citenamefont
  {Mir}}]{Andres-Carcasona:2024wqk}%
  \BibitemOpen
  \bibfield  {author} {\bibinfo {author} {\bibfnamefont {M.}~\bibnamefont
  {Andr\'es-Carcasona}}, \bibinfo {author} {\bibfnamefont {A.~J.}\ \bibnamefont
  {Iovino}}, \bibinfo {author} {\bibfnamefont {V.}~\bibnamefont {Vaskonen}},
  \bibinfo {author} {\bibfnamefont {H.}~\bibnamefont {Veerm\"ae}}, \bibinfo
  {author} {\bibfnamefont {M.}~\bibnamefont {Mart\'\i{}nez}}, \bibinfo {author}
  {\bibfnamefont {O.}~\bibnamefont {Pujol\`as}},\ and\ \bibinfo {author}
  {\bibfnamefont {L.~M.}\ \bibnamefont {Mir}},\ }\bibfield  {title} {\bibinfo
  {title} {{Constraints on primordial black holes from LIGO-Virgo-KAGRA O3
  events}},\ }\href {https://doi.org/10.1103/PhysRevD.110.023040} {\bibfield
  {journal} {\bibinfo  {journal} {Phys. Rev. D}\ }\textbf {\bibinfo {volume}
  {110}},\ \bibinfo {pages} {023040} (\bibinfo {year} {2024})},\ \Eprint
  {https://arxiv.org/abs/2405.05732} {arXiv:2405.05732 [astro-ph.CO]}
  \BibitemShut {NoStop}%
\bibitem [{\citenamefont {Young}\ and\ \citenamefont
  {Byrnes}(2013)}]{Young:2013oia}%
  \BibitemOpen
  \bibfield  {author} {\bibinfo {author} {\bibfnamefont {S.}~\bibnamefont
  {Young}}\ and\ \bibinfo {author} {\bibfnamefont {C.~T.}\ \bibnamefont
  {Byrnes}},\ }\bibfield  {title} {\bibinfo {title} {{Primordial black holes in
  non-Gaussian regimes}},\ }\href
  {https://doi.org/10.1088/1475-7516/2013/08/052} {\bibfield  {journal}
  {\bibinfo  {journal} {JCAP}\ }\textbf {\bibinfo {volume} {08}},\ \bibinfo
  {pages} {052}},\ \Eprint {https://arxiv.org/abs/1307.4995} {arXiv:1307.4995
  [astro-ph.CO]} \BibitemShut {NoStop}%
\bibitem [{\citenamefont {Bugaev}\ and\ \citenamefont
  {Klimai}(2013)}]{Bugaev:2013vba}%
  \BibitemOpen
  \bibfield  {author} {\bibinfo {author} {\bibfnamefont {E.~V.}\ \bibnamefont
  {Bugaev}}\ and\ \bibinfo {author} {\bibfnamefont {P.~A.}\ \bibnamefont
  {Klimai}},\ }\bibfield  {title} {\bibinfo {title} {{Primordial black hole
  constraints for curvaton models with predicted large non-Gaussianity}},\
  }\href {https://doi.org/10.1142/S021827181350034X} {\bibfield  {journal}
  {\bibinfo  {journal} {Int. J. Mod. Phys. D}\ }\textbf {\bibinfo {volume}
  {22}},\ \bibinfo {pages} {1350034} (\bibinfo {year} {2013})},\ \Eprint
  {https://arxiv.org/abs/1303.3146} {arXiv:1303.3146 [astro-ph.CO]}
  \BibitemShut {NoStop}%
\bibitem [{\citenamefont {Nakama}\ \emph {et~al.}(2017)\citenamefont {Nakama},
  \citenamefont {Silk},\ and\ \citenamefont {Kamionkowski}}]{Nakama:2016gzw}%
  \BibitemOpen
  \bibfield  {author} {\bibinfo {author} {\bibfnamefont {T.}~\bibnamefont
  {Nakama}}, \bibinfo {author} {\bibfnamefont {J.}~\bibnamefont {Silk}},\ and\
  \bibinfo {author} {\bibfnamefont {M.}~\bibnamefont {Kamionkowski}},\
  }\bibfield  {title} {\bibinfo {title} {{Stochastic gravitational waves
  associated with the formation of primordial black holes}},\ }\href
  {https://doi.org/10.1103/PhysRevD.95.043511} {\bibfield  {journal} {\bibinfo
  {journal} {Phys. Rev. D}\ }\textbf {\bibinfo {volume} {95}},\ \bibinfo
  {pages} {043511} (\bibinfo {year} {2017})},\ \Eprint
  {https://arxiv.org/abs/1612.06264} {arXiv:1612.06264 [astro-ph.CO]}
  \BibitemShut {NoStop}%
\bibitem [{\citenamefont {Byrnes}\ \emph {et~al.}(2012)\citenamefont {Byrnes},
  \citenamefont {Copeland},\ and\ \citenamefont {Green}}]{Byrnes:2012yx}%
  \BibitemOpen
  \bibfield  {author} {\bibinfo {author} {\bibfnamefont {C.~T.}\ \bibnamefont
  {Byrnes}}, \bibinfo {author} {\bibfnamefont {E.~J.}\ \bibnamefont
  {Copeland}},\ and\ \bibinfo {author} {\bibfnamefont {A.~M.}\ \bibnamefont
  {Green}},\ }\bibfield  {title} {\bibinfo {title} {{Primordial black holes as
  a tool for constraining non-Gaussianity}},\ }\href
  {https://doi.org/10.1103/PhysRevD.86.043512} {\bibfield  {journal} {\bibinfo
  {journal} {Phys. Rev. D}\ }\textbf {\bibinfo {volume} {86}},\ \bibinfo
  {pages} {043512} (\bibinfo {year} {2012})},\ \Eprint
  {https://arxiv.org/abs/1206.4188} {arXiv:1206.4188 [astro-ph.CO]}
  \BibitemShut {NoStop}%
\bibitem [{\citenamefont {Franciolini}\ \emph {et~al.}(2018)\citenamefont
  {Franciolini}, \citenamefont {Kehagias}, \citenamefont {Matarrese},\ and\
  \citenamefont {Riotto}}]{Franciolini:2018vbk}%
  \BibitemOpen
  \bibfield  {author} {\bibinfo {author} {\bibfnamefont {G.}~\bibnamefont
  {Franciolini}}, \bibinfo {author} {\bibfnamefont {A.}~\bibnamefont
  {Kehagias}}, \bibinfo {author} {\bibfnamefont {S.}~\bibnamefont
  {Matarrese}},\ and\ \bibinfo {author} {\bibfnamefont {A.}~\bibnamefont
  {Riotto}},\ }\bibfield  {title} {\bibinfo {title} {{Primordial Black Holes
  from Inflation and non-Gaussianity}},\ }\href
  {https://doi.org/10.1088/1475-7516/2018/03/016} {\bibfield  {journal}
  {\bibinfo  {journal} {JCAP}\ }\textbf {\bibinfo {volume} {03}},\ \bibinfo
  {pages} {016}},\ \Eprint {https://arxiv.org/abs/1801.09415} {arXiv:1801.09415
  [astro-ph.CO]} \BibitemShut {NoStop}%
\bibitem [{\citenamefont {Kawasaki}\ and\ \citenamefont
  {Nakatsuka}(2019)}]{Kawasaki:2019mbl}%
  \BibitemOpen
  \bibfield  {author} {\bibinfo {author} {\bibfnamefont {M.}~\bibnamefont
  {Kawasaki}}\ and\ \bibinfo {author} {\bibfnamefont {H.}~\bibnamefont
  {Nakatsuka}},\ }\bibfield  {title} {\bibinfo {title} {{Effect of nonlinearity
  between density and curvature perturbations on the primordial black hole
  formation}},\ }\href {https://doi.org/10.1103/PhysRevD.99.123501} {\bibfield
  {journal} {\bibinfo  {journal} {Phys. Rev. D}\ }\textbf {\bibinfo {volume}
  {99}},\ \bibinfo {pages} {123501} (\bibinfo {year} {2019})},\ \Eprint
  {https://arxiv.org/abs/1903.02994} {arXiv:1903.02994 [astro-ph.CO]}
  \BibitemShut {NoStop}%
\bibitem [{\citenamefont {Figueroa}\ \emph {et~al.}(2021)\citenamefont
  {Figueroa}, \citenamefont {Raatikainen}, \citenamefont {Rasanen},\ and\
  \citenamefont {Tomberg}}]{Figueroa:2020jkf}%
  \BibitemOpen
  \bibfield  {author} {\bibinfo {author} {\bibfnamefont {D.~G.}\ \bibnamefont
  {Figueroa}}, \bibinfo {author} {\bibfnamefont {S.}~\bibnamefont
  {Raatikainen}}, \bibinfo {author} {\bibfnamefont {S.}~\bibnamefont
  {Rasanen}},\ and\ \bibinfo {author} {\bibfnamefont {E.}~\bibnamefont
  {Tomberg}},\ }\bibfield  {title} {\bibinfo {title} {{Non-Gaussian Tail of the
  Curvature Perturbation in Stochastic Ultraslow-Roll Inflation: Implications
  for Primordial Black Hole Production}},\ }\href
  {https://doi.org/10.1103/PhysRevLett.127.101302} {\bibfield  {journal}
  {\bibinfo  {journal} {Phys. Rev. Lett.}\ }\textbf {\bibinfo {volume} {127}},\
  \bibinfo {pages} {101302} (\bibinfo {year} {2021})},\ \Eprint
  {https://arxiv.org/abs/2012.06551} {arXiv:2012.06551 [astro-ph.CO]}
  \BibitemShut {NoStop}%
\bibitem [{\citenamefont {Riccardi}\ \emph {et~al.}(2021)\citenamefont
  {Riccardi}, \citenamefont {Taoso},\ and\ \citenamefont
  {Urbano}}]{Riccardi:2021rlf}%
  \BibitemOpen
  \bibfield  {author} {\bibinfo {author} {\bibfnamefont {F.}~\bibnamefont
  {Riccardi}}, \bibinfo {author} {\bibfnamefont {M.}~\bibnamefont {Taoso}},\
  and\ \bibinfo {author} {\bibfnamefont {A.}~\bibnamefont {Urbano}},\
  }\bibfield  {title} {\bibinfo {title} {{Solving peak theory in the presence
  of local non-gaussianities}},\ }\href
  {https://doi.org/10.1088/1475-7516/2021/08/060} {\bibfield  {journal}
  {\bibinfo  {journal} {JCAP}\ }\textbf {\bibinfo {volume} {08}},\ \bibinfo
  {pages} {060}},\ \Eprint {https://arxiv.org/abs/2102.04084} {arXiv:2102.04084
  [astro-ph.CO]} \BibitemShut {NoStop}%
\bibitem [{\citenamefont {Taoso}\ and\ \citenamefont
  {Urbano}(2021)}]{Taoso:2021uvl}%
  \BibitemOpen
  \bibfield  {author} {\bibinfo {author} {\bibfnamefont {M.}~\bibnamefont
  {Taoso}}\ and\ \bibinfo {author} {\bibfnamefont {A.}~\bibnamefont {Urbano}},\
  }\bibfield  {title} {\bibinfo {title} {{Non-gaussianities for primordial
  black hole formation}},\ }\href
  {https://doi.org/10.1088/1475-7516/2021/08/016} {\bibfield  {journal}
  {\bibinfo  {journal} {JCAP}\ }\textbf {\bibinfo {volume} {08}},\ \bibinfo
  {pages} {016}},\ \Eprint {https://arxiv.org/abs/2102.03610} {arXiv:2102.03610
  [astro-ph.CO]} \BibitemShut {NoStop}%
\bibitem [{\citenamefont {Biagetti}\ \emph {et~al.}(2021)\citenamefont
  {Biagetti}, \citenamefont {De~Luca}, \citenamefont {Franciolini},
  \citenamefont {Kehagias},\ and\ \citenamefont {Riotto}}]{Biagetti:2021eep}%
  \BibitemOpen
  \bibfield  {author} {\bibinfo {author} {\bibfnamefont {M.}~\bibnamefont
  {Biagetti}}, \bibinfo {author} {\bibfnamefont {V.}~\bibnamefont {De~Luca}},
  \bibinfo {author} {\bibfnamefont {G.}~\bibnamefont {Franciolini}}, \bibinfo
  {author} {\bibfnamefont {A.}~\bibnamefont {Kehagias}},\ and\ \bibinfo
  {author} {\bibfnamefont {A.}~\bibnamefont {Riotto}},\ }\bibfield  {title}
  {\bibinfo {title} {{The formation probability of primordial black holes}},\
  }\href {https://doi.org/10.1016/j.physletb.2021.136602} {\bibfield  {journal}
  {\bibinfo  {journal} {Phys. Lett. B}\ }\textbf {\bibinfo {volume} {820}},\
  \bibinfo {pages} {136602} (\bibinfo {year} {2021})},\ \Eprint
  {https://arxiv.org/abs/2105.07810} {arXiv:2105.07810 [astro-ph.CO]}
  \BibitemShut {NoStop}%
\bibitem [{\citenamefont {Kitajima}\ \emph {et~al.}(2021)\citenamefont
  {Kitajima}, \citenamefont {Tada}, \citenamefont {Yokoyama},\ and\
  \citenamefont {Yoo}}]{Kitajima:2021fpq}%
  \BibitemOpen
  \bibfield  {author} {\bibinfo {author} {\bibfnamefont {N.}~\bibnamefont
  {Kitajima}}, \bibinfo {author} {\bibfnamefont {Y.}~\bibnamefont {Tada}},
  \bibinfo {author} {\bibfnamefont {S.}~\bibnamefont {Yokoyama}},\ and\
  \bibinfo {author} {\bibfnamefont {C.-M.}\ \bibnamefont {Yoo}},\ }\bibfield
  {title} {\bibinfo {title} {{Primordial black holes in peak theory with a
  non-Gaussian tail}},\ }\href {https://doi.org/10.1088/1475-7516/2021/10/053}
  {\bibfield  {journal} {\bibinfo  {journal} {JCAP}\ }\textbf {\bibinfo
  {volume} {10}},\ \bibinfo {pages} {053}},\ \Eprint
  {https://arxiv.org/abs/2109.00791} {arXiv:2109.00791 [astro-ph.CO]}
  \BibitemShut {NoStop}%
\bibitem [{\citenamefont {Hooshangi}\ \emph {et~al.}(2022)\citenamefont
  {Hooshangi}, \citenamefont {Namjoo},\ and\ \citenamefont
  {Noorbala}}]{Hooshangi:2021ubn}%
  \BibitemOpen
  \bibfield  {author} {\bibinfo {author} {\bibfnamefont {S.}~\bibnamefont
  {Hooshangi}}, \bibinfo {author} {\bibfnamefont {M.~H.}\ \bibnamefont
  {Namjoo}},\ and\ \bibinfo {author} {\bibfnamefont {M.}~\bibnamefont
  {Noorbala}},\ }\bibfield  {title} {\bibinfo {title} {{Rare events are
  nonperturbative: Primordial black holes from heavy-tailed distributions}},\
  }\href {https://doi.org/10.1016/j.physletb.2022.137400} {\bibfield  {journal}
  {\bibinfo  {journal} {Phys. Lett. B}\ }\textbf {\bibinfo {volume} {834}},\
  \bibinfo {pages} {137400} (\bibinfo {year} {2022})},\ \Eprint
  {https://arxiv.org/abs/2112.04520} {arXiv:2112.04520 [astro-ph.CO]}
  \BibitemShut {NoStop}%
\bibitem [{\citenamefont {Meng}\ \emph {et~al.}(2022)\citenamefont {Meng},
  \citenamefont {Yuan},\ and\ \citenamefont {Huang}}]{Meng:2022ixx}%
  \BibitemOpen
  \bibfield  {author} {\bibinfo {author} {\bibfnamefont {D.-S.}\ \bibnamefont
  {Meng}}, \bibinfo {author} {\bibfnamefont {C.}~\bibnamefont {Yuan}},\ and\
  \bibinfo {author} {\bibfnamefont {Q.-g.}\ \bibnamefont {Huang}},\ }\bibfield
  {title} {\bibinfo {title} {{One-loop correction to the enhanced curvature
  perturbation with local-type non-Gaussianity for the formation of primordial
  black holes}},\ }\href {https://doi.org/10.1103/PhysRevD.106.063508}
  {\bibfield  {journal} {\bibinfo  {journal} {Phys. Rev. D}\ }\textbf {\bibinfo
  {volume} {106}},\ \bibinfo {pages} {063508} (\bibinfo {year} {2022})},\
  \Eprint {https://arxiv.org/abs/2207.07668} {arXiv:2207.07668 [astro-ph.CO]}
  \BibitemShut {NoStop}%
\bibitem [{\citenamefont {Young}(2022)}]{Young:2022phe}%
  \BibitemOpen
  \bibfield  {author} {\bibinfo {author} {\bibfnamefont {S.}~\bibnamefont
  {Young}},\ }\bibfield  {title} {\bibinfo {title} {{Peaks and primordial black
  holes: the~effect of non-Gaussianity}},\ }\href
  {https://doi.org/10.1088/1475-7516/2022/05/037} {\bibfield  {journal}
  {\bibinfo  {journal} {JCAP}\ }\textbf {\bibinfo {volume} {05}}\bibfield
  {number} {\bibinfo  {number} { (05)},\ \bibinfo {pages} {037}},\ }\Eprint
  {https://arxiv.org/abs/2201.13345} {arXiv:2201.13345 [astro-ph.CO]}
  \BibitemShut {NoStop}%
\bibitem [{\citenamefont {Escriv{\`a}}\ \emph {et~al.}(2022)\citenamefont
  {Escriv{\`a}}, \citenamefont {Tada}, \citenamefont {Yokoyama},\ and\
  \citenamefont {Yoo}}]{Escriva:2022pnz}%
  \BibitemOpen
  \bibfield  {author} {\bibinfo {author} {\bibfnamefont {A.}~\bibnamefont
  {Escriv{\`a}}}, \bibinfo {author} {\bibfnamefont {Y.}~\bibnamefont {Tada}},
  \bibinfo {author} {\bibfnamefont {S.}~\bibnamefont {Yokoyama}},\ and\
  \bibinfo {author} {\bibfnamefont {C.-M.}\ \bibnamefont {Yoo}},\ }\bibfield
  {title} {\bibinfo {title} {{Simulation of primordial black holes with large
  negative non-Gaussianity}},\ }\href
  {https://doi.org/10.1088/1475-7516/2022/05/012} {\bibfield  {journal}
  {\bibinfo  {journal} {JCAP}\ }\textbf {\bibinfo {volume} {05}}\bibfield
  {number} {\bibinfo  {number} { (05)},\ \bibinfo {pages} {012}},\ }\Eprint
  {https://arxiv.org/abs/2202.01028} {arXiv:2202.01028 [astro-ph.CO]}
  \BibitemShut {NoStop}%
\bibitem [{\citenamefont {Hooshangi}\ \emph {et~al.}(2023)\citenamefont
  {Hooshangi}, \citenamefont {Namjoo},\ and\ \citenamefont
  {Noorbala}}]{Hooshangi:2023kss}%
  \BibitemOpen
  \bibfield  {author} {\bibinfo {author} {\bibfnamefont {S.}~\bibnamefont
  {Hooshangi}}, \bibinfo {author} {\bibfnamefont {M.~H.}\ \bibnamefont
  {Namjoo}},\ and\ \bibinfo {author} {\bibfnamefont {M.}~\bibnamefont
  {Noorbala}},\ }\bibfield  {title} {\bibinfo {title} {{Tail diversity from
  inflation}},\ }\href {https://doi.org/10.1088/1475-7516/2023/09/023}
  {\bibfield  {journal} {\bibinfo  {journal} {JCAP}\ }\textbf {\bibinfo
  {volume} {09}},\ \bibinfo {pages} {023}},\ \Eprint
  {https://arxiv.org/abs/2305.19257} {arXiv:2305.19257 [astro-ph.CO]}
  \BibitemShut {NoStop}%
\bibitem [{\citenamefont {Ianniccari}\ \emph
  {et~al.}(2024{\natexlab{b}})\citenamefont {Ianniccari}, \citenamefont
  {Iovino}, \citenamefont {Kehagias}, \citenamefont {Perrone},\ and\
  \citenamefont {Riotto}}]{Ianniccari:2024bkh}%
  \BibitemOpen
  \bibfield  {author} {\bibinfo {author} {\bibfnamefont {A.}~\bibnamefont
  {Ianniccari}}, \bibinfo {author} {\bibfnamefont {A.~J.}\ \bibnamefont
  {Iovino}}, \bibinfo {author} {\bibfnamefont {A.}~\bibnamefont {Kehagias}},
  \bibinfo {author} {\bibfnamefont {D.}~\bibnamefont {Perrone}},\ and\ \bibinfo
  {author} {\bibfnamefont {A.}~\bibnamefont {Riotto}},\ }\bibfield  {title}
  {\bibinfo {title} {{Primordial black hole abundance: The importance of
  broadness}},\ }\href {https://doi.org/10.1103/PhysRevD.109.123549} {\bibfield
   {journal} {\bibinfo  {journal} {Phys. Rev. D}\ }\textbf {\bibinfo {volume}
  {109}},\ \bibinfo {pages} {123549} (\bibinfo {year} {2024}{\natexlab{b}})},\
  \Eprint {https://arxiv.org/abs/2402.11033} {arXiv:2402.11033 [astro-ph.CO]}
  \BibitemShut {NoStop}%
\bibitem [{\citenamefont {Germani}\ and\ \citenamefont
  {Sheth}(2020)}]{Germani:2019zez}%
  \BibitemOpen
  \bibfield  {author} {\bibinfo {author} {\bibfnamefont {C.}~\bibnamefont
  {Germani}}\ and\ \bibinfo {author} {\bibfnamefont {R.~K.}\ \bibnamefont
  {Sheth}},\ }\bibfield  {title} {\bibinfo {title} {{Nonlinear statistics of
  primordial black holes from Gaussian curvature perturbations}},\ }\href
  {https://doi.org/10.1103/PhysRevD.101.063520} {\bibfield  {journal} {\bibinfo
   {journal} {Phys. Rev. D}\ }\textbf {\bibinfo {volume} {101}},\ \bibinfo
  {pages} {063520} (\bibinfo {year} {2020})},\ \Eprint
  {https://arxiv.org/abs/1912.07072} {arXiv:1912.07072 [astro-ph.CO]}
  \BibitemShut {NoStop}%
\bibitem [{\citenamefont {Sasaki}\ \emph {et~al.}(2006)\citenamefont {Sasaki},
  \citenamefont {Valiviita},\ and\ \citenamefont {Wands}}]{Sasaki:2006kq}%
  \BibitemOpen
  \bibfield  {author} {\bibinfo {author} {\bibfnamefont {M.}~\bibnamefont
  {Sasaki}}, \bibinfo {author} {\bibfnamefont {J.}~\bibnamefont {Valiviita}},\
  and\ \bibinfo {author} {\bibfnamefont {D.}~\bibnamefont {Wands}},\ }\bibfield
   {title} {\bibinfo {title} {{Non-Gaussianity of the primordial perturbation
  in the curvaton model}},\ }\href {https://doi.org/10.1103/PhysRevD.74.103003}
  {\bibfield  {journal} {\bibinfo  {journal} {Phys. Rev. D}\ }\textbf {\bibinfo
  {volume} {74}},\ \bibinfo {pages} {103003} (\bibinfo {year} {2006})},\
  \Eprint {https://arxiv.org/abs/astro-ph/0607627} {arXiv:astro-ph/0607627}
  \BibitemShut {NoStop}%
\bibitem [{\citenamefont {Pi}\ and\ \citenamefont {Sasaki}(2023)}]{Pi:2022ysn}%
  \BibitemOpen
  \bibfield  {author} {\bibinfo {author} {\bibfnamefont {S.}~\bibnamefont
  {Pi}}\ and\ \bibinfo {author} {\bibfnamefont {M.}~\bibnamefont {Sasaki}},\
  }\bibfield  {title} {\bibinfo {title} {{Logarithmic Duality of the Curvature
  Perturbation}},\ }\href {https://doi.org/10.1103/PhysRevLett.131.011002}
  {\bibfield  {journal} {\bibinfo  {journal} {Phys. Rev. Lett.}\ }\textbf
  {\bibinfo {volume} {131}},\ \bibinfo {pages} {011002} (\bibinfo {year}
  {2023})},\ \Eprint {https://arxiv.org/abs/2211.13932} {arXiv:2211.13932
  [astro-ph.CO]} \BibitemShut {NoStop}%
\bibitem [{\citenamefont {Atal}\ \emph {et~al.}(2019)\citenamefont {Atal},
  \citenamefont {Garriga},\ and\ \citenamefont
  {Marcos-Caballero}}]{Atal:2019cdz}%
  \BibitemOpen
  \bibfield  {author} {\bibinfo {author} {\bibfnamefont {V.}~\bibnamefont
  {Atal}}, \bibinfo {author} {\bibfnamefont {J.}~\bibnamefont {Garriga}},\ and\
  \bibinfo {author} {\bibfnamefont {A.}~\bibnamefont {Marcos-Caballero}},\
  }\bibfield  {title} {\bibinfo {title} {{Primordial black hole formation with
  non-Gaussian curvature perturbations}},\ }\href
  {https://doi.org/10.1088/1475-7516/2019/09/073} {\bibfield  {journal}
  {\bibinfo  {journal} {JCAP}\ }\textbf {\bibinfo {volume} {09}},\ \bibinfo
  {pages} {073}},\ \Eprint {https://arxiv.org/abs/1905.13202} {arXiv:1905.13202
  [astro-ph.CO]} \BibitemShut {NoStop}%
\bibitem [{\citenamefont {Tomberg}(2023)}]{Tomberg:2023kli}%
  \BibitemOpen
  \bibfield  {author} {\bibinfo {author} {\bibfnamefont {E.}~\bibnamefont
  {Tomberg}},\ }\bibfield  {title} {\bibinfo {title} {{Stochastic constant-roll
  inflation and primordial black holes}},\ }\href
  {https://doi.org/10.1103/PhysRevD.108.043502} {\bibfield  {journal} {\bibinfo
   {journal} {Phys. Rev. D}\ }\textbf {\bibinfo {volume} {108}},\ \bibinfo
  {pages} {043502} (\bibinfo {year} {2023})},\ \Eprint
  {https://arxiv.org/abs/2304.10903} {arXiv:2304.10903 [astro-ph.CO]}
  \BibitemShut {NoStop}%
\bibitem [{\citenamefont {Iovino}\ \emph {et~al.}(2026)\citenamefont {Iovino},
  \citenamefont {Matarrese}, \citenamefont {Perna}, \citenamefont
  {Ricciardone},\ and\ \citenamefont {Riotto}}]{Iovino:2024sgs}%
  \BibitemOpen
  \bibfield  {author} {\bibinfo {author} {\bibfnamefont {A.~J.}\ \bibnamefont
  {Iovino}}, \bibinfo {author} {\bibfnamefont {S.}~\bibnamefont {Matarrese}},
  \bibinfo {author} {\bibfnamefont {G.}~\bibnamefont {Perna}}, \bibinfo
  {author} {\bibfnamefont {A.}~\bibnamefont {Ricciardone}},\ and\ \bibinfo
  {author} {\bibfnamefont {A.}~\bibnamefont {Riotto}},\ }\bibfield  {title}
  {\bibinfo {title} {{How well do we know the scalar-induced gravitational
  waves?}},\ }\href {https://doi.org/10.1016/j.physletb.2025.140039} {\bibfield
   {journal} {\bibinfo  {journal} {Phys. Lett. B}\ }\textbf {\bibinfo {volume}
  {872}},\ \bibinfo {pages} {140039} (\bibinfo {year} {2026})},\ \Eprint
  {https://arxiv.org/abs/2412.06764} {arXiv:2412.06764 [astro-ph.CO]}
  \BibitemShut {NoStop}%
\bibitem [{\citenamefont {Kehagias}\ \emph {et~al.}(2019)\citenamefont
  {Kehagias}, \citenamefont {Musco},\ and\ \citenamefont
  {Riotto}}]{Kehagias:2019eil}%
  \BibitemOpen
  \bibfield  {author} {\bibinfo {author} {\bibfnamefont {A.}~\bibnamefont
  {Kehagias}}, \bibinfo {author} {\bibfnamefont {I.}~\bibnamefont {Musco}},\
  and\ \bibinfo {author} {\bibfnamefont {A.}~\bibnamefont {Riotto}},\
  }\bibfield  {title} {\bibinfo {title} {{Non-Gaussian Formation of Primordial
  Black Holes: Effects on the Threshold}},\ }\href
  {https://doi.org/10.1088/1475-7516/2019/12/029} {\bibfield  {journal}
  {\bibinfo  {journal} {JCAP}\ }\textbf {\bibinfo {volume} {12}},\ \bibinfo
  {pages} {029}},\ \Eprint {https://arxiv.org/abs/1906.07135} {arXiv:1906.07135
  [astro-ph.CO]} \BibitemShut {NoStop}%
\bibitem [{\citenamefont {Caravano}\ \emph {et~al.}(2025)\citenamefont
  {Caravano}, \citenamefont {Franciolini},\ and\ \citenamefont
  {Renaux-Petel}}]{Caravano:2025diq}%
  \BibitemOpen
  \bibfield  {author} {\bibinfo {author} {\bibfnamefont {A.}~\bibnamefont
  {Caravano}}, \bibinfo {author} {\bibfnamefont {G.}~\bibnamefont
  {Franciolini}},\ and\ \bibinfo {author} {\bibfnamefont {S.}~\bibnamefont
  {Renaux-Petel}},\ }\bibfield  {title} {\bibinfo {title} {{Ultraslow-roll
  inflation on the lattice. II. Nonperturbative curvature perturbation}},\
  }\href {https://doi.org/10.1103/39qd-gdfm} {\bibfield  {journal} {\bibinfo
  {journal} {Phys. Rev. D}\ }\textbf {\bibinfo {volume} {112}},\ \bibinfo
  {pages} {083508} (\bibinfo {year} {2025})},\ \Eprint
  {https://arxiv.org/abs/2506.11795} {arXiv:2506.11795 [astro-ph.CO]}
  \BibitemShut {NoStop}%
\bibitem [{\citenamefont {Atal}\ and\ \citenamefont
  {Germani}(2019)}]{Atal:2018neu}%
  \BibitemOpen
  \bibfield  {author} {\bibinfo {author} {\bibfnamefont {V.}~\bibnamefont
  {Atal}}\ and\ \bibinfo {author} {\bibfnamefont {C.}~\bibnamefont {Germani}},\
  }\bibfield  {title} {\bibinfo {title} {{The role of non-gaussianities in
  Primordial Black Hole formation}},\ }\href
  {https://doi.org/10.1016/j.dark.2019.100275} {\bibfield  {journal} {\bibinfo
  {journal} {Phys. Dark Univ.}\ }\textbf {\bibinfo {volume} {24}},\ \bibinfo
  {pages} {100275} (\bibinfo {year} {2019})},\ \Eprint
  {https://arxiv.org/abs/1811.07857} {arXiv:1811.07857 [astro-ph.CO]}
  \BibitemShut {NoStop}%
\bibitem [{\citenamefont {Firouzjahi}\ and\ \citenamefont
  {Riotto}(2023)}]{Firouzjahi:2023xke}%
  \BibitemOpen
  \bibfield  {author} {\bibinfo {author} {\bibfnamefont {H.}~\bibnamefont
  {Firouzjahi}}\ and\ \bibinfo {author} {\bibfnamefont {A.}~\bibnamefont
  {Riotto}},\ }\bibfield  {title} {\bibinfo {title} {{Sign of non-Gaussianity
  and the primordial black holes abundance}},\ }\href
  {https://doi.org/10.1103/PhysRevD.108.123504} {\bibfield  {journal} {\bibinfo
   {journal} {Phys. Rev. D}\ }\textbf {\bibinfo {volume} {108}},\ \bibinfo
  {pages} {123504} (\bibinfo {year} {2023})},\ \Eprint
  {https://arxiv.org/abs/2309.10536} {arXiv:2309.10536 [astro-ph.CO]}
  \BibitemShut {NoStop}%
\bibitem [{\citenamefont {Lewicki}\ \emph {et~al.}(2024)\citenamefont
  {Lewicki}, \citenamefont {Toczek},\ and\ \citenamefont
  {Vaskonen}}]{Lewicki:2024ghw}%
  \BibitemOpen
  \bibfield  {author} {\bibinfo {author} {\bibfnamefont {M.}~\bibnamefont
  {Lewicki}}, \bibinfo {author} {\bibfnamefont {P.}~\bibnamefont {Toczek}},\
  and\ \bibinfo {author} {\bibfnamefont {V.}~\bibnamefont {Vaskonen}},\
  }\bibfield  {title} {\bibinfo {title} {{Black Holes and Gravitational Waves
  from Slow First-Order Phase Transitions}},\ }\href
  {https://doi.org/10.1103/PhysRevLett.133.221003} {\bibfield  {journal}
  {\bibinfo  {journal} {Phys. Rev. Lett.}\ }\textbf {\bibinfo {volume} {133}},\
  \bibinfo {pages} {221003} (\bibinfo {year} {2024})},\ \Eprint
  {https://arxiv.org/abs/2402.04158} {arXiv:2402.04158 [astro-ph.CO]}
  \BibitemShut {NoStop}%
\bibitem [{\citenamefont {Franciolini}\ \emph {et~al.}(2023)\citenamefont
  {Franciolini}, \citenamefont {Iovino}, \citenamefont {Vaskonen},\ and\
  \citenamefont {Veermae}}]{Franciolini:2023pbf}%
  \BibitemOpen
  \bibfield  {author} {\bibinfo {author} {\bibfnamefont {G.}~\bibnamefont
  {Franciolini}}, \bibinfo {author} {\bibfnamefont {A.}~\bibnamefont {Iovino},
  \bibfnamefont {Junior.}}, \bibinfo {author} {\bibfnamefont {V.}~\bibnamefont
  {Vaskonen}},\ and\ \bibinfo {author} {\bibfnamefont {H.}~\bibnamefont
  {Veermae}},\ }\bibfield  {title} {\bibinfo {title} {{Recent Gravitational
  Wave Observation by Pulsar Timing Arrays and Primordial Black Holes: The
  Importance of Non-Gaussianities}},\ }\href
  {https://doi.org/10.1103/PhysRevLett.131.201401} {\bibfield  {journal}
  {\bibinfo  {journal} {Phys. Rev. Lett.}\ }\textbf {\bibinfo {volume} {131}},\
  \bibinfo {pages} {201401} (\bibinfo {year} {2023})},\ \Eprint
  {https://arxiv.org/abs/2306.17149} {arXiv:2306.17149 [astro-ph.CO]}
  \BibitemShut {NoStop}%
\bibitem [{\citenamefont {Flores}\ and\ \citenamefont
  {Kusenko}(2025)}]{Flores:2024eyy}%
  \BibitemOpen
  \bibfield  {author} {\bibinfo {author} {\bibfnamefont {M.~M.}\ \bibnamefont
  {Flores}}\ and\ \bibinfo {author} {\bibfnamefont {A.}~\bibnamefont
  {Kusenko}},\ }\bibinfo {title} {{New Ideas on~the~Formation and~Astrophysical
  Detection of~Primordial Black Holes}}\ (\bibinfo {year} {2025})\ \Eprint
  {https://arxiv.org/abs/2404.05430} {arXiv:2404.05430 [astro-ph.CO]}
  \BibitemShut {NoStop}%
\bibitem [{\citenamefont {Hawking}\ \emph {et~al.}(1982)\citenamefont
  {Hawking}, \citenamefont {Moss},\ and\ \citenamefont
  {Stewart}}]{Hawking:1982ga}%
  \BibitemOpen
  \bibfield  {author} {\bibinfo {author} {\bibfnamefont {S.~W.}\ \bibnamefont
  {Hawking}}, \bibinfo {author} {\bibfnamefont {I.~G.}\ \bibnamefont {Moss}},\
  and\ \bibinfo {author} {\bibfnamefont {J.~M.}\ \bibnamefont {Stewart}},\
  }\bibfield  {title} {\bibinfo {title} {{Bubble Collisions in the Very Early
  Universe}},\ }\href {https://doi.org/10.1103/PhysRevD.26.2681} {\bibfield
  {journal} {\bibinfo  {journal} {Phys. Rev. D}\ }\textbf {\bibinfo {volume}
  {26}},\ \bibinfo {pages} {2681} (\bibinfo {year} {1982})}\BibitemShut
  {NoStop}%
\bibitem [{\citenamefont {Jung}\ and\ \citenamefont
  {Okui}(2024)}]{Jung:2021mku}%
  \BibitemOpen
  \bibfield  {author} {\bibinfo {author} {\bibfnamefont {T.~H.}\ \bibnamefont
  {Jung}}\ and\ \bibinfo {author} {\bibfnamefont {T.}~\bibnamefont {Okui}},\
  }\bibfield  {title} {\bibinfo {title} {{Primordial black holes from bubble
  collisions during a first-order phase transition}},\ }\href
  {https://doi.org/10.1103/PhysRevD.110.115014} {\bibfield  {journal} {\bibinfo
   {journal} {Phys. Rev. D}\ }\textbf {\bibinfo {volume} {110}},\ \bibinfo
  {pages} {115014} (\bibinfo {year} {2024})},\ \Eprint
  {https://arxiv.org/abs/2110.04271} {arXiv:2110.04271 [hep-ph]} \BibitemShut
  {NoStop}%
\bibitem [{\citenamefont {Sato}\ \emph {et~al.}(1982)\citenamefont {Sato},
  \citenamefont {Kodama}, \citenamefont {Sasaki},\ and\ \citenamefont
  {Maeda}}]{Sato:1981gv}%
  \BibitemOpen
  \bibfield  {author} {\bibinfo {author} {\bibfnamefont {K.}~\bibnamefont
  {Sato}}, \bibinfo {author} {\bibfnamefont {H.}~\bibnamefont {Kodama}},
  \bibinfo {author} {\bibfnamefont {M.}~\bibnamefont {Sasaki}},\ and\ \bibinfo
  {author} {\bibfnamefont {K.-i.}\ \bibnamefont {Maeda}},\ }\bibfield  {title}
  {\bibinfo {title} {{Multiproduction of Universes by First Order Phase
  Transition of a Vacuum}},\ }\href
  {https://doi.org/10.1016/0370-2693(82)91152-2} {\bibfield  {journal}
  {\bibinfo  {journal} {Phys. Lett. B}\ }\textbf {\bibinfo {volume} {108}},\
  \bibinfo {pages} {103} (\bibinfo {year} {1982})}\BibitemShut {NoStop}%
\bibitem [{\citenamefont {Kodama}\ \emph {et~al.}(1982)\citenamefont {Kodama},
  \citenamefont {Sasaki},\ and\ \citenamefont {Sato}}]{Kodama:1982sf}%
  \BibitemOpen
  \bibfield  {author} {\bibinfo {author} {\bibfnamefont {H.}~\bibnamefont
  {Kodama}}, \bibinfo {author} {\bibfnamefont {M.}~\bibnamefont {Sasaki}},\
  and\ \bibinfo {author} {\bibfnamefont {K.}~\bibnamefont {Sato}},\ }\bibfield
  {title} {\bibinfo {title} {{Abundance of Primordial Holes Produced by
  Cosmological First Order Phase Transition}},\ }\href
  {https://doi.org/10.1143/PTP.68.1979} {\bibfield  {journal} {\bibinfo
  {journal} {Prog. Theor. Phys.}\ }\textbf {\bibinfo {volume} {68}},\ \bibinfo
  {pages} {1979} (\bibinfo {year} {1982})}\BibitemShut {NoStop}%
\bibitem [{\citenamefont {Gross}\ \emph {et~al.}(2021)\citenamefont {Gross},
  \citenamefont {Landini}, \citenamefont {Strumia},\ and\ \citenamefont
  {Teresi}}]{Gross:2021qgx}%
  \BibitemOpen
  \bibfield  {author} {\bibinfo {author} {\bibfnamefont {C.}~\bibnamefont
  {Gross}}, \bibinfo {author} {\bibfnamefont {G.}~\bibnamefont {Landini}},
  \bibinfo {author} {\bibfnamefont {A.}~\bibnamefont {Strumia}},\ and\ \bibinfo
  {author} {\bibfnamefont {D.}~\bibnamefont {Teresi}},\ }\bibfield  {title}
  {\bibinfo {title} {{Dark Matter as dark dwarfs and other macroscopic objects:
  multiverse relics?}},\ }\href {https://doi.org/10.1007/JHEP09(2021)033}
  {\bibfield  {journal} {\bibinfo  {journal} {JHEP}\ }\textbf {\bibinfo
  {volume} {09}},\ \bibinfo {pages} {033}},\ \Eprint
  {https://arxiv.org/abs/2105.02840} {arXiv:2105.02840 [hep-ph]} \BibitemShut
  {NoStop}%
\bibitem [{\citenamefont {Lewicki}\ \emph
  {et~al.}(2023{\natexlab{a}})\citenamefont {Lewicki}, \citenamefont {Toczek},\
  and\ \citenamefont {Vaskonen}}]{Lewicki:2023ioy}%
  \BibitemOpen
  \bibfield  {author} {\bibinfo {author} {\bibfnamefont {M.}~\bibnamefont
  {Lewicki}}, \bibinfo {author} {\bibfnamefont {P.}~\bibnamefont {Toczek}},\
  and\ \bibinfo {author} {\bibfnamefont {V.}~\bibnamefont {Vaskonen}},\
  }\bibfield  {title} {\bibinfo {title} {{Primordial black holes from strong
  first-order phase transitions}},\ }\href
  {https://doi.org/10.1007/JHEP09(2023)092} {\bibfield  {journal} {\bibinfo
  {journal} {JHEP}\ }\textbf {\bibinfo {volume} {09}},\ \bibinfo {pages}
  {092}},\ \Eprint {https://arxiv.org/abs/2305.04924} {arXiv:2305.04924
  [astro-ph.CO]} \BibitemShut {NoStop}%
\bibitem [{\citenamefont {Deng}\ and\ \citenamefont
  {Vilenkin}(2017)}]{Deng:2017uwc}%
  \BibitemOpen
  \bibfield  {author} {\bibinfo {author} {\bibfnamefont {H.}~\bibnamefont
  {Deng}}\ and\ \bibinfo {author} {\bibfnamefont {A.}~\bibnamefont
  {Vilenkin}},\ }\bibfield  {title} {\bibinfo {title} {{Primordial black hole
  formation by vacuum bubbles}},\ }\href
  {https://doi.org/10.1088/1475-7516/2017/12/044} {\bibfield  {journal}
  {\bibinfo  {journal} {JCAP}\ }\textbf {\bibinfo {volume} {12}},\ \bibinfo
  {pages} {044}},\ \Eprint {https://arxiv.org/abs/1710.02865} {arXiv:1710.02865
  [gr-qc]} \BibitemShut {NoStop}%
\bibitem [{\citenamefont {Kusenko}\ \emph {et~al.}(2020)\citenamefont
  {Kusenko}, \citenamefont {Sasaki}, \citenamefont {Sugiyama}, \citenamefont
  {Takada}, \citenamefont {Takhistov},\ and\ \citenamefont
  {Vitagliano}}]{Kusenko:2020pcg}%
  \BibitemOpen
  \bibfield  {author} {\bibinfo {author} {\bibfnamefont {A.}~\bibnamefont
  {Kusenko}}, \bibinfo {author} {\bibfnamefont {M.}~\bibnamefont {Sasaki}},
  \bibinfo {author} {\bibfnamefont {S.}~\bibnamefont {Sugiyama}}, \bibinfo
  {author} {\bibfnamefont {M.}~\bibnamefont {Takada}}, \bibinfo {author}
  {\bibfnamefont {V.}~\bibnamefont {Takhistov}},\ and\ \bibinfo {author}
  {\bibfnamefont {E.}~\bibnamefont {Vitagliano}},\ }\bibfield  {title}
  {\bibinfo {title} {{Exploring Primordial Black Holes from the Multiverse with
  Optical Telescopes}},\ }\href
  {https://doi.org/10.1103/PhysRevLett.125.181304} {\bibfield  {journal}
  {\bibinfo  {journal} {Phys. Rev. Lett.}\ }\textbf {\bibinfo {volume} {125}},\
  \bibinfo {pages} {181304} (\bibinfo {year} {2020})},\ \Eprint
  {https://arxiv.org/abs/2001.09160} {arXiv:2001.09160 [astro-ph.CO]}
  \BibitemShut {NoStop}%
\bibitem [{\citenamefont {Maeso}\ \emph {et~al.}(2022)\citenamefont {Maeso},
  \citenamefont {Marzola}, \citenamefont {Raidal}, \citenamefont {Vaskonen},\
  and\ \citenamefont {Veerm{\"a}e}}]{Maeso:2021xvl}%
  \BibitemOpen
  \bibfield  {author} {\bibinfo {author} {\bibfnamefont {D.~N.}\ \bibnamefont
  {Maeso}}, \bibinfo {author} {\bibfnamefont {L.}~\bibnamefont {Marzola}},
  \bibinfo {author} {\bibfnamefont {M.}~\bibnamefont {Raidal}}, \bibinfo
  {author} {\bibfnamefont {V.}~\bibnamefont {Vaskonen}},\ and\ \bibinfo
  {author} {\bibfnamefont {H.}~\bibnamefont {Veerm{\"a}e}},\ }\bibfield
  {title} {\bibinfo {title} {{Primordial black holes from spectator field
  bubbles}},\ }\href {https://doi.org/10.1088/1475-7516/2022/02/017} {\bibfield
   {journal} {\bibinfo  {journal} {JCAP}\ }\textbf {\bibinfo {volume}
  {02}}\bibfield  {number} {\bibinfo  {number} { (02)},\ \bibinfo {pages}
  {017}},\ }\Eprint {https://arxiv.org/abs/2112.01505} {arXiv:2112.01505
  [astro-ph.CO]} \BibitemShut {NoStop}%
\bibitem [{\citenamefont {Wang}\ \emph
  {et~al.}(2025{\natexlab{b}})\citenamefont {Wang}, \citenamefont {Zhang},\
  and\ \citenamefont {Suyama}}]{Wang:2025hwc}%
  \BibitemOpen
  \bibfield  {author} {\bibinfo {author} {\bibfnamefont {H.}~\bibnamefont
  {Wang}}, \bibinfo {author} {\bibfnamefont {Y.-l.}\ \bibnamefont {Zhang}},\
  and\ \bibinfo {author} {\bibfnamefont {T.}~\bibnamefont {Suyama}},\
  }\bibfield  {title} {\bibinfo {title} {{Nearly Monochromatic Primordial Black
  Holes as total Dark Matter from Bubble Collapse}},\ }\href@noop {} {\
  (\bibinfo {year} {2025}{\natexlab{b}})},\ \Eprint
  {https://arxiv.org/abs/2510.19233} {arXiv:2510.19233 [astro-ph.CO]}
  \BibitemShut {NoStop}%
\bibitem [{\citenamefont {Liu}\ \emph {et~al.}(2022)\citenamefont {Liu},
  \citenamefont {Bian}, \citenamefont {Cai}, \citenamefont {Guo},\ and\
  \citenamefont {Wang}}]{Liu:2021svg}%
  \BibitemOpen
  \bibfield  {author} {\bibinfo {author} {\bibfnamefont {J.}~\bibnamefont
  {Liu}}, \bibinfo {author} {\bibfnamefont {L.}~\bibnamefont {Bian}}, \bibinfo
  {author} {\bibfnamefont {R.-G.}\ \bibnamefont {Cai}}, \bibinfo {author}
  {\bibfnamefont {Z.-K.}\ \bibnamefont {Guo}},\ and\ \bibinfo {author}
  {\bibfnamefont {S.-J.}\ \bibnamefont {Wang}},\ }\bibfield  {title} {\bibinfo
  {title} {{Primordial black hole production during first-order phase
  transitions}},\ }\href {https://doi.org/10.1103/PhysRevD.105.L021303}
  {\bibfield  {journal} {\bibinfo  {journal} {Phys. Rev. D}\ }\textbf {\bibinfo
  {volume} {105}},\ \bibinfo {pages} {L021303} (\bibinfo {year} {2022})},\
  \Eprint {https://arxiv.org/abs/2106.05637} {arXiv:2106.05637 [astro-ph.CO]}
  \BibitemShut {NoStop}%
\bibitem [{\citenamefont {Kawana}\ \emph {et~al.}(2023)\citenamefont {Kawana},
  \citenamefont {Kim},\ and\ \citenamefont {Lu}}]{Kawana:2022olo}%
  \BibitemOpen
  \bibfield  {author} {\bibinfo {author} {\bibfnamefont {K.}~\bibnamefont
  {Kawana}}, \bibinfo {author} {\bibfnamefont {T.}~\bibnamefont {Kim}},\ and\
  \bibinfo {author} {\bibfnamefont {P.}~\bibnamefont {Lu}},\ }\bibfield
  {title} {\bibinfo {title} {{PBH formation from overdensities in delayed
  vacuum transitions}},\ }\href {https://doi.org/10.1103/PhysRevD.108.103531}
  {\bibfield  {journal} {\bibinfo  {journal} {Phys. Rev. D}\ }\textbf {\bibinfo
  {volume} {108}},\ \bibinfo {pages} {103531} (\bibinfo {year} {2023})},\
  \Eprint {https://arxiv.org/abs/2212.14037} {arXiv:2212.14037 [astro-ph.CO]}
  \BibitemShut {NoStop}%
\bibitem [{\citenamefont {Gouttenoire}\ and\ \citenamefont
  {Volansky}(2024)}]{Gouttenoire:2023naa}%
  \BibitemOpen
  \bibfield  {author} {\bibinfo {author} {\bibfnamefont {Y.}~\bibnamefont
  {Gouttenoire}}\ and\ \bibinfo {author} {\bibfnamefont {T.}~\bibnamefont
  {Volansky}},\ }\bibfield  {title} {\bibinfo {title} {{Primordial black holes
  from supercooled phase transitions}},\ }\href
  {https://doi.org/10.1103/PhysRevD.110.043514} {\bibfield  {journal} {\bibinfo
   {journal} {Phys. Rev. D}\ }\textbf {\bibinfo {volume} {110}},\ \bibinfo
  {pages} {043514} (\bibinfo {year} {2024})},\ \Eprint
  {https://arxiv.org/abs/2305.04942} {arXiv:2305.04942 [hep-ph]} \BibitemShut
  {NoStop}%
\bibitem [{\citenamefont {Baldes}\ and\ \citenamefont
  {Olea-Romacho}(2024)}]{Baldes:2023rqv}%
  \BibitemOpen
  \bibfield  {author} {\bibinfo {author} {\bibfnamefont {I.}~\bibnamefont
  {Baldes}}\ and\ \bibinfo {author} {\bibfnamefont {M.~O.}\ \bibnamefont
  {Olea-Romacho}},\ }\bibfield  {title} {\bibinfo {title} {{Primordial black
  holes as dark matter: interferometric tests of phase transition origin}},\
  }\href {https://doi.org/10.1007/JHEP01(2024)133} {\bibfield  {journal}
  {\bibinfo  {journal} {JHEP}\ }\textbf {\bibinfo {volume} {01}},\ \bibinfo
  {pages} {133}},\ \Eprint {https://arxiv.org/abs/2307.11639} {arXiv:2307.11639
  [hep-ph]} \BibitemShut {NoStop}%
\bibitem [{\citenamefont {Lewicki}\ \emph {et~al.}(2025)\citenamefont
  {Lewicki}, \citenamefont {Toczek},\ and\ \citenamefont
  {Vaskonen}}]{Lewicki:2024sfw}%
  \BibitemOpen
  \bibfield  {author} {\bibinfo {author} {\bibfnamefont {M.}~\bibnamefont
  {Lewicki}}, \bibinfo {author} {\bibfnamefont {P.}~\bibnamefont {Toczek}},\
  and\ \bibinfo {author} {\bibfnamefont {V.}~\bibnamefont {Vaskonen}},\
  }\bibfield  {title} {\bibinfo {title} {{Black holes and gravitational waves
  from phase transitions in realistic models}},\ }\href
  {https://doi.org/10.1016/j.dark.2025.102075} {\bibfield  {journal} {\bibinfo
  {journal} {Phys. Dark Univ.}\ }\textbf {\bibinfo {volume} {50}},\ \bibinfo
  {pages} {102075} (\bibinfo {year} {2025})},\ \Eprint
  {https://arxiv.org/abs/2412.10366} {arXiv:2412.10366 [astro-ph.CO]}
  \BibitemShut {NoStop}%
\bibitem [{\citenamefont {Franciolini}\ \emph {et~al.}(2025)\citenamefont
  {Franciolini}, \citenamefont {Gouttenoire},\ and\ \citenamefont
  {Jinno}}]{Franciolini:2025ztf}%
  \BibitemOpen
  \bibfield  {author} {\bibinfo {author} {\bibfnamefont {G.}~\bibnamefont
  {Franciolini}}, \bibinfo {author} {\bibfnamefont {Y.}~\bibnamefont
  {Gouttenoire}},\ and\ \bibinfo {author} {\bibfnamefont {R.}~\bibnamefont
  {Jinno}},\ }\bibfield  {title} {\bibinfo {title} {{Curvature Perturbations
  from First-Order Phase Transitions: Implications to Black Holes and
  Gravitational Waves}},\ }\href@noop {} {\  (\bibinfo {year} {2025})},\
  \Eprint {https://arxiv.org/abs/2503.01962} {arXiv:2503.01962 [hep-ph]}
  \BibitemShut {NoStop}%
\bibitem [{\citenamefont {Kierkla}\ \emph {et~al.}(2025)\citenamefont
  {Kierkla}, \citenamefont {Ramberg}, \citenamefont {Schicho},\ and\
  \citenamefont {Schmitt}}]{Kierkla:2025vwp}%
  \BibitemOpen
  \bibfield  {author} {\bibinfo {author} {\bibfnamefont {M.}~\bibnamefont
  {Kierkla}}, \bibinfo {author} {\bibfnamefont {N.}~\bibnamefont {Ramberg}},
  \bibinfo {author} {\bibfnamefont {P.}~\bibnamefont {Schicho}},\ and\ \bibinfo
  {author} {\bibfnamefont {D.}~\bibnamefont {Schmitt}},\ }\bibfield  {title}
  {\bibinfo {title} {{Theoretical uncertainties for primordial black holes from
  cosmological phase transitions}},\ }\href@noop {} {\  (\bibinfo {year}
  {2025})},\ \Eprint {https://arxiv.org/abs/2506.15496} {arXiv:2506.15496
  [hep-ph]} \BibitemShut {NoStop}%
\bibitem [{\citenamefont {Gouttenoire}\ \emph {et~al.}(2026)\citenamefont
  {Gouttenoire}, \citenamefont {Lewicki}, \citenamefont {Toczek},\ and\
  \citenamefont {Vaskonen}}]{Gouttenoire:2026xxx}%
  \BibitemOpen
  \bibfield  {author} {\bibinfo {author} {\bibfnamefont {Y.}~\bibnamefont
  {Gouttenoire}}, \bibinfo {author} {\bibfnamefont {M.}~\bibnamefont
  {Lewicki}}, \bibinfo {author} {\bibfnamefont {P.}~\bibnamefont {Toczek}},\
  and\ \bibinfo {author} {\bibfnamefont {V.}~\bibnamefont {Vaskonen}},\
  }\bibfield  {title} {\bibinfo {title} {{Resurrecting PBH formation in
  first-order phase transitions}},\ }\href@noop {} {\  (\bibinfo {year}
  {2026})},\ \Eprint {https://arxiv.org/abs/to appear} {to appear} \BibitemShut
  {NoStop}%
\bibitem [{\citenamefont {Dimopoulos}\ \emph {et~al.}(2019)\citenamefont
  {Dimopoulos}, \citenamefont {Markkanen}, \citenamefont {Racioppi},\ and\
  \citenamefont {Vaskonen}}]{Dimopoulos:2019wew}%
  \BibitemOpen
  \bibfield  {author} {\bibinfo {author} {\bibfnamefont {K.}~\bibnamefont
  {Dimopoulos}}, \bibinfo {author} {\bibfnamefont {T.}~\bibnamefont
  {Markkanen}}, \bibinfo {author} {\bibfnamefont {A.}~\bibnamefont
  {Racioppi}},\ and\ \bibinfo {author} {\bibfnamefont {V.}~\bibnamefont
  {Vaskonen}},\ }\bibfield  {title} {\bibinfo {title} {{Primordial Black Holes
  from Thermal Inflation}},\ }\href
  {https://doi.org/10.1088/1475-7516/2019/07/046} {\bibfield  {journal}
  {\bibinfo  {journal} {JCAP}\ }\textbf {\bibinfo {volume} {07}},\ \bibinfo
  {pages} {046}},\ \Eprint {https://arxiv.org/abs/1903.09598} {arXiv:1903.09598
  [astro-ph.CO]} \BibitemShut {NoStop}%
\bibitem [{\citenamefont {Bastero-Gil}\ \emph {et~al.}(2024)\citenamefont
  {Bastero-Gil}, \citenamefont {Gomes},\ and\ \citenamefont
  {Rosa}}]{Bastero-Gil:2023sub}%
  \BibitemOpen
  \bibfield  {author} {\bibinfo {author} {\bibfnamefont {M.}~\bibnamefont
  {Bastero-Gil}}, \bibinfo {author} {\bibfnamefont {J.~M.}\ \bibnamefont
  {Gomes}},\ and\ \bibinfo {author} {\bibfnamefont {J.~G.}\ \bibnamefont
  {Rosa}},\ }\bibfield  {title} {\bibinfo {title} {{Thermal curvature
  perturbations in thermal inflation}},\ }\href
  {https://doi.org/10.1088/1475-7516/2024/06/060} {\bibfield  {journal}
  {\bibinfo  {journal} {JCAP}\ }\textbf {\bibinfo {volume} {06}},\ \bibinfo
  {pages} {060}},\ \Eprint {https://arxiv.org/abs/2301.11666} {arXiv:2301.11666
  [hep-ph]} \BibitemShut {NoStop}%
\bibitem [{\citenamefont {Kitajima}\ and\ \citenamefont
  {Takahashi}(2020)}]{Kitajima:2020kig}%
  \BibitemOpen
  \bibfield  {author} {\bibinfo {author} {\bibfnamefont {N.}~\bibnamefont
  {Kitajima}}\ and\ \bibinfo {author} {\bibfnamefont {F.}~\bibnamefont
  {Takahashi}},\ }\bibfield  {title} {\bibinfo {title} {{Primordial Black Holes
  from QCD Axion Bubbles}},\ }\href
  {https://doi.org/10.1088/1475-7516/2020/11/060} {\bibfield  {journal}
  {\bibinfo  {journal} {JCAP}\ }\textbf {\bibinfo {volume} {11}},\ \bibinfo
  {pages} {060}},\ \Eprint {https://arxiv.org/abs/2006.13137} {arXiv:2006.13137
  [hep-ph]} \BibitemShut {NoStop}%
\bibitem [{\citenamefont {Baker}\ \emph
  {et~al.}(2025{\natexlab{a}})\citenamefont {Baker}, \citenamefont {Breitbach},
  \citenamefont {Kopp},\ and\ \citenamefont {Mittnacht}}]{Baker:2021nyl}%
  \BibitemOpen
  \bibfield  {author} {\bibinfo {author} {\bibfnamefont {M.~J.}\ \bibnamefont
  {Baker}}, \bibinfo {author} {\bibfnamefont {M.}~\bibnamefont {Breitbach}},
  \bibinfo {author} {\bibfnamefont {J.}~\bibnamefont {Kopp}},\ and\ \bibinfo
  {author} {\bibfnamefont {L.}~\bibnamefont {Mittnacht}},\ }\bibfield  {title}
  {\bibinfo {title} {{Primordial black holes from first-order cosmological
  phase transitions}},\ }\href {https://doi.org/10.1016/j.physletb.2025.139625}
  {\bibfield  {journal} {\bibinfo  {journal} {Phys. Lett. B}\ }\textbf
  {\bibinfo {volume} {868}},\ \bibinfo {pages} {139625} (\bibinfo {year}
  {2025}{\natexlab{a}})},\ \Eprint {https://arxiv.org/abs/2105.07481}
  {arXiv:2105.07481 [astro-ph.CO]} \BibitemShut {NoStop}%
\bibitem [{\citenamefont {Lewicki}\ \emph
  {et~al.}(2023{\natexlab{b}})\citenamefont {Lewicki}, \citenamefont
  {M{\"u}{\"u}rsepp}, \citenamefont {Pata}, \citenamefont {Vasar},
  \citenamefont {Vaskonen},\ and\ \citenamefont
  {Veerm{\"a}e}}]{Lewicki:2023mik}%
  \BibitemOpen
  \bibfield  {author} {\bibinfo {author} {\bibfnamefont {M.}~\bibnamefont
  {Lewicki}}, \bibinfo {author} {\bibfnamefont {K.}~\bibnamefont
  {M{\"u}{\"u}rsepp}}, \bibinfo {author} {\bibfnamefont {J.}~\bibnamefont
  {Pata}}, \bibinfo {author} {\bibfnamefont {M.}~\bibnamefont {Vasar}},
  \bibinfo {author} {\bibfnamefont {V.}~\bibnamefont {Vaskonen}},\ and\
  \bibinfo {author} {\bibfnamefont {H.}~\bibnamefont {Veerm{\"a}e}},\
  }\bibfield  {title} {\bibinfo {title} {{Dynamics of false vacuum bubbles with
  trapped particles}},\ }\href {https://doi.org/10.1103/PhysRevD.108.036023}
  {\bibfield  {journal} {\bibinfo  {journal} {Phys. Rev. D}\ }\textbf {\bibinfo
  {volume} {108}},\ \bibinfo {pages} {036023} (\bibinfo {year}
  {2023}{\natexlab{b}})},\ \Eprint {https://arxiv.org/abs/2305.07702}
  {arXiv:2305.07702 [hep-ph]} \BibitemShut {NoStop}%
\bibitem [{\citenamefont {Kawana}\ and\ \citenamefont
  {Xie}(2022)}]{Kawana:2021tde}%
  \BibitemOpen
  \bibfield  {author} {\bibinfo {author} {\bibfnamefont {K.}~\bibnamefont
  {Kawana}}\ and\ \bibinfo {author} {\bibfnamefont {K.-P.}\ \bibnamefont
  {Xie}},\ }\bibfield  {title} {\bibinfo {title} {{Primordial black holes from
  a cosmic phase transition: The collapse of Fermi-balls}},\ }\href
  {https://doi.org/10.1016/j.physletb.2021.136791} {\bibfield  {journal}
  {\bibinfo  {journal} {Phys. Lett. B}\ }\textbf {\bibinfo {volume} {824}},\
  \bibinfo {pages} {136791} (\bibinfo {year} {2022})},\ \Eprint
  {https://arxiv.org/abs/2106.00111} {arXiv:2106.00111 [astro-ph.CO]}
  \BibitemShut {NoStop}%
\bibitem [{\citenamefont {Vilenkin}\ and\ \citenamefont
  {Shellard}(2000)}]{Vilenkin:2000jqa}%
  \BibitemOpen
  \bibfield  {author} {\bibinfo {author} {\bibfnamefont {A.}~\bibnamefont
  {Vilenkin}}\ and\ \bibinfo {author} {\bibfnamefont {E.~P.~S.}\ \bibnamefont
  {Shellard}},\ }\href@noop {} {\emph {\bibinfo {title} {{Cosmic Strings and
  Other Topological Defects}}}}\ (\bibinfo  {publisher} {Cambridge University
  Press},\ \bibinfo {year} {2000})\BibitemShut {NoStop}%
\bibitem [{\citenamefont {Hawking}(1989)}]{Hawking:1987bn}%
  \BibitemOpen
  \bibfield  {author} {\bibinfo {author} {\bibfnamefont {S.~W.}\ \bibnamefont
  {Hawking}},\ }\bibfield  {title} {\bibinfo {title} {{Black Holes From Cosmic
  Strings}},\ }\href {https://doi.org/10.1016/0370-2693(89)90206-2} {\bibfield
  {journal} {\bibinfo  {journal} {Phys. Lett. B}\ }\textbf {\bibinfo {volume}
  {231}},\ \bibinfo {pages} {237} (\bibinfo {year} {1989})}\BibitemShut
  {NoStop}%
\bibitem [{\citenamefont {Polnarev}\ and\ \citenamefont
  {Zembowicz}(1991)}]{Polnarev:1988dh}%
  \BibitemOpen
  \bibfield  {author} {\bibinfo {author} {\bibfnamefont {A.}~\bibnamefont
  {Polnarev}}\ and\ \bibinfo {author} {\bibfnamefont {R.}~\bibnamefont
  {Zembowicz}},\ }\bibfield  {title} {\bibinfo {title} {{Formation of
  Primordial Black Holes by Cosmic Strings}},\ }\href
  {https://doi.org/10.1103/PhysRevD.43.1106} {\bibfield  {journal} {\bibinfo
  {journal} {Phys. Rev. D}\ }\textbf {\bibinfo {volume} {43}},\ \bibinfo
  {pages} {1106} (\bibinfo {year} {1991})}\BibitemShut {NoStop}%
\bibitem [{\citenamefont {Caldwell}\ and\ \citenamefont
  {Casper}(1996)}]{Caldwell:1995fu}%
  \BibitemOpen
  \bibfield  {author} {\bibinfo {author} {\bibfnamefont {R.~R.}\ \bibnamefont
  {Caldwell}}\ and\ \bibinfo {author} {\bibfnamefont {P.}~\bibnamefont
  {Casper}},\ }\bibfield  {title} {\bibinfo {title} {{Formation of black holes
  from collapsed cosmic string loops}},\ }\href
  {https://doi.org/10.1103/PhysRevD.53.3002} {\bibfield  {journal} {\bibinfo
  {journal} {Phys. Rev. D}\ }\textbf {\bibinfo {volume} {53}},\ \bibinfo
  {pages} {3002} (\bibinfo {year} {1996})},\ \Eprint
  {https://arxiv.org/abs/gr-qc/9509012} {arXiv:gr-qc/9509012} \BibitemShut
  {NoStop}%
\bibitem [{\citenamefont {MacGibbon}\ \emph {et~al.}(1998)\citenamefont
  {MacGibbon}, \citenamefont {Brandenberger},\ and\ \citenamefont
  {Wichoski}}]{MacGibbon:1997pu}%
  \BibitemOpen
  \bibfield  {author} {\bibinfo {author} {\bibfnamefont {J.~H.}\ \bibnamefont
  {MacGibbon}}, \bibinfo {author} {\bibfnamefont {R.~H.}\ \bibnamefont
  {Brandenberger}},\ and\ \bibinfo {author} {\bibfnamefont {U.~F.}\
  \bibnamefont {Wichoski}},\ }\bibfield  {title} {\bibinfo {title} {{Limits on
  black hole formation from cosmic string loops}},\ }\href
  {https://doi.org/10.1103/PhysRevD.57.2158} {\bibfield  {journal} {\bibinfo
  {journal} {Phys. Rev. D}\ }\textbf {\bibinfo {volume} {57}},\ \bibinfo
  {pages} {2158} (\bibinfo {year} {1998})},\ \Eprint
  {https://arxiv.org/abs/astro-ph/9707146} {arXiv:astro-ph/9707146}
  \BibitemShut {NoStop}%
\bibitem [{\citenamefont {Hansen}\ \emph {et~al.}(2000)\citenamefont {Hansen},
  \citenamefont {Christensen},\ and\ \citenamefont {Larsen}}]{Hansen:2000jv}%
  \BibitemOpen
  \bibfield  {author} {\bibinfo {author} {\bibfnamefont {R.~N.}\ \bibnamefont
  {Hansen}}, \bibinfo {author} {\bibfnamefont {M.}~\bibnamefont
  {Christensen}},\ and\ \bibinfo {author} {\bibfnamefont {A.~L.}\ \bibnamefont
  {Larsen}},\ }\bibfield  {title} {\bibinfo {title} {{Comment on `Formation of
  primordial black holes by cosmic strings'}},\ }\href
  {https://doi.org/10.1103/PhysRevD.61.108701} {\bibfield  {journal} {\bibinfo
  {journal} {Phys. Rev. D}\ }\textbf {\bibinfo {volume} {61}},\ \bibinfo
  {pages} {108701} (\bibinfo {year} {2000})},\ \Eprint
  {https://arxiv.org/abs/gr-qc/0005041} {arXiv:gr-qc/0005041} \BibitemShut
  {NoStop}%
\bibitem [{\citenamefont {Helfer}\ \emph {et~al.}(2019)\citenamefont {Helfer},
  \citenamefont {Aurrekoetxea},\ and\ \citenamefont {Lim}}]{Helfer:2018qgv}%
  \BibitemOpen
  \bibfield  {author} {\bibinfo {author} {\bibfnamefont {T.}~\bibnamefont
  {Helfer}}, \bibinfo {author} {\bibfnamefont {J.~C.}\ \bibnamefont
  {Aurrekoetxea}},\ and\ \bibinfo {author} {\bibfnamefont {E.~A.}\ \bibnamefont
  {Lim}},\ }\bibfield  {title} {\bibinfo {title} {{Cosmic String Loop Collapse
  in Full General Relativity}},\ }\href
  {https://doi.org/10.1103/PhysRevD.99.104028} {\bibfield  {journal} {\bibinfo
  {journal} {Phys. Rev. D}\ }\textbf {\bibinfo {volume} {99}},\ \bibinfo
  {pages} {104028} (\bibinfo {year} {2019})},\ \Eprint
  {https://arxiv.org/abs/1808.06678} {arXiv:1808.06678 [gr-qc]} \BibitemShut
  {NoStop}%
\bibitem [{\citenamefont {James-Turner}\ \emph {et~al.}(2020)\citenamefont
  {James-Turner}, \citenamefont {Weil}, \citenamefont {Green},\ and\
  \citenamefont {Copeland}}]{James-Turner:2019ssu}%
  \BibitemOpen
  \bibfield  {author} {\bibinfo {author} {\bibfnamefont {C.}~\bibnamefont
  {James-Turner}}, \bibinfo {author} {\bibfnamefont {D.~P.~B.}\ \bibnamefont
  {Weil}}, \bibinfo {author} {\bibfnamefont {A.~M.}\ \bibnamefont {Green}},\
  and\ \bibinfo {author} {\bibfnamefont {E.~J.}\ \bibnamefont {Copeland}},\
  }\bibfield  {title} {\bibinfo {title} {{Constraints on the cosmic string loop
  collapse fraction from primordial black holes}},\ }\href
  {https://doi.org/10.1103/PhysRevD.101.123526} {\bibfield  {journal} {\bibinfo
   {journal} {Phys. Rev. D}\ }\textbf {\bibinfo {volume} {101}},\ \bibinfo
  {pages} {123526} (\bibinfo {year} {2020})},\ \Eprint
  {https://arxiv.org/abs/1911.12658} {arXiv:1911.12658 [astro-ph.CO]}
  \BibitemShut {NoStop}%
\bibitem [{\citenamefont {Rubin}\ \emph {et~al.}(2000)\citenamefont {Rubin},
  \citenamefont {Khlopov},\ and\ \citenamefont {Sakharov}}]{Rubin:2000dq}%
  \BibitemOpen
  \bibfield  {author} {\bibinfo {author} {\bibfnamefont {S.~G.}\ \bibnamefont
  {Rubin}}, \bibinfo {author} {\bibfnamefont {M.~Y.}\ \bibnamefont {Khlopov}},\
  and\ \bibinfo {author} {\bibfnamefont {A.~S.}\ \bibnamefont {Sakharov}},\
  }\bibfield  {title} {\bibinfo {title} {{Primordial black holes from
  nonequilibrium second order phase transition}},\ }\href@noop {} {\bibfield
  {journal} {\bibinfo  {journal} {Grav. Cosmol.}\ }\textbf {\bibinfo {volume}
  {6}},\ \bibinfo {pages} {51} (\bibinfo {year} {2000})},\ \Eprint
  {https://arxiv.org/abs/hep-ph/0005271} {arXiv:hep-ph/0005271} \BibitemShut
  {NoStop}%
\bibitem [{\citenamefont {Ferrer}\ \emph {et~al.}(2019)\citenamefont {Ferrer},
  \citenamefont {Masso}, \citenamefont {Panico}, \citenamefont {Pujolas},\ and\
  \citenamefont {Rompineve}}]{Ferrer:2018uiu}%
  \BibitemOpen
  \bibfield  {author} {\bibinfo {author} {\bibfnamefont {F.}~\bibnamefont
  {Ferrer}}, \bibinfo {author} {\bibfnamefont {E.}~\bibnamefont {Masso}},
  \bibinfo {author} {\bibfnamefont {G.}~\bibnamefont {Panico}}, \bibinfo
  {author} {\bibfnamefont {O.}~\bibnamefont {Pujolas}},\ and\ \bibinfo {author}
  {\bibfnamefont {F.}~\bibnamefont {Rompineve}},\ }\bibfield  {title} {\bibinfo
  {title} {{Primordial Black Holes from the QCD axion}},\ }\href
  {https://doi.org/10.1103/PhysRevLett.122.101301} {\bibfield  {journal}
  {\bibinfo  {journal} {Phys. Rev. Lett.}\ }\textbf {\bibinfo {volume} {122}},\
  \bibinfo {pages} {101301} (\bibinfo {year} {2019})},\ \Eprint
  {https://arxiv.org/abs/1807.01707} {arXiv:1807.01707 [hep-ph]} \BibitemShut
  {NoStop}%
\bibitem [{\citenamefont {Gelmini}\ \emph {et~al.}(2023)\citenamefont
  {Gelmini}, \citenamefont {Hyman}, \citenamefont {Simpson},\ and\
  \citenamefont {Vitagliano}}]{Gelmini:2023ngs}%
  \BibitemOpen
  \bibfield  {author} {\bibinfo {author} {\bibfnamefont {G.~B.}\ \bibnamefont
  {Gelmini}}, \bibinfo {author} {\bibfnamefont {J.}~\bibnamefont {Hyman}},
  \bibinfo {author} {\bibfnamefont {A.}~\bibnamefont {Simpson}},\ and\ \bibinfo
  {author} {\bibfnamefont {E.}~\bibnamefont {Vitagliano}},\ }\bibfield  {title}
  {\bibinfo {title} {{Primordial black hole dark matter from catastrogenesis
  with unstable pseudo-Goldstone bosons}},\ }\href
  {https://doi.org/10.1088/1475-7516/2023/06/055} {\bibfield  {journal}
  {\bibinfo  {journal} {JCAP}\ }\textbf {\bibinfo {volume} {06}},\ \bibinfo
  {pages} {055}},\ \Eprint {https://arxiv.org/abs/2303.14107} {arXiv:2303.14107
  [hep-ph]} \BibitemShut {NoStop}%
\bibitem [{\citenamefont {Gouttenoire}\ and\ \citenamefont
  {Vitagliano}(2024{\natexlab{a}})}]{Gouttenoire:2023ftk}%
  \BibitemOpen
  \bibfield  {author} {\bibinfo {author} {\bibfnamefont {Y.}~\bibnamefont
  {Gouttenoire}}\ and\ \bibinfo {author} {\bibfnamefont {E.}~\bibnamefont
  {Vitagliano}},\ }\bibfield  {title} {\bibinfo {title} {{Domain wall
  interpretation of the PTA signal confronting black hole overproduction}},\
  }\href {https://doi.org/10.1103/PhysRevD.110.L061306} {\bibfield  {journal}
  {\bibinfo  {journal} {Phys. Rev. D}\ }\textbf {\bibinfo {volume} {110}},\
  \bibinfo {pages} {L061306} (\bibinfo {year} {2024}{\natexlab{a}})},\ \Eprint
  {https://arxiv.org/abs/2306.17841} {arXiv:2306.17841 [gr-qc]} \BibitemShut
  {NoStop}%
\bibitem [{\citenamefont {Gouttenoire}\ and\ \citenamefont
  {Vitagliano}(2024{\natexlab{b}})}]{Gouttenoire:2023gbn}%
  \BibitemOpen
  \bibfield  {author} {\bibinfo {author} {\bibfnamefont {Y.}~\bibnamefont
  {Gouttenoire}}\ and\ \bibinfo {author} {\bibfnamefont {E.}~\bibnamefont
  {Vitagliano}},\ }\bibfield  {title} {\bibinfo {title} {{Primordial black
  holes and wormholes from domain wall networks}},\ }\href
  {https://doi.org/10.1103/PhysRevD.109.123507} {\bibfield  {journal} {\bibinfo
   {journal} {Phys. Rev. D}\ }\textbf {\bibinfo {volume} {109}},\ \bibinfo
  {pages} {123507} (\bibinfo {year} {2024}{\natexlab{b}})},\ \Eprint
  {https://arxiv.org/abs/2311.07670} {arXiv:2311.07670 [hep-ph]} \BibitemShut
  {NoStop}%
\bibitem [{\citenamefont {Ferreira}\ \emph {et~al.}(2024)\citenamefont
  {Ferreira}, \citenamefont {Notari}, \citenamefont {Pujol{\`a}s},\ and\
  \citenamefont {Rompineve}}]{Ferreira:2024eru}%
  \BibitemOpen
  \bibfield  {author} {\bibinfo {author} {\bibfnamefont {R.~Z.}\ \bibnamefont
  {Ferreira}}, \bibinfo {author} {\bibfnamefont {A.}~\bibnamefont {Notari}},
  \bibinfo {author} {\bibfnamefont {O.}~\bibnamefont {Pujol{\`a}s}},\ and\
  \bibinfo {author} {\bibfnamefont {F.}~\bibnamefont {Rompineve}},\ }\bibfield
  {title} {\bibinfo {title} {{Collapsing domain wall networks: impact on pulsar
  timing arrays and primordial black holes}},\ }\href
  {https://doi.org/10.1088/1475-7516/2024/06/020} {\bibfield  {journal}
  {\bibinfo  {journal} {JCAP}\ }\textbf {\bibinfo {volume} {06}},\ \bibinfo
  {pages} {020}},\ \Eprint {https://arxiv.org/abs/2401.14331} {arXiv:2401.14331
  [astro-ph.CO]} \BibitemShut {NoStop}%
\bibitem [{\citenamefont {Dunsky}\ and\ \citenamefont
  {Kongsore}(2024)}]{Dunsky:2024zdo}%
  \BibitemOpen
  \bibfield  {author} {\bibinfo {author} {\bibfnamefont {D.~I.}\ \bibnamefont
  {Dunsky}}\ and\ \bibinfo {author} {\bibfnamefont {M.}~\bibnamefont
  {Kongsore}},\ }\bibfield  {title} {\bibinfo {title} {{Primordial black holes
  from axion domain wall collapse}},\ }\href
  {https://doi.org/10.1007/JHEP06(2024)198} {\bibfield  {journal} {\bibinfo
  {journal} {JHEP}\ }\textbf {\bibinfo {volume} {06}},\ \bibinfo {pages}
  {198}},\ \Eprint {https://arxiv.org/abs/2402.03426} {arXiv:2402.03426
  [hep-ph]} \BibitemShut {NoStop}%
\bibitem [{\citenamefont {Cotner}\ and\ \citenamefont
  {Kusenko}(2017{\natexlab{a}})}]{Cotner:2016cvr}%
  \BibitemOpen
  \bibfield  {author} {\bibinfo {author} {\bibfnamefont {E.}~\bibnamefont
  {Cotner}}\ and\ \bibinfo {author} {\bibfnamefont {A.}~\bibnamefont
  {Kusenko}},\ }\bibfield  {title} {\bibinfo {title} {{Primordial black holes
  from supersymmetry in the early universe}},\ }\href
  {https://doi.org/10.1103/PhysRevLett.119.031103} {\bibfield  {journal}
  {\bibinfo  {journal} {Phys. Rev. Lett.}\ }\textbf {\bibinfo {volume} {119}},\
  \bibinfo {pages} {031103} (\bibinfo {year} {2017}{\natexlab{a}})},\ \Eprint
  {https://arxiv.org/abs/1612.02529} {arXiv:1612.02529 [astro-ph.CO]}
  \BibitemShut {NoStop}%
\bibitem [{\citenamefont {Cotner}\ and\ \citenamefont
  {Kusenko}(2017{\natexlab{b}})}]{Cotner:2017tir}%
  \BibitemOpen
  \bibfield  {author} {\bibinfo {author} {\bibfnamefont {E.}~\bibnamefont
  {Cotner}}\ and\ \bibinfo {author} {\bibfnamefont {A.}~\bibnamefont
  {Kusenko}},\ }\bibfield  {title} {\bibinfo {title} {{Primordial black holes
  from scalar field evolution in the early universe}},\ }\href
  {https://doi.org/10.1103/PhysRevD.96.103002} {\bibfield  {journal} {\bibinfo
  {journal} {Phys. Rev. D}\ }\textbf {\bibinfo {volume} {96}},\ \bibinfo
  {pages} {103002} (\bibinfo {year} {2017}{\natexlab{b}})},\ \Eprint
  {https://arxiv.org/abs/1706.09003} {arXiv:1706.09003 [astro-ph.CO]}
  \BibitemShut {NoStop}%
\bibitem [{\citenamefont {Dvali}\ \emph {et~al.}(2021)\citenamefont {Dvali},
  \citenamefont {K{\"u}hnel},\ and\ \citenamefont
  {Zantedeschi}}]{Dvali:2021byy}%
  \BibitemOpen
  \bibfield  {author} {\bibinfo {author} {\bibfnamefont {G.}~\bibnamefont
  {Dvali}}, \bibinfo {author} {\bibfnamefont {F.}~\bibnamefont {K{\"u}hnel}},\
  and\ \bibinfo {author} {\bibfnamefont {M.}~\bibnamefont {Zantedeschi}},\
  }\bibfield  {title} {\bibinfo {title} {{Primordial black holes from
  confinement}},\ }\href {https://doi.org/10.1103/PhysRevD.104.123507}
  {\bibfield  {journal} {\bibinfo  {journal} {Phys. Rev. D}\ }\textbf {\bibinfo
  {volume} {104}},\ \bibinfo {pages} {123507} (\bibinfo {year} {2021})},\
  \Eprint {https://arxiv.org/abs/2108.09471} {arXiv:2108.09471 [hep-ph]}
  \BibitemShut {NoStop}%
\bibitem [{\citenamefont {Dolgov}\ and\ \citenamefont
  {Silk}(1993)}]{Dolgov:1992pu}%
  \BibitemOpen
  \bibfield  {author} {\bibinfo {author} {\bibfnamefont {A.}~\bibnamefont
  {Dolgov}}\ and\ \bibinfo {author} {\bibfnamefont {J.}~\bibnamefont {Silk}},\
  }\bibfield  {title} {\bibinfo {title} {{Baryon isocurvature fluctuations at
  small scales and baryonic dark matter}},\ }\href
  {https://doi.org/10.1103/PhysRevD.47.4244} {\bibfield  {journal} {\bibinfo
  {journal} {Phys. Rev. D}\ }\textbf {\bibinfo {volume} {47}},\ \bibinfo
  {pages} {4244} (\bibinfo {year} {1993})}\BibitemShut {NoStop}%
\bibitem [{\citenamefont {Hawking}(1975)}]{Hawking:1975vcx}%
  \BibitemOpen
  \bibfield  {author} {\bibinfo {author} {\bibfnamefont {S.~W.}\ \bibnamefont
  {Hawking}},\ }\bibfield  {title} {\bibinfo {title} {{Particle Creation by
  Black Holes}},\ }\href {https://doi.org/10.1007/BF02345020} {\bibfield
  {journal} {\bibinfo  {journal} {Commun. Math. Phys.}\ }\textbf {\bibinfo
  {volume} {43}},\ \bibinfo {pages} {199} (\bibinfo {year} {1975})},\ \bibinfo
  {note} {[Erratum: Commun.Math.Phys. 46, 206 (1976)]}\BibitemShut {NoStop}%
\bibitem [{\citenamefont {Page}(1976{\natexlab{a}})}]{Page:1976df}%
  \BibitemOpen
  \bibfield  {author} {\bibinfo {author} {\bibfnamefont {D.~N.}\ \bibnamefont
  {Page}},\ }\bibfield  {title} {\bibinfo {title} {{Particle Emission Rates
  from a Black Hole: Massless Particles from an Uncharged, Nonrotating Hole}},\
  }\href {https://doi.org/10.1103/PhysRevD.13.198} {\bibfield  {journal}
  {\bibinfo  {journal} {Phys. Rev. D}\ }\textbf {\bibinfo {volume} {13}},\
  \bibinfo {pages} {198} (\bibinfo {year} {1976}{\natexlab{a}})}\BibitemShut
  {NoStop}%
\bibitem [{\citenamefont {Page}(1976{\natexlab{b}})}]{Page:1976ki}%
  \BibitemOpen
  \bibfield  {author} {\bibinfo {author} {\bibfnamefont {D.~N.}\ \bibnamefont
  {Page}},\ }\bibfield  {title} {\bibinfo {title} {{Particle Emission Rates
  from a Black Hole. 2. Massless Particles from a Rotating Hole}},\ }\href
  {https://doi.org/10.1103/PhysRevD.14.3260} {\bibfield  {journal} {\bibinfo
  {journal} {Phys. Rev. D}\ }\textbf {\bibinfo {volume} {14}},\ \bibinfo
  {pages} {3260} (\bibinfo {year} {1976}{\natexlab{b}})}\BibitemShut {NoStop}%
\bibitem [{\citenamefont {Taylor}\ \emph {et~al.}(1998)\citenamefont {Taylor},
  \citenamefont {Chambers},\ and\ \citenamefont {Hiscock}}]{Taylor:1998dk}%
  \BibitemOpen
  \bibfield  {author} {\bibinfo {author} {\bibfnamefont {B.~E.}\ \bibnamefont
  {Taylor}}, \bibinfo {author} {\bibfnamefont {C.~M.}\ \bibnamefont
  {Chambers}},\ and\ \bibinfo {author} {\bibfnamefont {W.~A.}\ \bibnamefont
  {Hiscock}},\ }\bibfield  {title} {\bibinfo {title} {{Evaporation of a Kerr
  black hole by emission of scalar and higher spin particles}},\ }\href
  {https://doi.org/10.1103/PhysRevD.58.044012} {\bibfield  {journal} {\bibinfo
  {journal} {Phys. Rev. D}\ }\textbf {\bibinfo {volume} {58}},\ \bibinfo
  {pages} {044012} (\bibinfo {year} {1998})},\ \Eprint
  {https://arxiv.org/abs/gr-qc/9801044} {arXiv:gr-qc/9801044} \BibitemShut
  {NoStop}%
\bibitem [{\citenamefont {MacGibbon}\ and\ \citenamefont
  {Carr}(1991)}]{MacGibbon:1991vc}%
  \BibitemOpen
  \bibfield  {author} {\bibinfo {author} {\bibfnamefont {J.~H.}\ \bibnamefont
  {MacGibbon}}\ and\ \bibinfo {author} {\bibfnamefont {B.~J.}\ \bibnamefont
  {Carr}},\ }\bibfield  {title} {\bibinfo {title} {{Cosmic rays from primordial
  black holes}},\ }\href {https://doi.org/10.1086/169909} {\bibfield  {journal}
  {\bibinfo  {journal} {Astrophys. J.}\ }\textbf {\bibinfo {volume} {371}},\
  \bibinfo {pages} {447} (\bibinfo {year} {1991})}\BibitemShut {NoStop}%
\bibitem [{\citenamefont {MacGibbon}\ and\ \citenamefont
  {Webber}(1990)}]{MacGibbon:1990zk}%
  \BibitemOpen
  \bibfield  {author} {\bibinfo {author} {\bibfnamefont {J.~H.}\ \bibnamefont
  {MacGibbon}}\ and\ \bibinfo {author} {\bibfnamefont {B.~R.}\ \bibnamefont
  {Webber}},\ }\bibfield  {title} {\bibinfo {title} {{Quark and gluon jet
  emission from primordial black holes: The instantaneous spectra}},\ }\href
  {https://doi.org/10.1103/PhysRevD.41.3052} {\bibfield  {journal} {\bibinfo
  {journal} {Phys. Rev. D}\ }\textbf {\bibinfo {volume} {41}},\ \bibinfo
  {pages} {3052} (\bibinfo {year} {1990})}\BibitemShut {NoStop}%
\bibitem [{\citenamefont {Heckler}(1997)}]{Heckler:1997jv}%
  \BibitemOpen
  \bibfield  {author} {\bibinfo {author} {\bibfnamefont {A.~F.}\ \bibnamefont
  {Heckler}},\ }\bibfield  {title} {\bibinfo {title} {{Calculation of the
  emergent spectrum and observation of primordial black holes}},\ }\href
  {https://doi.org/10.1103/PhysRevLett.78.3430} {\bibfield  {journal} {\bibinfo
   {journal} {Phys. Rev. Lett.}\ }\textbf {\bibinfo {volume} {78}},\ \bibinfo
  {pages} {3430} (\bibinfo {year} {1997})},\ \Eprint
  {https://arxiv.org/abs/astro-ph/9702027} {arXiv:astro-ph/9702027}
  \BibitemShut {NoStop}%
\bibitem [{\citenamefont {De~Luca}\ \emph
  {et~al.}(2019{\natexlab{b}})\citenamefont {De~Luca}, \citenamefont
  {Desjacques}, \citenamefont {Franciolini}, \citenamefont {Malhotra},\ and\
  \citenamefont {Riotto}}]{DeLuca:2019buf}%
  \BibitemOpen
  \bibfield  {author} {\bibinfo {author} {\bibfnamefont {V.}~\bibnamefont
  {De~Luca}}, \bibinfo {author} {\bibfnamefont {V.}~\bibnamefont {Desjacques}},
  \bibinfo {author} {\bibfnamefont {G.}~\bibnamefont {Franciolini}}, \bibinfo
  {author} {\bibfnamefont {A.}~\bibnamefont {Malhotra}},\ and\ \bibinfo
  {author} {\bibfnamefont {A.}~\bibnamefont {Riotto}},\ }\bibfield  {title}
  {\bibinfo {title} {{The initial spin probability distribution of primordial
  black holes}},\ }\href {https://doi.org/10.1088/1475-7516/2019/05/018}
  {\bibfield  {journal} {\bibinfo  {journal} {JCAP}\ }\textbf {\bibinfo
  {volume} {05}},\ \bibinfo {pages} {018}},\ \Eprint
  {https://arxiv.org/abs/1903.01179} {arXiv:1903.01179 [astro-ph.CO]}
  \BibitemShut {NoStop}%
\bibitem [{\citenamefont {Arbey}\ \emph {et~al.}(2020)\citenamefont {Arbey},
  \citenamefont {Auffinger},\ and\ \citenamefont {Silk}}]{Arbey:2019jmj}%
  \BibitemOpen
  \bibfield  {author} {\bibinfo {author} {\bibfnamefont {A.}~\bibnamefont
  {Arbey}}, \bibinfo {author} {\bibfnamefont {J.}~\bibnamefont {Auffinger}},\
  and\ \bibinfo {author} {\bibfnamefont {J.}~\bibnamefont {Silk}},\ }\bibfield
  {title} {\bibinfo {title} {{Evolution of primordial black hole spin due to
  Hawking radiation}},\ }\href {https://doi.org/10.1093/mnras/staa765}
  {\bibfield  {journal} {\bibinfo  {journal} {Mon. Not. Roy. Astron. Soc.}\
  }\textbf {\bibinfo {volume} {494}},\ \bibinfo {pages} {1257} (\bibinfo {year}
  {2020})},\ \Eprint {https://arxiv.org/abs/1906.04196} {arXiv:1906.04196
  [astro-ph.CO]} \BibitemShut {NoStop}%
\bibitem [{\citenamefont {Cheek}\ \emph {et~al.}(2023)\citenamefont {Cheek},
  \citenamefont {Heurtier}, \citenamefont {Perez-Gonzalez},\ and\ \citenamefont
  {Turner}}]{Cheek:2022mmy}%
  \BibitemOpen
  \bibfield  {author} {\bibinfo {author} {\bibfnamefont {A.}~\bibnamefont
  {Cheek}}, \bibinfo {author} {\bibfnamefont {L.}~\bibnamefont {Heurtier}},
  \bibinfo {author} {\bibfnamefont {Y.~F.}\ \bibnamefont {Perez-Gonzalez}},\
  and\ \bibinfo {author} {\bibfnamefont {J.}~\bibnamefont {Turner}},\
  }\bibfield  {title} {\bibinfo {title} {{Evaporation of primordial black holes
  in the early Universe: Mass and spin distributions}},\ }\href
  {https://doi.org/10.1103/PhysRevD.108.015005} {\bibfield  {journal} {\bibinfo
   {journal} {Phys. Rev. D}\ }\textbf {\bibinfo {volume} {108}},\ \bibinfo
  {pages} {015005} (\bibinfo {year} {2023})},\ \Eprint
  {https://arxiv.org/abs/2212.03878} {arXiv:2212.03878 [hep-ph]} \BibitemShut
  {NoStop}%
\bibitem [{\citenamefont {Bell}\ and\ \citenamefont
  {Volkas}(1999)}]{Bell:1998jk}%
  \BibitemOpen
  \bibfield  {author} {\bibinfo {author} {\bibfnamefont {N.~F.}\ \bibnamefont
  {Bell}}\ and\ \bibinfo {author} {\bibfnamefont {R.~R.}\ \bibnamefont
  {Volkas}},\ }\bibfield  {title} {\bibinfo {title} {{Mirror matter and
  primordial black holes}},\ }\href
  {https://doi.org/10.1103/PhysRevD.59.107301} {\bibfield  {journal} {\bibinfo
  {journal} {Phys. Rev. D}\ }\textbf {\bibinfo {volume} {59}},\ \bibinfo
  {pages} {107301} (\bibinfo {year} {1999})},\ \Eprint
  {https://arxiv.org/abs/astro-ph/9812301} {arXiv:astro-ph/9812301}
  \BibitemShut {NoStop}%
\bibitem [{\citenamefont {Khlopov}\ \emph {et~al.}(2006)\citenamefont
  {Khlopov}, \citenamefont {Barrau},\ and\ \citenamefont
  {Grain}}]{Khlopov:2004tn}%
  \BibitemOpen
  \bibfield  {author} {\bibinfo {author} {\bibfnamefont {M.~Y.}\ \bibnamefont
  {Khlopov}}, \bibinfo {author} {\bibfnamefont {A.}~\bibnamefont {Barrau}},\
  and\ \bibinfo {author} {\bibfnamefont {J.}~\bibnamefont {Grain}},\ }\bibfield
   {title} {\bibinfo {title} {{Gravitino production by primordial black hole
  evaporation and constraints on the inhomogeneity of the early universe}},\
  }\href {https://doi.org/10.1088/0264-9381/23/6/004} {\bibfield  {journal}
  {\bibinfo  {journal} {Class. Quant. Grav.}\ }\textbf {\bibinfo {volume}
  {23}},\ \bibinfo {pages} {1875} (\bibinfo {year} {2006})},\ \Eprint
  {https://arxiv.org/abs/astro-ph/0406621} {arXiv:astro-ph/0406621}
  \BibitemShut {NoStop}%
\bibitem [{\citenamefont {Fujita}\ \emph {et~al.}(2014)\citenamefont {Fujita},
  \citenamefont {Kawasaki}, \citenamefont {Harigaya},\ and\ \citenamefont
  {Matsuda}}]{Fujita:2014hha}%
  \BibitemOpen
  \bibfield  {author} {\bibinfo {author} {\bibfnamefont {T.}~\bibnamefont
  {Fujita}}, \bibinfo {author} {\bibfnamefont {M.}~\bibnamefont {Kawasaki}},
  \bibinfo {author} {\bibfnamefont {K.}~\bibnamefont {Harigaya}},\ and\
  \bibinfo {author} {\bibfnamefont {R.}~\bibnamefont {Matsuda}},\ }\bibfield
  {title} {\bibinfo {title} {{Baryon asymmetry, dark matter, and density
  perturbation from primordial black holes}},\ }\href
  {https://doi.org/10.1103/PhysRevD.89.103501} {\bibfield  {journal} {\bibinfo
  {journal} {Phys. Rev. D}\ }\textbf {\bibinfo {volume} {89}},\ \bibinfo
  {pages} {103501} (\bibinfo {year} {2014})},\ \Eprint
  {https://arxiv.org/abs/1401.1909} {arXiv:1401.1909 [astro-ph.CO]}
  \BibitemShut {NoStop}%
\bibitem [{\citenamefont {Allahverdi}\ \emph {et~al.}(2018)\citenamefont
  {Allahverdi}, \citenamefont {Dent},\ and\ \citenamefont
  {Osinski}}]{Allahverdi:2017sks}%
  \BibitemOpen
  \bibfield  {author} {\bibinfo {author} {\bibfnamefont {R.}~\bibnamefont
  {Allahverdi}}, \bibinfo {author} {\bibfnamefont {J.}~\bibnamefont {Dent}},\
  and\ \bibinfo {author} {\bibfnamefont {J.}~\bibnamefont {Osinski}},\
  }\bibfield  {title} {\bibinfo {title} {{Nonthermal production of dark matter
  from primordial black holes}},\ }\href
  {https://doi.org/10.1103/PhysRevD.97.055013} {\bibfield  {journal} {\bibinfo
  {journal} {Phys. Rev. D}\ }\textbf {\bibinfo {volume} {97}},\ \bibinfo
  {pages} {055013} (\bibinfo {year} {2018})},\ \Eprint
  {https://arxiv.org/abs/1711.10511} {arXiv:1711.10511 [astro-ph.CO]}
  \BibitemShut {NoStop}%
\bibitem [{\citenamefont {Lennon}\ \emph {et~al.}(2018)\citenamefont {Lennon},
  \citenamefont {March-Russell}, \citenamefont {Petrossian-Byrne},\ and\
  \citenamefont {Tillim}}]{Lennon:2017tqq}%
  \BibitemOpen
  \bibfield  {author} {\bibinfo {author} {\bibfnamefont {O.}~\bibnamefont
  {Lennon}}, \bibinfo {author} {\bibfnamefont {J.}~\bibnamefont
  {March-Russell}}, \bibinfo {author} {\bibfnamefont {R.}~\bibnamefont
  {Petrossian-Byrne}},\ and\ \bibinfo {author} {\bibfnamefont {H.}~\bibnamefont
  {Tillim}},\ }\bibfield  {title} {\bibinfo {title} {{Black Hole Genesis of
  Dark Matter}},\ }\href {https://doi.org/10.1088/1475-7516/2018/04/009}
  {\bibfield  {journal} {\bibinfo  {journal} {JCAP}\ }\textbf {\bibinfo
  {volume} {04}},\ \bibinfo {pages} {009}},\ \Eprint
  {https://arxiv.org/abs/1712.07664} {arXiv:1712.07664 [hep-ph]} \BibitemShut
  {NoStop}%
\bibitem [{\citenamefont {Rasanen}\ and\ \citenamefont
  {Tomberg}(2019)}]{Rasanen:2018fom}%
  \BibitemOpen
  \bibfield  {author} {\bibinfo {author} {\bibfnamefont {S.}~\bibnamefont
  {Rasanen}}\ and\ \bibinfo {author} {\bibfnamefont {E.}~\bibnamefont
  {Tomberg}},\ }\bibfield  {title} {\bibinfo {title} {{Planck scale black hole
  dark matter from Higgs inflation}},\ }\href
  {https://doi.org/10.1088/1475-7516/2019/01/038} {\bibfield  {journal}
  {\bibinfo  {journal} {JCAP}\ }\textbf {\bibinfo {volume} {01}},\ \bibinfo
  {pages} {038}},\ \Eprint {https://arxiv.org/abs/1810.12608} {arXiv:1810.12608
  [astro-ph.CO]} \BibitemShut {NoStop}%
\bibitem [{\citenamefont {Morrison}\ \emph {et~al.}(2019)\citenamefont
  {Morrison}, \citenamefont {Profumo},\ and\ \citenamefont
  {Yu}}]{Morrison:2018xla}%
  \BibitemOpen
  \bibfield  {author} {\bibinfo {author} {\bibfnamefont {L.}~\bibnamefont
  {Morrison}}, \bibinfo {author} {\bibfnamefont {S.}~\bibnamefont {Profumo}},\
  and\ \bibinfo {author} {\bibfnamefont {Y.}~\bibnamefont {Yu}},\ }\bibfield
  {title} {\bibinfo {title} {{Melanopogenesis: Dark Matter of (almost) any Mass
  and Baryonic Matter from the Evaporation of Primordial Black Holes weighing a
  Ton (or less)}},\ }\href {https://doi.org/10.1088/1475-7516/2019/05/005}
  {\bibfield  {journal} {\bibinfo  {journal} {JCAP}\ }\textbf {\bibinfo
  {volume} {05}},\ \bibinfo {pages} {005}},\ \Eprint
  {https://arxiv.org/abs/1812.10606} {arXiv:1812.10606 [astro-ph.CO]}
  \BibitemShut {NoStop}%
\bibitem [{\citenamefont {Salvio}\ and\ \citenamefont
  {Veerm{\"a}e}(2020)}]{Salvio:2019llz}%
  \BibitemOpen
  \bibfield  {author} {\bibinfo {author} {\bibfnamefont {A.}~\bibnamefont
  {Salvio}}\ and\ \bibinfo {author} {\bibfnamefont {H.}~\bibnamefont
  {Veerm{\"a}e}},\ }\bibfield  {title} {\bibinfo {title} {{Horizonless
  ultracompact objects and dark matter in quadratic gravity}},\ }\href
  {https://doi.org/10.1088/1475-7516/2020/02/018} {\bibfield  {journal}
  {\bibinfo  {journal} {JCAP}\ }\textbf {\bibinfo {volume} {02}},\ \bibinfo
  {pages} {018}},\ \Eprint {https://arxiv.org/abs/1912.13333} {arXiv:1912.13333
  [gr-qc]} \BibitemShut {NoStop}%
\bibitem [{\citenamefont {Masina}(2020)}]{Masina:2020xhk}%
  \BibitemOpen
  \bibfield  {author} {\bibinfo {author} {\bibfnamefont {I.}~\bibnamefont
  {Masina}},\ }\bibfield  {title} {\bibinfo {title} {{Dark matter and dark
  radiation from evaporating primordial black holes}},\ }\href
  {https://doi.org/10.1140/epjp/s13360-020-00564-9} {\bibfield  {journal}
  {\bibinfo  {journal} {Eur. Phys. J. Plus}\ }\textbf {\bibinfo {volume}
  {135}},\ \bibinfo {pages} {552} (\bibinfo {year} {2020})},\ \Eprint
  {https://arxiv.org/abs/2004.04740} {arXiv:2004.04740 [hep-ph]} \BibitemShut
  {NoStop}%
\bibitem [{\citenamefont {Baldes}\ \emph {et~al.}(2020)\citenamefont {Baldes},
  \citenamefont {Decant}, \citenamefont {Hooper},\ and\ \citenamefont
  {Lopez-Honorez}}]{Baldes:2020nuv}%
  \BibitemOpen
  \bibfield  {author} {\bibinfo {author} {\bibfnamefont {I.}~\bibnamefont
  {Baldes}}, \bibinfo {author} {\bibfnamefont {Q.}~\bibnamefont {Decant}},
  \bibinfo {author} {\bibfnamefont {D.~C.}\ \bibnamefont {Hooper}},\ and\
  \bibinfo {author} {\bibfnamefont {L.}~\bibnamefont {Lopez-Honorez}},\
  }\bibfield  {title} {\bibinfo {title} {{Non-Cold Dark Matter from Primordial
  Black Hole Evaporation}},\ }\href
  {https://doi.org/10.1088/1475-7516/2020/08/045} {\bibfield  {journal}
  {\bibinfo  {journal} {JCAP}\ }\textbf {\bibinfo {volume} {08}},\ \bibinfo
  {pages} {045}},\ \Eprint {https://arxiv.org/abs/2004.14773} {arXiv:2004.14773
  [astro-ph.CO]} \BibitemShut {NoStop}%
\bibitem [{\citenamefont {MacGibbon}(1987)}]{MacGibbon:1987my}%
  \BibitemOpen
  \bibfield  {author} {\bibinfo {author} {\bibfnamefont {J.~H.}\ \bibnamefont
  {MacGibbon}},\ }\bibfield  {title} {\bibinfo {title} {{Can Planck-mass relics
  of evaporating black holes close the universe?}},\ }\href
  {https://doi.org/10.1038/329308a0} {\bibfield  {journal} {\bibinfo  {journal}
  {Nature}\ }\textbf {\bibinfo {volume} {329}},\ \bibinfo {pages} {308}
  (\bibinfo {year} {1987})}\BibitemShut {NoStop}%
\bibitem [{\citenamefont {Dvali}\ \emph {et~al.}(2020)\citenamefont {Dvali},
  \citenamefont {Eisemann}, \citenamefont {Michel},\ and\ \citenamefont
  {Zell}}]{Dvali:2020wft}%
  \BibitemOpen
  \bibfield  {author} {\bibinfo {author} {\bibfnamefont {G.}~\bibnamefont
  {Dvali}}, \bibinfo {author} {\bibfnamefont {L.}~\bibnamefont {Eisemann}},
  \bibinfo {author} {\bibfnamefont {M.}~\bibnamefont {Michel}},\ and\ \bibinfo
  {author} {\bibfnamefont {S.}~\bibnamefont {Zell}},\ }\bibfield  {title}
  {\bibinfo {title} {{Black hole metamorphosis and stabilization by memory
  burden}},\ }\href {https://doi.org/10.1103/PhysRevD.102.103523} {\bibfield
  {journal} {\bibinfo  {journal} {Phys. Rev. D}\ }\textbf {\bibinfo {volume}
  {102}},\ \bibinfo {pages} {103523} (\bibinfo {year} {2020})},\ \Eprint
  {https://arxiv.org/abs/2006.00011} {arXiv:2006.00011 [hep-th]} \BibitemShut
  {NoStop}%
\bibitem [{\citenamefont {Thoss}\ \emph {et~al.}(2024)\citenamefont {Thoss},
  \citenamefont {Burkert},\ and\ \citenamefont {Kohri}}]{Thoss:2024hsr}%
  \BibitemOpen
  \bibfield  {author} {\bibinfo {author} {\bibfnamefont {V.}~\bibnamefont
  {Thoss}}, \bibinfo {author} {\bibfnamefont {A.}~\bibnamefont {Burkert}},\
  and\ \bibinfo {author} {\bibfnamefont {K.}~\bibnamefont {Kohri}},\ }\bibfield
   {title} {\bibinfo {title} {{Breakdown of hawking evaporation opens new mass
  window for primordial black holes as dark matter candidate}},\ }\href
  {https://doi.org/10.1093/mnras/stae1098} {\bibfield  {journal} {\bibinfo
  {journal} {Mon. Not. Roy. Astron. Soc.}\ }\textbf {\bibinfo {volume} {532}},\
  \bibinfo {pages} {451} (\bibinfo {year} {2024})},\ \Eprint
  {https://arxiv.org/abs/2402.17823} {arXiv:2402.17823 [astro-ph.CO]}
  \BibitemShut {NoStop}%
\bibitem [{\citenamefont {Dvali}\ \emph {et~al.}(2024)\citenamefont {Dvali},
  \citenamefont {Valbuena-Berm{\'u}dez},\ and\ \citenamefont
  {Zantedeschi}}]{Dvali:2024hsb}%
  \BibitemOpen
  \bibfield  {author} {\bibinfo {author} {\bibfnamefont {G.}~\bibnamefont
  {Dvali}}, \bibinfo {author} {\bibfnamefont {J.~S.}\ \bibnamefont
  {Valbuena-Berm{\'u}dez}},\ and\ \bibinfo {author} {\bibfnamefont
  {M.}~\bibnamefont {Zantedeschi}},\ }\bibfield  {title} {\bibinfo {title}
  {{Memory burden effect in black holes and solitons: Implications for PBH}},\
  }\href {https://doi.org/10.1103/PhysRevD.110.056029} {\bibfield  {journal}
  {\bibinfo  {journal} {Phys. Rev. D}\ }\textbf {\bibinfo {volume} {110}},\
  \bibinfo {pages} {056029} (\bibinfo {year} {2024})},\ \Eprint
  {https://arxiv.org/abs/2405.13117} {arXiv:2405.13117 [hep-th]} \BibitemShut
  {NoStop}%
\bibitem [{\citenamefont {Dvali}\ \emph {et~al.}(2025)\citenamefont {Dvali},
  \citenamefont {Zantedeschi},\ and\ \citenamefont {Zell}}]{Dvali:2025ktz}%
  \BibitemOpen
  \bibfield  {author} {\bibinfo {author} {\bibfnamefont {G.}~\bibnamefont
  {Dvali}}, \bibinfo {author} {\bibfnamefont {M.}~\bibnamefont {Zantedeschi}},\
  and\ \bibinfo {author} {\bibfnamefont {S.}~\bibnamefont {Zell}},\ }\bibfield
  {title} {\bibinfo {title} {{Transitioning to Memory Burden: Detectable Small
  Primordial Black Holes as Dark Matter}},\ }\href@noop {} {\  (\bibinfo {year}
  {2025})},\ \Eprint {https://arxiv.org/abs/2503.21740} {arXiv:2503.21740
  [hep-ph]} \BibitemShut {NoStop}%
\bibitem [{\citenamefont {Dondarini}\ \emph {et~al.}(2025)\citenamefont
  {Dondarini}, \citenamefont {Marino}, \citenamefont {Panci},\ and\
  \citenamefont {Zantedeschi}}]{Dondarini:2025ktz}%
  \BibitemOpen
  \bibfield  {author} {\bibinfo {author} {\bibfnamefont {A.}~\bibnamefont
  {Dondarini}}, \bibinfo {author} {\bibfnamefont {G.}~\bibnamefont {Marino}},
  \bibinfo {author} {\bibfnamefont {P.}~\bibnamefont {Panci}},\ and\ \bibinfo
  {author} {\bibfnamefont {M.}~\bibnamefont {Zantedeschi}},\ }\bibfield
  {title} {\bibinfo {title} {{The fast, the slow and the merging: probes of
  evaporating memory burdened PBHs}},\ }\href
  {https://doi.org/10.1088/1475-7516/2025/11/006} {\bibfield  {journal}
  {\bibinfo  {journal} {JCAP}\ }\textbf {\bibinfo {volume} {11}},\ \bibinfo
  {pages} {006}},\ \Eprint {https://arxiv.org/abs/2506.13861} {arXiv:2506.13861
  [hep-ph]} \BibitemShut {NoStop}%
\bibitem [{\citenamefont {Anantua}\ \emph {et~al.}(2009)\citenamefont
  {Anantua}, \citenamefont {Easther},\ and\ \citenamefont
  {Giblin}}]{Anantua:2008am}%
  \BibitemOpen
  \bibfield  {author} {\bibinfo {author} {\bibfnamefont {R.}~\bibnamefont
  {Anantua}}, \bibinfo {author} {\bibfnamefont {R.}~\bibnamefont {Easther}},\
  and\ \bibinfo {author} {\bibfnamefont {J.~T.}\ \bibnamefont {Giblin}},\
  }\bibfield  {title} {\bibinfo {title} {{GUT-Scale Primordial Black Holes:
  Consequences and Constraints}},\ }\href
  {https://doi.org/10.1103/PhysRevLett.103.111303} {\bibfield  {journal}
  {\bibinfo  {journal} {Phys. Rev. Lett.}\ }\textbf {\bibinfo {volume} {103}},\
  \bibinfo {pages} {111303} (\bibinfo {year} {2009})},\ \Eprint
  {https://arxiv.org/abs/0812.0825} {arXiv:0812.0825 [astro-ph]} \BibitemShut
  {NoStop}%
\bibitem [{\citenamefont {Dolgov}\ and\ \citenamefont
  {Ejlli}(2011)}]{Dolgov:2011cq}%
  \BibitemOpen
  \bibfield  {author} {\bibinfo {author} {\bibfnamefont {A.~D.}\ \bibnamefont
  {Dolgov}}\ and\ \bibinfo {author} {\bibfnamefont {D.}~\bibnamefont {Ejlli}},\
  }\bibfield  {title} {\bibinfo {title} {{Relic gravitational waves from light
  primordial black holes}},\ }\href
  {https://doi.org/10.1103/PhysRevD.84.024028} {\bibfield  {journal} {\bibinfo
  {journal} {Phys. Rev. D}\ }\textbf {\bibinfo {volume} {84}},\ \bibinfo
  {pages} {024028} (\bibinfo {year} {2011})},\ \Eprint
  {https://arxiv.org/abs/1105.2303} {arXiv:1105.2303 [astro-ph.CO]}
  \BibitemShut {NoStop}%
\bibitem [{\citenamefont {Dong}\ \emph {et~al.}(2016)\citenamefont {Dong},
  \citenamefont {Kinney},\ and\ \citenamefont {Stojkovic}}]{Dong:2015yjs}%
  \BibitemOpen
  \bibfield  {author} {\bibinfo {author} {\bibfnamefont {R.}~\bibnamefont
  {Dong}}, \bibinfo {author} {\bibfnamefont {W.~H.}\ \bibnamefont {Kinney}},\
  and\ \bibinfo {author} {\bibfnamefont {D.}~\bibnamefont {Stojkovic}},\
  }\bibfield  {title} {\bibinfo {title} {{Gravitational wave production by
  Hawking radiation from rotating primordial black holes}},\ }\href
  {https://doi.org/10.1088/1475-7516/2016/10/034} {\bibfield  {journal}
  {\bibinfo  {journal} {JCAP}\ }\textbf {\bibinfo {volume} {10}},\ \bibinfo
  {pages} {034}},\ \Eprint {https://arxiv.org/abs/1511.05642} {arXiv:1511.05642
  [astro-ph.CO]} \BibitemShut {NoStop}%
\bibitem [{\citenamefont {Ireland}\ \emph {et~al.}(2023)\citenamefont
  {Ireland}, \citenamefont {Profumo},\ and\ \citenamefont
  {Scharnhorst}}]{Ireland:2023avg}%
  \BibitemOpen
  \bibfield  {author} {\bibinfo {author} {\bibfnamefont {A.}~\bibnamefont
  {Ireland}}, \bibinfo {author} {\bibfnamefont {S.}~\bibnamefont {Profumo}},\
  and\ \bibinfo {author} {\bibfnamefont {J.}~\bibnamefont {Scharnhorst}},\
  }\bibfield  {title} {\bibinfo {title} {{Primordial gravitational waves from
  black hole evaporation in standard and nonstandard cosmologies}},\ }\href
  {https://doi.org/10.1103/PhysRevD.107.104021} {\bibfield  {journal} {\bibinfo
   {journal} {Phys. Rev. D}\ }\textbf {\bibinfo {volume} {107}},\ \bibinfo
  {pages} {104021} (\bibinfo {year} {2023})},\ \Eprint
  {https://arxiv.org/abs/2302.10188} {arXiv:2302.10188 [gr-qc]} \BibitemShut
  {NoStop}%
\bibitem [{\citenamefont {Gehrman}\ \emph {et~al.}(2023)\citenamefont
  {Gehrman}, \citenamefont {Shams Es~Haghi}, \citenamefont {Sinha},\ and\
  \citenamefont {Xu}}]{Gehrman:2023esa}%
  \BibitemOpen
  \bibfield  {author} {\bibinfo {author} {\bibfnamefont {T.~C.}\ \bibnamefont
  {Gehrman}}, \bibinfo {author} {\bibfnamefont {B.}~\bibnamefont {Shams
  Es~Haghi}}, \bibinfo {author} {\bibfnamefont {K.}~\bibnamefont {Sinha}},\
  and\ \bibinfo {author} {\bibfnamefont {T.}~\bibnamefont {Xu}},\ }\bibfield
  {title} {\bibinfo {title} {{The primordial black holes that disappeared:
  connections to dark matter and MHz-GHz gravitational Waves}},\ }\href
  {https://doi.org/10.1088/1475-7516/2023/10/001} {\bibfield  {journal}
  {\bibinfo  {journal} {JCAP}\ }\textbf {\bibinfo {volume} {10}},\ \bibinfo
  {pages} {001}},\ \Eprint {https://arxiv.org/abs/2304.09194} {arXiv:2304.09194
  [hep-ph]} \BibitemShut {NoStop}%
\bibitem [{\citenamefont {Zagorac}\ \emph {et~al.}(2019)\citenamefont
  {Zagorac}, \citenamefont {Easther},\ and\ \citenamefont
  {Padmanabhan}}]{Zagorac:2019ekv}%
  \BibitemOpen
  \bibfield  {author} {\bibinfo {author} {\bibfnamefont {J.~L.}\ \bibnamefont
  {Zagorac}}, \bibinfo {author} {\bibfnamefont {R.}~\bibnamefont {Easther}},\
  and\ \bibinfo {author} {\bibfnamefont {N.}~\bibnamefont {Padmanabhan}},\
  }\bibfield  {title} {\bibinfo {title} {{GUT-Scale Primordial Black Holes:
  Mergers and Gravitational Waves}},\ }\href
  {https://doi.org/10.1088/1475-7516/2019/06/052} {\bibfield  {journal}
  {\bibinfo  {journal} {JCAP}\ }\textbf {\bibinfo {volume} {06}},\ \bibinfo
  {pages} {052}},\ \Eprint {https://arxiv.org/abs/1903.05053} {arXiv:1903.05053
  [astro-ph.CO]} \BibitemShut {NoStop}%
\bibitem [{\citenamefont {Hooper}\ \emph {et~al.}(2020)\citenamefont {Hooper},
  \citenamefont {Krnjaic}, \citenamefont {March-Russell}, \citenamefont
  {McDermott},\ and\ \citenamefont {Petrossian-Byrne}}]{Hooper:2020evu}%
  \BibitemOpen
  \bibfield  {author} {\bibinfo {author} {\bibfnamefont {D.}~\bibnamefont
  {Hooper}}, \bibinfo {author} {\bibfnamefont {G.}~\bibnamefont {Krnjaic}},
  \bibinfo {author} {\bibfnamefont {J.}~\bibnamefont {March-Russell}}, \bibinfo
  {author} {\bibfnamefont {S.~D.}\ \bibnamefont {McDermott}},\ and\ \bibinfo
  {author} {\bibfnamefont {R.}~\bibnamefont {Petrossian-Byrne}},\ }\bibfield
  {title} {\bibinfo {title} {{Hot Gravitons and Gravitational Waves From Kerr
  Black Holes in the Early Universe}},\ }\href@noop {} {\  (\bibinfo {year}
  {2020})},\ \Eprint {https://arxiv.org/abs/2004.00618} {arXiv:2004.00618
  [astro-ph.CO]} \BibitemShut {NoStop}%
\bibitem [{\citenamefont {Inomata}\ \emph {et~al.}(2020)\citenamefont
  {Inomata}, \citenamefont {Kawasaki}, \citenamefont {Mukaida}, \citenamefont
  {Terada},\ and\ \citenamefont {Yanagida}}]{Inomata:2020lmk}%
  \BibitemOpen
  \bibfield  {author} {\bibinfo {author} {\bibfnamefont {K.}~\bibnamefont
  {Inomata}}, \bibinfo {author} {\bibfnamefont {M.}~\bibnamefont {Kawasaki}},
  \bibinfo {author} {\bibfnamefont {K.}~\bibnamefont {Mukaida}}, \bibinfo
  {author} {\bibfnamefont {T.}~\bibnamefont {Terada}},\ and\ \bibinfo {author}
  {\bibfnamefont {T.~T.}\ \bibnamefont {Yanagida}},\ }\bibfield  {title}
  {\bibinfo {title} {{Gravitational Wave Production right after a Primordial
  Black Hole Evaporation}},\ }\href
  {https://doi.org/10.1103/PhysRevD.101.123533} {\bibfield  {journal} {\bibinfo
   {journal} {Phys. Rev. D}\ }\textbf {\bibinfo {volume} {101}},\ \bibinfo
  {pages} {123533} (\bibinfo {year} {2020})},\ \Eprint
  {https://arxiv.org/abs/2003.10455} {arXiv:2003.10455 [astro-ph.CO]}
  \BibitemShut {NoStop}%
\bibitem [{\citenamefont {Papanikolaou}\ \emph {et~al.}(2021)\citenamefont
  {Papanikolaou}, \citenamefont {Vennin},\ and\ \citenamefont
  {Langlois}}]{Papanikolaou:2020qtd}%
  \BibitemOpen
  \bibfield  {author} {\bibinfo {author} {\bibfnamefont {T.}~\bibnamefont
  {Papanikolaou}}, \bibinfo {author} {\bibfnamefont {V.}~\bibnamefont
  {Vennin}},\ and\ \bibinfo {author} {\bibfnamefont {D.}~\bibnamefont
  {Langlois}},\ }\bibfield  {title} {\bibinfo {title} {{Gravitational waves
  from a universe filled with primordial black holes}},\ }\href
  {https://doi.org/10.1088/1475-7516/2021/03/053} {\bibfield  {journal}
  {\bibinfo  {journal} {JCAP}\ }\textbf {\bibinfo {volume} {03}},\ \bibinfo
  {pages} {053}},\ \Eprint {https://arxiv.org/abs/2010.11573} {arXiv:2010.11573
  [astro-ph.CO]} \BibitemShut {NoStop}%
\bibitem [{\citenamefont {Dom{\`e}nech}\ \emph
  {et~al.}(2021{\natexlab{a}})\citenamefont {Dom{\`e}nech}, \citenamefont
  {Lin},\ and\ \citenamefont {Sasaki}}]{Domenech:2020ssp}%
  \BibitemOpen
  \bibfield  {author} {\bibinfo {author} {\bibfnamefont {G.}~\bibnamefont
  {Dom{\`e}nech}}, \bibinfo {author} {\bibfnamefont {C.}~\bibnamefont {Lin}},\
  and\ \bibinfo {author} {\bibfnamefont {M.}~\bibnamefont {Sasaki}},\
  }\bibfield  {title} {\bibinfo {title} {{Gravitational wave constraints on the
  primordial black hole dominated early universe}},\ }\href
  {https://doi.org/10.1088/1475-7516/2021/11/E01} {\bibfield  {journal}
  {\bibinfo  {journal} {JCAP}\ }\textbf {\bibinfo {volume} {04}},\ \bibinfo
  {pages} {062}},\ \bibinfo {note} {[Erratum: JCAP 11, E01 (2021)]},\ \Eprint
  {https://arxiv.org/abs/2012.08151} {arXiv:2012.08151 [gr-qc]} \BibitemShut
  {NoStop}%
\bibitem [{\citenamefont {Dom{\`e}nech}\ \emph
  {et~al.}(2021{\natexlab{b}})\citenamefont {Dom{\`e}nech}, \citenamefont
  {Takhistov},\ and\ \citenamefont {Sasaki}}]{Domenech:2021wkk}%
  \BibitemOpen
  \bibfield  {author} {\bibinfo {author} {\bibfnamefont {G.}~\bibnamefont
  {Dom{\`e}nech}}, \bibinfo {author} {\bibfnamefont {V.}~\bibnamefont
  {Takhistov}},\ and\ \bibinfo {author} {\bibfnamefont {M.}~\bibnamefont
  {Sasaki}},\ }\bibfield  {title} {\bibinfo {title} {{Exploring evaporating
  primordial black holes with gravitational waves}},\ }\href
  {https://doi.org/10.1016/j.physletb.2021.136722} {\bibfield  {journal}
  {\bibinfo  {journal} {Phys. Lett. B}\ }\textbf {\bibinfo {volume} {823}},\
  \bibinfo {pages} {136722} (\bibinfo {year} {2021}{\natexlab{b}})},\ \Eprint
  {https://arxiv.org/abs/2105.06816} {arXiv:2105.06816 [astro-ph.CO]}
  \BibitemShut {NoStop}%
\bibitem [{\citenamefont {Papanikolaou}(2022)}]{Papanikolaou:2022chm}%
  \BibitemOpen
  \bibfield  {author} {\bibinfo {author} {\bibfnamefont {T.}~\bibnamefont
  {Papanikolaou}},\ }\bibfield  {title} {\bibinfo {title} {{Gravitational waves
  induced from primordial black hole fluctuations: the~effect of an extended
  mass function}},\ }\href {https://doi.org/10.1088/1475-7516/2022/10/089}
  {\bibfield  {journal} {\bibinfo  {journal} {JCAP}\ }\textbf {\bibinfo
  {volume} {10}},\ \bibinfo {pages} {089}},\ \Eprint
  {https://arxiv.org/abs/2207.11041} {arXiv:2207.11041 [astro-ph.CO]}
  \BibitemShut {NoStop}%
\bibitem [{\citenamefont {Bhaumik}\ \emph {et~al.}(2022)\citenamefont
  {Bhaumik}, \citenamefont {Ghoshal},\ and\ \citenamefont
  {Lewicki}}]{Bhaumik:2022pil}%
  \BibitemOpen
  \bibfield  {author} {\bibinfo {author} {\bibfnamefont {N.}~\bibnamefont
  {Bhaumik}}, \bibinfo {author} {\bibfnamefont {A.}~\bibnamefont {Ghoshal}},\
  and\ \bibinfo {author} {\bibfnamefont {M.}~\bibnamefont {Lewicki}},\
  }\bibfield  {title} {\bibinfo {title} {{Doubly peaked induced stochastic
  gravitational wave background: testing baryogenesis from primordial black
  holes}},\ }\href {https://doi.org/10.1007/JHEP07(2022)130} {\bibfield
  {journal} {\bibinfo  {journal} {JHEP}\ }\textbf {\bibinfo {volume} {07}},\
  \bibinfo {pages} {130}},\ \Eprint {https://arxiv.org/abs/2205.06260}
  {arXiv:2205.06260 [astro-ph.CO]} \BibitemShut {NoStop}%
\bibitem [{\citenamefont {Auffinger}(2023)}]{Auffinger:2022khh}%
  \BibitemOpen
  \bibfield  {author} {\bibinfo {author} {\bibfnamefont {J.}~\bibnamefont
  {Auffinger}},\ }\bibfield  {title} {\bibinfo {title} {{Primordial black hole
  constraints with Hawking radiation\textemdash{}A review}},\ }\href
  {https://doi.org/10.1016/j.ppnp.2023.104040} {\bibfield  {journal} {\bibinfo
  {journal} {Prog. Part. Nucl. Phys.}\ }\textbf {\bibinfo {volume} {131}},\
  \bibinfo {pages} {104040} (\bibinfo {year} {2023})},\ \Eprint
  {https://arxiv.org/abs/2206.02672} {arXiv:2206.02672 [astro-ph.CO]}
  \BibitemShut {NoStop}%
\bibitem [{\citenamefont {Lehmann}\ \emph {et~al.}(2019)\citenamefont
  {Lehmann}, \citenamefont {Johnson}, \citenamefont {Profumo},\ and\
  \citenamefont {Schwemberger}}]{Lehmann:2019zgt}%
  \BibitemOpen
  \bibfield  {author} {\bibinfo {author} {\bibfnamefont {B.~V.}\ \bibnamefont
  {Lehmann}}, \bibinfo {author} {\bibfnamefont {C.}~\bibnamefont {Johnson}},
  \bibinfo {author} {\bibfnamefont {S.}~\bibnamefont {Profumo}},\ and\ \bibinfo
  {author} {\bibfnamefont {T.}~\bibnamefont {Schwemberger}},\ }\bibfield
  {title} {\bibinfo {title} {{Direct detection of primordial black hole relics
  as dark matter}},\ }\href {https://doi.org/10.1088/1475-7516/2019/10/046}
  {\bibfield  {journal} {\bibinfo  {journal} {JCAP}\ }\textbf {\bibinfo
  {volume} {10}},\ \bibinfo {pages} {046}},\ \Eprint
  {https://arxiv.org/abs/1906.06348} {arXiv:1906.06348 [hep-ph]} \BibitemShut
  {NoStop}%
\bibitem [{\citenamefont {Bai}\ and\ \citenamefont
  {Orlofsky}(2020)}]{Bai:2019zcd}%
  \BibitemOpen
  \bibfield  {author} {\bibinfo {author} {\bibfnamefont {Y.}~\bibnamefont
  {Bai}}\ and\ \bibinfo {author} {\bibfnamefont {N.}~\bibnamefont {Orlofsky}},\
  }\bibfield  {title} {\bibinfo {title} {{Primordial Extremal Black Holes as
  Dark Matter}},\ }\href {https://doi.org/10.1103/PhysRevD.101.055006}
  {\bibfield  {journal} {\bibinfo  {journal} {Phys. Rev. D}\ }\textbf {\bibinfo
  {volume} {101}},\ \bibinfo {pages} {055006} (\bibinfo {year} {2020})},\
  \Eprint {https://arxiv.org/abs/1906.04858} {arXiv:1906.04858 [hep-ph]}
  \BibitemShut {NoStop}%
\bibitem [{\citenamefont {Carr}\ \emph {et~al.}(2010)\citenamefont {Carr},
  \citenamefont {Kohri}, \citenamefont {Sendouda},\ and\ \citenamefont
  {Yokoyama}}]{Carr:2009jm}%
  \BibitemOpen
  \bibfield  {author} {\bibinfo {author} {\bibfnamefont {B.~J.}\ \bibnamefont
  {Carr}}, \bibinfo {author} {\bibfnamefont {K.}~\bibnamefont {Kohri}},
  \bibinfo {author} {\bibfnamefont {Y.}~\bibnamefont {Sendouda}},\ and\
  \bibinfo {author} {\bibfnamefont {J.}~\bibnamefont {Yokoyama}},\ }\bibfield
  {title} {\bibinfo {title} {{New cosmological constraints on primordial black
  holes}},\ }\href {https://doi.org/10.1103/PhysRevD.81.104019} {\bibfield
  {journal} {\bibinfo  {journal} {Phys. Rev. D}\ }\textbf {\bibinfo {volume}
  {81}},\ \bibinfo {pages} {104019} (\bibinfo {year} {2010})},\ \Eprint
  {https://arxiv.org/abs/0912.5297} {arXiv:0912.5297 [astro-ph.CO]}
  \BibitemShut {NoStop}%
\bibitem [{\citenamefont {Keith}\ \emph {et~al.}(2020)\citenamefont {Keith},
  \citenamefont {Hooper}, \citenamefont {Blinov},\ and\ \citenamefont
  {McDermott}}]{Keith:2020jww}%
  \BibitemOpen
  \bibfield  {author} {\bibinfo {author} {\bibfnamefont {C.}~\bibnamefont
  {Keith}}, \bibinfo {author} {\bibfnamefont {D.}~\bibnamefont {Hooper}},
  \bibinfo {author} {\bibfnamefont {N.}~\bibnamefont {Blinov}},\ and\ \bibinfo
  {author} {\bibfnamefont {S.~D.}\ \bibnamefont {McDermott}},\ }\bibfield
  {title} {\bibinfo {title} {{Constraints on Primordial Black Holes From Big
  Bang Nucleosynthesis Revisited}},\ }\href
  {https://doi.org/10.1103/PhysRevD.102.103512} {\bibfield  {journal} {\bibinfo
   {journal} {Phys. Rev. D}\ }\textbf {\bibinfo {volume} {102}},\ \bibinfo
  {pages} {103512} (\bibinfo {year} {2020})},\ \Eprint
  {https://arxiv.org/abs/2006.03608} {arXiv:2006.03608 [astro-ph.CO]}
  \BibitemShut {NoStop}%
\bibitem [{\citenamefont {Kohri}\ and\ \citenamefont
  {Yokoyama}(2000)}]{Kohri:1999ex}%
  \BibitemOpen
  \bibfield  {author} {\bibinfo {author} {\bibfnamefont {K.}~\bibnamefont
  {Kohri}}\ and\ \bibinfo {author} {\bibfnamefont {J.}~\bibnamefont
  {Yokoyama}},\ }\bibfield  {title} {\bibinfo {title} {{Primordial black holes
  and primordial nucleosynthesis. 1. Effects of hadron injection from low mass
  holes}},\ }\href {https://doi.org/10.1103/PhysRevD.61.023501} {\bibfield
  {journal} {\bibinfo  {journal} {Phys. Rev. D}\ }\textbf {\bibinfo {volume}
  {61}},\ \bibinfo {pages} {023501} (\bibinfo {year} {2000})},\ \Eprint
  {https://arxiv.org/abs/astro-ph/9908160} {arXiv:astro-ph/9908160}
  \BibitemShut {NoStop}%
\bibitem [{\citenamefont {Boccia}\ \emph {et~al.}(2025)\citenamefont {Boccia},
  \citenamefont {Iocco},\ and\ \citenamefont {Visinelli}}]{Boccia:2024nly}%
  \BibitemOpen
  \bibfield  {author} {\bibinfo {author} {\bibfnamefont {A.}~\bibnamefont
  {Boccia}}, \bibinfo {author} {\bibfnamefont {F.}~\bibnamefont {Iocco}},\ and\
  \bibinfo {author} {\bibfnamefont {L.}~\bibnamefont {Visinelli}},\ }\bibfield
  {title} {\bibinfo {title} {{Constraining the primordial black hole abundance
  through big-bang nucleosynthesis}},\ }\href
  {https://doi.org/10.1103/PhysRevD.111.063508} {\bibfield  {journal} {\bibinfo
   {journal} {Phys. Rev. D}\ }\textbf {\bibinfo {volume} {111}},\ \bibinfo
  {pages} {063508} (\bibinfo {year} {2025})},\ \Eprint
  {https://arxiv.org/abs/2405.18493} {arXiv:2405.18493 [astro-ph.CO]}
  \BibitemShut {NoStop}%
\bibitem [{\citenamefont {Wu}\ and\ \citenamefont {Xu}(2025)}]{Wu:2025ovd}%
  \BibitemOpen
  \bibfield  {author} {\bibinfo {author} {\bibfnamefont {Q.-f.}\ \bibnamefont
  {Wu}}\ and\ \bibinfo {author} {\bibfnamefont {X.-J.}\ \bibnamefont {Xu}},\
  }\bibfield  {title} {\bibinfo {title} {{Primordial Black Holes Evaporating
  before Big Bang Nucleosynthesis}},\ }\href@noop {} {\  (\bibinfo {year}
  {2025})},\ \Eprint {https://arxiv.org/abs/2509.05618} {arXiv:2509.05618
  [astro-ph.CO]} \BibitemShut {NoStop}%
\bibitem [{\citenamefont {Acharya}\ and\ \citenamefont
  {Khatri}(2020)}]{Acharya:2020jbv}%
  \BibitemOpen
  \bibfield  {author} {\bibinfo {author} {\bibfnamefont {S.~K.}\ \bibnamefont
  {Acharya}}\ and\ \bibinfo {author} {\bibfnamefont {R.}~\bibnamefont
  {Khatri}},\ }\bibfield  {title} {\bibinfo {title} {{CMB and BBN constraints
  on evaporating primordial black holes revisited}},\ }\href
  {https://doi.org/10.1088/1475-7516/2020/06/018} {\bibfield  {journal}
  {\bibinfo  {journal} {JCAP}\ }\textbf {\bibinfo {volume} {06}},\ \bibinfo
  {pages} {018}},\ \Eprint {https://arxiv.org/abs/2002.00898} {arXiv:2002.00898
  [astro-ph.CO]} \BibitemShut {NoStop}%
\bibitem [{\citenamefont {Boudaud}\ and\ \citenamefont
  {Cirelli}(2019)}]{Boudaud:2018hqb}%
  \BibitemOpen
  \bibfield  {author} {\bibinfo {author} {\bibfnamefont {M.}~\bibnamefont
  {Boudaud}}\ and\ \bibinfo {author} {\bibfnamefont {M.}~\bibnamefont
  {Cirelli}},\ }\bibfield  {title} {\bibinfo {title} {{Voyager 1 $e^\pm$
  Further Constrain Primordial Black Holes as Dark Matter}},\ }\href
  {https://doi.org/10.1103/PhysRevLett.122.041104} {\bibfield  {journal}
  {\bibinfo  {journal} {Phys. Rev. Lett.}\ }\textbf {\bibinfo {volume} {122}},\
  \bibinfo {pages} {041104} (\bibinfo {year} {2019})},\ \Eprint
  {https://arxiv.org/abs/1807.03075} {arXiv:1807.03075 [astro-ph.HE]}
  \BibitemShut {NoStop}%
\bibitem [{\citenamefont {Clark}\ \emph {et~al.}(2017)\citenamefont {Clark},
  \citenamefont {Dutta}, \citenamefont {Gao}, \citenamefont {Strigari},\ and\
  \citenamefont {Watson}}]{Clark:2016nst}%
  \BibitemOpen
  \bibfield  {author} {\bibinfo {author} {\bibfnamefont {S.}~\bibnamefont
  {Clark}}, \bibinfo {author} {\bibfnamefont {B.}~\bibnamefont {Dutta}},
  \bibinfo {author} {\bibfnamefont {Y.}~\bibnamefont {Gao}}, \bibinfo {author}
  {\bibfnamefont {L.~E.}\ \bibnamefont {Strigari}},\ and\ \bibinfo {author}
  {\bibfnamefont {S.}~\bibnamefont {Watson}},\ }\bibfield  {title} {\bibinfo
  {title} {{Planck Constraint on Relic Primordial Black Holes}},\ }\href
  {https://doi.org/10.1103/PhysRevD.95.083006} {\bibfield  {journal} {\bibinfo
  {journal} {Phys. Rev. D}\ }\textbf {\bibinfo {volume} {95}},\ \bibinfo
  {pages} {083006} (\bibinfo {year} {2017})},\ \Eprint
  {https://arxiv.org/abs/1612.07738} {arXiv:1612.07738 [astro-ph.CO]}
  \BibitemShut {NoStop}%
\bibitem [{\citenamefont {DeRocco}\ and\ \citenamefont
  {Graham}(2019)}]{DeRocco:2019fjq}%
  \BibitemOpen
  \bibfield  {author} {\bibinfo {author} {\bibfnamefont {W.}~\bibnamefont
  {DeRocco}}\ and\ \bibinfo {author} {\bibfnamefont {P.~W.}\ \bibnamefont
  {Graham}},\ }\bibfield  {title} {\bibinfo {title} {{Constraining Primordial
  Black Hole Abundance with the Galactic 511 keV Line}},\ }\href
  {https://doi.org/10.1103/PhysRevLett.123.251102} {\bibfield  {journal}
  {\bibinfo  {journal} {Phys. Rev. Lett.}\ }\textbf {\bibinfo {volume} {123}},\
  \bibinfo {pages} {251102} (\bibinfo {year} {2019})},\ \Eprint
  {https://arxiv.org/abs/1906.07740} {arXiv:1906.07740 [astro-ph.CO]}
  \BibitemShut {NoStop}%
\bibitem [{\citenamefont {Dasgupta}\ \emph {et~al.}(2020)\citenamefont
  {Dasgupta}, \citenamefont {Laha},\ and\ \citenamefont
  {Ray}}]{Dasgupta:2019cae}%
  \BibitemOpen
  \bibfield  {author} {\bibinfo {author} {\bibfnamefont {B.}~\bibnamefont
  {Dasgupta}}, \bibinfo {author} {\bibfnamefont {R.}~\bibnamefont {Laha}},\
  and\ \bibinfo {author} {\bibfnamefont {A.}~\bibnamefont {Ray}},\ }\bibfield
  {title} {\bibinfo {title} {{Neutrino and positron constraints on spinning
  primordial black hole dark matter}},\ }\href
  {https://doi.org/10.1103/PhysRevLett.125.101101} {\bibfield  {journal}
  {\bibinfo  {journal} {Phys. Rev. Lett.}\ }\textbf {\bibinfo {volume} {125}},\
  \bibinfo {pages} {101101} (\bibinfo {year} {2020})},\ \Eprint
  {https://arxiv.org/abs/1912.01014} {arXiv:1912.01014 [hep-ph]} \BibitemShut
  {NoStop}%
\bibitem [{\citenamefont {Laha}(2019)}]{Laha:2019ssq}%
  \BibitemOpen
  \bibfield  {author} {\bibinfo {author} {\bibfnamefont {R.}~\bibnamefont
  {Laha}},\ }\bibfield  {title} {\bibinfo {title} {{Primordial Black Holes as a
  Dark Matter Candidate Are Severely Constrained by the Galactic Center 511 keV
  $\gamma$ -Ray Line}},\ }\href
  {https://doi.org/10.1103/PhysRevLett.123.251101} {\bibfield  {journal}
  {\bibinfo  {journal} {Phys. Rev. Lett.}\ }\textbf {\bibinfo {volume} {123}},\
  \bibinfo {pages} {251101} (\bibinfo {year} {2019})},\ \Eprint
  {https://arxiv.org/abs/1906.09994} {arXiv:1906.09994 [astro-ph.HE]}
  \BibitemShut {NoStop}%
\bibitem [{\citenamefont {Laha}\ \emph {et~al.}(2020)\citenamefont {Laha},
  \citenamefont {Mu{\~n}oz},\ and\ \citenamefont {Slatyer}}]{Laha:2020ivk}%
  \BibitemOpen
  \bibfield  {author} {\bibinfo {author} {\bibfnamefont {R.}~\bibnamefont
  {Laha}}, \bibinfo {author} {\bibfnamefont {J.~B.}\ \bibnamefont
  {Mu{\~n}oz}},\ and\ \bibinfo {author} {\bibfnamefont {T.~R.}\ \bibnamefont
  {Slatyer}},\ }\bibfield  {title} {\bibinfo {title} {{INTEGRAL constraints on
  primordial black holes and particle dark matter}},\ }\href
  {https://doi.org/10.1103/PhysRevD.101.123514} {\bibfield  {journal} {\bibinfo
   {journal} {Phys. Rev. D}\ }\textbf {\bibinfo {volume} {101}},\ \bibinfo
  {pages} {123514} (\bibinfo {year} {2020})},\ \Eprint
  {https://arxiv.org/abs/2004.00627} {arXiv:2004.00627 [astro-ph.CO]}
  \BibitemShut {NoStop}%
\bibitem [{\citenamefont {Saha}\ and\ \citenamefont
  {Laha}(2022)}]{Saha:2021pqf}%
  \BibitemOpen
  \bibfield  {author} {\bibinfo {author} {\bibfnamefont {A.~K.}\ \bibnamefont
  {Saha}}\ and\ \bibinfo {author} {\bibfnamefont {R.}~\bibnamefont {Laha}},\
  }\bibfield  {title} {\bibinfo {title} {{Sensitivities on nonspinning and
  spinning primordial black hole dark matter with global 21-cm troughs}},\
  }\href {https://doi.org/10.1103/PhysRevD.105.103026} {\bibfield  {journal}
  {\bibinfo  {journal} {Phys. Rev. D}\ }\textbf {\bibinfo {volume} {105}},\
  \bibinfo {pages} {103026} (\bibinfo {year} {2022})},\ \Eprint
  {https://arxiv.org/abs/2112.10794} {arXiv:2112.10794 [astro-ph.CO]}
  \BibitemShut {NoStop}%
\bibitem [{\citenamefont {Mittal}\ \emph {et~al.}(2022)\citenamefont {Mittal},
  \citenamefont {Ray}, \citenamefont {Kulkarni},\ and\ \citenamefont
  {Dasgupta}}]{Mittal:2021egv}%
  \BibitemOpen
  \bibfield  {author} {\bibinfo {author} {\bibfnamefont {S.}~\bibnamefont
  {Mittal}}, \bibinfo {author} {\bibfnamefont {A.}~\bibnamefont {Ray}},
  \bibinfo {author} {\bibfnamefont {G.}~\bibnamefont {Kulkarni}},\ and\
  \bibinfo {author} {\bibfnamefont {B.}~\bibnamefont {Dasgupta}},\ }\bibfield
  {title} {\bibinfo {title} {{Constraining primordial black holes as dark
  matter using the global 21-cm signal with X-ray heating and excess radio
  background}},\ }\href {https://doi.org/10.1088/1475-7516/2022/03/030}
  {\bibfield  {journal} {\bibinfo  {journal} {JCAP}\ }\textbf {\bibinfo
  {volume} {03}},\ \bibinfo {pages} {030}},\ \Eprint
  {https://arxiv.org/abs/2107.02190} {arXiv:2107.02190 [astro-ph.CO]}
  \BibitemShut {NoStop}%
\bibitem [{\citenamefont {Calz{\`a}}\ \emph {et~al.}(2024)\citenamefont
  {Calz{\`a}}, \citenamefont {March-Russell},\ and\ \citenamefont
  {Rosa}}]{Calza:2021czr}%
  \BibitemOpen
  \bibfield  {author} {\bibinfo {author} {\bibfnamefont {M.}~\bibnamefont
  {Calz{\`a}}}, \bibinfo {author} {\bibfnamefont {J.}~\bibnamefont
  {March-Russell}},\ and\ \bibinfo {author} {\bibfnamefont {J.~G.}\
  \bibnamefont {Rosa}},\ }\bibfield  {title} {\bibinfo {title} {{Evaporating
  Primordial Black Holes, the String Axiverse, and Hot Dark Radiation}},\
  }\href {https://doi.org/10.1103/PhysRevLett.133.261003} {\bibfield  {journal}
  {\bibinfo  {journal} {Phys. Rev. Lett.}\ }\textbf {\bibinfo {volume} {133}},\
  \bibinfo {pages} {261003} (\bibinfo {year} {2024})},\ \Eprint
  {https://arxiv.org/abs/2110.13602} {arXiv:2110.13602 [astro-ph.CO]}
  \BibitemShut {NoStop}%
\bibitem [{\citenamefont {Berteaud}\ \emph {et~al.}(2022)\citenamefont
  {Berteaud}, \citenamefont {Calore}, \citenamefont {Iguaz}, \citenamefont
  {Serpico},\ and\ \citenamefont {Siegert}}]{Berteaud:2022tws}%
  \BibitemOpen
  \bibfield  {author} {\bibinfo {author} {\bibfnamefont {J.}~\bibnamefont
  {Berteaud}}, \bibinfo {author} {\bibfnamefont {F.}~\bibnamefont {Calore}},
  \bibinfo {author} {\bibfnamefont {J.}~\bibnamefont {Iguaz}}, \bibinfo
  {author} {\bibfnamefont {P.~D.}\ \bibnamefont {Serpico}},\ and\ \bibinfo
  {author} {\bibfnamefont {T.}~\bibnamefont {Siegert}},\ }\bibfield  {title}
  {\bibinfo {title} {{Strong constraints on primordial black hole dark matter
  from 16~years of INTEGRAL/SPI observations}},\ }\href
  {https://doi.org/10.1103/PhysRevD.106.023030} {\bibfield  {journal} {\bibinfo
   {journal} {Phys. Rev. D}\ }\textbf {\bibinfo {volume} {106}},\ \bibinfo
  {pages} {023030} (\bibinfo {year} {2022})},\ \Eprint
  {https://arxiv.org/abs/2202.07483} {arXiv:2202.07483 [astro-ph.HE]}
  \BibitemShut {NoStop}%
\bibitem [{\citenamefont {De~la Torre~Luque}\ \emph {et~al.}(2024)\citenamefont
  {De~la Torre~Luque}, \citenamefont {Koechler},\ and\ \citenamefont
  {Balaji}}]{DelaTorreLuque:2024qms}%
  \BibitemOpen
  \bibfield  {author} {\bibinfo {author} {\bibfnamefont {P.}~\bibnamefont
  {De~la Torre~Luque}}, \bibinfo {author} {\bibfnamefont {J.}~\bibnamefont
  {Koechler}},\ and\ \bibinfo {author} {\bibfnamefont {S.}~\bibnamefont
  {Balaji}},\ }\bibfield  {title} {\bibinfo {title} {{Refining Galactic
  primordial black hole evaporation constraints}},\ }\href
  {https://doi.org/10.1103/PhysRevD.110.123022} {\bibfield  {journal} {\bibinfo
   {journal} {Phys. Rev. D}\ }\textbf {\bibinfo {volume} {110}},\ \bibinfo
  {pages} {123022} (\bibinfo {year} {2024})},\ \Eprint
  {https://arxiv.org/abs/2406.11949} {arXiv:2406.11949 [astro-ph.HE]}
  \BibitemShut {NoStop}%
\bibitem [{\citenamefont {Saha}\ \emph {et~al.}(2025)\citenamefont {Saha},
  \citenamefont {Singh}, \citenamefont {Parashari},\ and\ \citenamefont
  {Laha}}]{Saha:2024ies}%
  \BibitemOpen
  \bibfield  {author} {\bibinfo {author} {\bibfnamefont {A.~K.}\ \bibnamefont
  {Saha}}, \bibinfo {author} {\bibfnamefont {A.}~\bibnamefont {Singh}},
  \bibinfo {author} {\bibfnamefont {P.}~\bibnamefont {Parashari}},\ and\
  \bibinfo {author} {\bibfnamefont {R.}~\bibnamefont {Laha}},\ }\bibfield
  {title} {\bibinfo {title} {{Hunting primordial black hole dark matter in the
  Lyman-$\alpha $ forest}},\ }\href
  {https://doi.org/10.1140/epjc/s10052-025-14827-1} {\bibfield  {journal}
  {\bibinfo  {journal} {Eur. Phys. J. C}\ }\textbf {\bibinfo {volume} {85}},\
  \bibinfo {pages} {1117} (\bibinfo {year} {2025})},\ \Eprint
  {https://arxiv.org/abs/2409.10617} {arXiv:2409.10617 [astro-ph.CO]}
  \BibitemShut {NoStop}%
\bibitem [{\citenamefont {Su}\ \emph {et~al.}(2024)\citenamefont {Su},
  \citenamefont {Pan}, \citenamefont {Wang}, \citenamefont {Zu}, \citenamefont
  {Yang},\ and\ \citenamefont {Feng}}]{Su:2024hrp}%
  \BibitemOpen
  \bibfield  {author} {\bibinfo {author} {\bibfnamefont {B.-Y.}\ \bibnamefont
  {Su}}, \bibinfo {author} {\bibfnamefont {X.}~\bibnamefont {Pan}}, \bibinfo
  {author} {\bibfnamefont {G.-S.}\ \bibnamefont {Wang}}, \bibinfo {author}
  {\bibfnamefont {L.}~\bibnamefont {Zu}}, \bibinfo {author} {\bibfnamefont
  {Y.}~\bibnamefont {Yang}},\ and\ \bibinfo {author} {\bibfnamefont
  {L.}~\bibnamefont {Feng}},\ }\bibfield  {title} {\bibinfo {title}
  {{Constraining primordial black holes as dark matter using AMS-02 data}},\
  }\href {https://doi.org/10.1140/epjc/s10052-024-12773-y} {\bibfield
  {journal} {\bibinfo  {journal} {Eur. Phys. J. C}\ }\textbf {\bibinfo {volume}
  {84}},\ \bibinfo {pages} {606} (\bibinfo {year} {2024})},\ \bibinfo {note}
  {[Erratum: Eur.Phys.J.C 84, 768 (2024)]},\ \Eprint
  {https://arxiv.org/abs/2403.04988} {arXiv:2403.04988 [astro-ph.HE]}
  \BibitemShut {NoStop}%
\bibitem [{\citenamefont {Khan}\ \emph {et~al.}(2025)\citenamefont {Khan},
  \citenamefont {Ray}, \citenamefont {Kulkarni},\ and\ \citenamefont
  {Dasgupta}}]{Khan:2025kag}%
  \BibitemOpen
  \bibfield  {author} {\bibinfo {author} {\bibfnamefont {N.~K.}\ \bibnamefont
  {Khan}}, \bibinfo {author} {\bibfnamefont {A.}~\bibnamefont {Ray}}, \bibinfo
  {author} {\bibfnamefont {G.}~\bibnamefont {Kulkarni}},\ and\ \bibinfo
  {author} {\bibfnamefont {B.}~\bibnamefont {Dasgupta}},\ }\bibfield  {title}
  {\bibinfo {title} {{Stronger Constraints on Primordial Black Holes as Dark
  Matter Derived from the Thermal Evolution of the Intergalactic Medium over
  the Last Twelve Billion Years}},\ }\href@noop {} {\  (\bibinfo {year}
  {2025})},\ \Eprint {https://arxiv.org/abs/2503.15595} {arXiv:2503.15595
  [astro-ph.CO]} \BibitemShut {NoStop}%
\bibitem [{\citenamefont {Dimopoulos}\ and\ \citenamefont
  {Landsberg}(2001)}]{Dimopoulos:2001hw}%
  \BibitemOpen
  \bibfield  {author} {\bibinfo {author} {\bibfnamefont {S.}~\bibnamefont
  {Dimopoulos}}\ and\ \bibinfo {author} {\bibfnamefont {G.~L.}\ \bibnamefont
  {Landsberg}},\ }\bibfield  {title} {\bibinfo {title} {{Black holes at the
  LHC}},\ }\href {https://doi.org/10.1103/PhysRevLett.87.161602} {\bibfield
  {journal} {\bibinfo  {journal} {Phys. Rev. Lett.}\ }\textbf {\bibinfo
  {volume} {87}},\ \bibinfo {pages} {161602} (\bibinfo {year} {2001})},\
  \Eprint {https://arxiv.org/abs/hep-ph/0106295} {arXiv:hep-ph/0106295}
  \BibitemShut {NoStop}%
\bibitem [{\citenamefont {Giddings}\ and\ \citenamefont
  {Thomas}(2002)}]{Giddings:2001bu}%
  \BibitemOpen
  \bibfield  {author} {\bibinfo {author} {\bibfnamefont {S.~B.}\ \bibnamefont
  {Giddings}}\ and\ \bibinfo {author} {\bibfnamefont {S.~D.}\ \bibnamefont
  {Thomas}},\ }\bibfield  {title} {\bibinfo {title} {{High-energy colliders as
  black hole factories: The End of short distance physics}},\ }\href
  {https://doi.org/10.1103/PhysRevD.65.056010} {\bibfield  {journal} {\bibinfo
  {journal} {Phys. Rev. D}\ }\textbf {\bibinfo {volume} {65}},\ \bibinfo
  {pages} {056010} (\bibinfo {year} {2002})},\ \Eprint
  {https://arxiv.org/abs/hep-ph/0106219} {arXiv:hep-ph/0106219} \BibitemShut
  {NoStop}%
\bibitem [{\citenamefont {Raidal}\ \emph {et~al.}(2018)\citenamefont {Raidal},
  \citenamefont {Solodukhin}, \citenamefont {Vaskonen},\ and\ \citenamefont
  {Veerm{\"a}e}}]{Raidal:2018eoo}%
  \BibitemOpen
  \bibfield  {author} {\bibinfo {author} {\bibfnamefont {M.}~\bibnamefont
  {Raidal}}, \bibinfo {author} {\bibfnamefont {S.}~\bibnamefont {Solodukhin}},
  \bibinfo {author} {\bibfnamefont {V.}~\bibnamefont {Vaskonen}},\ and\
  \bibinfo {author} {\bibfnamefont {H.}~\bibnamefont {Veerm{\"a}e}},\
  }\bibfield  {title} {\bibinfo {title} {{Light Primordial Exotic Compact
  Objects as All Dark Matter}},\ }\href
  {https://doi.org/10.1103/PhysRevD.97.123520} {\bibfield  {journal} {\bibinfo
  {journal} {Phys. Rev. D}\ }\textbf {\bibinfo {volume} {97}},\ \bibinfo
  {pages} {123520} (\bibinfo {year} {2018})},\ \Eprint
  {https://arxiv.org/abs/1802.07728} {arXiv:1802.07728 [astro-ph.CO]}
  \BibitemShut {NoStop}%
\bibitem [{\citenamefont {Boluna}\ \emph {et~al.}(2024)\citenamefont {Boluna},
  \citenamefont {Profumo}, \citenamefont {Bl\'e},\ and\ \citenamefont
  {Hennings}}]{Boluna:2023jlo}%
  \BibitemOpen
  \bibfield  {author} {\bibinfo {author} {\bibfnamefont {X.}~\bibnamefont
  {Boluna}}, \bibinfo {author} {\bibfnamefont {S.}~\bibnamefont {Profumo}},
  \bibinfo {author} {\bibfnamefont {J.}~\bibnamefont {Bl\'e}},\ and\ \bibinfo
  {author} {\bibfnamefont {D.}~\bibnamefont {Hennings}},\ }\bibfield  {title}
  {\bibinfo {title} {{Searching for Exploding black holes}},\ }\href
  {https://doi.org/10.1088/1475-7516/2024/04/024} {\bibfield  {journal}
  {\bibinfo  {journal} {JCAP}\ }\textbf {\bibinfo {volume} {04}},\ \bibinfo
  {pages} {024}},\ \Eprint {https://arxiv.org/abs/2307.06467} {arXiv:2307.06467
  [astro-ph.HE]} \BibitemShut {NoStop}%
\bibitem [{\citenamefont {Baker}\ \emph
  {et~al.}(2025{\natexlab{b}})\citenamefont {Baker}, \citenamefont
  {Iguaz~Juan}, \citenamefont {Symons},\ and\ \citenamefont
  {Thamm}}]{Baker:2025zxm}%
  \BibitemOpen
  \bibfield  {author} {\bibinfo {author} {\bibfnamefont {M.~J.}\ \bibnamefont
  {Baker}}, \bibinfo {author} {\bibfnamefont {J.}~\bibnamefont {Iguaz~Juan}},
  \bibinfo {author} {\bibfnamefont {A.}~\bibnamefont {Symons}},\ and\ \bibinfo
  {author} {\bibfnamefont {A.}~\bibnamefont {Thamm}},\ }\bibfield  {title}
  {\bibinfo {title} {{Could We Observe an Exploding Black Hole in the Near
  Future?}},\ }\href {https://doi.org/10.1103/nwgd-g3zl} {\bibfield  {journal}
  {\bibinfo  {journal} {Phys. Rev. Lett.}\ }\textbf {\bibinfo {volume} {135}},\
  \bibinfo {pages} {111002} (\bibinfo {year} {2025}{\natexlab{b}})},\ \Eprint
  {https://arxiv.org/abs/2503.10755} {arXiv:2503.10755 [hep-ph]} \BibitemShut
  {NoStop}%
\bibitem [{\citenamefont {Ukwatta}\ \emph {et~al.}(2010)\citenamefont
  {Ukwatta}, \citenamefont {MacGibbon}, \citenamefont {Parke}, \citenamefont
  {Dhuga}, \citenamefont {Rhodes}, \citenamefont {Eskandarian}, \citenamefont
  {Gehrels}, \citenamefont {Maximon},\ and\ \citenamefont
  {Morris}}]{Ukwatta:2010zn}%
  \BibitemOpen
  \bibfield  {author} {\bibinfo {author} {\bibfnamefont {T.~N.}\ \bibnamefont
  {Ukwatta}}, \bibinfo {author} {\bibfnamefont {J.~H.}\ \bibnamefont
  {MacGibbon}}, \bibinfo {author} {\bibfnamefont {W.~C.}\ \bibnamefont
  {Parke}}, \bibinfo {author} {\bibfnamefont {K.~S.}\ \bibnamefont {Dhuga}},
  \bibinfo {author} {\bibfnamefont {S.}~\bibnamefont {Rhodes}}, \bibinfo
  {author} {\bibfnamefont {A.}~\bibnamefont {Eskandarian}}, \bibinfo {author}
  {\bibfnamefont {N.}~\bibnamefont {Gehrels}}, \bibinfo {author} {\bibfnamefont
  {L.}~\bibnamefont {Maximon}},\ and\ \bibinfo {author} {\bibfnamefont {D.~C.}\
  \bibnamefont {Morris}},\ }\bibfield  {title} {\bibinfo {title} {{Sensitivity
  of the FERMI Detectors to Gamma-Ray Bursts from Evaporating Primordial Black
  Holes (PBHs)}},\ }in\ \href {https://doi.org/10.1142/9789814374552_0278}
  {\emph {\bibinfo {booktitle} {{12th Marcel Grossmann Meeting on General
  Relativity}}}}\ (\bibinfo {year} {2010})\ pp.\ \bibinfo {pages}
  {1588--1590},\ \Eprint {https://arxiv.org/abs/1003.4515} {arXiv:1003.4515
  [astro-ph.HE]} \BibitemShut {NoStop}%
\bibitem [{\citenamefont {Cline}(1995)}]{Cline:1995jf}%
  \BibitemOpen
  \bibfield  {author} {\bibinfo {author} {\bibfnamefont {D.~B.}\ \bibnamefont
  {Cline}},\ }\bibfield  {title} {\bibinfo {title} {{The Search for GRB from
  primordial black hole evaporation}},\ }\href
  {https://doi.org/10.1007/BF00658656} {\bibfield  {journal} {\bibinfo
  {journal} {Astrophys. Space Sci.}\ }\textbf {\bibinfo {volume} {231}},\
  \bibinfo {pages} {393} (\bibinfo {year} {1995})}\BibitemShut {NoStop}%
\bibitem [{\citenamefont {Ukwatta}\ \emph {et~al.}(2016)\citenamefont {Ukwatta}
  \emph {et~al.}}]{Ukwatta:2015mfb}%
  \BibitemOpen
  \bibfield  {author} {\bibinfo {author} {\bibfnamefont {T.~N.}\ \bibnamefont
  {Ukwatta}} \emph {et~al.},\ }\bibfield  {title} {\bibinfo {title}
  {{Investigation of Primordial Black Hole Bursts using Interplanetary Network
  Gamma-ray Bursts}},\ }\href {https://doi.org/10.3847/0004-637X/826/1/98}
  {\bibfield  {journal} {\bibinfo  {journal} {Astrophys. J.}\ }\textbf
  {\bibinfo {volume} {826}},\ \bibinfo {pages} {98} (\bibinfo {year} {2016})},\
  \Eprint {https://arxiv.org/abs/1512.01264} {arXiv:1512.01264 [astro-ph.HE]}
  \BibitemShut {NoStop}%
\bibitem [{\citenamefont {Wright}(1996)}]{Wright:1995bi}%
  \BibitemOpen
  \bibfield  {author} {\bibinfo {author} {\bibfnamefont {E.~L.}\ \bibnamefont
  {Wright}},\ }\bibfield  {title} {\bibinfo {title} {{On the density of pbh's
  in the galactic halo}},\ }\href {https://doi.org/10.1086/176910} {\bibfield
  {journal} {\bibinfo  {journal} {Astrophys. J.}\ }\textbf {\bibinfo {volume}
  {459}},\ \bibinfo {pages} {487} (\bibinfo {year} {1996})},\ \Eprint
  {https://arxiv.org/abs/astro-ph/9509074} {arXiv:astro-ph/9509074}
  \BibitemShut {NoStop}%
\bibitem [{\citenamefont {Lehoucq}\ \emph {et~al.}(2009)\citenamefont
  {Lehoucq}, \citenamefont {Casse}, \citenamefont {Casandjian},\ and\
  \citenamefont {Grenier}}]{Lehoucq:2009ge}%
  \BibitemOpen
  \bibfield  {author} {\bibinfo {author} {\bibfnamefont {R.}~\bibnamefont
  {Lehoucq}}, \bibinfo {author} {\bibfnamefont {M.}~\bibnamefont {Casse}},
  \bibinfo {author} {\bibfnamefont {J.~M.}\ \bibnamefont {Casandjian}},\ and\
  \bibinfo {author} {\bibfnamefont {I.}~\bibnamefont {Grenier}},\ }\bibfield
  {title} {\bibinfo {title} {{New constraints on the primordial black hole
  number density from Galactic gamma-ray astronomy}},\ }\href
  {https://doi.org/10.1051/0004-6361/200911961} {\bibfield  {journal} {\bibinfo
   {journal} {Astron. Astrophys.}\ }\textbf {\bibinfo {volume} {502}},\
  \bibinfo {pages} {37} (\bibinfo {year} {2009})},\ \Eprint
  {https://arxiv.org/abs/0906.1648} {arXiv:0906.1648 [astro-ph.HE]}
  \BibitemShut {NoStop}%
\bibitem [{\citenamefont {Carr}\ \emph
  {et~al.}(2016{\natexlab{b}})\citenamefont {Carr}, \citenamefont {Kohri},
  \citenamefont {Sendouda},\ and\ \citenamefont {Yokoyama}}]{Carr:2016hva}%
  \BibitemOpen
  \bibfield  {author} {\bibinfo {author} {\bibfnamefont {B.~J.}\ \bibnamefont
  {Carr}}, \bibinfo {author} {\bibfnamefont {K.}~\bibnamefont {Kohri}},
  \bibinfo {author} {\bibfnamefont {Y.}~\bibnamefont {Sendouda}},\ and\
  \bibinfo {author} {\bibfnamefont {J.}~\bibnamefont {Yokoyama}},\ }\bibfield
  {title} {\bibinfo {title} {{Constraints on primordial black holes from the
  Galactic gamma-ray background}},\ }\href
  {https://doi.org/10.1103/PhysRevD.94.044029} {\bibfield  {journal} {\bibinfo
  {journal} {Phys. Rev. D}\ }\textbf {\bibinfo {volume} {94}},\ \bibinfo
  {pages} {044029} (\bibinfo {year} {2016}{\natexlab{b}})},\ \Eprint
  {https://arxiv.org/abs/1604.05349} {arXiv:1604.05349 [astro-ph.CO]}
  \BibitemShut {NoStop}%
\bibitem [{\citenamefont {Adriani}\ \emph {et~al.}(2025)\citenamefont {Adriani}
  \emph {et~al.}}]{KM3NeT:2025ccp}%
  \BibitemOpen
  \bibfield  {author} {\bibinfo {author} {\bibfnamefont {O.}~\bibnamefont
  {Adriani}} \emph {et~al.} (\bibinfo {collaboration} {KM3NeT}),\ }\bibfield
  {title} {\bibinfo {title} {{Ultrahigh-Energy Event KM3-230213A within the
  Global Neutrino Landscape}},\ }\href {https://doi.org/10.1103/yypk-zmb8}
  {\bibfield  {journal} {\bibinfo  {journal} {Phys. Rev. X}\ }\textbf {\bibinfo
  {volume} {15}},\ \bibinfo {pages} {031016} (\bibinfo {year} {2025})},\
  \Eprint {https://arxiv.org/abs/2502.08173} {arXiv:2502.08173 [astro-ph.HE]}
  \BibitemShut {NoStop}%
\bibitem [{\citenamefont {Boccia}\ and\ \citenamefont
  {Iocco}(2025)}]{Boccia:2025hpm}%
  \BibitemOpen
  \bibfield  {author} {\bibinfo {author} {\bibfnamefont {A.}~\bibnamefont
  {Boccia}}\ and\ \bibinfo {author} {\bibfnamefont {F.}~\bibnamefont {Iocco}},\
  }\bibfield  {title} {\bibinfo {title} {{Could the KM3{\textendash}230213A
  event be caused by an evaporating primordial black hole?}},\ }\href
  {https://doi.org/10.1103/qxcj-fpwn} {\bibfield  {journal} {\bibinfo
  {journal} {Phys. Rev. D}\ }\textbf {\bibinfo {volume} {112}},\ \bibinfo
  {pages} {063045} (\bibinfo {year} {2025})},\ \Eprint
  {https://arxiv.org/abs/2502.19245} {arXiv:2502.19245 [astro-ph.HE]}
  \BibitemShut {NoStop}%
\bibitem [{\citenamefont {Baker}\ \emph
  {et~al.}(2025{\natexlab{c}})\citenamefont {Baker}, \citenamefont
  {Iguaz~Juan}, \citenamefont {Symons},\ and\ \citenamefont
  {Thamm}}]{Baker:2025cff}%
  \BibitemOpen
  \bibfield  {author} {\bibinfo {author} {\bibfnamefont {M.~J.}\ \bibnamefont
  {Baker}}, \bibinfo {author} {\bibfnamefont {J.}~\bibnamefont {Iguaz~Juan}},
  \bibinfo {author} {\bibfnamefont {A.}~\bibnamefont {Symons}},\ and\ \bibinfo
  {author} {\bibfnamefont {A.}~\bibnamefont {Thamm}},\ }\bibfield  {title}
  {\bibinfo {title} {{Explaining the PeV Neutrino Fluxes at KM3NeT and IceCube
  with Quasi-Extremal Primordial Black Holes}},\ }\href@noop {} {\  (\bibinfo
  {year} {2025}{\natexlab{c}})},\ \Eprint {https://arxiv.org/abs/2505.22722}
  {arXiv:2505.22722 [hep-ph]} \BibitemShut {NoStop}%
\bibitem [{\citenamefont {Aldecoa-Tamayo}\ \emph {et~al.}(2025)\citenamefont
  {Aldecoa-Tamayo}, \citenamefont {Byrnes},\ and\ \citenamefont
  {Seery}}]{Aldecoa-Tamayo:2025dqe}%
  \BibitemOpen
  \bibfield  {author} {\bibinfo {author} {\bibfnamefont {I.}~\bibnamefont
  {Aldecoa-Tamayo}}, \bibinfo {author} {\bibfnamefont {C.~T.}\ \bibnamefont
  {Byrnes}},\ and\ \bibinfo {author} {\bibfnamefont {D.}~\bibnamefont
  {Seery}},\ }\bibfield  {title} {\bibinfo {title} {{Primordial black holes in
  Randall-Sundrum: Cosmological signatures}},\ }\href@noop {} {\  (\bibinfo
  {year} {2025})},\ \Eprint {https://arxiv.org/abs/2509.13409}
  {arXiv:2509.13409 [astro-ph.CO]} \BibitemShut {NoStop}%
\bibitem [{\citenamefont {Smirnov}\ \emph {et~al.}(2024)\citenamefont
  {Smirnov}, \citenamefont {Goobar}, \citenamefont {Linden},\ and\
  \citenamefont {M\"ortsell}}]{Smirnov:2022zip}%
  \BibitemOpen
  \bibfield  {author} {\bibinfo {author} {\bibfnamefont {J.}~\bibnamefont
  {Smirnov}}, \bibinfo {author} {\bibfnamefont {A.}~\bibnamefont {Goobar}},
  \bibinfo {author} {\bibfnamefont {T.}~\bibnamefont {Linden}},\ and\ \bibinfo
  {author} {\bibfnamefont {E.}~\bibnamefont {M\"ortsell}},\ }\bibfield  {title}
  {\bibinfo {title} {{White Dwarfs in Dwarf Spheroidal Galaxies: A New Class of
  Compact-Dark-Matter Detectors}},\ }\href
  {https://doi.org/10.1103/PhysRevLett.132.151401} {\bibfield  {journal}
  {\bibinfo  {journal} {Phys. Rev. Lett.}\ }\textbf {\bibinfo {volume} {132}},\
  \bibinfo {pages} {151401} (\bibinfo {year} {2024})},\ \Eprint
  {https://arxiv.org/abs/2211.00013} {arXiv:2211.00013 [astro-ph.CO]}
  \BibitemShut {NoStop}%
\bibitem [{\citenamefont {Paczynski}(1986)}]{Paczynski:1985jf}%
  \BibitemOpen
  \bibfield  {author} {\bibinfo {author} {\bibfnamefont {B.}~\bibnamefont
  {Paczynski}},\ }\bibfield  {title} {\bibinfo {title} {{Gravitational
  microlensing by the galactic halo}},\ }\href {https://doi.org/10.1086/164140}
  {\bibfield  {journal} {\bibinfo  {journal} {Astrophys. J.}\ }\textbf
  {\bibinfo {volume} {304}},\ \bibinfo {pages} {1} (\bibinfo {year}
  {1986})}\BibitemShut {NoStop}%
\bibitem [{\citenamefont {Allsman}\ \emph {et~al.}(2001)\citenamefont {Allsman}
  \emph {et~al.}}]{Macho:2000nvd}%
  \BibitemOpen
  \bibfield  {author} {\bibinfo {author} {\bibfnamefont {R.~A.}\ \bibnamefont
  {Allsman}} \emph {et~al.} (\bibinfo {collaboration} {Macho}),\ }\bibfield
  {title} {\bibinfo {title} {{MACHO project limits on black hole dark matter in
  the 1-30 solar mass range}},\ }\href {https://doi.org/10.1086/319636}
  {\bibfield  {journal} {\bibinfo  {journal} {Astrophys. J. Lett.}\ }\textbf
  {\bibinfo {volume} {550}},\ \bibinfo {pages} {L169} (\bibinfo {year}
  {2001})},\ \Eprint {https://arxiv.org/abs/astro-ph/0011506}
  {arXiv:astro-ph/0011506} \BibitemShut {NoStop}%
\bibitem [{\citenamefont {Tisserand}\ \emph {et~al.}(2007)\citenamefont
  {Tisserand} \emph {et~al.}}]{EROS-2:2006ryy}%
  \BibitemOpen
  \bibfield  {author} {\bibinfo {author} {\bibfnamefont {P.}~\bibnamefont
  {Tisserand}} \emph {et~al.} (\bibinfo {collaboration} {EROS-2}),\ }\bibfield
  {title} {\bibinfo {title} {{Limits on the Macho Content of the Galactic Halo
  from the EROS-2 Survey of the Magellanic Clouds}},\ }\href
  {https://doi.org/10.1051/0004-6361:20066017} {\bibfield  {journal} {\bibinfo
  {journal} {Astron. Astrophys.}\ }\textbf {\bibinfo {volume} {469}},\ \bibinfo
  {pages} {387} (\bibinfo {year} {2007})},\ \Eprint
  {https://arxiv.org/abs/astro-ph/0607207} {arXiv:astro-ph/0607207}
  \BibitemShut {NoStop}%
\bibitem [{\citenamefont {Griest}\ \emph {et~al.}(2014)\citenamefont {Griest},
  \citenamefont {Cieplak},\ and\ \citenamefont {Lehner}}]{Griest:2013aaa}%
  \BibitemOpen
  \bibfield  {author} {\bibinfo {author} {\bibfnamefont {K.}~\bibnamefont
  {Griest}}, \bibinfo {author} {\bibfnamefont {A.~M.}\ \bibnamefont
  {Cieplak}},\ and\ \bibinfo {author} {\bibfnamefont {M.~J.}\ \bibnamefont
  {Lehner}},\ }\bibfield  {title} {\bibinfo {title} {{Experimental Limits on
  Primordial Black Hole Dark Matter from the First 2 yr of Kepler Data}},\
  }\href {https://doi.org/10.1088/0004-637X/786/2/158} {\bibfield  {journal}
  {\bibinfo  {journal} {Astrophys. J.}\ }\textbf {\bibinfo {volume} {786}},\
  \bibinfo {pages} {158} (\bibinfo {year} {2014})},\ \Eprint
  {https://arxiv.org/abs/1307.5798} {arXiv:1307.5798 [astro-ph.CO]}
  \BibitemShut {NoStop}%
\bibitem [{\citenamefont {Smyth}\ \emph {et~al.}(2020)\citenamefont {Smyth},
  \citenamefont {Profumo}, \citenamefont {English}, \citenamefont {Jeltema},
  \citenamefont {McKinnon},\ and\ \citenamefont
  {Guhathakurta}}]{Smyth:2019whb}%
  \BibitemOpen
  \bibfield  {author} {\bibinfo {author} {\bibfnamefont {N.}~\bibnamefont
  {Smyth}}, \bibinfo {author} {\bibfnamefont {S.}~\bibnamefont {Profumo}},
  \bibinfo {author} {\bibfnamefont {S.}~\bibnamefont {English}}, \bibinfo
  {author} {\bibfnamefont {T.}~\bibnamefont {Jeltema}}, \bibinfo {author}
  {\bibfnamefont {K.}~\bibnamefont {McKinnon}},\ and\ \bibinfo {author}
  {\bibfnamefont {P.}~\bibnamefont {Guhathakurta}},\ }\bibfield  {title}
  {\bibinfo {title} {{Updated Constraints on Asteroid-Mass Primordial Black
  Holes as Dark Matter}},\ }\href {https://doi.org/10.1103/PhysRevD.101.063005}
  {\bibfield  {journal} {\bibinfo  {journal} {Phys. Rev. D}\ }\textbf {\bibinfo
  {volume} {101}},\ \bibinfo {pages} {063005} (\bibinfo {year} {2020})},\
  \Eprint {https://arxiv.org/abs/1910.01285} {arXiv:1910.01285 [astro-ph.CO]}
  \BibitemShut {NoStop}%
\bibitem [{\citenamefont {Mr\'oz}\ \emph {et~al.}(2024)\citenamefont {Mr\'oz}
  \emph {et~al.}}]{Mroz:2024mse}%
  \BibitemOpen
  \bibfield  {author} {\bibinfo {author} {\bibfnamefont {P.}~\bibnamefont
  {Mr\'oz}} \emph {et~al.},\ }\bibfield  {title} {\bibinfo {title} {{No massive
  black holes in the Milky Way halo}},\ }\href
  {https://doi.org/10.1038/s41586-024-07704-6} {\bibfield  {journal} {\bibinfo
  {journal} {Nature}\ }\textbf {\bibinfo {volume} {632}},\ \bibinfo {pages}
  {749} (\bibinfo {year} {2024})},\ \Eprint {https://arxiv.org/abs/2403.02386}
  {arXiv:2403.02386 [astro-ph.GA]} \BibitemShut {NoStop}%
\bibitem [{\citenamefont {Mr{\'o}z}\ \emph {et~al.}(2024)\citenamefont
  {Mr{\'o}z} \emph {et~al.}}]{Mroz:2024wia}%
  \BibitemOpen
  \bibfield  {author} {\bibinfo {author} {\bibfnamefont {P.}~\bibnamefont
  {Mr{\'o}z}} \emph {et~al.},\ }\bibfield  {title} {\bibinfo {title} {{Limits
  on Planetary-mass Primordial Black Holes from the OGLE High-cadence Survey of
  the Magellanic Clouds}},\ }\href {https://doi.org/10.3847/2041-8213/ad8e68}
  {\bibfield  {journal} {\bibinfo  {journal} {Astrophys. J. Lett.}\ }\textbf
  {\bibinfo {volume} {976}},\ \bibinfo {pages} {L19} (\bibinfo {year}
  {2024})},\ \Eprint {https://arxiv.org/abs/2410.06251} {arXiv:2410.06251
  [astro-ph.CO]} \BibitemShut {NoStop}%
\bibitem [{\citenamefont {Mr{\'o}z}\ \emph
  {et~al.}(2025{\natexlab{a}})\citenamefont {Mr{\'o}z} \emph
  {et~al.}}]{Mroz:2025xbl}%
  \BibitemOpen
  \bibfield  {author} {\bibinfo {author} {\bibfnamefont {P.}~\bibnamefont
  {Mr{\'o}z}} \emph {et~al.},\ }\bibfield  {title} {\bibinfo {title}
  {{Microlensing Optical Depth, Event Rate, and Limits on Compact Objects in
  Dark Matter Based on 20 Yr of OGLE Observations of the Small Magellanic
  Cloud}},\ }\href {https://doi.org/10.3847/1538-4365/adf842} {\bibfield
  {journal} {\bibinfo  {journal} {Astrophys. J. Suppl.}\ }\textbf {\bibinfo
  {volume} {280}},\ \bibinfo {pages} {49} (\bibinfo {year}
  {2025}{\natexlab{a}})},\ \Eprint {https://arxiv.org/abs/2507.13794}
  {arXiv:2507.13794 [astro-ph.GA]} \BibitemShut {NoStop}%
\bibitem [{\citenamefont {Hawkins}\ and\ \citenamefont
  {Garc{\'\i}a-Bellido}(2025)}]{Hawkins:2025mlo}%
  \BibitemOpen
  \bibfield  {author} {\bibinfo {author} {\bibfnamefont {M.~R.~S.}\
  \bibnamefont {Hawkins}}\ and\ \bibinfo {author} {\bibfnamefont
  {J.}~\bibnamefont {Garc{\'\i}a-Bellido}},\ }\bibfield  {title} {\bibinfo
  {title} {{A critical analysis of the recent OGLE limits on stellar mass
  primordial black holes in the halo of the Milky Way}},\ }\href@noop {} {\
  (\bibinfo {year} {2025})},\ \Eprint {https://arxiv.org/abs/2509.05400}
  {arXiv:2509.05400 [astro-ph.GA]} \BibitemShut {NoStop}%
\bibitem [{\citenamefont {Mr{\'o}z}\ \emph
  {et~al.}(2025{\natexlab{b}})\citenamefont {Mr{\'o}z} \emph
  {et~al.}}]{Mroz:2025aor}%
  \BibitemOpen
  \bibfield  {author} {\bibinfo {author} {\bibfnamefont {P.}~\bibnamefont
  {Mr{\'o}z}} \emph {et~al.},\ }\bibfield  {title} {\bibinfo {title} {{Eppur
  non si trovano: Comments on the Primordial Black Hole Limits in the Galactic
  Halo}},\ }\href@noop {} {\  (\bibinfo {year} {2025}{\natexlab{b}})},\ \Eprint
  {https://arxiv.org/abs/2510.20005} {arXiv:2510.20005 [astro-ph.CO]}
  \BibitemShut {NoStop}%
\bibitem [{\citenamefont {Zumalacarregui}\ and\ \citenamefont
  {Seljak}(2018)}]{Zumalacarregui:2017qqd}%
  \BibitemOpen
  \bibfield  {author} {\bibinfo {author} {\bibfnamefont {M.}~\bibnamefont
  {Zumalacarregui}}\ and\ \bibinfo {author} {\bibfnamefont {U.}~\bibnamefont
  {Seljak}},\ }\bibfield  {title} {\bibinfo {title} {{Limits on stellar-mass
  compact objects as dark matter from gravitational lensing of type Ia
  supernovae}},\ }\href {https://doi.org/10.1103/PhysRevLett.121.141101}
  {\bibfield  {journal} {\bibinfo  {journal} {Phys. Rev. Lett.}\ }\textbf
  {\bibinfo {volume} {121}},\ \bibinfo {pages} {141101} (\bibinfo {year}
  {2018})},\ \Eprint {https://arxiv.org/abs/1712.02240} {arXiv:1712.02240
  [astro-ph.CO]} \BibitemShut {NoStop}%
\bibitem [{\citenamefont {Urrutia}\ \emph {et~al.}(2023)\citenamefont
  {Urrutia}, \citenamefont {Vaskonen},\ and\ \citenamefont
  {Veerm\"ae}}]{Urrutia:2023mtk}%
  \BibitemOpen
  \bibfield  {author} {\bibinfo {author} {\bibfnamefont {J.}~\bibnamefont
  {Urrutia}}, \bibinfo {author} {\bibfnamefont {V.}~\bibnamefont {Vaskonen}},\
  and\ \bibinfo {author} {\bibfnamefont {H.}~\bibnamefont {Veerm\"ae}},\
  }\bibfield  {title} {\bibinfo {title} {{Gravitational wave microlensing by
  dressed primordial black holes}},\ }\href
  {https://doi.org/10.1103/PhysRevD.108.023507} {\bibfield  {journal} {\bibinfo
   {journal} {Phys. Rev. D}\ }\textbf {\bibinfo {volume} {108}},\ \bibinfo
  {pages} {023507} (\bibinfo {year} {2023})},\ \Eprint
  {https://arxiv.org/abs/2303.17601} {arXiv:2303.17601 [astro-ph.CO]}
  \BibitemShut {NoStop}%
\bibitem [{\citenamefont {Peta{\v{c}}}\ \emph {et~al.}(2022)\citenamefont
  {Peta{\v{c}}}, \citenamefont {Lavalle},\ and\ \citenamefont
  {Jedamzik}}]{Petac:2022rio}%
  \BibitemOpen
  \bibfield  {author} {\bibinfo {author} {\bibfnamefont {M.}~\bibnamefont
  {Peta{\v{c}}}}, \bibinfo {author} {\bibfnamefont {J.}~\bibnamefont
  {Lavalle}},\ and\ \bibinfo {author} {\bibfnamefont {K.}~\bibnamefont
  {Jedamzik}},\ }\bibfield  {title} {\bibinfo {title} {{Microlensing
  constraints on clustered primordial black holes}},\ }\href
  {https://doi.org/10.1103/PhysRevD.105.083520} {\bibfield  {journal} {\bibinfo
   {journal} {Phys. Rev. D}\ }\textbf {\bibinfo {volume} {105}},\ \bibinfo
  {pages} {083520} (\bibinfo {year} {2022})},\ \Eprint
  {https://arxiv.org/abs/2201.02521} {arXiv:2201.02521 [astro-ph.CO]}
  \BibitemShut {NoStop}%
\bibitem [{\citenamefont {Gorton}\ and\ \citenamefont
  {Green}(2022)}]{Gorton:2022fyb}%
  \BibitemOpen
  \bibfield  {author} {\bibinfo {author} {\bibfnamefont {M.}~\bibnamefont
  {Gorton}}\ and\ \bibinfo {author} {\bibfnamefont {A.~M.}\ \bibnamefont
  {Green}},\ }\bibfield  {title} {\bibinfo {title} {{Effect of clustering on
  primordial black hole microlensing constraints}},\ }\href
  {https://doi.org/10.1088/1475-7516/2022/08/035} {\bibfield  {journal}
  {\bibinfo  {journal} {JCAP}\ }\textbf {\bibinfo {volume} {08}}\bibfield
  {number} {\bibinfo  {number} { (08)},\ \bibinfo {pages} {035}},\ }\Eprint
  {https://arxiv.org/abs/2203.04209} {arXiv:2203.04209 [astro-ph.CO]}
  \BibitemShut {NoStop}%
\bibitem [{\citenamefont {De~Luca}\ \emph
  {et~al.}(2022{\natexlab{a}})\citenamefont {De~Luca}, \citenamefont
  {Franciolini}, \citenamefont {Riotto},\ and\ \citenamefont
  {Veerm{\"a}e}}]{DeLuca:2022uvz}%
  \BibitemOpen
  \bibfield  {author} {\bibinfo {author} {\bibfnamefont {V.}~\bibnamefont
  {De~Luca}}, \bibinfo {author} {\bibfnamefont {G.}~\bibnamefont
  {Franciolini}}, \bibinfo {author} {\bibfnamefont {A.}~\bibnamefont
  {Riotto}},\ and\ \bibinfo {author} {\bibfnamefont {H.}~\bibnamefont
  {Veerm{\"a}e}},\ }\bibfield  {title} {\bibinfo {title} {{Ruling Out Initially
  Clustered Primordial Black Holes as Dark Matter}},\ }\href
  {https://doi.org/10.1103/PhysRevLett.129.191302} {\bibfield  {journal}
  {\bibinfo  {journal} {Phys. Rev. Lett.}\ }\textbf {\bibinfo {volume} {129}},\
  \bibinfo {pages} {191302} (\bibinfo {year} {2022}{\natexlab{a}})},\ \Eprint
  {https://arxiv.org/abs/2208.01683} {arXiv:2208.01683 [astro-ph.CO]}
  \BibitemShut {NoStop}%
\bibitem [{\citenamefont {Green}(2017)}]{Green:2017qoa}%
  \BibitemOpen
  \bibfield  {author} {\bibinfo {author} {\bibfnamefont {A.~M.}\ \bibnamefont
  {Green}},\ }\bibfield  {title} {\bibinfo {title} {{Astrophysical
  uncertainties on stellar microlensing constraints on multi-Solar mass
  primordial black hole dark matter}},\ }\href
  {https://doi.org/10.1103/PhysRevD.96.043020} {\bibfield  {journal} {\bibinfo
  {journal} {Phys. Rev. D}\ }\textbf {\bibinfo {volume} {96}},\ \bibinfo
  {pages} {043020} (\bibinfo {year} {2017})},\ \Eprint
  {https://arxiv.org/abs/1705.10818} {arXiv:1705.10818 [astro-ph.CO]}
  \BibitemShut {NoStop}%
\bibitem [{\citenamefont {Green}(2025)}]{Green:2025dut}%
  \BibitemOpen
  \bibfield  {author} {\bibinfo {author} {\bibfnamefont {A.~M.}\ \bibnamefont
  {Green}},\ }\bibfield  {title} {\bibinfo {title} {{Primordial black hole
  stellar microlensing constraints: understanding their dependence on the
  density and velocity distributions}},\ }\href
  {https://doi.org/10.1088/1475-7516/2025/04/023} {\bibfield  {journal}
  {\bibinfo  {journal} {JCAP}\ }\textbf {\bibinfo {volume} {04}},\ \bibinfo
  {pages} {023}},\ \Eprint {https://arxiv.org/abs/2501.02610} {arXiv:2501.02610
  [astro-ph.GA]} \BibitemShut {NoStop}%
\bibitem [{\citenamefont {Li}\ \emph {et~al.}(2025)\citenamefont {Li},
  \citenamefont {Tang}, \citenamefont {Huang},\ and\ \citenamefont
  {Liu}}]{Li:2025mqx}%
  \BibitemOpen
  \bibfield  {author} {\bibinfo {author} {\bibfnamefont {B.}~\bibnamefont
  {Li}}, \bibinfo {author} {\bibfnamefont {C.-Y.}\ \bibnamefont {Tang}},
  \bibinfo {author} {\bibfnamefont {Z.-R.}\ \bibnamefont {Huang}},\ and\
  \bibinfo {author} {\bibfnamefont {L.-H.}\ \bibnamefont {Liu}},\ }\bibfield
  {title} {\bibinfo {title} {{Microlensing of dark matter models in the Milky
  Way}},\ }\href {https://doi.org/10.1088/1475-7516/2025/12/008} {\bibfield
  {journal} {\bibinfo  {journal} {JCAP}\ }\textbf {\bibinfo {volume} {12}},\
  \bibinfo {pages} {008}},\ \Eprint {https://arxiv.org/abs/2507.00770}
  {arXiv:2507.00770 [astro-ph.GA]} \BibitemShut {NoStop}%
\bibitem [{\citenamefont {Alcock}\ \emph {et~al.}(2000)\citenamefont {Alcock}
  \emph {et~al.}}]{MACHO:2000qbb}%
  \BibitemOpen
  \bibfield  {author} {\bibinfo {author} {\bibfnamefont {C.}~\bibnamefont
  {Alcock}} \emph {et~al.} (\bibinfo {collaboration} {MACHO}),\ }\bibfield
  {title} {\bibinfo {title} {{The MACHO project: Microlensing results from 5.7
  years of LMC observations}},\ }\href {https://doi.org/10.1086/309512}
  {\bibfield  {journal} {\bibinfo  {journal} {Astrophys. J.}\ }\textbf
  {\bibinfo {volume} {542}},\ \bibinfo {pages} {281} (\bibinfo {year}
  {2000})},\ \Eprint {https://arxiv.org/abs/astro-ph/0001272}
  {arXiv:astro-ph/0001272} \BibitemShut {NoStop}%
\bibitem [{\citenamefont {Niikura}\ \emph
  {et~al.}(2019{\natexlab{b}})\citenamefont {Niikura} \emph
  {et~al.}}]{Niikura:2017zjd}%
  \BibitemOpen
  \bibfield  {author} {\bibinfo {author} {\bibfnamefont {H.}~\bibnamefont
  {Niikura}} \emph {et~al.},\ }\bibfield  {title} {\bibinfo {title}
  {{Microlensing constraints on primordial black holes with Subaru/HSC
  Andromeda observations}},\ }\href {https://doi.org/10.1038/s41550-019-0723-1}
  {\bibfield  {journal} {\bibinfo  {journal} {Nature Astron.}\ }\textbf
  {\bibinfo {volume} {3}},\ \bibinfo {pages} {524} (\bibinfo {year}
  {2019}{\natexlab{b}})},\ \Eprint {https://arxiv.org/abs/1701.02151}
  {arXiv:1701.02151 [astro-ph.CO]} \BibitemShut {NoStop}%
\bibitem [{\citenamefont {{Schild}}\ and\ \citenamefont
  {{Smith}}(1991)}]{1991AJ....101..813S}%
  \BibitemOpen
  \bibfield  {author} {\bibinfo {author} {\bibfnamefont {R.~E.}\ \bibnamefont
  {{Schild}}}\ and\ \bibinfo {author} {\bibfnamefont {R.~C.}\ \bibnamefont
  {{Smith}}},\ }\bibfield  {title} {\bibinfo {title} {{Microlensing in the
  Q0957+561 Gravitational Mirage}},\ }\href {https://doi.org/10.1086/115724} {\
  \textbf {\bibinfo {volume} {101}},\ \bibinfo {pages} {813} (\bibinfo {year}
  {1991})}\BibitemShut {NoStop}%
\bibitem [{\citenamefont {Hawkins}(2020)}]{Hawkins:2020zie}%
  \BibitemOpen
  \bibfield  {author} {\bibinfo {author} {\bibfnamefont {M.~R.~S.}\
  \bibnamefont {Hawkins}},\ }\bibfield  {title} {\bibinfo {title} {{The
  signature of primordial black holes in the dark matter halos of galaxies}},\
  }\href {https://doi.org/10.1051/0004-6361/201936462} {\bibfield  {journal}
  {\bibinfo  {journal} {Astron. Astrophys.}\ }\textbf {\bibinfo {volume}
  {633}},\ \bibinfo {pages} {A107} (\bibinfo {year} {2020})},\ \Eprint
  {https://arxiv.org/abs/2001.07633} {arXiv:2001.07633 [astro-ph.GA]}
  \BibitemShut {NoStop}%
\bibitem [{\citenamefont {Hawkins}(2022)}]{Hawkins:2022vqo}%
  \BibitemOpen
  \bibfield  {author} {\bibinfo {author} {\bibfnamefont {M.~R.~S.}\
  \bibnamefont {Hawkins}},\ }\bibfield  {title} {\bibinfo {title} {{New
  evidence for a cosmological distribution of stellar mass primordial black
  holes}},\ }\href {https://doi.org/10.1093/mnras/stac863} {\bibfield
  {journal} {\bibinfo  {journal} {Mon. Not. Roy. Astron. Soc.}\ }\textbf
  {\bibinfo {volume} {512}},\ \bibinfo {pages} {5706} (\bibinfo {year}
  {2022})},\ \Eprint {https://arxiv.org/abs/2204.09143} {arXiv:2204.09143
  [astro-ph.CO]} \BibitemShut {NoStop}%
\bibitem [{\citenamefont {Nakamura}\ \emph {et~al.}(1997)\citenamefont
  {Nakamura}, \citenamefont {Sasaki}, \citenamefont {Tanaka},\ and\
  \citenamefont {Thorne}}]{Nakamura:1997sm}%
  \BibitemOpen
  \bibfield  {author} {\bibinfo {author} {\bibfnamefont {T.}~\bibnamefont
  {Nakamura}}, \bibinfo {author} {\bibfnamefont {M.}~\bibnamefont {Sasaki}},
  \bibinfo {author} {\bibfnamefont {T.}~\bibnamefont {Tanaka}},\ and\ \bibinfo
  {author} {\bibfnamefont {K.~S.}\ \bibnamefont {Thorne}},\ }\bibfield  {title}
  {\bibinfo {title} {{Gravitational waves from coalescing black hole MACHO
  binaries}},\ }\href {https://doi.org/10.1086/310886} {\bibfield  {journal}
  {\bibinfo  {journal} {Astrophys. J. Lett.}\ }\textbf {\bibinfo {volume}
  {487}},\ \bibinfo {pages} {L139} (\bibinfo {year} {1997})},\ \Eprint
  {https://arxiv.org/abs/astro-ph/9708060} {arXiv:astro-ph/9708060}
  \BibitemShut {NoStop}%
\bibitem [{\citenamefont {Raidal}\ \emph {et~al.}(2025)\citenamefont {Raidal},
  \citenamefont {Vaskonen},\ and\ \citenamefont {Veerm\"ae}}]{Raidal:2024bmm}%
  \BibitemOpen
  \bibfield  {author} {\bibinfo {author} {\bibfnamefont {M.}~\bibnamefont
  {Raidal}}, \bibinfo {author} {\bibfnamefont {V.}~\bibnamefont {Vaskonen}},\
  and\ \bibinfo {author} {\bibfnamefont {H.}~\bibnamefont {Veerm\"ae}},\
  }\bibinfo {title} {{Formation of~Primordial Black Hole Binaries and~Their
  Merger Rates}},\ in\ \href {https://doi.org/10.1007/978-981-97-8887-3_16}
  {\emph {\bibinfo {booktitle} {{Primordial Black Holes}}}},\ \bibinfo {editor}
  {edited by\ \bibinfo {editor} {\bibfnamefont {C.}~\bibnamefont {Byrnes}},
  \bibinfo {editor} {\bibfnamefont {G.}~\bibnamefont {Franciolini}}, \bibinfo
  {editor} {\bibfnamefont {T.}~\bibnamefont {Harada}}, \bibinfo {editor}
  {\bibfnamefont {P.}~\bibnamefont {Pani}},\ and\ \bibinfo {editor}
  {\bibfnamefont {M.}~\bibnamefont {Sasaki}}}\ (\bibinfo {year} {2025})\
  \Eprint {https://arxiv.org/abs/2404.08416} {arXiv:2404.08416 [astro-ph.CO]}
  \BibitemShut {NoStop}%
\bibitem [{\citenamefont {Ali-Ha{\"\i}moud}\ \emph {et~al.}(2017)\citenamefont
  {Ali-Ha{\"\i}moud}, \citenamefont {Kovetz},\ and\ \citenamefont
  {Kamionkowski}}]{Ali-Haimoud:2017rtz}%
  \BibitemOpen
  \bibfield  {author} {\bibinfo {author} {\bibfnamefont {Y.}~\bibnamefont
  {Ali-Ha{\"\i}moud}}, \bibinfo {author} {\bibfnamefont {E.~D.}\ \bibnamefont
  {Kovetz}},\ and\ \bibinfo {author} {\bibfnamefont {M.}~\bibnamefont
  {Kamionkowski}},\ }\bibfield  {title} {\bibinfo {title} {{Merger rate of
  primordial black-hole binaries}},\ }\href
  {https://doi.org/10.1103/PhysRevD.96.123523} {\bibfield  {journal} {\bibinfo
  {journal} {Phys. Rev. D}\ }\textbf {\bibinfo {volume} {96}},\ \bibinfo
  {pages} {123523} (\bibinfo {year} {2017})},\ \Eprint
  {https://arxiv.org/abs/1709.06576} {arXiv:1709.06576 [astro-ph.CO]}
  \BibitemShut {NoStop}%
\bibitem [{\citenamefont {Raidal}\ \emph {et~al.}(2019)\citenamefont {Raidal},
  \citenamefont {Spethmann}, \citenamefont {Vaskonen},\ and\ \citenamefont
  {Veerm\"ae}}]{Raidal:2018bbj}%
  \BibitemOpen
  \bibfield  {author} {\bibinfo {author} {\bibfnamefont {M.}~\bibnamefont
  {Raidal}}, \bibinfo {author} {\bibfnamefont {C.}~\bibnamefont {Spethmann}},
  \bibinfo {author} {\bibfnamefont {V.}~\bibnamefont {Vaskonen}},\ and\
  \bibinfo {author} {\bibfnamefont {H.}~\bibnamefont {Veerm\"ae}},\ }\bibfield
  {title} {\bibinfo {title} {{Formation and Evolution of Primordial Black Hole
  Binaries in the Early Universe}},\ }\href
  {https://doi.org/10.1088/1475-7516/2019/02/018} {\bibfield  {journal}
  {\bibinfo  {journal} {JCAP}\ }\textbf {\bibinfo {volume} {02}},\ \bibinfo
  {pages} {018}},\ \Eprint {https://arxiv.org/abs/1812.01930} {arXiv:1812.01930
  [astro-ph.CO]} \BibitemShut {NoStop}%
\bibitem [{\citenamefont {Vaskonen}\ and\ \citenamefont
  {Veerm\"ae}(2020)}]{Vaskonen:2019jpv}%
  \BibitemOpen
  \bibfield  {author} {\bibinfo {author} {\bibfnamefont {V.}~\bibnamefont
  {Vaskonen}}\ and\ \bibinfo {author} {\bibfnamefont {H.}~\bibnamefont
  {Veerm\"ae}},\ }\bibfield  {title} {\bibinfo {title} {{Lower bound on the
  primordial black hole merger rate}},\ }\href
  {https://doi.org/10.1103/PhysRevD.101.043015} {\bibfield  {journal} {\bibinfo
   {journal} {Phys. Rev. D}\ }\textbf {\bibinfo {volume} {101}},\ \bibinfo
  {pages} {043015} (\bibinfo {year} {2020})},\ \Eprint
  {https://arxiv.org/abs/1908.09752} {arXiv:1908.09752 [astro-ph.CO]}
  \BibitemShut {NoStop}%
\bibitem [{\citenamefont {Delos}\ \emph {et~al.}(2024)\citenamefont {Delos},
  \citenamefont {Rantala}, \citenamefont {Young},\ and\ \citenamefont
  {Schmidt}}]{Delos:2024poq}%
  \BibitemOpen
  \bibfield  {author} {\bibinfo {author} {\bibfnamefont {M.~S.}\ \bibnamefont
  {Delos}}, \bibinfo {author} {\bibfnamefont {A.}~\bibnamefont {Rantala}},
  \bibinfo {author} {\bibfnamefont {S.}~\bibnamefont {Young}},\ and\ \bibinfo
  {author} {\bibfnamefont {F.}~\bibnamefont {Schmidt}},\ }\bibfield  {title}
  {\bibinfo {title} {{Structure formation with primordial black holes:
  collisional dynamics, binaries, and gravitational waves}},\ }\href
  {https://doi.org/10.1088/1475-7516/2024/12/005} {\bibfield  {journal}
  {\bibinfo  {journal} {JCAP}\ }\textbf {\bibinfo {volume} {12}},\ \bibinfo
  {pages} {005}},\ \Eprint {https://arxiv.org/abs/2410.01876} {arXiv:2410.01876
  [astro-ph.CO]} \BibitemShut {NoStop}%
\bibitem [{\citenamefont {Franciolini}\ \emph
  {et~al.}(2022{\natexlab{b}})\citenamefont {Franciolini}, \citenamefont
  {Kritos}, \citenamefont {Berti},\ and\ \citenamefont
  {Silk}}]{Franciolini:2022ewd}%
  \BibitemOpen
  \bibfield  {author} {\bibinfo {author} {\bibfnamefont {G.}~\bibnamefont
  {Franciolini}}, \bibinfo {author} {\bibfnamefont {K.}~\bibnamefont {Kritos}},
  \bibinfo {author} {\bibfnamefont {E.}~\bibnamefont {Berti}},\ and\ \bibinfo
  {author} {\bibfnamefont {J.}~\bibnamefont {Silk}},\ }\bibfield  {title}
  {\bibinfo {title} {{Primordial black hole mergers from three-body
  interactions}},\ }\href {https://doi.org/10.1103/PhysRevD.106.083529}
  {\bibfield  {journal} {\bibinfo  {journal} {Phys. Rev. D}\ }\textbf {\bibinfo
  {volume} {106}},\ \bibinfo {pages} {083529} (\bibinfo {year}
  {2022}{\natexlab{b}})},\ \Eprint {https://arxiv.org/abs/2205.15340}
  {arXiv:2205.15340 [astro-ph.CO]} \BibitemShut {NoStop}%
\bibitem [{\citenamefont {Abbott}\ \emph
  {et~al.}(2019{\natexlab{a}})\citenamefont {Abbott} \emph
  {et~al.}}]{LIGOScientific:2018mvr}%
  \BibitemOpen
  \bibfield  {author} {\bibinfo {author} {\bibfnamefont {B.~P.}\ \bibnamefont
  {Abbott}} \emph {et~al.} (\bibinfo {collaboration} {LIGO Scientific,
  Virgo}),\ }\bibfield  {title} {\bibinfo {title} {{GWTC-1: A
  Gravitational-Wave Transient Catalog of Compact Binary Mergers Observed by
  LIGO and Virgo during the First and Second Observing Runs}},\ }\href
  {https://doi.org/10.1103/PhysRevX.9.031040} {\bibfield  {journal} {\bibinfo
  {journal} {Phys. Rev. X}\ }\textbf {\bibinfo {volume} {9}},\ \bibinfo {pages}
  {031040} (\bibinfo {year} {2019}{\natexlab{a}})},\ \Eprint
  {https://arxiv.org/abs/1811.12907} {arXiv:1811.12907 [astro-ph.HE]}
  \BibitemShut {NoStop}%
\bibitem [{\citenamefont {Abbott}\ \emph {et~al.}(2021)\citenamefont {Abbott}
  \emph {et~al.}}]{LIGOScientific:2020ibl}%
  \BibitemOpen
  \bibfield  {author} {\bibinfo {author} {\bibfnamefont {R.}~\bibnamefont
  {Abbott}} \emph {et~al.} (\bibinfo {collaboration} {LIGO Scientific,
  Virgo}),\ }\bibfield  {title} {\bibinfo {title} {{GWTC-2: Compact Binary
  Coalescences Observed by LIGO and Virgo During the First Half of the Third
  Observing Run}},\ }\href {https://doi.org/10.1103/PhysRevX.11.021053}
  {\bibfield  {journal} {\bibinfo  {journal} {Phys. Rev. X}\ }\textbf {\bibinfo
  {volume} {11}},\ \bibinfo {pages} {021053} (\bibinfo {year} {2021})},\
  \Eprint {https://arxiv.org/abs/2010.14527} {arXiv:2010.14527 [gr-qc]}
  \BibitemShut {NoStop}%
\bibitem [{\citenamefont {Abbott}\ \emph
  {et~al.}(2023{\natexlab{a}})\citenamefont {Abbott} \emph
  {et~al.}}]{KAGRA:2021vkt}%
  \BibitemOpen
  \bibfield  {author} {\bibinfo {author} {\bibfnamefont {R.}~\bibnamefont
  {Abbott}} \emph {et~al.} (\bibinfo {collaboration} {KAGRA, VIRGO, LIGO
  Scientific}),\ }\bibfield  {title} {\bibinfo {title} {{GWTC-3: Compact Binary
  Coalescences Observed by LIGO and Virgo during the Second Part of the Third
  Observing Run}},\ }\href {https://doi.org/10.1103/PhysRevX.13.041039}
  {\bibfield  {journal} {\bibinfo  {journal} {Phys. Rev. X}\ }\textbf {\bibinfo
  {volume} {13}},\ \bibinfo {pages} {041039} (\bibinfo {year}
  {2023}{\natexlab{a}})},\ \Eprint {https://arxiv.org/abs/2111.03606}
  {arXiv:2111.03606 [gr-qc]} \BibitemShut {NoStop}%
\bibitem [{\citenamefont {Abac}\ \emph
  {et~al.}(2025{\natexlab{a}})\citenamefont {Abac} \emph
  {et~al.}}]{LIGOScientific:2025slb}%
  \BibitemOpen
  \bibfield  {author} {\bibinfo {author} {\bibfnamefont {A.~G.}\ \bibnamefont
  {Abac}} \emph {et~al.} (\bibinfo {collaboration} {LIGO Scientific, VIRGO,
  KAGRA}),\ }\bibfield  {title} {\bibinfo {title} {{GWTC-4.0: Updating the
  Gravitational-Wave Transient Catalog with Observations from the First Part of
  the Fourth LIGO-Virgo-KAGRA Observing Run}},\ }\href@noop {} {\  (\bibinfo
  {year} {2025}{\natexlab{a}})},\ \Eprint {https://arxiv.org/abs/2508.18082}
  {arXiv:2508.18082 [gr-qc]} \BibitemShut {NoStop}%
\bibitem [{\citenamefont {Abbott}\ \emph
  {et~al.}(2019{\natexlab{b}})\citenamefont {Abbott} \emph
  {et~al.}}]{LIGOScientific:2019kan}%
  \BibitemOpen
  \bibfield  {author} {\bibinfo {author} {\bibfnamefont {B.~P.}\ \bibnamefont
  {Abbott}} \emph {et~al.} (\bibinfo {collaboration} {LIGO Scientific,
  Virgo}),\ }\bibfield  {title} {\bibinfo {title} {{Search for Subsolar Mass
  Ultracompact Binaries in Advanced LIGO{\textquoteright}s Second Observing
  Run}},\ }\href {https://doi.org/10.1103/PhysRevLett.123.161102} {\bibfield
  {journal} {\bibinfo  {journal} {Phys. Rev. Lett.}\ }\textbf {\bibinfo
  {volume} {123}},\ \bibinfo {pages} {161102} (\bibinfo {year}
  {2019}{\natexlab{b}})},\ \Eprint {https://arxiv.org/abs/1904.08976}
  {arXiv:1904.08976 [astro-ph.CO]} \BibitemShut {NoStop}%
\bibitem [{\citenamefont {Kavanagh}\ \emph {et~al.}(2018)\citenamefont
  {Kavanagh}, \citenamefont {Gaggero},\ and\ \citenamefont
  {Bertone}}]{Kavanagh:2018ggo}%
  \BibitemOpen
  \bibfield  {author} {\bibinfo {author} {\bibfnamefont {B.~J.}\ \bibnamefont
  {Kavanagh}}, \bibinfo {author} {\bibfnamefont {D.}~\bibnamefont {Gaggero}},\
  and\ \bibinfo {author} {\bibfnamefont {G.}~\bibnamefont {Bertone}},\
  }\bibfield  {title} {\bibinfo {title} {{Merger rate of a subdominant
  population of primordial black holes}},\ }\href
  {https://doi.org/10.1103/PhysRevD.98.023536} {\bibfield  {journal} {\bibinfo
  {journal} {Phys. Rev. D}\ }\textbf {\bibinfo {volume} {98}},\ \bibinfo
  {pages} {023536} (\bibinfo {year} {2018})},\ \Eprint
  {https://arxiv.org/abs/1805.09034} {arXiv:1805.09034 [astro-ph.CO]}
  \BibitemShut {NoStop}%
\bibitem [{\citenamefont {Wong}\ \emph {et~al.}(2021)\citenamefont {Wong},
  \citenamefont {Franciolini}, \citenamefont {De~Luca}, \citenamefont
  {Baibhav}, \citenamefont {Berti}, \citenamefont {Pani},\ and\ \citenamefont
  {Riotto}}]{Wong:2020yig}%
  \BibitemOpen
  \bibfield  {author} {\bibinfo {author} {\bibfnamefont {K.~W.~K.}\
  \bibnamefont {Wong}}, \bibinfo {author} {\bibfnamefont {G.}~\bibnamefont
  {Franciolini}}, \bibinfo {author} {\bibfnamefont {V.}~\bibnamefont
  {De~Luca}}, \bibinfo {author} {\bibfnamefont {V.}~\bibnamefont {Baibhav}},
  \bibinfo {author} {\bibfnamefont {E.}~\bibnamefont {Berti}}, \bibinfo
  {author} {\bibfnamefont {P.}~\bibnamefont {Pani}},\ and\ \bibinfo {author}
  {\bibfnamefont {A.}~\bibnamefont {Riotto}},\ }\bibfield  {title} {\bibinfo
  {title} {{Constraining the primordial black hole scenario with Bayesian
  inference and machine learning: the GWTC-2 gravitational wave catalog}},\
  }\href {https://doi.org/10.1103/PhysRevD.103.023026} {\bibfield  {journal}
  {\bibinfo  {journal} {Phys. Rev. D}\ }\textbf {\bibinfo {volume} {103}},\
  \bibinfo {pages} {023026} (\bibinfo {year} {2021})},\ \Eprint
  {https://arxiv.org/abs/2011.01865} {arXiv:2011.01865 [gr-qc]} \BibitemShut
  {NoStop}%
\bibitem [{\citenamefont {H\"utsi}\ \emph {et~al.}(2021)\citenamefont
  {H\"utsi}, \citenamefont {Raidal}, \citenamefont {Vaskonen},\ and\
  \citenamefont {Veerm\"ae}}]{Hutsi:2020sol}%
  \BibitemOpen
  \bibfield  {author} {\bibinfo {author} {\bibfnamefont {G.}~\bibnamefont
  {H\"utsi}}, \bibinfo {author} {\bibfnamefont {M.}~\bibnamefont {Raidal}},
  \bibinfo {author} {\bibfnamefont {V.}~\bibnamefont {Vaskonen}},\ and\
  \bibinfo {author} {\bibfnamefont {H.}~\bibnamefont {Veerm\"ae}},\ }\bibfield
  {title} {\bibinfo {title} {{Two populations of LIGO-Virgo black holes}},\
  }\href {https://doi.org/10.1088/1475-7516/2021/03/068} {\bibfield  {journal}
  {\bibinfo  {journal} {JCAP}\ }\textbf {\bibinfo {volume} {03}},\ \bibinfo
  {pages} {068}},\ \Eprint {https://arxiv.org/abs/2012.02786} {arXiv:2012.02786
  [astro-ph.CO]} \BibitemShut {NoStop}%
\bibitem [{\citenamefont {De~Luca}\ \emph
  {et~al.}(2021{\natexlab{b}})\citenamefont {De~Luca}, \citenamefont
  {Franciolini}, \citenamefont {Pani},\ and\ \citenamefont
  {Riotto}}]{DeLuca:2021wjr}%
  \BibitemOpen
  \bibfield  {author} {\bibinfo {author} {\bibfnamefont {V.}~\bibnamefont
  {De~Luca}}, \bibinfo {author} {\bibfnamefont {G.}~\bibnamefont
  {Franciolini}}, \bibinfo {author} {\bibfnamefont {P.}~\bibnamefont {Pani}},\
  and\ \bibinfo {author} {\bibfnamefont {A.}~\bibnamefont {Riotto}},\
  }\bibfield  {title} {\bibinfo {title} {{Bayesian Evidence for Both
  Astrophysical and Primordial Black Holes: Mapping the GWTC-2 Catalog to
  Third-Generation Detectors}},\ }\href
  {https://doi.org/10.1088/1475-7516/2021/05/003} {\bibfield  {journal}
  {\bibinfo  {journal} {JCAP}\ }\textbf {\bibinfo {volume} {05}},\ \bibinfo
  {pages} {003}},\ \Eprint {https://arxiv.org/abs/2102.03809} {arXiv:2102.03809
  [astro-ph.CO]} \BibitemShut {NoStop}%
\bibitem [{\citenamefont {Gasparotto}\ \emph {et~al.}(2025)\citenamefont
  {Gasparotto}, \citenamefont {Franciolini},\ and\ \citenamefont
  {Domcke}}]{Gasparotto:2025wok}%
  \BibitemOpen
  \bibfield  {author} {\bibinfo {author} {\bibfnamefont {S.}~\bibnamefont
  {Gasparotto}}, \bibinfo {author} {\bibfnamefont {G.}~\bibnamefont
  {Franciolini}},\ and\ \bibinfo {author} {\bibfnamefont {V.}~\bibnamefont
  {Domcke}},\ }\bibfield  {title} {\bibinfo {title} {{Gravitational wave memory
  of primordial black hole mergers}},\ }\href
  {https://doi.org/10.1103/ngfw-dvwz} {\bibfield  {journal} {\bibinfo
  {journal} {Phys. Rev. D}\ }\textbf {\bibinfo {volume} {112}},\ \bibinfo
  {pages} {103021} (\bibinfo {year} {2025})},\ \Eprint
  {https://arxiv.org/abs/2505.01356} {arXiv:2505.01356 [astro-ph.CO]}
  \BibitemShut {NoStop}%
\bibitem [{\citenamefont {Boybeyi}\ \emph {et~al.}(2025)\citenamefont
  {Boybeyi}, \citenamefont {Clesse}, \citenamefont {Kuroyanagi},\ and\
  \citenamefont {Sakellariadou}}]{Boybeyi:2024mhp}%
  \BibitemOpen
  \bibfield  {author} {\bibinfo {author} {\bibfnamefont {T.}~\bibnamefont
  {Boybeyi}}, \bibinfo {author} {\bibfnamefont {S.}~\bibnamefont {Clesse}},
  \bibinfo {author} {\bibfnamefont {S.}~\bibnamefont {Kuroyanagi}},\ and\
  \bibinfo {author} {\bibfnamefont {M.}~\bibnamefont {Sakellariadou}},\
  }\bibfield  {title} {\bibinfo {title} {{Search for a gravitational wave
  background from primordial black hole binaries using data from the first
  three LIGO-Virgo-KAGRA observing runs}},\ }\href
  {https://doi.org/10.1103/zphk-3ld9} {\bibfield  {journal} {\bibinfo
  {journal} {Phys. Rev. D}\ }\textbf {\bibinfo {volume} {112}},\ \bibinfo
  {pages} {023551} (\bibinfo {year} {2025})},\ \Eprint
  {https://arxiv.org/abs/2412.18318} {arXiv:2412.18318 [astro-ph.CO]}
  \BibitemShut {NoStop}%
\bibitem [{\citenamefont {Abbott}\ \emph
  {et~al.}(2023{\natexlab{b}})\citenamefont {Abbott} \emph
  {et~al.}}]{KAGRA:2021duu}%
  \BibitemOpen
  \bibfield  {author} {\bibinfo {author} {\bibfnamefont {R.}~\bibnamefont
  {Abbott}} \emph {et~al.} (\bibinfo {collaboration} {KAGRA, VIRGO, LIGO
  Scientific}),\ }\bibfield  {title} {\bibinfo {title} {{Population of Merging
  Compact Binaries Inferred Using Gravitational Waves through GWTC-3}},\ }\href
  {https://doi.org/10.1103/PhysRevX.13.011048} {\bibfield  {journal} {\bibinfo
  {journal} {Phys. Rev. X}\ }\textbf {\bibinfo {volume} {13}},\ \bibinfo
  {pages} {011048} (\bibinfo {year} {2023}{\natexlab{b}})},\ \Eprint
  {https://arxiv.org/abs/2111.03634} {arXiv:2111.03634 [astro-ph.HE]}
  \BibitemShut {NoStop}%
\bibitem [{\citenamefont {Abac}\ \emph
  {et~al.}(2025{\natexlab{b}})\citenamefont {Abac} \emph
  {et~al.}}]{LIGOScientific:2025pvj}%
  \BibitemOpen
  \bibfield  {author} {\bibinfo {author} {\bibfnamefont {A.~G.}\ \bibnamefont
  {Abac}} \emph {et~al.} (\bibinfo {collaboration} {LIGO Scientific, VIRGO,
  KAGRA}),\ }\bibfield  {title} {\bibinfo {title} {{GWTC-4.0: Population
  Properties of Merging Compact Binaries}},\ }\href@noop {} {\  (\bibinfo
  {year} {2025}{\natexlab{b}})},\ \Eprint {https://arxiv.org/abs/2508.18083}
  {arXiv:2508.18083 [astro-ph.HE]} \BibitemShut {NoStop}%
\bibitem [{\citenamefont {{Fowler}}\ and\ \citenamefont
  {{Hoyle}}(1964)}]{1964ApJS....9..201F}%
  \BibitemOpen
  \bibfield  {author} {\bibinfo {author} {\bibfnamefont {W.~A.}\ \bibnamefont
  {{Fowler}}}\ and\ \bibinfo {author} {\bibfnamefont {F.}~\bibnamefont
  {{Hoyle}}},\ }\bibfield  {title} {\bibinfo {title} {{Neutrino Processes and
  Pair Formation in Massive Stars and Supernovae.}},\ }\href
  {https://doi.org/10.1086/190103} {\bibfield  {journal} {\bibinfo  {journal}
  {\apjs}\ }\textbf {\bibinfo {volume} {9}},\ \bibinfo {pages} {201} (\bibinfo
  {year} {1964})}\BibitemShut {NoStop}%
\bibitem [{\citenamefont {Woosley}(2017)}]{Woosley:2016hmi}%
  \BibitemOpen
  \bibfield  {author} {\bibinfo {author} {\bibfnamefont {S.~E.}\ \bibnamefont
  {Woosley}},\ }\bibfield  {title} {\bibinfo {title} {{Pulsational
  Pair-Instability Supernovae}},\ }\href
  {https://doi.org/10.3847/1538-4357/836/2/244} {\bibfield  {journal} {\bibinfo
   {journal} {Astrophys. J.}\ }\textbf {\bibinfo {volume} {836}},\ \bibinfo
  {pages} {244} (\bibinfo {year} {2017})},\ \Eprint
  {https://arxiv.org/abs/1608.08939} {arXiv:1608.08939 [astro-ph.HE]}
  \BibitemShut {NoStop}%
\bibitem [{\citenamefont {Bailyn}\ \emph {et~al.}(1998)\citenamefont {Bailyn},
  \citenamefont {Jain}, \citenamefont {Coppi},\ and\ \citenamefont
  {Orosz}}]{Bailyn:1997xt}%
  \BibitemOpen
  \bibfield  {author} {\bibinfo {author} {\bibfnamefont {C.~D.}\ \bibnamefont
  {Bailyn}}, \bibinfo {author} {\bibfnamefont {R.~K.}\ \bibnamefont {Jain}},
  \bibinfo {author} {\bibfnamefont {P.}~\bibnamefont {Coppi}},\ and\ \bibinfo
  {author} {\bibfnamefont {J.~A.}\ \bibnamefont {Orosz}},\ }\bibfield  {title}
  {\bibinfo {title} {{The Mass distribution of stellar black holes}},\ }\href
  {https://doi.org/10.1086/305614} {\bibfield  {journal} {\bibinfo  {journal}
  {Astrophys. J.}\ }\textbf {\bibinfo {volume} {499}},\ \bibinfo {pages} {367}
  (\bibinfo {year} {1998})},\ \Eprint {https://arxiv.org/abs/astro-ph/9708032}
  {arXiv:astro-ph/9708032} \BibitemShut {NoStop}%
\bibitem [{\citenamefont {Farr}\ \emph {et~al.}(2011)\citenamefont {Farr},
  \citenamefont {Sravan}, \citenamefont {Cantrell}, \citenamefont {Kreidberg},
  \citenamefont {Bailyn}, \citenamefont {Mandel},\ and\ \citenamefont
  {Kalogera}}]{Farr:2010tu}%
  \BibitemOpen
  \bibfield  {author} {\bibinfo {author} {\bibfnamefont {W.~M.}\ \bibnamefont
  {Farr}}, \bibinfo {author} {\bibfnamefont {N.}~\bibnamefont {Sravan}},
  \bibinfo {author} {\bibfnamefont {A.}~\bibnamefont {Cantrell}}, \bibinfo
  {author} {\bibfnamefont {L.}~\bibnamefont {Kreidberg}}, \bibinfo {author}
  {\bibfnamefont {C.~D.}\ \bibnamefont {Bailyn}}, \bibinfo {author}
  {\bibfnamefont {I.}~\bibnamefont {Mandel}},\ and\ \bibinfo {author}
  {\bibfnamefont {V.}~\bibnamefont {Kalogera}},\ }\bibfield  {title} {\bibinfo
  {title} {{The Mass Distribution of Stellar-Mass Black Holes}},\ }\href
  {https://doi.org/10.1088/0004-637X/741/2/103} {\bibfield  {journal} {\bibinfo
   {journal} {Astrophys. J.}\ }\textbf {\bibinfo {volume} {741}},\ \bibinfo
  {pages} {103} (\bibinfo {year} {2011})},\ \Eprint
  {https://arxiv.org/abs/1011.1459} {arXiv:1011.1459 [astro-ph.GA]}
  \BibitemShut {NoStop}%
\bibitem [{\citenamefont {Yuan}\ \emph
  {et~al.}(2025{\natexlab{a}})\citenamefont {Yuan}, \citenamefont {Chen},\ and\
  \citenamefont {Liu}}]{Yuan:2025avq}%
  \BibitemOpen
  \bibfield  {author} {\bibinfo {author} {\bibfnamefont {C.}~\bibnamefont
  {Yuan}}, \bibinfo {author} {\bibfnamefont {Z.-C.}\ \bibnamefont {Chen}},\
  and\ \bibinfo {author} {\bibfnamefont {L.}~\bibnamefont {Liu}},\ }\bibfield
  {title} {\bibinfo {title} {{GW231123 mass gap event and the primordial black
  hole scenario}},\ }\href {https://doi.org/10.1103/2vfn-48kh} {\bibfield
  {journal} {\bibinfo  {journal} {Phys. Rev. D}\ }\textbf {\bibinfo {volume}
  {112}},\ \bibinfo {pages} {L081306} (\bibinfo {year} {2025}{\natexlab{a}})},\
  \Eprint {https://arxiv.org/abs/2507.15701} {arXiv:2507.15701 [astro-ph.CO]}
  \BibitemShut {NoStop}%
\bibitem [{\citenamefont {De~Luca}\ \emph
  {et~al.}(2025{\natexlab{a}})\citenamefont {De~Luca}, \citenamefont
  {Franciolini},\ and\ \citenamefont {Riotto}}]{DeLuca:2025fln}%
  \BibitemOpen
  \bibfield  {author} {\bibinfo {author} {\bibfnamefont {V.}~\bibnamefont
  {De~Luca}}, \bibinfo {author} {\bibfnamefont {G.}~\bibnamefont
  {Franciolini}},\ and\ \bibinfo {author} {\bibfnamefont {A.}~\bibnamefont
  {Riotto}},\ }\bibfield  {title} {\bibinfo {title} {{GW231123: a Possible
  Primordial Black Hole Origin}},\ }\href@noop {} {\  (\bibinfo {year}
  {2025}{\natexlab{a}})},\ \Eprint {https://arxiv.org/abs/2508.09965}
  {arXiv:2508.09965 [astro-ph.CO]} \BibitemShut {NoStop}%
\bibitem [{\citenamefont {Cannon}\ \emph {et~al.}(2020)\citenamefont {Cannon}
  \emph {et~al.}}]{Cannon:2020qnf}%
  \BibitemOpen
  \bibfield  {author} {\bibinfo {author} {\bibfnamefont {K.}~\bibnamefont
  {Cannon}} \emph {et~al.},\ }\bibfield  {title} {\bibinfo {title} {{GstLAL: A
  software framework for gravitational wave discovery}},\ }\href@noop {} {\
  (\bibinfo {year} {2020})},\ \Eprint {https://arxiv.org/abs/2010.05082}
  {arXiv:2010.05082 [astro-ph.IM]} \BibitemShut {NoStop}%
\bibitem [{\citenamefont {Prunier}\ \emph {et~al.}(2024)\citenamefont
  {Prunier}, \citenamefont {Morr{\'a}s}, \citenamefont {Siles}, \citenamefont
  {Clesse}, \citenamefont {Garc{\'\i}a-Bellido},\ and\ \citenamefont
  {Ruiz~Morales}}]{Prunier:2023uoo}%
  \BibitemOpen
  \bibfield  {author} {\bibinfo {author} {\bibfnamefont {M.}~\bibnamefont
  {Prunier}}, \bibinfo {author} {\bibfnamefont {G.}~\bibnamefont {Morr{\'a}s}},
  \bibinfo {author} {\bibfnamefont {J.~F.~N.}\ \bibnamefont {Siles}}, \bibinfo
  {author} {\bibfnamefont {S.}~\bibnamefont {Clesse}}, \bibinfo {author}
  {\bibfnamefont {J.}~\bibnamefont {Garc{\'\i}a-Bellido}},\ and\ \bibinfo
  {author} {\bibfnamefont {E.}~\bibnamefont {Ruiz~Morales}},\ }\bibfield
  {title} {\bibinfo {title} {{Analysis of the subsolar-mass black hole
  candidate SSM200308 from the second part of the third observing run of
  Advanced LIGO-Virgo}},\ }\href {https://doi.org/10.1016/j.dark.2024.101582}
  {\bibfield  {journal} {\bibinfo  {journal} {Phys. Dark Univ.}\ }\textbf
  {\bibinfo {volume} {46}},\ \bibinfo {pages} {101582} (\bibinfo {year}
  {2024})},\ \Eprint {https://arxiv.org/abs/2311.16085} {arXiv:2311.16085
  [gr-qc]} \BibitemShut {NoStop}%
\bibitem [{\citenamefont {De~Luca}\ \emph
  {et~al.}(2025{\natexlab{b}})\citenamefont {De~Luca}, \citenamefont {Iovino},\
  and\ \citenamefont {Riotto}}]{DeLuca:2025uov}%
  \BibitemOpen
  \bibfield  {author} {\bibinfo {author} {\bibfnamefont {V.}~\bibnamefont
  {De~Luca}}, \bibinfo {author} {\bibfnamefont {A.~J.}\ \bibnamefont
  {Iovino}},\ and\ \bibinfo {author} {\bibfnamefont {A.}~\bibnamefont
  {Riotto}},\ }\bibfield  {title} {\bibinfo {title} {{Primordial Black Hole
  Ringdown: the Irreducible Stochastic Gravitational Wave Background}},\
  }\href@noop {} {\  (\bibinfo {year} {2025}{\natexlab{b}})},\ \Eprint
  {https://arxiv.org/abs/2507.04083} {arXiv:2507.04083 [gr-qc]} \BibitemShut
  {NoStop}%
\bibitem [{\citenamefont {Yuan}\ \emph
  {et~al.}(2025{\natexlab{b}})\citenamefont {Yuan}, \citenamefont {Zhong},\
  and\ \citenamefont {Huang}}]{Yuan:2025bdp}%
  \BibitemOpen
  \bibfield  {author} {\bibinfo {author} {\bibfnamefont {C.}~\bibnamefont
  {Yuan}}, \bibinfo {author} {\bibfnamefont {Z.}~\bibnamefont {Zhong}},\ and\
  \bibinfo {author} {\bibfnamefont {Q.-G.}\ \bibnamefont {Huang}},\ }\bibfield
  {title} {\bibinfo {title} {{Whispers from the Early Universe: The Ringdown of
  Primordial Black Holes}},\ }\href@noop {} {\  (\bibinfo {year}
  {2025}{\natexlab{b}})},\ \Eprint {https://arxiv.org/abs/2507.07665}
  {arXiv:2507.07665 [astro-ph.CO]} \BibitemShut {NoStop}%
\bibitem [{\citenamefont {Kogut}\ \emph {et~al.}(2011)\citenamefont {Kogut}
  \emph {et~al.}}]{Kogut:2011xw}%
  \BibitemOpen
  \bibfield  {author} {\bibinfo {author} {\bibfnamefont {A.}~\bibnamefont
  {Kogut}} \emph {et~al.},\ }\bibfield  {title} {\bibinfo {title} {{The
  Primordial Inflation Explorer (PIXIE): A Nulling Polarimeter for Cosmic
  Microwave Background Observations}},\ }\href
  {https://doi.org/10.1088/1475-7516/2011/07/025} {\bibfield  {journal}
  {\bibinfo  {journal} {JCAP}\ }\textbf {\bibinfo {volume} {07}},\ \bibinfo
  {pages} {025}},\ \Eprint {https://arxiv.org/abs/1105.2044} {arXiv:1105.2044
  [astro-ph.CO]} \BibitemShut {NoStop}%
\bibitem [{\citenamefont {Kogut}\ \emph {et~al.}(2019)\citenamefont {Kogut},
  \citenamefont {Abitbol}, \citenamefont {Chluba}, \citenamefont
  {Delabrouille}, \citenamefont {Fixsen}, \citenamefont {Hill}, \citenamefont
  {Patil},\ and\ \citenamefont {Rotti}}]{Kogut:2019vqh}%
  \BibitemOpen
  \bibfield  {author} {\bibinfo {author} {\bibfnamefont {A.}~\bibnamefont
  {Kogut}}, \bibinfo {author} {\bibfnamefont {M.~H.}\ \bibnamefont {Abitbol}},
  \bibinfo {author} {\bibfnamefont {J.}~\bibnamefont {Chluba}}, \bibinfo
  {author} {\bibfnamefont {J.}~\bibnamefont {Delabrouille}}, \bibinfo {author}
  {\bibfnamefont {D.}~\bibnamefont {Fixsen}}, \bibinfo {author} {\bibfnamefont
  {J.~C.}\ \bibnamefont {Hill}}, \bibinfo {author} {\bibfnamefont {S.~P.}\
  \bibnamefont {Patil}},\ and\ \bibinfo {author} {\bibfnamefont
  {A.}~\bibnamefont {Rotti}},\ }\bibfield  {title} {\bibinfo {title} {{CMB
  Spectral Distortions: Status and Prospects}},\ }\href@noop {} {\bibfield
  {journal} {\bibinfo  {journal} {Bull. Am. Astron. Soc.}\ }\textbf {\bibinfo
  {volume} {51}},\ \bibinfo {pages} {113} (\bibinfo {year} {2019})},\ \Eprint
  {https://arxiv.org/abs/1907.13195} {arXiv:1907.13195 [astro-ph.CO]}
  \BibitemShut {NoStop}%
\bibitem [{\citenamefont {Chluba}\ \emph {et~al.}(2021)\citenamefont {Chluba}
  \emph {et~al.}}]{Chluba:2019nxa}%
  \BibitemOpen
  \bibfield  {author} {\bibinfo {author} {\bibfnamefont {J.}~\bibnamefont
  {Chluba}} \emph {et~al.},\ }\bibfield  {title} {\bibinfo {title} {{New
  horizons in cosmology with spectral distortions of the cosmic microwave
  background}},\ }\href {https://doi.org/10.1007/s10686-021-09729-5} {\bibfield
   {journal} {\bibinfo  {journal} {Exper. Astron.}\ }\textbf {\bibinfo {volume}
  {51}},\ \bibinfo {pages} {1515} (\bibinfo {year} {2021})},\ \Eprint
  {https://arxiv.org/abs/1909.01593} {arXiv:1909.01593 [astro-ph.CO]}
  \BibitemShut {NoStop}%
\bibitem [{\citenamefont {Carr}\ and\ \citenamefont
  {Sakellariadou}(1999)}]{Carr:1997cn}%
  \BibitemOpen
  \bibfield  {author} {\bibinfo {author} {\bibfnamefont {B.~J.}\ \bibnamefont
  {Carr}}\ and\ \bibinfo {author} {\bibfnamefont {M.}~\bibnamefont
  {Sakellariadou}},\ }\bibfield  {title} {\bibinfo {title} {{Dynamical
  constraints on dark compact objects}},\ }\href
  {https://doi.org/10.1086/307071} {\bibfield  {journal} {\bibinfo  {journal}
  {Astrophys. J.}\ }\textbf {\bibinfo {volume} {516}},\ \bibinfo {pages} {195}
  (\bibinfo {year} {1999})}\BibitemShut {NoStop}%
\bibitem [{\citenamefont {Brandt}(2016)}]{Brandt:2016aco}%
  \BibitemOpen
  \bibfield  {author} {\bibinfo {author} {\bibfnamefont {T.~D.}\ \bibnamefont
  {Brandt}},\ }\bibfield  {title} {\bibinfo {title} {{Constraints on MACHO Dark
  Matter from Compact Stellar Systems in Ultra-Faint Dwarf Galaxies}},\ }\href
  {https://doi.org/10.3847/2041-8205/824/2/L31} {\bibfield  {journal} {\bibinfo
   {journal} {Astrophys. J. Lett.}\ }\textbf {\bibinfo {volume} {824}},\
  \bibinfo {pages} {L31} (\bibinfo {year} {2016})},\ \Eprint
  {https://arxiv.org/abs/1605.03665} {arXiv:1605.03665 [astro-ph.GA]}
  \BibitemShut {NoStop}%
\bibitem [{\citenamefont {Koushiappas}\ and\ \citenamefont
  {Loeb}(2017)}]{Koushiappas:2017chw}%
  \BibitemOpen
  \bibfield  {author} {\bibinfo {author} {\bibfnamefont {S.~M.}\ \bibnamefont
  {Koushiappas}}\ and\ \bibinfo {author} {\bibfnamefont {A.}~\bibnamefont
  {Loeb}},\ }\bibfield  {title} {\bibinfo {title} {{Dynamics of Dwarf Galaxies
  Disfavor Stellar-Mass Black Holes as Dark Matter}},\ }\href
  {https://doi.org/10.1103/PhysRevLett.119.041102} {\bibfield  {journal}
  {\bibinfo  {journal} {Phys. Rev. Lett.}\ }\textbf {\bibinfo {volume} {119}},\
  \bibinfo {pages} {041102} (\bibinfo {year} {2017})},\ \Eprint
  {https://arxiv.org/abs/1704.01668} {arXiv:1704.01668 [astro-ph.GA]}
  \BibitemShut {NoStop}%
\bibitem [{\citenamefont {Monroy-Rodr\'\i{}guez}\ and\ \citenamefont
  {Allen}(2014)}]{Monroy-Rodriguez:2014ula}%
  \BibitemOpen
  \bibfield  {author} {\bibinfo {author} {\bibfnamefont {M.~A.}\ \bibnamefont
  {Monroy-Rodr\'\i{}guez}}\ and\ \bibinfo {author} {\bibfnamefont
  {C.}~\bibnamefont {Allen}},\ }\bibfield  {title} {\bibinfo {title} {{The end
  of the MACHO era- revisited: new limits on MACHO masses from halo wide
  binaries}},\ }\href {https://doi.org/10.1088/0004-637X/790/2/159} {\bibfield
  {journal} {\bibinfo  {journal} {Astrophys. J.}\ }\textbf {\bibinfo {volume}
  {790}},\ \bibinfo {pages} {159} (\bibinfo {year} {2014})},\ \Eprint
  {https://arxiv.org/abs/1406.5169} {arXiv:1406.5169 [astro-ph.GA]}
  \BibitemShut {NoStop}%
\bibitem [{\citenamefont {Carr}\ \emph
  {et~al.}(2021{\natexlab{c}})\citenamefont {Carr}, \citenamefont {Kuhnel},\
  and\ \citenamefont {Visinelli}}]{Carr:2020erq}%
  \BibitemOpen
  \bibfield  {author} {\bibinfo {author} {\bibfnamefont {B.}~\bibnamefont
  {Carr}}, \bibinfo {author} {\bibfnamefont {F.}~\bibnamefont {Kuhnel}},\ and\
  \bibinfo {author} {\bibfnamefont {L.}~\bibnamefont {Visinelli}},\ }\bibfield
  {title} {\bibinfo {title} {{Constraints on Stupendously Large Black Holes}},\
  }\href {https://doi.org/10.1093/mnras/staa3651} {\bibfield  {journal}
  {\bibinfo  {journal} {Mon. Not. Roy. Astron. Soc.}\ }\textbf {\bibinfo
  {volume} {501}},\ \bibinfo {pages} {2029} (\bibinfo {year}
  {2021}{\natexlab{c}})},\ \Eprint {https://arxiv.org/abs/2008.08077}
  {arXiv:2008.08077 [astro-ph.CO]} \BibitemShut {NoStop}%
\bibitem [{\citenamefont {Murgia}\ \emph {et~al.}(2019)\citenamefont {Murgia},
  \citenamefont {Scelfo}, \citenamefont {Viel},\ and\ \citenamefont
  {Raccanelli}}]{Murgia:2019duy}%
  \BibitemOpen
  \bibfield  {author} {\bibinfo {author} {\bibfnamefont {R.}~\bibnamefont
  {Murgia}}, \bibinfo {author} {\bibfnamefont {G.}~\bibnamefont {Scelfo}},
  \bibinfo {author} {\bibfnamefont {M.}~\bibnamefont {Viel}},\ and\ \bibinfo
  {author} {\bibfnamefont {A.}~\bibnamefont {Raccanelli}},\ }\bibfield  {title}
  {\bibinfo {title} {{Lyman-\ensuremath{\alpha} Forest Constraints on
  Primordial Black Holes as Dark Matter}},\ }\href
  {https://doi.org/10.1103/PhysRevLett.123.071102} {\bibfield  {journal}
  {\bibinfo  {journal} {Phys. Rev. Lett.}\ }\textbf {\bibinfo {volume} {123}},\
  \bibinfo {pages} {071102} (\bibinfo {year} {2019})},\ \Eprint
  {https://arxiv.org/abs/1903.10509} {arXiv:1903.10509 [astro-ph.CO]}
  \BibitemShut {NoStop}%
\bibitem [{\citenamefont {Gouttenoire}\ \emph {et~al.}(2024)\citenamefont
  {Gouttenoire}, \citenamefont {Trifinopoulos}, \citenamefont {Valogiannis},\
  and\ \citenamefont {Vanvlasselaer}}]{Gouttenoire:2023nzr}%
  \BibitemOpen
  \bibfield  {author} {\bibinfo {author} {\bibfnamefont {Y.}~\bibnamefont
  {Gouttenoire}}, \bibinfo {author} {\bibfnamefont {S.}~\bibnamefont
  {Trifinopoulos}}, \bibinfo {author} {\bibfnamefont {G.}~\bibnamefont
  {Valogiannis}},\ and\ \bibinfo {author} {\bibfnamefont {M.}~\bibnamefont
  {Vanvlasselaer}},\ }\bibfield  {title} {\bibinfo {title} {{Scrutinizing the
  primordial black hole interpretation of PTA gravitational waves and JWST
  early galaxies}},\ }\href {https://doi.org/10.1103/PhysRevD.109.123002}
  {\bibfield  {journal} {\bibinfo  {journal} {Phys. Rev. D}\ }\textbf {\bibinfo
  {volume} {109}},\ \bibinfo {pages} {123002} (\bibinfo {year} {2024})},\
  \Eprint {https://arxiv.org/abs/2307.01457} {arXiv:2307.01457 [astro-ph.CO]}
  \BibitemShut {NoStop}%
\bibitem [{\citenamefont {Ellis}\ \emph {et~al.}(2025)\citenamefont {Ellis},
  \citenamefont {Fairbairn}, \citenamefont {Urrutia},\ and\ \citenamefont
  {Vaskonen}}]{Ellis:2025xju}%
  \BibitemOpen
  \bibfield  {author} {\bibinfo {author} {\bibfnamefont {J.}~\bibnamefont
  {Ellis}}, \bibinfo {author} {\bibfnamefont {M.}~\bibnamefont {Fairbairn}},
  \bibinfo {author} {\bibfnamefont {J.}~\bibnamefont {Urrutia}},\ and\ \bibinfo
  {author} {\bibfnamefont {V.}~\bibnamefont {Vaskonen}},\ }\bibfield  {title}
  {\bibinfo {title} {{Starlight from JWST: Implications for star formation and
  dark matter models}},\ }\href@noop {} {\  (\bibinfo {year} {2025})},\ \Eprint
  {https://arxiv.org/abs/2504.20043} {arXiv:2504.20043 [astro-ph.CO]}
  \BibitemShut {NoStop}%
\bibitem [{\citenamefont {Ivanov}\ and\ \citenamefont
  {Trifinopoulos}(2025)}]{Ivanov:2025pbu}%
  \BibitemOpen
  \bibfield  {author} {\bibinfo {author} {\bibfnamefont {M.~M.}\ \bibnamefont
  {Ivanov}}\ and\ \bibinfo {author} {\bibfnamefont {S.}~\bibnamefont
  {Trifinopoulos}},\ }\bibfield  {title} {\bibinfo {title} {{Effective Field
  Theory Constraints on Primordial Black Holes from the High-Redshift
  Lyman-$\alpha$ Forest}},\ }\href@noop {} {\  (\bibinfo {year} {2025})},\
  \Eprint {https://arxiv.org/abs/2508.04767} {arXiv:2508.04767 [astro-ph.CO]}
  \BibitemShut {NoStop}%
\bibitem [{\citenamefont {Gerlach}\ \emph {et~al.}(2025)\citenamefont
  {Gerlach}, \citenamefont {Gouttenoire}, \citenamefont {Iovino},\ and\
  \citenamefont {Leister}}]{Gerlach:2025vco}%
  \BibitemOpen
  \bibfield  {author} {\bibinfo {author} {\bibfnamefont {C.}~\bibnamefont
  {Gerlach}}, \bibinfo {author} {\bibfnamefont {Y.}~\bibnamefont
  {Gouttenoire}}, \bibinfo {author} {\bibfnamefont {A.~J.}\ \bibnamefont
  {Iovino}},\ and\ \bibinfo {author} {\bibfnamefont {N.}~\bibnamefont
  {Leister}},\ }\bibfield  {title} {\bibinfo {title} {{Closing the Mass Window
  for Stupendously Large Black Holes}},\ }\href@noop {} {\  (\bibinfo {year}
  {2025})},\ \Eprint {https://arxiv.org/abs/2508.08238} {arXiv:2508.08238
  [astro-ph.CO]} \BibitemShut {NoStop}%
\bibitem [{\citenamefont {Fuller}\ \emph {et~al.}(2017)\citenamefont {Fuller},
  \citenamefont {Kusenko},\ and\ \citenamefont {Takhistov}}]{Fuller:2017uyd}%
  \BibitemOpen
  \bibfield  {author} {\bibinfo {author} {\bibfnamefont {G.~M.}\ \bibnamefont
  {Fuller}}, \bibinfo {author} {\bibfnamefont {A.}~\bibnamefont {Kusenko}},\
  and\ \bibinfo {author} {\bibfnamefont {V.}~\bibnamefont {Takhistov}},\
  }\bibfield  {title} {\bibinfo {title} {{Primordial Black Holes and
  $r$-Process Nucleosynthesis}},\ }\href
  {https://doi.org/10.1103/PhysRevLett.119.061101} {\bibfield  {journal}
  {\bibinfo  {journal} {Phys. Rev. Lett.}\ }\textbf {\bibinfo {volume} {119}},\
  \bibinfo {pages} {061101} (\bibinfo {year} {2017})},\ \Eprint
  {https://arxiv.org/abs/1704.01129} {arXiv:1704.01129 [astro-ph.HE]}
  \BibitemShut {NoStop}%
\bibitem [{\citenamefont {Abramowicz}\ \emph {et~al.}(2018)\citenamefont
  {Abramowicz}, \citenamefont {Bejger},\ and\ \citenamefont
  {Wielgus}}]{Abramowicz:2017zbp}%
  \BibitemOpen
  \bibfield  {author} {\bibinfo {author} {\bibfnamefont {M.~A.}\ \bibnamefont
  {Abramowicz}}, \bibinfo {author} {\bibfnamefont {M.}~\bibnamefont {Bejger}},\
  and\ \bibinfo {author} {\bibfnamefont {M.}~\bibnamefont {Wielgus}},\
  }\bibfield  {title} {\bibinfo {title} {{Collisions of neutron stars with
  primordial black holes as fast radio bursts engines}},\ }\href
  {https://doi.org/10.3847/1538-4357/aae64a} {\bibfield  {journal} {\bibinfo
  {journal} {Astrophys. J.}\ }\textbf {\bibinfo {volume} {868}},\ \bibinfo
  {pages} {17} (\bibinfo {year} {2018})},\ \Eprint
  {https://arxiv.org/abs/1704.05931} {arXiv:1704.05931 [astro-ph.HE]}
  \BibitemShut {NoStop}%
\bibitem [{\citenamefont {Kainulainen}\ \emph {et~al.}(2021)\citenamefont
  {Kainulainen}, \citenamefont {Nurmi}, \citenamefont {Schiappacasse},\ and\
  \citenamefont {Yanagida}}]{Kainulainen:2021rbg}%
  \BibitemOpen
  \bibfield  {author} {\bibinfo {author} {\bibfnamefont {K.}~\bibnamefont
  {Kainulainen}}, \bibinfo {author} {\bibfnamefont {S.}~\bibnamefont {Nurmi}},
  \bibinfo {author} {\bibfnamefont {E.~D.}\ \bibnamefont {Schiappacasse}},\
  and\ \bibinfo {author} {\bibfnamefont {T.~T.}\ \bibnamefont {Yanagida}},\
  }\bibfield  {title} {\bibinfo {title} {{Can primordial black holes as all
  dark matter explain fast radio bursts?}},\ }\href
  {https://doi.org/10.1103/PhysRevD.104.123033} {\bibfield  {journal} {\bibinfo
   {journal} {Phys. Rev. D}\ }\textbf {\bibinfo {volume} {104}},\ \bibinfo
  {pages} {123033} (\bibinfo {year} {2021})},\ \Eprint
  {https://arxiv.org/abs/2108.08717} {arXiv:2108.08717 [astro-ph.HE]}
  \BibitemShut {NoStop}%
\bibitem [{\citenamefont {Amaral}\ and\ \citenamefont
  {Schiappacasse}(2024)}]{Amaral:2023ekd}%
  \BibitemOpen
  \bibfield  {author} {\bibinfo {author} {\bibfnamefont {D.~W.~P.}\
  \bibnamefont {Amaral}}\ and\ \bibinfo {author} {\bibfnamefont {E.~D.}\
  \bibnamefont {Schiappacasse}},\ }\bibfield  {title} {\bibinfo {title}
  {{Rescuing the primordial black holes all-dark matter hypothesis from the
  fast radio bursts tension}},\ }\href
  {https://doi.org/10.1103/PhysRevD.110.083532} {\bibfield  {journal} {\bibinfo
   {journal} {Phys. Rev. D}\ }\textbf {\bibinfo {volume} {110}},\ \bibinfo
  {pages} {083532} (\bibinfo {year} {2024})},\ \Eprint
  {https://arxiv.org/abs/2312.09285} {arXiv:2312.09285 [hep-ph]} \BibitemShut
  {NoStop}%
\bibitem [{\citenamefont {Cang}\ \emph {et~al.}(2023)\citenamefont {Cang},
  \citenamefont {Ma},\ and\ \citenamefont {Gao}}]{Cang:2022jyc}%
  \BibitemOpen
  \bibfield  {author} {\bibinfo {author} {\bibfnamefont {J.}~\bibnamefont
  {Cang}}, \bibinfo {author} {\bibfnamefont {Y.-Z.}\ \bibnamefont {Ma}},\ and\
  \bibinfo {author} {\bibfnamefont {Y.}~\bibnamefont {Gao}},\ }\bibfield
  {title} {\bibinfo {title} {{Implications for Primordial Black Holes from
  Cosmological Constraints on Scalar-induced Gravitational Waves}},\ }\href
  {https://doi.org/10.3847/1538-4357/acc949} {\bibfield  {journal} {\bibinfo
  {journal} {Astrophys. J.}\ }\textbf {\bibinfo {volume} {949}},\ \bibinfo
  {pages} {64} (\bibinfo {year} {2023})},\ \Eprint
  {https://arxiv.org/abs/2210.03476} {arXiv:2210.03476 [astro-ph.CO]}
  \BibitemShut {NoStop}%
\bibitem [{\citenamefont {Boldrini}\ \emph {et~al.}(2020)\citenamefont
  {Boldrini}, \citenamefont {Miki}, \citenamefont {Wagner}, \citenamefont
  {Mohayaee}, \citenamefont {Silk},\ and\ \citenamefont
  {Arbey}}]{Boldrini:2019isx}%
  \BibitemOpen
  \bibfield  {author} {\bibinfo {author} {\bibfnamefont {P.}~\bibnamefont
  {Boldrini}}, \bibinfo {author} {\bibfnamefont {Y.}~\bibnamefont {Miki}},
  \bibinfo {author} {\bibfnamefont {A.~Y.}\ \bibnamefont {Wagner}}, \bibinfo
  {author} {\bibfnamefont {R.}~\bibnamefont {Mohayaee}}, \bibinfo {author}
  {\bibfnamefont {J.}~\bibnamefont {Silk}},\ and\ \bibinfo {author}
  {\bibfnamefont {A.}~\bibnamefont {Arbey}},\ }\bibfield  {title} {\bibinfo
  {title} {{Cusp-to-core transition in low-mass dwarf galaxies induced by
  dynamical heating of cold dark matter by primordial black holes}},\ }\href
  {https://doi.org/10.1093/mnras/staa150} {\bibfield  {journal} {\bibinfo
  {journal} {Mon. Not. Roy. Astron. Soc.}\ }\textbf {\bibinfo {volume} {492}},\
  \bibinfo {pages} {5218} (\bibinfo {year} {2020})},\ \Eprint
  {https://arxiv.org/abs/1909.07395} {arXiv:1909.07395 [astro-ph.CO]}
  \BibitemShut {NoStop}%
\bibitem [{\citenamefont {Cappelluti}\ \emph {et~al.}(2022)\citenamefont
  {Cappelluti}, \citenamefont {Hasinger},\ and\ \citenamefont
  {Natarajan}}]{Cappelluti:2021usg}%
  \BibitemOpen
  \bibfield  {author} {\bibinfo {author} {\bibfnamefont {N.}~\bibnamefont
  {Cappelluti}}, \bibinfo {author} {\bibfnamefont {G.}~\bibnamefont
  {Hasinger}},\ and\ \bibinfo {author} {\bibfnamefont {P.}~\bibnamefont
  {Natarajan}},\ }\bibfield  {title} {\bibinfo {title} {{Exploring the
  High-redshift PBH-{\ensuremath{\Lambda}}CDM Universe: Early Black Hole
  Seeding, the First Stars and Cosmic Radiation Backgrounds}},\ }\href
  {https://doi.org/10.3847/1538-4357/ac332d} {\bibfield  {journal} {\bibinfo
  {journal} {Astrophys. J.}\ }\textbf {\bibinfo {volume} {926}},\ \bibinfo
  {pages} {205} (\bibinfo {year} {2022})},\ \Eprint
  {https://arxiv.org/abs/2109.08701} {arXiv:2109.08701 [astro-ph.CO]}
  \BibitemShut {NoStop}%
\bibitem [{\citenamefont {H\"utsi}\ \emph {et~al.}(2023)\citenamefont
  {H\"utsi}, \citenamefont {Raidal}, \citenamefont {Urrutia}, \citenamefont
  {Vaskonen},\ and\ \citenamefont {Veerm\"ae}}]{Hutsi:2022fzw}%
  \BibitemOpen
  \bibfield  {author} {\bibinfo {author} {\bibfnamefont {G.}~\bibnamefont
  {H\"utsi}}, \bibinfo {author} {\bibfnamefont {M.}~\bibnamefont {Raidal}},
  \bibinfo {author} {\bibfnamefont {J.}~\bibnamefont {Urrutia}}, \bibinfo
  {author} {\bibfnamefont {V.}~\bibnamefont {Vaskonen}},\ and\ \bibinfo
  {author} {\bibfnamefont {H.}~\bibnamefont {Veerm\"ae}},\ }\bibfield  {title}
  {\bibinfo {title} {{Did JWST observe imprints of axion miniclusters or
  primordial black holes?}},\ }\href
  {https://doi.org/10.1103/PhysRevD.107.043502} {\bibfield  {journal} {\bibinfo
   {journal} {Phys. Rev. D}\ }\textbf {\bibinfo {volume} {107}},\ \bibinfo
  {pages} {043502} (\bibinfo {year} {2023})},\ \Eprint
  {https://arxiv.org/abs/2211.02651} {arXiv:2211.02651 [astro-ph.CO]}
  \BibitemShut {NoStop}%
\bibitem [{\citenamefont {Colazo}\ \emph {et~al.}(2024)\citenamefont {Colazo},
  \citenamefont {Stasyszyn},\ and\ \citenamefont {Padilla}}]{Colazo:2024jmz}%
  \BibitemOpen
  \bibfield  {author} {\bibinfo {author} {\bibfnamefont {P.~E.}\ \bibnamefont
  {Colazo}}, \bibinfo {author} {\bibfnamefont {F.}~\bibnamefont {Stasyszyn}},\
  and\ \bibinfo {author} {\bibfnamefont {N.}~\bibnamefont {Padilla}},\
  }\bibfield  {title} {\bibinfo {title} {{Structure formation with primordial
  black holes to alleviate early star formation tension revealed by JWST}},\
  }\href {https://doi.org/10.1051/0004-6361/202449565} {\bibfield  {journal}
  {\bibinfo  {journal} {Astron. Astrophys.}\ }\textbf {\bibinfo {volume}
  {685}},\ \bibinfo {pages} {L8} (\bibinfo {year} {2024})},\ \Eprint
  {https://arxiv.org/abs/2404.13110} {arXiv:2404.13110 [astro-ph.CO]}
  \BibitemShut {NoStop}%
\bibitem [{\citenamefont {Zhang}\ \emph {et~al.}(2024)\citenamefont {Zhang},
  \citenamefont {Bromm},\ and\ \citenamefont {Liu}}]{Zhang:2024ytf}%
  \BibitemOpen
  \bibfield  {author} {\bibinfo {author} {\bibfnamefont {S.}~\bibnamefont
  {Zhang}}, \bibinfo {author} {\bibfnamefont {V.}~\bibnamefont {Bromm}},\ and\
  \bibinfo {author} {\bibfnamefont {B.}~\bibnamefont {Liu}},\ }\bibfield
  {title} {\bibinfo {title} {{How Do Primordial Black Holes Change the Halo
  Mass Function and Structure?}},\ }\href
  {https://doi.org/10.3847/1538-4357/ad7b0d} {\bibfield  {journal} {\bibinfo
  {journal} {Astrophys. J.}\ }\textbf {\bibinfo {volume} {975}},\ \bibinfo
  {pages} {139} (\bibinfo {year} {2024})},\ \Eprint
  {https://arxiv.org/abs/2405.11381} {arXiv:2405.11381 [astro-ph.CO]}
  \BibitemShut {NoStop}%
\bibitem [{\citenamefont {Dayal}(2024)}]{Dayal:2024zwq}%
  \BibitemOpen
  \bibfield  {author} {\bibinfo {author} {\bibfnamefont {P.}~\bibnamefont
  {Dayal}},\ }\bibfield  {title} {\bibinfo {title} {{Exploring a primordial
  solution for early black holes detected with JWST}},\ }\href
  {https://doi.org/10.1051/0004-6361/202451481} {\bibfield  {journal} {\bibinfo
   {journal} {Astron. Astrophys.}\ }\textbf {\bibinfo {volume} {690}},\
  \bibinfo {pages} {A182} (\bibinfo {year} {2024})},\ \Eprint
  {https://arxiv.org/abs/2407.07162} {arXiv:2407.07162 [astro-ph.GA]}
  \BibitemShut {NoStop}%
\bibitem [{\citenamefont {Dayal}\ and\ \citenamefont
  {maiolino}(2025)}]{Dayal:2025aiv}%
  \BibitemOpen
  \bibfield  {author} {\bibinfo {author} {\bibfnamefont {P.}~\bibnamefont
  {Dayal}}\ and\ \bibinfo {author} {\bibfnamefont {R.}~\bibnamefont
  {maiolino}},\ }\bibfield  {title} {\bibinfo {title} {{The properties of
  primordially-seeded black holes and their hosts in the first billion years:
  implications for JWST}},\ }\href@noop {} {\  (\bibinfo {year} {2025})},\
  \Eprint {https://arxiv.org/abs/2506.08116} {arXiv:2506.08116 [astro-ph.GA]}
  \BibitemShut {NoStop}%
\bibitem [{\citenamefont {Prole}\ \emph {et~al.}(2025)\citenamefont {Prole},
  \citenamefont {Regan}, \citenamefont {Mehta}, \citenamefont {Coles},\ and\
  \citenamefont {Dayal}}]{Prole:2025snf}%
  \BibitemOpen
  \bibfield  {author} {\bibinfo {author} {\bibfnamefont {L.~R.}\ \bibnamefont
  {Prole}}, \bibinfo {author} {\bibfnamefont {J.~A.}\ \bibnamefont {Regan}},
  \bibinfo {author} {\bibfnamefont {D.}~\bibnamefont {Mehta}}, \bibinfo
  {author} {\bibfnamefont {P.}~\bibnamefont {Coles}},\ and\ \bibinfo {author}
  {\bibfnamefont {P.}~\bibnamefont {Dayal}},\ }\bibfield  {title} {\bibinfo
  {title} {{Primordial black holes in cosmological simulations: growth
  prospects for supermassive black holes}}\ }\href
  {https://doi.org/10.33232/001c.143643} {10.33232/001c.143643} (\bibinfo
  {year} {2025}),\ \Eprint {https://arxiv.org/abs/2506.11233} {arXiv:2506.11233
  [astro-ph.GA]} \BibitemShut {NoStop}%
\bibitem [{\citenamefont {Zhang}\ \emph
  {et~al.}(2025{\natexlab{a}})\citenamefont {Zhang}, \citenamefont {Liu},
  \citenamefont {Bromm}, \citenamefont {Jeon}, \citenamefont {Boylan-Kolchin},\
  and\ \citenamefont {Kuhnel}}]{Zhang:2025asq}%
  \BibitemOpen
  \bibfield  {author} {\bibinfo {author} {\bibfnamefont {S.}~\bibnamefont
  {Zhang}}, \bibinfo {author} {\bibfnamefont {B.}~\bibnamefont {Liu}}, \bibinfo
  {author} {\bibfnamefont {V.}~\bibnamefont {Bromm}}, \bibinfo {author}
  {\bibfnamefont {J.}~\bibnamefont {Jeon}}, \bibinfo {author} {\bibfnamefont
  {M.}~\bibnamefont {Boylan-Kolchin}},\ and\ \bibinfo {author} {\bibfnamefont
  {F.}~\bibnamefont {Kuhnel}},\ }\bibfield  {title} {\bibinfo {title} {{How do
  Massive Primordial Black Holes Impact the Formation of the First Stars and
  Galaxies?}},\ }\href@noop {} {\  (\bibinfo {year} {2025}{\natexlab{a}})},\
  \Eprint {https://arxiv.org/abs/2503.17585} {arXiv:2503.17585 [astro-ph.GA]}
  \BibitemShut {NoStop}%
\bibitem [{\citenamefont {Matteri}\ \emph {et~al.}(2025)\citenamefont
  {Matteri}, \citenamefont {Pallottini},\ and\ \citenamefont
  {Ferrara}}]{Matteri:2025klg}%
  \BibitemOpen
  \bibfield  {author} {\bibinfo {author} {\bibfnamefont {A.}~\bibnamefont
  {Matteri}}, \bibinfo {author} {\bibfnamefont {A.}~\bibnamefont
  {Pallottini}},\ and\ \bibinfo {author} {\bibfnamefont {A.}~\bibnamefont
  {Ferrara}},\ }\bibfield  {title} {\bibinfo {title} {{Can primordial black
  holes explain the overabundance of bright super-early galaxies?}},\ }\href
  {https://doi.org/10.1051/0004-6361/202553701} {\bibfield  {journal} {\bibinfo
   {journal} {Astron. Astrophys.}\ }\textbf {\bibinfo {volume} {697}},\
  \bibinfo {pages} {A65} (\bibinfo {year} {2025})},\ \Eprint
  {https://arxiv.org/abs/2503.01968} {arXiv:2503.01968 [astro-ph.GA]}
  \BibitemShut {NoStop}%
\bibitem [{\citenamefont {Zhang}\ \emph
  {et~al.}(2025{\natexlab{b}})\citenamefont {Zhang}, \citenamefont {Liu},
  \citenamefont {Bromm},\ and\ \citenamefont {K{\"u}hnel}}]{Zhang:2025oyl}%
  \BibitemOpen
  \bibfield  {author} {\bibinfo {author} {\bibfnamefont {S.}~\bibnamefont
  {Zhang}}, \bibinfo {author} {\bibfnamefont {B.}~\bibnamefont {Liu}}, \bibinfo
  {author} {\bibfnamefont {V.}~\bibnamefont {Bromm}},\ and\ \bibinfo {author}
  {\bibfnamefont {F.}~\bibnamefont {K{\"u}hnel}},\ }\bibfield  {title}
  {\bibinfo {title} {{Primordial Black Holes as Seeds for Extremely Overmassive
  AGN Observed by JWST}},\ }\href@noop {} {\  (\bibinfo {year}
  {2025}{\natexlab{b}})},\ \Eprint {https://arxiv.org/abs/2512.14066}
  {arXiv:2512.14066 [astro-ph.GA]} \BibitemShut {NoStop}%
\bibitem [{\citenamefont {Maiolino}\ \emph {et~al.}(2025)\citenamefont
  {Maiolino} \emph {et~al.}}]{Maiolino:2025tih}%
  \BibitemOpen
  \bibfield  {author} {\bibinfo {author} {\bibfnamefont {R.}~\bibnamefont
  {Maiolino}} \emph {et~al.},\ }\bibfield  {title} {\bibinfo {title} {{A black
  hole in a near-pristine galaxy 700 million years after the Big Bang}},\
  }\href@noop {} {\  (\bibinfo {year} {2025})},\ \Eprint
  {https://arxiv.org/abs/2505.22567} {arXiv:2505.22567 [astro-ph.GA]}
  \BibitemShut {NoStop}%
\bibitem [{\citenamefont {Huang}\ \emph
  {et~al.}(2024{\natexlab{a}})\citenamefont {Huang}, \citenamefont {Jiang},
  \citenamefont {He}, \citenamefont {Wang},\ and\ \citenamefont
  {Piao}}]{Hai-LongHuang:2024gtx}%
  \BibitemOpen
  \bibfield  {author} {\bibinfo {author} {\bibfnamefont {H.-L.}\ \bibnamefont
  {Huang}}, \bibinfo {author} {\bibfnamefont {J.-Q.}\ \bibnamefont {Jiang}},
  \bibinfo {author} {\bibfnamefont {J.}~\bibnamefont {He}}, \bibinfo {author}
  {\bibfnamefont {Y.-T.}\ \bibnamefont {Wang}},\ and\ \bibinfo {author}
  {\bibfnamefont {Y.-S.}\ \bibnamefont {Piao}},\ }\bibfield  {title} {\bibinfo
  {title} {{Sub-Eddington accreting supermassive primordial black holes explain
  Little Red Dots}},\ }\href@noop {} {\  (\bibinfo {year}
  {2024}{\natexlab{a}})},\ \Eprint {https://arxiv.org/abs/2410.20663}
  {arXiv:2410.20663 [astro-ph.GA]} \BibitemShut {NoStop}%
\bibitem [{\citenamefont {Zhang}\ \emph
  {et~al.}(2025{\natexlab{c}})\citenamefont {Zhang}, \citenamefont {Feng},\
  and\ \citenamefont {An}}]{Zhang:2025tgm}%
  \BibitemOpen
  \bibfield  {author} {\bibinfo {author} {\bibfnamefont {B.}~\bibnamefont
  {Zhang}}, \bibinfo {author} {\bibfnamefont {W.-X.}\ \bibnamefont {Feng}},\
  and\ \bibinfo {author} {\bibfnamefont {H.}~\bibnamefont {An}},\ }\bibfield
  {title} {\bibinfo {title} {{Little Red Dots from Small-Scale Primordial Black
  Hole Clustering}},\ }\href@noop {} {\  (\bibinfo {year}
  {2025}{\natexlab{c}})},\ \Eprint {https://arxiv.org/abs/2507.07171}
  {arXiv:2507.07171 [astro-ph.CO]} \BibitemShut {NoStop}%
\bibitem [{\citenamefont {De~Luca}\ \emph
  {et~al.}(2025{\natexlab{c}})\citenamefont {De~Luca}, \citenamefont
  {Del~Grosso}, \citenamefont {Franciolini}, \citenamefont {Kritos},
  \citenamefont {Berti}, \citenamefont {D'Orazio},\ and\ \citenamefont
  {Silk}}]{DeLuca:2025nao}%
  \BibitemOpen
  \bibfield  {author} {\bibinfo {author} {\bibfnamefont {V.}~\bibnamefont
  {De~Luca}}, \bibinfo {author} {\bibfnamefont {L.}~\bibnamefont {Del~Grosso}},
  \bibinfo {author} {\bibfnamefont {G.}~\bibnamefont {Franciolini}}, \bibinfo
  {author} {\bibfnamefont {K.}~\bibnamefont {Kritos}}, \bibinfo {author}
  {\bibfnamefont {E.}~\bibnamefont {Berti}}, \bibinfo {author} {\bibfnamefont
  {D.}~\bibnamefont {D'Orazio}},\ and\ \bibinfo {author} {\bibfnamefont
  {J.}~\bibnamefont {Silk}},\ }\bibfield  {title} {\bibinfo {title} {{A
  cosmologist's take on Little Red Dots}},\ }\href@noop {} {\  (\bibinfo {year}
  {2025}{\natexlab{c}})},\ \Eprint {https://arxiv.org/abs/2512.19666}
  {arXiv:2512.19666 [astro-ph.CO]} \BibitemShut {NoStop}%
\bibitem [{\citenamefont {Liu}\ and\ \citenamefont
  {Bromm}(2022)}]{Liu:2022bvr}%
  \BibitemOpen
  \bibfield  {author} {\bibinfo {author} {\bibfnamefont {B.}~\bibnamefont
  {Liu}}\ and\ \bibinfo {author} {\bibfnamefont {V.}~\bibnamefont {Bromm}},\
  }\bibfield  {title} {\bibinfo {title} {{Accelerating Early Massive Galaxy
  Formation with Primordial Black Holes}},\ }\href
  {https://doi.org/10.3847/2041-8213/ac927f} {\bibfield  {journal} {\bibinfo
  {journal} {Astrophys. J. Lett.}\ }\textbf {\bibinfo {volume} {937}},\
  \bibinfo {pages} {L30} (\bibinfo {year} {2022})},\ \Eprint
  {https://arxiv.org/abs/2208.13178} {arXiv:2208.13178 [astro-ph.CO]}
  \BibitemShut {NoStop}%
\bibitem [{\citenamefont {Ziparo}\ \emph {et~al.}(2025)\citenamefont {Ziparo},
  \citenamefont {Gallerani},\ and\ \citenamefont {Ferrara}}]{Ziparo:2024nwh}%
  \BibitemOpen
  \bibfield  {author} {\bibinfo {author} {\bibfnamefont {F.}~\bibnamefont
  {Ziparo}}, \bibinfo {author} {\bibfnamefont {S.}~\bibnamefont {Gallerani}},\
  and\ \bibinfo {author} {\bibfnamefont {A.}~\bibnamefont {Ferrara}},\
  }\bibfield  {title} {\bibinfo {title} {{Primordial black holes as
  supermassive black hole seeds}},\ }\href
  {https://doi.org/10.1088/1475-7516/2025/04/040} {\bibfield  {journal}
  {\bibinfo  {journal} {JCAP}\ }\textbf {\bibinfo {volume} {04}},\ \bibinfo
  {pages} {040}},\ \Eprint {https://arxiv.org/abs/2411.03448} {arXiv:2411.03448
  [astro-ph.CO]} \BibitemShut {NoStop}%
\bibitem [{\citenamefont {Huang}\ \emph
  {et~al.}(2024{\natexlab{b}})\citenamefont {Huang}, \citenamefont {Wang},\
  and\ \citenamefont {Piao}}]{Hai-LongHuang:2024vvz}%
  \BibitemOpen
  \bibfield  {author} {\bibinfo {author} {\bibfnamefont {H.-L.}\ \bibnamefont
  {Huang}}, \bibinfo {author} {\bibfnamefont {Y.-T.}\ \bibnamefont {Wang}},\
  and\ \bibinfo {author} {\bibfnamefont {Y.-S.}\ \bibnamefont {Piao}},\
  }\bibfield  {title} {\bibinfo {title} {{Supermassive primordial black holes
  for the GHZ9 and UHZ1 observed by the JWST}},\ }\href@noop {} {\  (\bibinfo
  {year} {2024}{\natexlab{b}})},\ \Eprint {https://arxiv.org/abs/2410.05891}
  {arXiv:2410.05891 [astro-ph.GA]} \BibitemShut {NoStop}%
\bibitem [{\citenamefont {Zhang}\ \emph
  {et~al.}(2025{\natexlab{d}})\citenamefont {Zhang}, \citenamefont {Liu},\ and\
  \citenamefont {Bromm}}]{Zhang:2025grn}%
  \BibitemOpen
  \bibfield  {author} {\bibinfo {author} {\bibfnamefont {S.}~\bibnamefont
  {Zhang}}, \bibinfo {author} {\bibfnamefont {B.}~\bibnamefont {Liu}},\ and\
  \bibinfo {author} {\bibfnamefont {V.}~\bibnamefont {Bromm}},\ }\bibfield
  {title} {\bibinfo {title} {{A Novel Formation Channel for Supermassive Black
  Hole Binaries in the Early Universe via Primordial Black Holes}},\ }\href
  {https://doi.org/10.3847/1538-4357/ae061c} {\bibfield  {journal} {\bibinfo
  {journal} {Astrophys. J.}\ }\textbf {\bibinfo {volume} {992}},\ \bibinfo
  {pages} {136} (\bibinfo {year} {2025}{\natexlab{d}})},\ \Eprint
  {https://arxiv.org/abs/2508.00774} {arXiv:2508.00774 [astro-ph.GA]}
  \BibitemShut {NoStop}%
\bibitem [{\citenamefont {Ricotti}\ \emph {et~al.}(2008)\citenamefont
  {Ricotti}, \citenamefont {Ostriker},\ and\ \citenamefont
  {Mack}}]{Ricotti:2007au}%
  \BibitemOpen
  \bibfield  {author} {\bibinfo {author} {\bibfnamefont {M.}~\bibnamefont
  {Ricotti}}, \bibinfo {author} {\bibfnamefont {J.~P.}\ \bibnamefont
  {Ostriker}},\ and\ \bibinfo {author} {\bibfnamefont {K.~J.}\ \bibnamefont
  {Mack}},\ }\bibfield  {title} {\bibinfo {title} {{Effect of Primordial Black
  Holes on the Cosmic Microwave Background and Cosmological Parameter
  Estimates}},\ }\href {https://doi.org/10.1086/587831} {\bibfield  {journal}
  {\bibinfo  {journal} {Astrophys. J.}\ }\textbf {\bibinfo {volume} {680}},\
  \bibinfo {pages} {829} (\bibinfo {year} {2008})},\ \Eprint
  {https://arxiv.org/abs/0709.0524} {arXiv:0709.0524 [astro-ph]} \BibitemShut
  {NoStop}%
\bibitem [{\citenamefont {Serpico}\ \emph {et~al.}(2020)\citenamefont
  {Serpico}, \citenamefont {Poulin}, \citenamefont {Inman},\ and\ \citenamefont
  {Kohri}}]{Serpico:2020ehh}%
  \BibitemOpen
  \bibfield  {author} {\bibinfo {author} {\bibfnamefont {P.~D.}\ \bibnamefont
  {Serpico}}, \bibinfo {author} {\bibfnamefont {V.}~\bibnamefont {Poulin}},
  \bibinfo {author} {\bibfnamefont {D.}~\bibnamefont {Inman}},\ and\ \bibinfo
  {author} {\bibfnamefont {K.}~\bibnamefont {Kohri}},\ }\bibfield  {title}
  {\bibinfo {title} {{Cosmic microwave background bounds on primordial black
  holes including dark matter halo accretion}},\ }\href
  {https://doi.org/10.1103/PhysRevResearch.2.023204} {\bibfield  {journal}
  {\bibinfo  {journal} {Phys. Rev. Res.}\ }\textbf {\bibinfo {volume} {2}},\
  \bibinfo {pages} {023204} (\bibinfo {year} {2020})},\ \Eprint
  {https://arxiv.org/abs/2002.10771} {arXiv:2002.10771 [astro-ph.CO]}
  \BibitemShut {NoStop}%
\bibitem [{\citenamefont {Inoue}\ and\ \citenamefont
  {Kusenko}(2017)}]{Inoue:2017csr}%
  \BibitemOpen
  \bibfield  {author} {\bibinfo {author} {\bibfnamefont {Y.}~\bibnamefont
  {Inoue}}\ and\ \bibinfo {author} {\bibfnamefont {A.}~\bibnamefont
  {Kusenko}},\ }\bibfield  {title} {\bibinfo {title} {{New X-ray bound on
  density of primordial black holes}},\ }\href
  {https://doi.org/10.1088/1475-7516/2017/10/034} {\bibfield  {journal}
  {\bibinfo  {journal} {JCAP}\ }\textbf {\bibinfo {volume} {10}},\ \bibinfo
  {pages} {034}},\ \Eprint {https://arxiv.org/abs/1705.00791} {arXiv:1705.00791
  [astro-ph.CO]} \BibitemShut {NoStop}%
\bibitem [{\citenamefont {Hektor}\ \emph {et~al.}(2018)\citenamefont {Hektor},
  \citenamefont {H{\"u}tsi}, \citenamefont {Marzola}, \citenamefont {Raidal},
  \citenamefont {Vaskonen},\ and\ \citenamefont
  {Veerm{\"a}e}}]{Hektor:2018qqw}%
  \BibitemOpen
  \bibfield  {author} {\bibinfo {author} {\bibfnamefont {A.}~\bibnamefont
  {Hektor}}, \bibinfo {author} {\bibfnamefont {G.}~\bibnamefont {H{\"u}tsi}},
  \bibinfo {author} {\bibfnamefont {L.}~\bibnamefont {Marzola}}, \bibinfo
  {author} {\bibfnamefont {M.}~\bibnamefont {Raidal}}, \bibinfo {author}
  {\bibfnamefont {V.}~\bibnamefont {Vaskonen}},\ and\ \bibinfo {author}
  {\bibfnamefont {H.}~\bibnamefont {Veerm{\"a}e}},\ }\bibfield  {title}
  {\bibinfo {title} {{Constraining Primordial Black Holes with the EDGES 21-cm
  Absorption Signal}},\ }\href {https://doi.org/10.1103/PhysRevD.98.023503}
  {\bibfield  {journal} {\bibinfo  {journal} {Phys. Rev. D}\ }\textbf {\bibinfo
  {volume} {98}},\ \bibinfo {pages} {023503} (\bibinfo {year} {2018})},\
  \Eprint {https://arxiv.org/abs/1803.09697} {arXiv:1803.09697 [astro-ph.CO]}
  \BibitemShut {NoStop}%
\bibitem [{\citenamefont {Cang}\ \emph {et~al.}(2022)\citenamefont {Cang},
  \citenamefont {Gao},\ and\ \citenamefont {Ma}}]{Cang:2021owu}%
  \BibitemOpen
  \bibfield  {author} {\bibinfo {author} {\bibfnamefont {J.}~\bibnamefont
  {Cang}}, \bibinfo {author} {\bibfnamefont {Y.}~\bibnamefont {Gao}},\ and\
  \bibinfo {author} {\bibfnamefont {Y.-Z.}\ \bibnamefont {Ma}},\ }\bibfield
  {title} {\bibinfo {title} {{21-cm constraints on spinning primordial black
  holes}},\ }\href {https://doi.org/10.1088/1475-7516/2022/03/012} {\bibfield
  {journal} {\bibinfo  {journal} {JCAP}\ }\textbf {\bibinfo {volume}
  {03}}\bibfield  {number} {\bibinfo  {number} { (03)},\ \bibinfo {pages}
  {012}},\ }\Eprint {https://arxiv.org/abs/2108.13256} {arXiv:2108.13256
  [astro-ph.CO]} \BibitemShut {NoStop}%
\bibitem [{\citenamefont {Agius}\ \emph {et~al.}(2025)\citenamefont {Agius},
  \citenamefont {Essig}, \citenamefont {Gaggero}, \citenamefont
  {Palomares-Ruiz}, \citenamefont {Suczewski},\ and\ \citenamefont
  {Valli}}]{Agius:2025xbj}%
  \BibitemOpen
  \bibfield  {author} {\bibinfo {author} {\bibfnamefont {D.}~\bibnamefont
  {Agius}}, \bibinfo {author} {\bibfnamefont {R.}~\bibnamefont {Essig}},
  \bibinfo {author} {\bibfnamefont {D.}~\bibnamefont {Gaggero}}, \bibinfo
  {author} {\bibfnamefont {S.}~\bibnamefont {Palomares-Ruiz}}, \bibinfo
  {author} {\bibfnamefont {G.}~\bibnamefont {Suczewski}},\ and\ \bibinfo
  {author} {\bibfnamefont {M.}~\bibnamefont {Valli}},\ }\bibfield  {title}
  {\bibinfo {title} {{Astrophysical uncertainties challenge 21-cm forecasts: A
  primordial black hole case study}},\ }\href@noop {} {\  (\bibinfo {year}
  {2025})},\ \Eprint {https://arxiv.org/abs/2510.14877} {arXiv:2510.14877
  [astro-ph.CO]} \BibitemShut {NoStop}%
\bibitem [{\citenamefont {Agius}\ \emph {et~al.}(2024)\citenamefont {Agius},
  \citenamefont {Essig}, \citenamefont {Gaggero}, \citenamefont {Scarcella},
  \citenamefont {Suczewski},\ and\ \citenamefont {Valli}}]{Agius:2024ecw}%
  \BibitemOpen
  \bibfield  {author} {\bibinfo {author} {\bibfnamefont {D.}~\bibnamefont
  {Agius}}, \bibinfo {author} {\bibfnamefont {R.}~\bibnamefont {Essig}},
  \bibinfo {author} {\bibfnamefont {D.}~\bibnamefont {Gaggero}}, \bibinfo
  {author} {\bibfnamefont {F.}~\bibnamefont {Scarcella}}, \bibinfo {author}
  {\bibfnamefont {G.}~\bibnamefont {Suczewski}},\ and\ \bibinfo {author}
  {\bibfnamefont {M.}~\bibnamefont {Valli}},\ }\bibfield  {title} {\bibinfo
  {title} {{Feedback in the dark: a critical examination of CMB bounds on
  primordial black holes}},\ }\href
  {https://doi.org/10.1088/1475-7516/2024/07/003} {\bibfield  {journal}
  {\bibinfo  {journal} {JCAP}\ }\textbf {\bibinfo {volume} {07}},\ \bibinfo
  {pages} {003}},\ \Eprint {https://arxiv.org/abs/2403.18895} {arXiv:2403.18895
  [hep-ph]} \BibitemShut {NoStop}%
\bibitem [{\citenamefont {Facchinetti}\ \emph {et~al.}(2023)\citenamefont
  {Facchinetti}, \citenamefont {Lucca},\ and\ \citenamefont
  {Clesse}}]{Facchinetti:2022kbg}%
  \BibitemOpen
  \bibfield  {author} {\bibinfo {author} {\bibfnamefont {G.}~\bibnamefont
  {Facchinetti}}, \bibinfo {author} {\bibfnamefont {M.}~\bibnamefont {Lucca}},\
  and\ \bibinfo {author} {\bibfnamefont {S.}~\bibnamefont {Clesse}},\
  }\bibfield  {title} {\bibinfo {title} {{Relaxing CMB bounds on primordial
  black holes: The role of ionization fronts}},\ }\href
  {https://doi.org/10.1103/PhysRevD.107.043537} {\bibfield  {journal} {\bibinfo
   {journal} {Phys. Rev. D}\ }\textbf {\bibinfo {volume} {107}},\ \bibinfo
  {pages} {043537} (\bibinfo {year} {2023})},\ \Eprint
  {https://arxiv.org/abs/2212.07969} {arXiv:2212.07969 [astro-ph.CO]}
  \BibitemShut {NoStop}%
\bibitem [{\citenamefont {De~Luca}\ \emph
  {et~al.}(2020{\natexlab{b}})\citenamefont {De~Luca}, \citenamefont
  {Franciolini}, \citenamefont {Pani},\ and\ \citenamefont
  {Riotto}}]{DeLuca:2020fpg}%
  \BibitemOpen
  \bibfield  {author} {\bibinfo {author} {\bibfnamefont {V.}~\bibnamefont
  {De~Luca}}, \bibinfo {author} {\bibfnamefont {G.}~\bibnamefont
  {Franciolini}}, \bibinfo {author} {\bibfnamefont {P.}~\bibnamefont {Pani}},\
  and\ \bibinfo {author} {\bibfnamefont {A.}~\bibnamefont {Riotto}},\
  }\bibfield  {title} {\bibinfo {title} {{Constraints on Primordial Black
  Holes: the Importance of Accretion}},\ }\href
  {https://doi.org/10.1103/PhysRevD.102.043505} {\bibfield  {journal} {\bibinfo
   {journal} {Phys. Rev. D}\ }\textbf {\bibinfo {volume} {102}},\ \bibinfo
  {pages} {043505} (\bibinfo {year} {2020}{\natexlab{b}})},\ \Eprint
  {https://arxiv.org/abs/2003.12589} {arXiv:2003.12589 [astro-ph.CO]}
  \BibitemShut {NoStop}%
\bibitem [{\citenamefont {H{\"u}tsi}\ \emph {et~al.}(2019)\citenamefont
  {H{\"u}tsi}, \citenamefont {Raidal},\ and\ \citenamefont
  {Veerm{\"a}e}}]{Hutsi:2019hlw}%
  \BibitemOpen
  \bibfield  {author} {\bibinfo {author} {\bibfnamefont {G.}~\bibnamefont
  {H{\"u}tsi}}, \bibinfo {author} {\bibfnamefont {M.}~\bibnamefont {Raidal}},\
  and\ \bibinfo {author} {\bibfnamefont {H.}~\bibnamefont {Veerm{\"a}e}},\
  }\bibfield  {title} {\bibinfo {title} {{Small-scale structure of primordial
  black hole dark matter and its implications for accretion}},\ }\href
  {https://doi.org/10.1103/PhysRevD.100.083016} {\bibfield  {journal} {\bibinfo
   {journal} {Phys. Rev. D}\ }\textbf {\bibinfo {volume} {100}},\ \bibinfo
  {pages} {083016} (\bibinfo {year} {2019})},\ \Eprint
  {https://arxiv.org/abs/1907.06533} {arXiv:1907.06533 [astro-ph.CO]}
  \BibitemShut {NoStop}%
\bibitem [{\citenamefont {Kashlinsky}(2016)}]{Kashlinsky:2016sdv}%
  \BibitemOpen
  \bibfield  {author} {\bibinfo {author} {\bibfnamefont {A.}~\bibnamefont
  {Kashlinsky}},\ }\bibfield  {title} {\bibinfo {title} {{LIGO gravitational
  wave detection, primordial black holes and the near-IR cosmic infrared
  background anisotropies}},\ }\href
  {https://doi.org/10.3847/2041-8205/823/2/L25} {\bibfield  {journal} {\bibinfo
   {journal} {Astrophys. J. Lett.}\ }\textbf {\bibinfo {volume} {823}},\
  \bibinfo {pages} {L25} (\bibinfo {year} {2016})},\ \Eprint
  {https://arxiv.org/abs/1605.04023} {arXiv:1605.04023 [astro-ph.CO]}
  \BibitemShut {NoStop}%
\bibitem [{\citenamefont {Kashlinsky}\ \emph {et~al.}(2018)\citenamefont
  {Kashlinsky}, \citenamefont {Arendt}, \citenamefont {Atrio-Barandela},
  \citenamefont {Cappelluti}, \citenamefont {Ferrara},\ and\ \citenamefont
  {Hasinger}}]{Kashlinsky:2018mnu}%
  \BibitemOpen
  \bibfield  {author} {\bibinfo {author} {\bibfnamefont {A.}~\bibnamefont
  {Kashlinsky}}, \bibinfo {author} {\bibfnamefont {R.~G.}\ \bibnamefont
  {Arendt}}, \bibinfo {author} {\bibfnamefont {F.}~\bibnamefont
  {Atrio-Barandela}}, \bibinfo {author} {\bibfnamefont {N.}~\bibnamefont
  {Cappelluti}}, \bibinfo {author} {\bibfnamefont {A.}~\bibnamefont
  {Ferrara}},\ and\ \bibinfo {author} {\bibfnamefont {G.}~\bibnamefont
  {Hasinger}},\ }\bibfield  {title} {\bibinfo {title} {{Looking at cosmic
  near-infrared background radiation anisotropies}},\ }\href
  {https://doi.org/10.1103/RevModPhys.90.025006} {\bibfield  {journal}
  {\bibinfo  {journal} {Rev. Mod. Phys.}\ }\textbf {\bibinfo {volume} {90}},\
  \bibinfo {pages} {025006} (\bibinfo {year} {2018})},\ \Eprint
  {https://arxiv.org/abs/1802.07774} {arXiv:1802.07774 [astro-ph.CO]}
  \BibitemShut {NoStop}%
\bibitem [{\citenamefont {Capela}\ \emph
  {et~al.}(2013{\natexlab{a}})\citenamefont {Capela}, \citenamefont
  {Pshirkov},\ and\ \citenamefont {Tinyakov}}]{Capela:2013yf}%
  \BibitemOpen
  \bibfield  {author} {\bibinfo {author} {\bibfnamefont {F.}~\bibnamefont
  {Capela}}, \bibinfo {author} {\bibfnamefont {M.}~\bibnamefont {Pshirkov}},\
  and\ \bibinfo {author} {\bibfnamefont {P.}~\bibnamefont {Tinyakov}},\
  }\bibfield  {title} {\bibinfo {title} {{Constraints on primordial black holes
  as dark matter candidates from capture by neutron stars}},\ }\href
  {https://doi.org/10.1103/PhysRevD.87.123524} {\bibfield  {journal} {\bibinfo
  {journal} {Phys. Rev. D}\ }\textbf {\bibinfo {volume} {87}},\ \bibinfo
  {pages} {123524} (\bibinfo {year} {2013}{\natexlab{a}})},\ \Eprint
  {https://arxiv.org/abs/1301.4984} {arXiv:1301.4984 [astro-ph.CO]}
  \BibitemShut {NoStop}%
\bibitem [{\citenamefont {Montero-Camacho}\ \emph {et~al.}(2019)\citenamefont
  {Montero-Camacho}, \citenamefont {Fang}, \citenamefont {Vasquez},
  \citenamefont {Silva},\ and\ \citenamefont
  {Hirata}}]{Montero-Camacho:2019jte}%
  \BibitemOpen
  \bibfield  {author} {\bibinfo {author} {\bibfnamefont {P.}~\bibnamefont
  {Montero-Camacho}}, \bibinfo {author} {\bibfnamefont {X.}~\bibnamefont
  {Fang}}, \bibinfo {author} {\bibfnamefont {G.}~\bibnamefont {Vasquez}},
  \bibinfo {author} {\bibfnamefont {M.}~\bibnamefont {Silva}},\ and\ \bibinfo
  {author} {\bibfnamefont {C.~M.}\ \bibnamefont {Hirata}},\ }\bibfield  {title}
  {\bibinfo {title} {{Revisiting constraints on asteroid-mass primordial black
  holes as dark matter candidates}},\ }\href
  {https://doi.org/10.1088/1475-7516/2019/08/031} {\bibfield  {journal}
  {\bibinfo  {journal} {JCAP}\ }\textbf {\bibinfo {volume} {08}},\ \bibinfo
  {pages} {031}},\ \Eprint {https://arxiv.org/abs/1906.05950} {arXiv:1906.05950
  [astro-ph.CO]} \BibitemShut {NoStop}%
\bibitem [{\citenamefont {Caiozzo}\ \emph {et~al.}(2024)\citenamefont
  {Caiozzo}, \citenamefont {Bertone},\ and\ \citenamefont
  {K{\"u}hnel}}]{Caiozzo:2024flz}%
  \BibitemOpen
  \bibfield  {author} {\bibinfo {author} {\bibfnamefont {R.}~\bibnamefont
  {Caiozzo}}, \bibinfo {author} {\bibfnamefont {G.}~\bibnamefont {Bertone}},\
  and\ \bibinfo {author} {\bibfnamefont {F.}~\bibnamefont {K{\"u}hnel}},\
  }\bibfield  {title} {\bibinfo {title} {{Revisiting primordial black hole
  capture by neutron~stars}},\ }\href
  {https://doi.org/10.1088/1475-7516/2024/07/091} {\bibfield  {journal}
  {\bibinfo  {journal} {JCAP}\ }\textbf {\bibinfo {volume} {07}},\ \bibinfo
  {pages} {091}},\ \Eprint {https://arxiv.org/abs/2404.08057} {arXiv:2404.08057
  [astro-ph.HE]} \BibitemShut {NoStop}%
\bibitem [{\citenamefont {Holst}\ \emph {et~al.}(2025)\citenamefont {Holst},
  \citenamefont {G{\'e}nolini},\ and\ \citenamefont {Serpico}}]{Holst:2025vgk}%
  \BibitemOpen
  \bibfield  {author} {\bibinfo {author} {\bibfnamefont {I.}~\bibnamefont
  {Holst}}, \bibinfo {author} {\bibfnamefont {Y.}~\bibnamefont
  {G{\'e}nolini}},\ and\ \bibinfo {author} {\bibfnamefont {P.~D.}\ \bibnamefont
  {Serpico}},\ }\bibfield  {title} {\bibinfo {title} {{Tidal effects on
  primordial black hole capture in neutron stars}},\ }\href
  {https://doi.org/10.1088/1475-7516/2025/09/050} {\bibfield  {journal}
  {\bibinfo  {journal} {JCAP}\ }\textbf {\bibinfo {volume} {09}},\ \bibinfo
  {pages} {050}},\ \Eprint {https://arxiv.org/abs/2505.04709} {arXiv:2505.04709
  [astro-ph.HE]} \BibitemShut {NoStop}%
\bibitem [{\citenamefont {Capela}\ \emph
  {et~al.}(2013{\natexlab{b}})\citenamefont {Capela}, \citenamefont
  {Pshirkov},\ and\ \citenamefont {Tinyakov}}]{Capela:2012jz}%
  \BibitemOpen
  \bibfield  {author} {\bibinfo {author} {\bibfnamefont {F.}~\bibnamefont
  {Capela}}, \bibinfo {author} {\bibfnamefont {M.}~\bibnamefont {Pshirkov}},\
  and\ \bibinfo {author} {\bibfnamefont {P.}~\bibnamefont {Tinyakov}},\
  }\bibfield  {title} {\bibinfo {title} {{Constraints on Primordial Black Holes
  as Dark Matter Candidates from Star Formation}},\ }\href
  {https://doi.org/10.1103/PhysRevD.87.023507} {\bibfield  {journal} {\bibinfo
  {journal} {Phys. Rev. D}\ }\textbf {\bibinfo {volume} {87}},\ \bibinfo
  {pages} {023507} (\bibinfo {year} {2013}{\natexlab{b}})},\ \Eprint
  {https://arxiv.org/abs/1209.6021} {arXiv:1209.6021 [astro-ph.CO]}
  \BibitemShut {NoStop}%
\bibitem [{\citenamefont {Capela}\ \emph {et~al.}(2014)\citenamefont {Capela},
  \citenamefont {Pshirkov},\ and\ \citenamefont {Tinyakov}}]{Capela:2014ita}%
  \BibitemOpen
  \bibfield  {author} {\bibinfo {author} {\bibfnamefont {F.}~\bibnamefont
  {Capela}}, \bibinfo {author} {\bibfnamefont {M.}~\bibnamefont {Pshirkov}},\
  and\ \bibinfo {author} {\bibfnamefont {P.}~\bibnamefont {Tinyakov}},\
  }\bibfield  {title} {\bibinfo {title} {{Adiabatic contraction revisited:
  implications for primordial black holes}},\ }\href
  {https://doi.org/10.1103/PhysRevD.90.083507} {\bibfield  {journal} {\bibinfo
  {journal} {Phys. Rev. D}\ }\textbf {\bibinfo {volume} {90}},\ \bibinfo
  {pages} {083507} (\bibinfo {year} {2014})},\ \Eprint
  {https://arxiv.org/abs/1403.7098} {arXiv:1403.7098 [astro-ph.CO]}
  \BibitemShut {NoStop}%
\bibitem [{\citenamefont {Oncins}\ \emph {et~al.}(2022)\citenamefont {Oncins},
  \citenamefont {Miralda-Escud{\'e}}, \citenamefont {Guti{\'e}rrez},\ and\
  \citenamefont {Gil-Pons}}]{Oncins:2022djq}%
  \BibitemOpen
  \bibfield  {author} {\bibinfo {author} {\bibfnamefont {M.}~\bibnamefont
  {Oncins}}, \bibinfo {author} {\bibfnamefont {J.}~\bibnamefont
  {Miralda-Escud{\'e}}}, \bibinfo {author} {\bibfnamefont {J.~L.}\ \bibnamefont
  {Guti{\'e}rrez}},\ and\ \bibinfo {author} {\bibfnamefont {P.}~\bibnamefont
  {Gil-Pons}},\ }\bibfield  {title} {\bibinfo {title} {{Primordial black holes
  capture by stars and induced collapse to low-mass stellar black holes}},\
  }\href {https://doi.org/10.1093/mnras/stac2647} {\bibfield  {journal}
  {\bibinfo  {journal} {Mon. Not. Roy. Astron. Soc.}\ }\textbf {\bibinfo
  {volume} {517}},\ \bibinfo {pages} {28} (\bibinfo {year} {2022})},\ \Eprint
  {https://arxiv.org/abs/2205.13003} {arXiv:2205.13003 [astro-ph.GA]}
  \BibitemShut {NoStop}%
\bibitem [{\citenamefont {Esser}\ and\ \citenamefont
  {Tinyakov}(2023)}]{Esser:2022owk}%
  \BibitemOpen
  \bibfield  {author} {\bibinfo {author} {\bibfnamefont {N.}~\bibnamefont
  {Esser}}\ and\ \bibinfo {author} {\bibfnamefont {P.}~\bibnamefont
  {Tinyakov}},\ }\bibfield  {title} {\bibinfo {title} {{Constraints on
  primordial black holes from observation of stars in dwarf galaxies}},\ }\href
  {https://doi.org/10.1103/PhysRevD.107.103052} {\bibfield  {journal} {\bibinfo
   {journal} {Phys. Rev. D}\ }\textbf {\bibinfo {volume} {107}},\ \bibinfo
  {pages} {103052} (\bibinfo {year} {2023})},\ \Eprint
  {https://arxiv.org/abs/2207.07412} {arXiv:2207.07412 [astro-ph.HE]}
  \BibitemShut {NoStop}%
\bibitem [{\citenamefont {Esser}\ \emph {et~al.}(2024)\citenamefont {Esser},
  \citenamefont {De~Rijcke},\ and\ \citenamefont {Tinyakov}}]{Esser:2023yut}%
  \BibitemOpen
  \bibfield  {author} {\bibinfo {author} {\bibfnamefont {N.}~\bibnamefont
  {Esser}}, \bibinfo {author} {\bibfnamefont {S.}~\bibnamefont {De~Rijcke}},\
  and\ \bibinfo {author} {\bibfnamefont {P.}~\bibnamefont {Tinyakov}},\
  }\bibfield  {title} {\bibinfo {title} {{The impact of primordial black holes
  on the stellar mass function of ultra-faint dwarf galaxies}},\ }\href
  {https://doi.org/10.1093/mnras/stae147} {\bibfield  {journal} {\bibinfo
  {journal} {Mon. Not. Roy. Astron. Soc.}\ }\textbf {\bibinfo {volume} {529}},\
  \bibinfo {pages} {32} (\bibinfo {year} {2024})},\ \Eprint
  {https://arxiv.org/abs/2311.12658} {arXiv:2311.12658 [astro-ph.GA]}
  \BibitemShut {NoStop}%
\bibitem [{\citenamefont {Esser}\ \emph {et~al.}(2025)\citenamefont {Esser},
  \citenamefont {Filion}, \citenamefont {De~Rijcke}, \citenamefont
  {Kallivayalil}, \citenamefont {Richstein}, \citenamefont {Tinyakov},\ and\
  \citenamefont {Wyse}}]{Esser:2025pnt}%
  \BibitemOpen
  \bibfield  {author} {\bibinfo {author} {\bibfnamefont {N.}~\bibnamefont
  {Esser}}, \bibinfo {author} {\bibfnamefont {C.}~\bibnamefont {Filion}},
  \bibinfo {author} {\bibfnamefont {S.}~\bibnamefont {De~Rijcke}}, \bibinfo
  {author} {\bibfnamefont {N.}~\bibnamefont {Kallivayalil}}, \bibinfo {author}
  {\bibfnamefont {H.}~\bibnamefont {Richstein}}, \bibinfo {author}
  {\bibfnamefont {P.}~\bibnamefont {Tinyakov}},\ and\ \bibinfo {author}
  {\bibfnamefont {R.~F.~G.}\ \bibnamefont {Wyse}},\ }\bibfield  {title}
  {\bibinfo {title} {{Constraints on asteroid-mass primordial black holes in
  dwarf galaxies using Hubble Space Telescope photometry}},\ }\href
  {https://doi.org/10.1051/0004-6361/202554687} {\bibfield  {journal} {\bibinfo
   {journal} {Astron. Astrophys.}\ }\textbf {\bibinfo {volume} {698}},\
  \bibinfo {pages} {A290} (\bibinfo {year} {2025})},\ \Eprint
  {https://arxiv.org/abs/2503.03352} {arXiv:2503.03352 [astro-ph.GA]}
  \BibitemShut {NoStop}%
\bibitem [{\citenamefont {Baumgarte}\ and\ \citenamefont
  {Shapiro}(2021)}]{Baumgarte:2021thx}%
  \BibitemOpen
  \bibfield  {author} {\bibinfo {author} {\bibfnamefont {T.~W.}\ \bibnamefont
  {Baumgarte}}\ and\ \bibinfo {author} {\bibfnamefont {S.~L.}\ \bibnamefont
  {Shapiro}},\ }\bibfield  {title} {\bibinfo {title} {{Neutron Stars Harboring
  a Primordial Black Hole: Maximum Survival Time}},\ }\href
  {https://doi.org/10.1103/PhysRevD.103.L081303} {\bibfield  {journal}
  {\bibinfo  {journal} {Phys. Rev. D}\ }\textbf {\bibinfo {volume} {103}},\
  \bibinfo {pages} {L081303} (\bibinfo {year} {2021})},\ \Eprint
  {https://arxiv.org/abs/2101.12220} {arXiv:2101.12220 [astro-ph.HE]}
  \BibitemShut {NoStop}%
\bibitem [{\citenamefont {Bellinger}\ \emph {et~al.}(2023)\citenamefont
  {Bellinger}, \citenamefont {Caplan}, \citenamefont {Ryu}, \citenamefont
  {Bollimpalli}, \citenamefont {Ball}, \citenamefont {K{\"u}hnel},
  \citenamefont {Farmer}, \citenamefont {de~Mink},\ and\ \citenamefont
  {Christensen-Dalsgaard}}]{Bellinger:2023wou}%
  \BibitemOpen
  \bibfield  {author} {\bibinfo {author} {\bibfnamefont {E.~P.}\ \bibnamefont
  {Bellinger}}, \bibinfo {author} {\bibfnamefont {M.~E.}\ \bibnamefont
  {Caplan}}, \bibinfo {author} {\bibfnamefont {T.}~\bibnamefont {Ryu}},
  \bibinfo {author} {\bibfnamefont {D.}~\bibnamefont {Bollimpalli}}, \bibinfo
  {author} {\bibfnamefont {W.~H.}\ \bibnamefont {Ball}}, \bibinfo {author}
  {\bibfnamefont {F.}~\bibnamefont {K{\"u}hnel}}, \bibinfo {author}
  {\bibfnamefont {R.}~\bibnamefont {Farmer}}, \bibinfo {author} {\bibfnamefont
  {S.~E.}\ \bibnamefont {de~Mink}},\ and\ \bibinfo {author} {\bibfnamefont
  {J.}~\bibnamefont {Christensen-Dalsgaard}},\ }\bibfield  {title} {\bibinfo
  {title} {{Solar Evolution Models with a Central Black Hole}},\ }\href
  {https://doi.org/10.3847/1538-4357/ad04de} {\bibfield  {journal} {\bibinfo
  {journal} {Astrophys. J.}\ }\textbf {\bibinfo {volume} {959}},\ \bibinfo
  {pages} {113} (\bibinfo {year} {2023})},\ \Eprint
  {https://arxiv.org/abs/2312.06782} {arXiv:2312.06782 [astro-ph.SR]}
  \BibitemShut {NoStop}%
\bibitem [{\citenamefont {Graham}\ \emph {et~al.}(2015)\citenamefont {Graham},
  \citenamefont {Rajendran},\ and\ \citenamefont {Varela}}]{Graham:2015apa}%
  \BibitemOpen
  \bibfield  {author} {\bibinfo {author} {\bibfnamefont {P.~W.}\ \bibnamefont
  {Graham}}, \bibinfo {author} {\bibfnamefont {S.}~\bibnamefont {Rajendran}},\
  and\ \bibinfo {author} {\bibfnamefont {J.}~\bibnamefont {Varela}},\
  }\bibfield  {title} {\bibinfo {title} {{Dark Matter Triggers of
  Supernovae}},\ }\href {https://doi.org/10.1103/PhysRevD.92.063007} {\bibfield
   {journal} {\bibinfo  {journal} {Phys. Rev. D}\ }\textbf {\bibinfo {volume}
  {92}},\ \bibinfo {pages} {063007} (\bibinfo {year} {2015})},\ \Eprint
  {https://arxiv.org/abs/1505.04444} {arXiv:1505.04444 [hep-ph]} \BibitemShut
  {NoStop}%
\bibitem [{\citenamefont {Barnacka}\ \emph {et~al.}(2012)\citenamefont
  {Barnacka}, \citenamefont {Glicenstein},\ and\ \citenamefont
  {Moderski}}]{Barnacka:2012bm}%
  \BibitemOpen
  \bibfield  {author} {\bibinfo {author} {\bibfnamefont {A.}~\bibnamefont
  {Barnacka}}, \bibinfo {author} {\bibfnamefont {J.~F.}\ \bibnamefont
  {Glicenstein}},\ and\ \bibinfo {author} {\bibfnamefont {R.}~\bibnamefont
  {Moderski}},\ }\bibfield  {title} {\bibinfo {title} {{New constraints on
  primordial black holes abundance from femtolensing of gamma-ray bursts}},\
  }\href {https://doi.org/10.1103/PhysRevD.86.043001} {\bibfield  {journal}
  {\bibinfo  {journal} {Phys. Rev. D}\ }\textbf {\bibinfo {volume} {86}},\
  \bibinfo {pages} {043001} (\bibinfo {year} {2012})},\ \Eprint
  {https://arxiv.org/abs/1204.2056} {arXiv:1204.2056 [astro-ph.CO]}
  \BibitemShut {NoStop}%
\bibitem [{\citenamefont {Katz}\ \emph {et~al.}(2018)\citenamefont {Katz},
  \citenamefont {Kopp}, \citenamefont {Sibiryakov},\ and\ \citenamefont
  {Xue}}]{Katz:2018zrn}%
  \BibitemOpen
  \bibfield  {author} {\bibinfo {author} {\bibfnamefont {A.}~\bibnamefont
  {Katz}}, \bibinfo {author} {\bibfnamefont {J.}~\bibnamefont {Kopp}}, \bibinfo
  {author} {\bibfnamefont {S.}~\bibnamefont {Sibiryakov}},\ and\ \bibinfo
  {author} {\bibfnamefont {W.}~\bibnamefont {Xue}},\ }\bibfield  {title}
  {\bibinfo {title} {{Femtolensing by Dark Matter Revisited}},\ }\href
  {https://doi.org/10.1088/1475-7516/2018/12/005} {\bibfield  {journal}
  {\bibinfo  {journal} {JCAP}\ }\textbf {\bibinfo {volume} {12}},\ \bibinfo
  {pages} {005}},\ \Eprint {https://arxiv.org/abs/1807.11495} {arXiv:1807.11495
  [astro-ph.CO]} \BibitemShut {NoStop}%
\bibitem [{\citenamefont {Bai}\ and\ \citenamefont
  {Orlofsky}(2019)}]{Bai:2018bej}%
  \BibitemOpen
  \bibfield  {author} {\bibinfo {author} {\bibfnamefont {Y.}~\bibnamefont
  {Bai}}\ and\ \bibinfo {author} {\bibfnamefont {N.}~\bibnamefont {Orlofsky}},\
  }\bibfield  {title} {\bibinfo {title} {{Microlensing of X-ray Pulsars: a
  Method to Detect Primordial Black Hole Dark Matter}},\ }\href
  {https://doi.org/10.1103/PhysRevD.99.123019} {\bibfield  {journal} {\bibinfo
  {journal} {Phys. Rev. D}\ }\textbf {\bibinfo {volume} {99}},\ \bibinfo
  {pages} {123019} (\bibinfo {year} {2019})},\ \Eprint
  {https://arxiv.org/abs/1812.01427} {arXiv:1812.01427 [astro-ph.HE]}
  \BibitemShut {NoStop}%
\bibitem [{\citenamefont {Tamta}\ \emph {et~al.}(2025)\citenamefont {Tamta},
  \citenamefont {Raj},\ and\ \citenamefont {Sharma}}]{Tamta:2024pow}%
  \BibitemOpen
  \bibfield  {author} {\bibinfo {author} {\bibfnamefont {M.}~\bibnamefont
  {Tamta}}, \bibinfo {author} {\bibfnamefont {N.}~\bibnamefont {Raj}},\ and\
  \bibinfo {author} {\bibfnamefont {P.}~\bibnamefont {Sharma}},\ }\bibfield
  {title} {\bibinfo {title} {{Entering the window of primordial black hole dark
  matter with x-ray microlensing}},\ }\href
  {https://doi.org/10.1103/PhysRevD.111.043043} {\bibfield  {journal} {\bibinfo
   {journal} {Phys. Rev. D}\ }\textbf {\bibinfo {volume} {111}},\ \bibinfo
  {pages} {043043} (\bibinfo {year} {2025})},\ \Eprint
  {https://arxiv.org/abs/2405.20365} {arXiv:2405.20365 [astro-ph.HE]}
  \BibitemShut {NoStop}%
\bibitem [{\citenamefont {Anchordoqui}\ \emph {et~al.}(2024)\citenamefont
  {Anchordoqui}, \citenamefont {Antoniadis}, \citenamefont {Lust},\ and\
  \citenamefont {Castillo}}]{Anchordoqui:2024tdj}%
  \BibitemOpen
  \bibfield  {author} {\bibinfo {author} {\bibfnamefont {L.~A.}\ \bibnamefont
  {Anchordoqui}}, \bibinfo {author} {\bibfnamefont {I.}~\bibnamefont
  {Antoniadis}}, \bibinfo {author} {\bibfnamefont {D.}~\bibnamefont {Lust}},\
  and\ \bibinfo {author} {\bibfnamefont {K.~P.~n.}\ \bibnamefont {Castillo}},\
  }\bibfield  {title} {\bibinfo {title} {{Through the looking glass into the
  dark dimension: Searching for bulk black hole dark matter with microlensing
  of X-ray pulsars}},\ }\href {https://doi.org/10.1016/j.dark.2024.101681}
  {\bibfield  {journal} {\bibinfo  {journal} {Phys. Dark Univ.}\ }\textbf
  {\bibinfo {volume} {46}},\ \bibinfo {pages} {101681} (\bibinfo {year}
  {2024})},\ \Eprint {https://arxiv.org/abs/2409.12904} {arXiv:2409.12904
  [hep-ph]} \BibitemShut {NoStop}%
\bibitem [{\citenamefont {Tran}\ \emph {et~al.}(2024)\citenamefont {Tran},
  \citenamefont {Geller}, \citenamefont {Lehmann},\ and\ \citenamefont
  {Kaiser}}]{Tran:2023jci}%
  \BibitemOpen
  \bibfield  {author} {\bibinfo {author} {\bibfnamefont {T.~X.}\ \bibnamefont
  {Tran}}, \bibinfo {author} {\bibfnamefont {S.~R.}\ \bibnamefont {Geller}},
  \bibinfo {author} {\bibfnamefont {B.~V.}\ \bibnamefont {Lehmann}},\ and\
  \bibinfo {author} {\bibfnamefont {D.~I.}\ \bibnamefont {Kaiser}},\ }\bibfield
   {title} {\bibinfo {title} {{Close encounters of the primordial kind: A new
  observable for primordial black holes as dark matter}},\ }\href
  {https://doi.org/10.1103/PhysRevD.110.063533} {\bibfield  {journal} {\bibinfo
   {journal} {Phys. Rev. D}\ }\textbf {\bibinfo {volume} {110}},\ \bibinfo
  {pages} {063533} (\bibinfo {year} {2024})},\ \Eprint
  {https://arxiv.org/abs/2312.17217} {arXiv:2312.17217 [astro-ph.CO]}
  \BibitemShut {NoStop}%
\bibitem [{\citenamefont {Belotsky}\ \emph {et~al.}(2014)\citenamefont
  {Belotsky}, \citenamefont {Dmitriev}, \citenamefont {Esipova}, \citenamefont
  {Gani}, \citenamefont {Grobov}, \citenamefont {Khlopov}, \citenamefont
  {Kirillov}, \citenamefont {Rubin},\ and\ \citenamefont
  {Svadkovsky}}]{Belotsky:2014kca}%
  \BibitemOpen
  \bibfield  {author} {\bibinfo {author} {\bibfnamefont {K.~M.}\ \bibnamefont
  {Belotsky}}, \bibinfo {author} {\bibfnamefont {A.~D.}\ \bibnamefont
  {Dmitriev}}, \bibinfo {author} {\bibfnamefont {E.~A.}\ \bibnamefont
  {Esipova}}, \bibinfo {author} {\bibfnamefont {V.~A.}\ \bibnamefont {Gani}},
  \bibinfo {author} {\bibfnamefont {A.~V.}\ \bibnamefont {Grobov}}, \bibinfo
  {author} {\bibfnamefont {M.~Y.}\ \bibnamefont {Khlopov}}, \bibinfo {author}
  {\bibfnamefont {A.~A.}\ \bibnamefont {Kirillov}}, \bibinfo {author}
  {\bibfnamefont {S.~G.}\ \bibnamefont {Rubin}},\ and\ \bibinfo {author}
  {\bibfnamefont {I.~V.}\ \bibnamefont {Svadkovsky}},\ }\bibfield  {title}
  {\bibinfo {title} {{Signatures of primordial black hole dark matter}},\
  }\href {https://doi.org/10.1142/S0217732314400057} {\bibfield  {journal}
  {\bibinfo  {journal} {Mod. Phys. Lett. A}\ }\textbf {\bibinfo {volume}
  {29}},\ \bibinfo {pages} {1440005} (\bibinfo {year} {2014})},\ \Eprint
  {https://arxiv.org/abs/1410.0203} {arXiv:1410.0203 [astro-ph.CO]}
  \BibitemShut {NoStop}%
\bibitem [{\citenamefont {K{\"u}hnel}\ and\ \citenamefont
  {Freese}(2017)}]{Kuhnel:2017pwq}%
  \BibitemOpen
  \bibfield  {author} {\bibinfo {author} {\bibfnamefont {F.}~\bibnamefont
  {K{\"u}hnel}}\ and\ \bibinfo {author} {\bibfnamefont {K.}~\bibnamefont
  {Freese}},\ }\bibfield  {title} {\bibinfo {title} {{Constraints on Primordial
  Black Holes with Extended Mass Functions}},\ }\href
  {https://doi.org/10.1103/PhysRevD.95.083508} {\bibfield  {journal} {\bibinfo
  {journal} {Phys. Rev. D}\ }\textbf {\bibinfo {volume} {95}},\ \bibinfo
  {pages} {083508} (\bibinfo {year} {2017})},\ \Eprint
  {https://arxiv.org/abs/1701.07223} {arXiv:1701.07223 [astro-ph.CO]}
  \BibitemShut {NoStop}%
\bibitem [{\citenamefont {De~Luca}\ \emph
  {et~al.}(2022{\natexlab{b}})\citenamefont {De~Luca}, \citenamefont
  {Franciolini}, \citenamefont {Kehagias}, \citenamefont {Pani},\ and\
  \citenamefont {Riotto}}]{DeLuca:2021pls}%
  \BibitemOpen
  \bibfield  {author} {\bibinfo {author} {\bibfnamefont {V.}~\bibnamefont
  {De~Luca}}, \bibinfo {author} {\bibfnamefont {G.}~\bibnamefont
  {Franciolini}}, \bibinfo {author} {\bibfnamefont {A.}~\bibnamefont
  {Kehagias}}, \bibinfo {author} {\bibfnamefont {P.}~\bibnamefont {Pani}},\
  and\ \bibinfo {author} {\bibfnamefont {A.}~\bibnamefont {Riotto}},\
  }\bibfield  {title} {\bibinfo {title} {{Primordial black holes in
  matter-dominated eras: The role of accretion}},\ }\href
  {https://doi.org/10.1016/j.physletb.2022.137265} {\bibfield  {journal}
  {\bibinfo  {journal} {Phys. Lett. B}\ }\textbf {\bibinfo {volume} {832}},\
  \bibinfo {pages} {137265} (\bibinfo {year} {2022}{\natexlab{b}})},\ \Eprint
  {https://arxiv.org/abs/2112.02534} {arXiv:2112.02534 [astro-ph.CO]}
  \BibitemShut {NoStop}%
\bibitem [{\citenamefont {Tomita}(1975)}]{Tomita:1975kj}%
  \BibitemOpen
  \bibfield  {author} {\bibinfo {author} {\bibfnamefont {K.}~\bibnamefont
  {Tomita}},\ }\bibfield  {title} {\bibinfo {title} {{Evolution of
  Irregularities in a Chaotic Early Universe}},\ }\href
  {https://doi.org/10.1143/PTP.54.730} {\bibfield  {journal} {\bibinfo
  {journal} {Prog. Theor. Phys.}\ }\textbf {\bibinfo {volume} {54}},\ \bibinfo
  {pages} {730} (\bibinfo {year} {1975})}\BibitemShut {NoStop}%
\bibitem [{\citenamefont {Matarrese}\ \emph {et~al.}(1993)\citenamefont
  {Matarrese}, \citenamefont {Pantano},\ and\ \citenamefont
  {Saez}}]{Matarrese:1992rp}%
  \BibitemOpen
  \bibfield  {author} {\bibinfo {author} {\bibfnamefont {S.}~\bibnamefont
  {Matarrese}}, \bibinfo {author} {\bibfnamefont {O.}~\bibnamefont {Pantano}},\
  and\ \bibinfo {author} {\bibfnamefont {D.}~\bibnamefont {Saez}},\ }\bibfield
  {title} {\bibinfo {title} {{A General relativistic approach to the nonlinear
  evolution of collisionless matter}},\ }\href
  {https://doi.org/10.1103/PhysRevD.47.1311} {\bibfield  {journal} {\bibinfo
  {journal} {Phys. Rev. D}\ }\textbf {\bibinfo {volume} {47}},\ \bibinfo
  {pages} {1311} (\bibinfo {year} {1993})}\BibitemShut {NoStop}%
\bibitem [{\citenamefont {Matarrese}\ \emph {et~al.}(1994)\citenamefont
  {Matarrese}, \citenamefont {Pantano},\ and\ \citenamefont
  {Saez}}]{Matarrese:1993zf}%
  \BibitemOpen
  \bibfield  {author} {\bibinfo {author} {\bibfnamefont {S.}~\bibnamefont
  {Matarrese}}, \bibinfo {author} {\bibfnamefont {O.}~\bibnamefont {Pantano}},\
  and\ \bibinfo {author} {\bibfnamefont {D.}~\bibnamefont {Saez}},\ }\bibfield
  {title} {\bibinfo {title} {{General relativistic dynamics of irrotational
  dust: Cosmological implications}},\ }\href
  {https://doi.org/10.1103/PhysRevLett.72.320} {\bibfield  {journal} {\bibinfo
  {journal} {Phys. Rev. Lett.}\ }\textbf {\bibinfo {volume} {72}},\ \bibinfo
  {pages} {320} (\bibinfo {year} {1994})},\ \Eprint
  {https://arxiv.org/abs/astro-ph/9310036} {arXiv:astro-ph/9310036}
  \BibitemShut {NoStop}%
\bibitem [{\citenamefont {Matarrese}\ \emph {et~al.}(1998)\citenamefont
  {Matarrese}, \citenamefont {Mollerach},\ and\ \citenamefont
  {Bruni}}]{Matarrese:1997ay}%
  \BibitemOpen
  \bibfield  {author} {\bibinfo {author} {\bibfnamefont {S.}~\bibnamefont
  {Matarrese}}, \bibinfo {author} {\bibfnamefont {S.}~\bibnamefont
  {Mollerach}},\ and\ \bibinfo {author} {\bibfnamefont {M.}~\bibnamefont
  {Bruni}},\ }\bibfield  {title} {\bibinfo {title} {{Second order perturbations
  of the Einstein-de Sitter universe}},\ }\href
  {https://doi.org/10.1103/PhysRevD.58.043504} {\bibfield  {journal} {\bibinfo
  {journal} {Phys. Rev. D}\ }\textbf {\bibinfo {volume} {58}},\ \bibinfo
  {pages} {043504} (\bibinfo {year} {1998})},\ \Eprint
  {https://arxiv.org/abs/astro-ph/9707278} {arXiv:astro-ph/9707278}
  \BibitemShut {NoStop}%
\bibitem [{\citenamefont {Acquaviva}\ \emph {et~al.}(2003)\citenamefont
  {Acquaviva}, \citenamefont {Bartolo}, \citenamefont {Matarrese},\ and\
  \citenamefont {Riotto}}]{Acquaviva:2002ud}%
  \BibitemOpen
  \bibfield  {author} {\bibinfo {author} {\bibfnamefont {V.}~\bibnamefont
  {Acquaviva}}, \bibinfo {author} {\bibfnamefont {N.}~\bibnamefont {Bartolo}},
  \bibinfo {author} {\bibfnamefont {S.}~\bibnamefont {Matarrese}},\ and\
  \bibinfo {author} {\bibfnamefont {A.}~\bibnamefont {Riotto}},\ }\bibfield
  {title} {\bibinfo {title} {{Second order cosmological perturbations from
  inflation}},\ }\href {https://doi.org/10.1016/S0550-3213(03)00550-9}
  {\bibfield  {journal} {\bibinfo  {journal} {Nucl. Phys. B}\ }\textbf
  {\bibinfo {volume} {667}},\ \bibinfo {pages} {119} (\bibinfo {year}
  {2003})},\ \Eprint {https://arxiv.org/abs/astro-ph/0209156}
  {arXiv:astro-ph/0209156} \BibitemShut {NoStop}%
\bibitem [{\citenamefont {Mollerach}\ \emph {et~al.}(2004)\citenamefont
  {Mollerach}, \citenamefont {Harari},\ and\ \citenamefont
  {Matarrese}}]{Mollerach:2003nq}%
  \BibitemOpen
  \bibfield  {author} {\bibinfo {author} {\bibfnamefont {S.}~\bibnamefont
  {Mollerach}}, \bibinfo {author} {\bibfnamefont {D.}~\bibnamefont {Harari}},\
  and\ \bibinfo {author} {\bibfnamefont {S.}~\bibnamefont {Matarrese}},\
  }\bibfield  {title} {\bibinfo {title} {{CMB polarization from secondary
  vector and tensor modes}},\ }\href
  {https://doi.org/10.1103/PhysRevD.69.063002} {\bibfield  {journal} {\bibinfo
  {journal} {Phys. Rev. D}\ }\textbf {\bibinfo {volume} {69}},\ \bibinfo
  {pages} {063002} (\bibinfo {year} {2004})},\ \Eprint
  {https://arxiv.org/abs/astro-ph/0310711} {arXiv:astro-ph/0310711}
  \BibitemShut {NoStop}%
\bibitem [{\citenamefont {Carbone}\ and\ \citenamefont
  {Matarrese}(2005)}]{Carbone:2004iv}%
  \BibitemOpen
  \bibfield  {author} {\bibinfo {author} {\bibfnamefont {C.}~\bibnamefont
  {Carbone}}\ and\ \bibinfo {author} {\bibfnamefont {S.}~\bibnamefont
  {Matarrese}},\ }\bibfield  {title} {\bibinfo {title} {{A Unified treatment of
  cosmological perturbations from super-horizon to small scales}},\ }\href
  {https://doi.org/10.1103/PhysRevD.71.043508} {\bibfield  {journal} {\bibinfo
  {journal} {Phys. Rev. D}\ }\textbf {\bibinfo {volume} {71}},\ \bibinfo
  {pages} {043508} (\bibinfo {year} {2005})},\ \Eprint
  {https://arxiv.org/abs/astro-ph/0407611} {arXiv:astro-ph/0407611}
  \BibitemShut {NoStop}%
\bibitem [{\citenamefont {Ananda}\ \emph {et~al.}(2007)\citenamefont {Ananda},
  \citenamefont {Clarkson},\ and\ \citenamefont {Wands}}]{Ananda:2006af}%
  \BibitemOpen
  \bibfield  {author} {\bibinfo {author} {\bibfnamefont {K.~N.}\ \bibnamefont
  {Ananda}}, \bibinfo {author} {\bibfnamefont {C.}~\bibnamefont {Clarkson}},\
  and\ \bibinfo {author} {\bibfnamefont {D.}~\bibnamefont {Wands}},\ }\bibfield
   {title} {\bibinfo {title} {{The Cosmological gravitational wave background
  from primordial density perturbations}},\ }\href
  {https://doi.org/10.1103/PhysRevD.75.123518} {\bibfield  {journal} {\bibinfo
  {journal} {Phys. Rev. D}\ }\textbf {\bibinfo {volume} {75}},\ \bibinfo
  {pages} {123518} (\bibinfo {year} {2007})},\ \Eprint
  {https://arxiv.org/abs/gr-qc/0612013} {arXiv:gr-qc/0612013} \BibitemShut
  {NoStop}%
\bibitem [{\citenamefont {Baumann}\ \emph {et~al.}(2007)\citenamefont
  {Baumann}, \citenamefont {Steinhardt}, \citenamefont {Takahashi},\ and\
  \citenamefont {Ichiki}}]{Baumann:2007zm}%
  \BibitemOpen
  \bibfield  {author} {\bibinfo {author} {\bibfnamefont {D.}~\bibnamefont
  {Baumann}}, \bibinfo {author} {\bibfnamefont {P.~J.}\ \bibnamefont
  {Steinhardt}}, \bibinfo {author} {\bibfnamefont {K.}~\bibnamefont
  {Takahashi}},\ and\ \bibinfo {author} {\bibfnamefont {K.}~\bibnamefont
  {Ichiki}},\ }\bibfield  {title} {\bibinfo {title} {{Gravitational Wave
  Spectrum Induced by Primordial Scalar Perturbations}},\ }\href
  {https://doi.org/10.1103/PhysRevD.76.084019} {\bibfield  {journal} {\bibinfo
  {journal} {Phys. Rev. D}\ }\textbf {\bibinfo {volume} {76}},\ \bibinfo
  {pages} {084019} (\bibinfo {year} {2007})},\ \Eprint
  {https://arxiv.org/abs/hep-th/0703290} {arXiv:hep-th/0703290} \BibitemShut
  {NoStop}%
\bibitem [{\citenamefont {Dom\`enech}(2021)}]{Domenech:2021ztg}%
  \BibitemOpen
  \bibfield  {author} {\bibinfo {author} {\bibfnamefont {G.}~\bibnamefont
  {Dom\`enech}},\ }\bibfield  {title} {\bibinfo {title} {{Scalar Induced
  Gravitational Waves Review}},\ }\href
  {https://doi.org/10.3390/universe7110398} {\bibfield  {journal} {\bibinfo
  {journal} {Universe}\ }\textbf {\bibinfo {volume} {7}},\ \bibinfo {pages}
  {398} (\bibinfo {year} {2021})},\ \Eprint {https://arxiv.org/abs/2109.01398}
  {arXiv:2109.01398 [gr-qc]} \BibitemShut {NoStop}%
\bibitem [{\citenamefont {Saito}\ and\ \citenamefont
  {Yokoyama}(2009)}]{Saito:2008jc}%
  \BibitemOpen
  \bibfield  {author} {\bibinfo {author} {\bibfnamefont {R.}~\bibnamefont
  {Saito}}\ and\ \bibinfo {author} {\bibfnamefont {J.}~\bibnamefont
  {Yokoyama}},\ }\bibfield  {title} {\bibinfo {title} {{Gravitational wave
  background as a probe of the primordial black hole abundance}},\ }\href
  {https://doi.org/10.1103/PhysRevLett.102.161101} {\bibfield  {journal}
  {\bibinfo  {journal} {Phys. Rev. Lett.}\ }\textbf {\bibinfo {volume} {102}},\
  \bibinfo {pages} {161101} (\bibinfo {year} {2009})},\ \bibinfo {note}
  {[Erratum: Phys.Rev.Lett. 107, 069901 (2011)]},\ \Eprint
  {https://arxiv.org/abs/0812.4339} {arXiv:0812.4339 [astro-ph]} \BibitemShut
  {NoStop}%
\bibitem [{\citenamefont {Zheng}\ \emph {et~al.}(2022)\citenamefont {Zheng},
  \citenamefont {Shi},\ and\ \citenamefont {Qiu}}]{ZhengRuiFeng:2021zoz}%
  \BibitemOpen
  \bibfield  {author} {\bibinfo {author} {\bibfnamefont {R.}~\bibnamefont
  {Zheng}}, \bibinfo {author} {\bibfnamefont {J.}~\bibnamefont {Shi}},\ and\
  \bibinfo {author} {\bibfnamefont {T.}~\bibnamefont {Qiu}},\ }\bibfield
  {title} {\bibinfo {title} {{On primordial black holes and secondary
  gravitational waves generated from inflation with solo/multi-bumpy potential
  *}},\ }\href {https://doi.org/10.1088/1674-1137/ac42bd} {\bibfield  {journal}
  {\bibinfo  {journal} {Chin. Phys. C}\ }\textbf {\bibinfo {volume} {46}},\
  \bibinfo {pages} {045103} (\bibinfo {year} {2022})},\ \Eprint
  {https://arxiv.org/abs/2106.04303} {arXiv:2106.04303 [astro-ph.CO]}
  \BibitemShut {NoStop}%
\bibitem [{\citenamefont {Zhang}\ \emph {et~al.}(2021)\citenamefont {Zhang},
  \citenamefont {Lin},\ and\ \citenamefont {Lu}}]{Zhang:2021vak}%
  \BibitemOpen
  \bibfield  {author} {\bibinfo {author} {\bibfnamefont {F.}~\bibnamefont
  {Zhang}}, \bibinfo {author} {\bibfnamefont {J.}~\bibnamefont {Lin}},\ and\
  \bibinfo {author} {\bibfnamefont {Y.}~\bibnamefont {Lu}},\ }\bibfield
  {title} {\bibinfo {title} {{Double-peaked inflation model: Scalar induced
  gravitational waves and primordial-black-hole suppression from primordial
  non-Gaussianity}},\ }\href {https://doi.org/10.1103/PhysRevD.104.063515}
  {\bibfield  {journal} {\bibinfo  {journal} {Phys. Rev. D}\ }\textbf {\bibinfo
  {volume} {104}},\ \bibinfo {pages} {063515} (\bibinfo {year} {2021})},\
  \bibinfo {note} {[Erratum: Phys.Rev.D 104, 129902 (2021)]},\ \Eprint
  {https://arxiv.org/abs/2106.10792} {arXiv:2106.10792 [gr-qc]} \BibitemShut
  {NoStop}%
\bibitem [{\citenamefont {Yuan}\ and\ \citenamefont
  {Huang}(2021)}]{Yuan:2021qgz}%
  \BibitemOpen
  \bibfield  {author} {\bibinfo {author} {\bibfnamefont {C.}~\bibnamefont
  {Yuan}}\ and\ \bibinfo {author} {\bibfnamefont {Q.-G.}\ \bibnamefont
  {Huang}},\ }\bibfield  {title} {\bibinfo {title} {{A topic review on probing
  primordial black hole dark matter with scalar induced gravitational waves}},\
  }\href {https://doi.org/10.1016/j.isci.2021.102860} {\bibfield  {journal}
  {\bibinfo  {journal} {iScience}\ }\textbf {\bibinfo {volume} {24}},\ \bibinfo
  {pages} {102860} (\bibinfo {year} {2021})},\ \Eprint
  {https://arxiv.org/abs/2103.04739} {arXiv:2103.04739 [astro-ph.GA]}
  \BibitemShut {NoStop}%
\bibitem [{\citenamefont {Yi}\ and\ \citenamefont {Fei}(2023)}]{Yi:2022ymw}%
  \BibitemOpen
  \bibfield  {author} {\bibinfo {author} {\bibfnamefont {Z.}~\bibnamefont
  {Yi}}\ and\ \bibinfo {author} {\bibfnamefont {Q.}~\bibnamefont {Fei}},\
  }\bibfield  {title} {\bibinfo {title} {{Constraints on primordial curvature
  spectrum from primordial black holes and scalar-induced gravitational
  waves}},\ }\href {https://doi.org/10.1140/epjc/s10052-023-11233-3} {\bibfield
   {journal} {\bibinfo  {journal} {Eur. Phys. J. C}\ }\textbf {\bibinfo
  {volume} {83}},\ \bibinfo {pages} {82} (\bibinfo {year} {2023})},\ \Eprint
  {https://arxiv.org/abs/2210.03641} {arXiv:2210.03641 [astro-ph.CO]}
  \BibitemShut {NoStop}%
\bibitem [{\citenamefont {Zhao}\ \emph {et~al.}(2023)\citenamefont {Zhao},
  \citenamefont {Liu},\ and\ \citenamefont {Li}}]{Zhao:2023xnh}%
  \BibitemOpen
  \bibfield  {author} {\bibinfo {author} {\bibfnamefont {J.-X.}\ \bibnamefont
  {Zhao}}, \bibinfo {author} {\bibfnamefont {X.-H.}\ \bibnamefont {Liu}},\ and\
  \bibinfo {author} {\bibfnamefont {N.}~\bibnamefont {Li}},\ }\bibfield
  {title} {\bibinfo {title} {{Primordial black holes and scalar-induced
  gravitational waves from the perturbations on the inflaton potential in peak
  theory}},\ }\href {https://doi.org/10.1103/PhysRevD.107.043515} {\bibfield
  {journal} {\bibinfo  {journal} {Phys. Rev. D}\ }\textbf {\bibinfo {volume}
  {107}},\ \bibinfo {pages} {043515} (\bibinfo {year} {2023})},\ \Eprint
  {https://arxiv.org/abs/2302.06886} {arXiv:2302.06886 [astro-ph.CO]}
  \BibitemShut {NoStop}%
\bibitem [{\citenamefont {Gouttenoire}\ \emph {et~al.}(2025)\citenamefont
  {Gouttenoire}, \citenamefont {Trifinopoulos},\ and\ \citenamefont
  {Vanvlasselaer}}]{Gouttenoire:2025jxe}%
  \BibitemOpen
  \bibfield  {author} {\bibinfo {author} {\bibfnamefont {Y.}~\bibnamefont
  {Gouttenoire}}, \bibinfo {author} {\bibfnamefont {S.}~\bibnamefont
  {Trifinopoulos}},\ and\ \bibinfo {author} {\bibfnamefont {M.}~\bibnamefont
  {Vanvlasselaer}},\ }\bibfield  {title} {\bibinfo {title} {{Implications for
  Pulsar Timing Arrays of Sub-solar Black Hole Detections: From LVK to Einstein
  Telescope and Cosmic Explorer}},\ }\href@noop {} {\  (\bibinfo {year}
  {2025})},\ \Eprint {https://arxiv.org/abs/2508.19328} {arXiv:2508.19328
  [astro-ph.CO]} \BibitemShut {NoStop}%
\bibitem [{\citenamefont {Bari}\ \emph {et~al.}(2024)\citenamefont {Bari},
  \citenamefont {Bartolo}, \citenamefont {Dom\`enech},\ and\ \citenamefont
  {Matarrese}}]{Bari:2023rcw}%
  \BibitemOpen
  \bibfield  {author} {\bibinfo {author} {\bibfnamefont {P.}~\bibnamefont
  {Bari}}, \bibinfo {author} {\bibfnamefont {N.}~\bibnamefont {Bartolo}},
  \bibinfo {author} {\bibfnamefont {G.}~\bibnamefont {Dom\`enech}},\ and\
  \bibinfo {author} {\bibfnamefont {S.}~\bibnamefont {Matarrese}},\ }\bibfield
  {title} {\bibinfo {title} {{Gravitational waves induced by scalar-tensor
  mixing}},\ }\href {https://doi.org/10.1103/PhysRevD.109.023509} {\bibfield
  {journal} {\bibinfo  {journal} {Phys. Rev. D}\ }\textbf {\bibinfo {volume}
  {109}},\ \bibinfo {pages} {023509} (\bibinfo {year} {2024})},\ \Eprint
  {https://arxiv.org/abs/2307.05404} {arXiv:2307.05404 [astro-ph.CO]}
  \BibitemShut {NoStop}%
\bibitem [{\citenamefont {Picard}\ and\ \citenamefont
  {Malik}(2024)}]{Picard:2023sbz}%
  \BibitemOpen
  \bibfield  {author} {\bibinfo {author} {\bibfnamefont {R.}~\bibnamefont
  {Picard}}\ and\ \bibinfo {author} {\bibfnamefont {K.~A.}\ \bibnamefont
  {Malik}},\ }\bibfield  {title} {\bibinfo {title} {{Induced gravitational
  waves: the effect of first order tensor perturbations}},\ }\href
  {https://doi.org/10.1088/1475-7516/2024/10/010} {\bibfield  {journal}
  {\bibinfo  {journal} {JCAP}\ }\textbf {\bibinfo {volume} {10}},\ \bibinfo
  {pages} {010}},\ \Eprint {https://arxiv.org/abs/2311.14513} {arXiv:2311.14513
  [astro-ph.CO]} \BibitemShut {NoStop}%
\bibitem [{\citenamefont {Iovino}\ \emph
  {et~al.}(2025{\natexlab{a}})\citenamefont {Iovino}, \citenamefont {Perna},
  \citenamefont {Perrone}, \citenamefont {Racco},\ and\ \citenamefont
  {Riotto}}]{Iovino:2025xkq}%
  \BibitemOpen
  \bibfield  {author} {\bibinfo {author} {\bibfnamefont {A.~J.}\ \bibnamefont
  {Iovino}}, \bibinfo {author} {\bibfnamefont {G.}~\bibnamefont {Perna}},
  \bibinfo {author} {\bibfnamefont {D.}~\bibnamefont {Perrone}}, \bibinfo
  {author} {\bibfnamefont {D.}~\bibnamefont {Racco}},\ and\ \bibinfo {author}
  {\bibfnamefont {A.}~\bibnamefont {Riotto}},\ }\bibfield  {title} {\bibinfo
  {title} {{Understanding the Nature of Scalar-Induced Gravitational Waves}},\
  }\href@noop {} {\  (\bibinfo {year} {2025}{\natexlab{a}})},\ \Eprint
  {https://arxiv.org/abs/2509.24774} {arXiv:2509.24774 [gr-qc]} \BibitemShut
  {NoStop}%
\bibitem [{\citenamefont {Espinosa}\ \emph {et~al.}(2018)\citenamefont
  {Espinosa}, \citenamefont {Racco},\ and\ \citenamefont
  {Riotto}}]{Espinosa:2018eve}%
  \BibitemOpen
  \bibfield  {author} {\bibinfo {author} {\bibfnamefont {J.~R.}\ \bibnamefont
  {Espinosa}}, \bibinfo {author} {\bibfnamefont {D.}~\bibnamefont {Racco}},\
  and\ \bibinfo {author} {\bibfnamefont {A.}~\bibnamefont {Riotto}},\
  }\bibfield  {title} {\bibinfo {title} {{A Cosmological Signature of the SM
  Higgs Instability: Gravitational Waves}},\ }\href
  {https://doi.org/10.1088/1475-7516/2018/09/012} {\bibfield  {journal}
  {\bibinfo  {journal} {JCAP}\ }\textbf {\bibinfo {volume} {09}},\ \bibinfo
  {pages} {012}},\ \Eprint {https://arxiv.org/abs/1804.07732} {arXiv:1804.07732
  [hep-ph]} \BibitemShut {NoStop}%
\bibitem [{\citenamefont {Kohri}\ and\ \citenamefont
  {Terada}(2018)}]{Kohri:2018awv}%
  \BibitemOpen
  \bibfield  {author} {\bibinfo {author} {\bibfnamefont {K.}~\bibnamefont
  {Kohri}}\ and\ \bibinfo {author} {\bibfnamefont {T.}~\bibnamefont {Terada}},\
  }\bibfield  {title} {\bibinfo {title} {{Semianalytic calculation of
  gravitational wave spectrum nonlinearly induced from primordial curvature
  perturbations}},\ }\href {https://doi.org/10.1103/PhysRevD.97.123532}
  {\bibfield  {journal} {\bibinfo  {journal} {Phys. Rev. D}\ }\textbf {\bibinfo
  {volume} {97}},\ \bibinfo {pages} {123532} (\bibinfo {year} {2018})},\
  \Eprint {https://arxiv.org/abs/1804.08577} {arXiv:1804.08577 [gr-qc]}
  \BibitemShut {NoStop}%
\bibitem [{\citenamefont {Adshead}\ \emph {et~al.}(2021)\citenamefont
  {Adshead}, \citenamefont {Lozanov},\ and\ \citenamefont
  {Weiner}}]{Adshead:2021hnm}%
  \BibitemOpen
  \bibfield  {author} {\bibinfo {author} {\bibfnamefont {P.}~\bibnamefont
  {Adshead}}, \bibinfo {author} {\bibfnamefont {K.~D.}\ \bibnamefont
  {Lozanov}},\ and\ \bibinfo {author} {\bibfnamefont {Z.~J.}\ \bibnamefont
  {Weiner}},\ }\bibfield  {title} {\bibinfo {title} {{Non-Gaussianity and the
  induced gravitational wave background}},\ }\href
  {https://doi.org/10.1088/1475-7516/2021/10/080} {\bibfield  {journal}
  {\bibinfo  {journal} {JCAP}\ }\textbf {\bibinfo {volume} {10}},\ \bibinfo
  {pages} {080}},\ \Eprint {https://arxiv.org/abs/2105.01659} {arXiv:2105.01659
  [astro-ph.CO]} \BibitemShut {NoStop}%
\bibitem [{\citenamefont {Perna}\ \emph {et~al.}(2024)\citenamefont {Perna},
  \citenamefont {Testini}, \citenamefont {Ricciardone},\ and\ \citenamefont
  {Matarrese}}]{Perna:2024ehx}%
  \BibitemOpen
  \bibfield  {author} {\bibinfo {author} {\bibfnamefont {G.}~\bibnamefont
  {Perna}}, \bibinfo {author} {\bibfnamefont {C.}~\bibnamefont {Testini}},
  \bibinfo {author} {\bibfnamefont {A.}~\bibnamefont {Ricciardone}},\ and\
  \bibinfo {author} {\bibfnamefont {S.}~\bibnamefont {Matarrese}},\ }\bibfield
  {title} {\bibinfo {title} {{Fully non-Gaussian Scalar-Induced Gravitational
  Waves}},\ }\href {https://doi.org/10.1088/1475-7516/2024/05/086} {\bibfield
  {journal} {\bibinfo  {journal} {JCAP}\ }\textbf {\bibinfo {volume} {05}},\
  \bibinfo {pages} {086}},\ \Eprint {https://arxiv.org/abs/2403.06962}
  {arXiv:2403.06962 [astro-ph.CO]} \BibitemShut {NoStop}%
\bibitem [{\citenamefont {Gammal}\ \emph {et~al.}(2025)\citenamefont {Gammal}
  \emph {et~al.}}]{LISACosmologyWorkingGroup:2025vdz}%
  \BibitemOpen
  \bibfield  {author} {\bibinfo {author} {\bibfnamefont {J.~E.}\ \bibnamefont
  {Gammal}} \emph {et~al.} (\bibinfo {collaboration} {LISA Cosmology Working
  Group}),\ }\bibfield  {title} {\bibinfo {title} {{Reconstructing primordial
  curvature perturbations via scalar-induced gravitational waves with LISA}},\
  }\href {https://doi.org/10.1088/1475-7516/2025/05/062} {\bibfield  {journal}
  {\bibinfo  {journal} {JCAP}\ }\textbf {\bibinfo {volume} {05}},\ \bibinfo
  {pages} {062}},\ \Eprint {https://arxiv.org/abs/2501.11320} {arXiv:2501.11320
  [astro-ph.CO]} \BibitemShut {NoStop}%
\bibitem [{\citenamefont {Gangui}\ \emph {et~al.}(1994)\citenamefont {Gangui},
  \citenamefont {Lucchin}, \citenamefont {Matarrese},\ and\ \citenamefont
  {Mollerach}}]{Gangui:1993tt}%
  \BibitemOpen
  \bibfield  {author} {\bibinfo {author} {\bibfnamefont {A.}~\bibnamefont
  {Gangui}}, \bibinfo {author} {\bibfnamefont {F.}~\bibnamefont {Lucchin}},
  \bibinfo {author} {\bibfnamefont {S.}~\bibnamefont {Matarrese}},\ and\
  \bibinfo {author} {\bibfnamefont {S.}~\bibnamefont {Mollerach}},\ }\bibfield
  {title} {\bibinfo {title} {{The Three point correlation function of the
  cosmic microwave background in inflationary models}},\ }\href
  {https://doi.org/10.1086/174421} {\bibfield  {journal} {\bibinfo  {journal}
  {Astrophys. J.}\ }\textbf {\bibinfo {volume} {430}},\ \bibinfo {pages} {447}
  (\bibinfo {year} {1994})},\ \Eprint {https://arxiv.org/abs/astro-ph/9312033}
  {arXiv:astro-ph/9312033} \BibitemShut {NoStop}%
\bibitem [{\citenamefont {Matarrese}\ \emph {et~al.}(2000)\citenamefont
  {Matarrese}, \citenamefont {Verde},\ and\ \citenamefont
  {Jimenez}}]{Matarrese:2000iz}%
  \BibitemOpen
  \bibfield  {author} {\bibinfo {author} {\bibfnamefont {S.}~\bibnamefont
  {Matarrese}}, \bibinfo {author} {\bibfnamefont {L.}~\bibnamefont {Verde}},\
  and\ \bibinfo {author} {\bibfnamefont {R.}~\bibnamefont {Jimenez}},\
  }\bibfield  {title} {\bibinfo {title} {{The Abundance of high-redshift
  objects as a probe of non-Gaussian initial conditions}},\ }\href
  {https://doi.org/10.1086/309412} {\bibfield  {journal} {\bibinfo  {journal}
  {Astrophys. J.}\ }\textbf {\bibinfo {volume} {541}},\ \bibinfo {pages} {10}
  (\bibinfo {year} {2000})},\ \Eprint {https://arxiv.org/abs/astro-ph/0001366}
  {arXiv:astro-ph/0001366} \BibitemShut {NoStop}%
\bibitem [{\citenamefont {Bartolo}\ \emph {et~al.}(2002)\citenamefont
  {Bartolo}, \citenamefont {Matarrese},\ and\ \citenamefont
  {Riotto}}]{Bartolo:2001cw}%
  \BibitemOpen
  \bibfield  {author} {\bibinfo {author} {\bibfnamefont {N.}~\bibnamefont
  {Bartolo}}, \bibinfo {author} {\bibfnamefont {S.}~\bibnamefont {Matarrese}},\
  and\ \bibinfo {author} {\bibfnamefont {A.}~\bibnamefont {Riotto}},\
  }\bibfield  {title} {\bibinfo {title} {{Nongaussianity from inflation}},\
  }\href {https://doi.org/10.1103/PhysRevD.65.103505} {\bibfield  {journal}
  {\bibinfo  {journal} {Phys. Rev. D}\ }\textbf {\bibinfo {volume} {65}},\
  \bibinfo {pages} {103505} (\bibinfo {year} {2002})},\ \Eprint
  {https://arxiv.org/abs/hep-ph/0112261} {arXiv:hep-ph/0112261} \BibitemShut
  {NoStop}%
\bibitem [{\citenamefont {Maldacena}(2003)}]{Maldacena:2002vr}%
  \BibitemOpen
  \bibfield  {author} {\bibinfo {author} {\bibfnamefont {J.~M.}\ \bibnamefont
  {Maldacena}},\ }\bibfield  {title} {\bibinfo {title} {{Non-Gaussian features
  of primordial fluctuations in single field inflationary models}},\ }\href
  {https://doi.org/10.1088/1126-6708/2003/05/013} {\bibfield  {journal}
  {\bibinfo  {journal} {JHEP}\ }\textbf {\bibinfo {volume} {05}},\ \bibinfo
  {pages} {013}},\ \Eprint {https://arxiv.org/abs/astro-ph/0210603}
  {arXiv:astro-ph/0210603} \BibitemShut {NoStop}%
\bibitem [{\citenamefont {Bartolo}\ \emph {et~al.}(2004)\citenamefont
  {Bartolo}, \citenamefont {Komatsu}, \citenamefont {Matarrese},\ and\
  \citenamefont {Riotto}}]{Bartolo:2004if}%
  \BibitemOpen
  \bibfield  {author} {\bibinfo {author} {\bibfnamefont {N.}~\bibnamefont
  {Bartolo}}, \bibinfo {author} {\bibfnamefont {E.}~\bibnamefont {Komatsu}},
  \bibinfo {author} {\bibfnamefont {S.}~\bibnamefont {Matarrese}},\ and\
  \bibinfo {author} {\bibfnamefont {A.}~\bibnamefont {Riotto}},\ }\bibfield
  {title} {\bibinfo {title} {{Non-Gaussianity from inflation: Theory and
  observations}},\ }\href {https://doi.org/10.1016/j.physrep.2004.08.022}
  {\bibfield  {journal} {\bibinfo  {journal} {Phys. Rept.}\ }\textbf {\bibinfo
  {volume} {402}},\ \bibinfo {pages} {103} (\bibinfo {year} {2004})},\ \Eprint
  {https://arxiv.org/abs/astro-ph/0406398} {arXiv:astro-ph/0406398}
  \BibitemShut {NoStop}%
\bibitem [{\citenamefont {Chen}(2010)}]{Chen:2010xka}%
  \BibitemOpen
  \bibfield  {author} {\bibinfo {author} {\bibfnamefont {X.}~\bibnamefont
  {Chen}},\ }\bibfield  {title} {\bibinfo {title} {{Primordial
  Non-Gaussianities from Inflation Models}},\ }\href
  {https://doi.org/10.1155/2010/638979} {\bibfield  {journal} {\bibinfo
  {journal} {Adv. Astron.}\ }\textbf {\bibinfo {volume} {2010}},\ \bibinfo
  {pages} {638979} (\bibinfo {year} {2010})},\ \Eprint
  {https://arxiv.org/abs/1002.1416} {arXiv:1002.1416 [astro-ph.CO]}
  \BibitemShut {NoStop}%
\bibitem [{\citenamefont {Byrnes}\ and\ \citenamefont
  {Choi}(2010)}]{Byrnes:2010em}%
  \BibitemOpen
  \bibfield  {author} {\bibinfo {author} {\bibfnamefont {C.~T.}\ \bibnamefont
  {Byrnes}}\ and\ \bibinfo {author} {\bibfnamefont {K.-Y.}\ \bibnamefont
  {Choi}},\ }\bibfield  {title} {\bibinfo {title} {{Review of local
  non-Gaussianity from multi-field inflation}},\ }\href
  {https://doi.org/10.1155/2010/724525} {\bibfield  {journal} {\bibinfo
  {journal} {Adv. Astron.}\ }\textbf {\bibinfo {volume} {2010}},\ \bibinfo
  {pages} {724525} (\bibinfo {year} {2010})},\ \Eprint
  {https://arxiv.org/abs/1002.3110} {arXiv:1002.3110 [astro-ph.CO]}
  \BibitemShut {NoStop}%
\bibitem [{\citenamefont {Wands}(2010)}]{Wands:2010af}%
  \BibitemOpen
  \bibfield  {author} {\bibinfo {author} {\bibfnamefont {D.}~\bibnamefont
  {Wands}},\ }\bibfield  {title} {\bibinfo {title} {{Local non-Gaussianity from
  inflation}},\ }\href {https://doi.org/10.1088/0264-9381/27/12/124002}
  {\bibfield  {journal} {\bibinfo  {journal} {Class. Quant. Grav.}\ }\textbf
  {\bibinfo {volume} {27}},\ \bibinfo {pages} {124002} (\bibinfo {year}
  {2010})},\ \Eprint {https://arxiv.org/abs/1004.0818} {arXiv:1004.0818
  [astro-ph.CO]} \BibitemShut {NoStop}%
\bibitem [{\citenamefont {Renaux-Petel}(2015)}]{Renaux-Petel:2015bja}%
  \BibitemOpen
  \bibfield  {author} {\bibinfo {author} {\bibfnamefont {S.}~\bibnamefont
  {Renaux-Petel}},\ }\bibfield  {title} {\bibinfo {title} {{Primordial
  non-Gaussianities after Planck 2015: an introductory review}},\ }\href
  {https://doi.org/10.1016/j.crhy.2015.08.003} {\bibfield  {journal} {\bibinfo
  {journal} {Comptes Rendus Physique}\ }\textbf {\bibinfo {volume} {16}},\
  \bibinfo {pages} {969} (\bibinfo {year} {2015})},\ \Eprint
  {https://arxiv.org/abs/1508.06740} {arXiv:1508.06740 [astro-ph.CO]}
  \BibitemShut {NoStop}%
\bibitem [{\citenamefont {Ach{\'u}carro}\ \emph {et~al.}(2022)\citenamefont
  {Ach{\'u}carro} \emph {et~al.}}]{Achucarro:2022qrl}%
  \BibitemOpen
  \bibfield  {author} {\bibinfo {author} {\bibfnamefont {A.}~\bibnamefont
  {Ach{\'u}carro}} \emph {et~al.},\ }\bibfield  {title} {\bibinfo {title}
  {{Inflation: Theory and Observations}},\ }\href@noop {} {\  (\bibinfo {year}
  {2022})},\ \Eprint {https://arxiv.org/abs/2203.08128} {arXiv:2203.08128
  [astro-ph.CO]} \BibitemShut {NoStop}%
\bibitem [{\citenamefont {Cai}\ \emph {et~al.}(2019)\citenamefont {Cai},
  \citenamefont {Pi},\ and\ \citenamefont {Sasaki}}]{Cai:2018dig}%
  \BibitemOpen
  \bibfield  {author} {\bibinfo {author} {\bibfnamefont {R.-g.}\ \bibnamefont
  {Cai}}, \bibinfo {author} {\bibfnamefont {S.}~\bibnamefont {Pi}},\ and\
  \bibinfo {author} {\bibfnamefont {M.}~\bibnamefont {Sasaki}},\ }\bibfield
  {title} {\bibinfo {title} {{Gravitational Waves Induced by non-Gaussian
  Scalar Perturbations}},\ }\href
  {https://doi.org/10.1103/PhysRevLett.122.201101} {\bibfield  {journal}
  {\bibinfo  {journal} {Phys. Rev. Lett.}\ }\textbf {\bibinfo {volume} {122}},\
  \bibinfo {pages} {201101} (\bibinfo {year} {2019})},\ \Eprint
  {https://arxiv.org/abs/1810.11000} {arXiv:1810.11000 [astro-ph.CO]}
  \BibitemShut {NoStop}%
\bibitem [{\citenamefont {Unal}(2019)}]{Unal:2018yaa}%
  \BibitemOpen
  \bibfield  {author} {\bibinfo {author} {\bibfnamefont {C.}~\bibnamefont
  {Unal}},\ }\bibfield  {title} {\bibinfo {title} {{Imprints of Primordial
  Non-Gaussianity on Gravitational Wave Spectrum}},\ }\href
  {https://doi.org/10.1103/PhysRevD.99.041301} {\bibfield  {journal} {\bibinfo
  {journal} {Phys. Rev. D}\ }\textbf {\bibinfo {volume} {99}},\ \bibinfo
  {pages} {041301} (\bibinfo {year} {2019})},\ \Eprint
  {https://arxiv.org/abs/1811.09151} {arXiv:1811.09151 [astro-ph.CO]}
  \BibitemShut {NoStop}%
\bibitem [{\citenamefont {Garcia-Saenz}\ \emph {et~al.}(2023)\citenamefont
  {Garcia-Saenz}, \citenamefont {Pinol}, \citenamefont {Renaux-Petel},\ and\
  \citenamefont {Werth}}]{Garcia-Saenz:2022tzu}%
  \BibitemOpen
  \bibfield  {author} {\bibinfo {author} {\bibfnamefont {S.}~\bibnamefont
  {Garcia-Saenz}}, \bibinfo {author} {\bibfnamefont {L.}~\bibnamefont {Pinol}},
  \bibinfo {author} {\bibfnamefont {S.}~\bibnamefont {Renaux-Petel}},\ and\
  \bibinfo {author} {\bibfnamefont {D.}~\bibnamefont {Werth}},\ }\bibfield
  {title} {\bibinfo {title} {{No-go theorem for scalar-trispectrum-induced
  gravitational waves}},\ }\href
  {https://doi.org/10.1088/1475-7516/2023/03/057} {\bibfield  {journal}
  {\bibinfo  {journal} {JCAP}\ }\textbf {\bibinfo {volume} {03}},\ \bibinfo
  {pages} {057}},\ \Eprint {https://arxiv.org/abs/2207.14267} {arXiv:2207.14267
  [astro-ph.CO]} \BibitemShut {NoStop}%
\bibitem [{\citenamefont {Abac}\ \emph
  {et~al.}(2025{\natexlab{c}})\citenamefont {Abac} \emph
  {et~al.}}]{ET:2025xjr}%
  \BibitemOpen
  \bibfield  {author} {\bibinfo {author} {\bibfnamefont {A.}~\bibnamefont
  {Abac}} \emph {et~al.} (\bibinfo {collaboration} {ET}),\ }\bibfield  {title}
  {\bibinfo {title} {{The Science of the Einstein Telescope}},\ }\href@noop {}
  {\  (\bibinfo {year} {2025}{\natexlab{c}})},\ \Eprint
  {https://arxiv.org/abs/2503.12263} {arXiv:2503.12263 [gr-qc]} \BibitemShut
  {NoStop}%
\bibitem [{\citenamefont {Agazie}\ \emph {et~al.}(2023)\citenamefont {Agazie}
  \emph {et~al.}}]{NANOGrav:2023gor}%
  \BibitemOpen
  \bibfield  {author} {\bibinfo {author} {\bibfnamefont {G.}~\bibnamefont
  {Agazie}} \emph {et~al.} (\bibinfo {collaboration} {NANOGrav}),\ }\bibfield
  {title} {\bibinfo {title} {{The NANOGrav 15 yr Data Set: Evidence for a
  Gravitational-wave Background}},\ }\href
  {https://doi.org/10.3847/2041-8213/acdac6} {\bibfield  {journal} {\bibinfo
  {journal} {Astrophys. J. Lett.}\ }\textbf {\bibinfo {volume} {951}},\
  \bibinfo {pages} {L8} (\bibinfo {year} {2023})},\ \Eprint
  {https://arxiv.org/abs/2306.16213} {arXiv:2306.16213 [astro-ph.HE]}
  \BibitemShut {NoStop}%
\bibitem [{\citenamefont {Antoniadis}\ \emph {et~al.}(2023)\citenamefont
  {Antoniadis} \emph {et~al.}}]{EPTA:2023fyk}%
  \BibitemOpen
  \bibfield  {author} {\bibinfo {author} {\bibfnamefont {J.}~\bibnamefont
  {Antoniadis}} \emph {et~al.} (\bibinfo {collaboration} {EPTA, InPTA:}),\
  }\bibfield  {title} {\bibinfo {title} {{The second data release from the
  European Pulsar Timing Array - III. Search for gravitational wave signals}},\
  }\href {https://doi.org/10.1051/0004-6361/202346844} {\bibfield  {journal}
  {\bibinfo  {journal} {Astron. Astrophys.}\ }\textbf {\bibinfo {volume}
  {678}},\ \bibinfo {pages} {A50} (\bibinfo {year} {2023})},\ \Eprint
  {https://arxiv.org/abs/2306.16214} {arXiv:2306.16214 [astro-ph.HE]}
  \BibitemShut {NoStop}%
\bibitem [{\citenamefont {Reardon}\ \emph {et~al.}(2023)\citenamefont {Reardon}
  \emph {et~al.}}]{Reardon:2023gzh}%
  \BibitemOpen
  \bibfield  {author} {\bibinfo {author} {\bibfnamefont {D.~J.}\ \bibnamefont
  {Reardon}} \emph {et~al.},\ }\bibfield  {title} {\bibinfo {title} {{Search
  for an Isotropic Gravitational-wave Background with the Parkes Pulsar Timing
  Array}},\ }\href {https://doi.org/10.3847/2041-8213/acdd02} {\bibfield
  {journal} {\bibinfo  {journal} {Astrophys. J. Lett.}\ }\textbf {\bibinfo
  {volume} {951}},\ \bibinfo {pages} {L6} (\bibinfo {year} {2023})},\ \Eprint
  {https://arxiv.org/abs/2306.16215} {arXiv:2306.16215 [astro-ph.HE]}
  \BibitemShut {NoStop}%
\bibitem [{\citenamefont {Xu}\ \emph {et~al.}(2023)\citenamefont {Xu} \emph
  {et~al.}}]{Xu:2023wog}%
  \BibitemOpen
  \bibfield  {author} {\bibinfo {author} {\bibfnamefont {H.}~\bibnamefont {Xu}}
  \emph {et~al.},\ }\bibfield  {title} {\bibinfo {title} {{Searching for the
  Nano-Hertz Stochastic Gravitational Wave Background with the Chinese Pulsar
  Timing Array Data Release I}},\ }\href
  {https://doi.org/10.1088/1674-4527/acdfa5} {\bibfield  {journal} {\bibinfo
  {journal} {Res. Astron. Astrophys.}\ }\textbf {\bibinfo {volume} {23}},\
  \bibinfo {pages} {075024} (\bibinfo {year} {2023})},\ \Eprint
  {https://arxiv.org/abs/2306.16216} {arXiv:2306.16216 [astro-ph.HE]}
  \BibitemShut {NoStop}%
\bibitem [{\citenamefont {Amaro-Seoane}\ \emph {et~al.}(2017)\citenamefont
  {Amaro-Seoane} \emph {et~al.}}]{LISA:2017pwj}%
  \BibitemOpen
  \bibfield  {author} {\bibinfo {author} {\bibfnamefont {P.}~\bibnamefont
  {Amaro-Seoane}} \emph {et~al.} (\bibinfo {collaboration} {LISA}),\ }\bibfield
   {title} {\bibinfo {title} {{Laser Interferometer Space Antenna}},\
  }\href@noop {} {\  (\bibinfo {year} {2017})},\ \Eprint
  {https://arxiv.org/abs/1702.00786} {arXiv:1702.00786 [astro-ph.IM]}
  \BibitemShut {NoStop}%
\bibitem [{\citenamefont {Bartolo}\ \emph {et~al.}(2019)\citenamefont
  {Bartolo}, \citenamefont {De~Luca}, \citenamefont {Franciolini},
  \citenamefont {Lewis}, \citenamefont {Peloso},\ and\ \citenamefont
  {Riotto}}]{Bartolo:2018evs}%
  \BibitemOpen
  \bibfield  {author} {\bibinfo {author} {\bibfnamefont {N.}~\bibnamefont
  {Bartolo}}, \bibinfo {author} {\bibfnamefont {V.}~\bibnamefont {De~Luca}},
  \bibinfo {author} {\bibfnamefont {G.}~\bibnamefont {Franciolini}}, \bibinfo
  {author} {\bibfnamefont {A.}~\bibnamefont {Lewis}}, \bibinfo {author}
  {\bibfnamefont {M.}~\bibnamefont {Peloso}},\ and\ \bibinfo {author}
  {\bibfnamefont {A.}~\bibnamefont {Riotto}},\ }\bibfield  {title} {\bibinfo
  {title} {{Primordial Black Hole Dark Matter: LISA Serendipity}},\ }\href
  {https://doi.org/10.1103/PhysRevLett.122.211301} {\bibfield  {journal}
  {\bibinfo  {journal} {Phys. Rev. Lett.}\ }\textbf {\bibinfo {volume} {122}},\
  \bibinfo {pages} {211301} (\bibinfo {year} {2019})},\ \Eprint
  {https://arxiv.org/abs/1810.12218} {arXiv:1810.12218 [astro-ph.CO]}
  \BibitemShut {NoStop}%
\bibitem [{\citenamefont {Iovino}\ \emph
  {et~al.}(2025{\natexlab{b}})\citenamefont {Iovino}, \citenamefont {Perna},\
  and\ \citenamefont {Veerm{\"a}e}}]{Iovino:2025cdy}%
  \BibitemOpen
  \bibfield  {author} {\bibinfo {author} {\bibfnamefont {A.}~\bibnamefont
  {Iovino}, \bibfnamefont {Junior.}}, \bibinfo {author} {\bibfnamefont
  {G.}~\bibnamefont {Perna}},\ and\ \bibinfo {author} {\bibfnamefont
  {H.}~\bibnamefont {Veerm{\"a}e}},\ }\bibfield  {title} {\bibinfo {title}
  {{The impact of non-Gaussianity when searching for Primordial Black Holes
  with LISA}},\ }\href@noop {} {\  (\bibinfo {year} {2025}{\natexlab{b}})},\
  \Eprint {https://arxiv.org/abs/2512.13648} {arXiv:2512.13648 [astro-ph.CO]}
  \BibitemShut {NoStop}%
\bibitem [{\citenamefont {El-Neaj}\ \emph {et~al.}(2020)\citenamefont {El-Neaj}
  \emph {et~al.}}]{AEDGE:2019nxb}%
  \BibitemOpen
  \bibfield  {author} {\bibinfo {author} {\bibfnamefont {Y.~A.}\ \bibnamefont
  {El-Neaj}} \emph {et~al.} (\bibinfo {collaboration} {AEDGE}),\ }\bibfield
  {title} {\bibinfo {title} {{AEDGE: Atomic Experiment for Dark Matter and
  Gravity Exploration in Space}},\ }\href
  {https://doi.org/10.1140/epjqt/s40507-020-0080-0} {\bibfield  {journal}
  {\bibinfo  {journal} {EPJ Quant. Technol.}\ }\textbf {\bibinfo {volume}
  {7}},\ \bibinfo {pages} {6} (\bibinfo {year} {2020})},\ \Eprint
  {https://arxiv.org/abs/1908.00802} {arXiv:1908.00802 [gr-qc]} \BibitemShut
  {NoStop}%
\bibitem [{\citenamefont {Madau}\ and\ \citenamefont
  {Fragos}(2017)}]{Madau:2016jbv}%
  \BibitemOpen
  \bibfield  {author} {\bibinfo {author} {\bibfnamefont {P.}~\bibnamefont
  {Madau}}\ and\ \bibinfo {author} {\bibfnamefont {T.}~\bibnamefont {Fragos}},\
  }\bibfield  {title} {\bibinfo {title} {{Radiation Backgrounds at Cosmic Dawn:
  X-Rays from Compact Binaries}},\ }\href
  {https://doi.org/10.3847/1538-4357/aa6af9} {\bibfield  {journal} {\bibinfo
  {journal} {Astrophys. J.}\ }\textbf {\bibinfo {volume} {840}},\ \bibinfo
  {pages} {39} (\bibinfo {year} {2017})},\ \Eprint
  {https://arxiv.org/abs/1606.07887} {arXiv:1606.07887 [astro-ph.GA]}
  \BibitemShut {NoStop}%
\bibitem [{\citenamefont {Ng}\ \emph {et~al.}(2022)\citenamefont {Ng},
  \citenamefont {Franciolini}, \citenamefont {Berti}, \citenamefont {Pani},
  \citenamefont {Riotto},\ and\ \citenamefont {Vitale}}]{Ng:2022agi}%
  \BibitemOpen
  \bibfield  {author} {\bibinfo {author} {\bibfnamefont {K.~K.~Y.}\
  \bibnamefont {Ng}}, \bibinfo {author} {\bibfnamefont {G.}~\bibnamefont
  {Franciolini}}, \bibinfo {author} {\bibfnamefont {E.}~\bibnamefont {Berti}},
  \bibinfo {author} {\bibfnamefont {P.}~\bibnamefont {Pani}}, \bibinfo {author}
  {\bibfnamefont {A.}~\bibnamefont {Riotto}},\ and\ \bibinfo {author}
  {\bibfnamefont {S.}~\bibnamefont {Vitale}},\ }\bibfield  {title} {\bibinfo
  {title} {{Constraining High-redshift Stellar-mass Primordial Black Holes with
  Next-generation Ground-based Gravitational-wave Detectors}},\ }\href
  {https://doi.org/10.3847/2041-8213/ac7aae} {\bibfield  {journal} {\bibinfo
  {journal} {Astrophys. J. Lett.}\ }\textbf {\bibinfo {volume} {933}},\
  \bibinfo {pages} {L41} (\bibinfo {year} {2022})},\ \Eprint
  {https://arxiv.org/abs/2204.11864} {arXiv:2204.11864 [astro-ph.CO]}
  \BibitemShut {NoStop}%
\bibitem [{\citenamefont {Badurina}\ \emph {et~al.}(2020)\citenamefont
  {Badurina} \emph {et~al.}}]{Badurina:2019hst}%
  \BibitemOpen
  \bibfield  {author} {\bibinfo {author} {\bibfnamefont {L.}~\bibnamefont
  {Badurina}} \emph {et~al.},\ }\bibfield  {title} {\bibinfo {title} {{AION: An
  Atom Interferometer Observatory and Network}},\ }\href
  {https://doi.org/10.1088/1475-7516/2020/05/011} {\bibfield  {journal}
  {\bibinfo  {journal} {JCAP}\ }\textbf {\bibinfo {volume} {05}},\ \bibinfo
  {pages} {011}},\ \Eprint {https://arxiv.org/abs/1911.11755} {arXiv:1911.11755
  [astro-ph.CO]} \BibitemShut {NoStop}%
\bibitem [{\citenamefont {Pujolas}\ \emph {et~al.}(2021)\citenamefont
  {Pujolas}, \citenamefont {Vaskonen},\ and\ \citenamefont
  {Veerm{\"a}e}}]{Pujolas:2021yaw}%
  \BibitemOpen
  \bibfield  {author} {\bibinfo {author} {\bibfnamefont {O.}~\bibnamefont
  {Pujolas}}, \bibinfo {author} {\bibfnamefont {V.}~\bibnamefont {Vaskonen}},\
  and\ \bibinfo {author} {\bibfnamefont {H.}~\bibnamefont {Veerm{\"a}e}},\
  }\bibfield  {title} {\bibinfo {title} {{Prospects for probing gravitational
  waves from primordial black hole binaries}},\ }\href
  {https://doi.org/10.1103/PhysRevD.104.083521} {\bibfield  {journal} {\bibinfo
   {journal} {Phys. Rev. D}\ }\textbf {\bibinfo {volume} {104}},\ \bibinfo
  {pages} {083521} (\bibinfo {year} {2021})},\ \Eprint
  {https://arxiv.org/abs/2107.03379} {arXiv:2107.03379 [astro-ph.CO]}
  \BibitemShut {NoStop}%
\bibitem [{\citenamefont {Takahashi}\ and\ \citenamefont
  {Nakamura}(2003)}]{Takahashi:2003ix}%
  \BibitemOpen
  \bibfield  {author} {\bibinfo {author} {\bibfnamefont {R.}~\bibnamefont
  {Takahashi}}\ and\ \bibinfo {author} {\bibfnamefont {T.}~\bibnamefont
  {Nakamura}},\ }\bibfield  {title} {\bibinfo {title} {{Wave effects in
  gravitational lensing of gravitational waves from chirping binaries}},\
  }\href {https://doi.org/10.1086/377430} {\bibfield  {journal} {\bibinfo
  {journal} {Astrophys. J.}\ }\textbf {\bibinfo {volume} {595}},\ \bibinfo
  {pages} {1039} (\bibinfo {year} {2003})},\ \Eprint
  {https://arxiv.org/abs/astro-ph/0305055} {arXiv:astro-ph/0305055}
  \BibitemShut {NoStop}%
\bibitem [{\citenamefont {Abac}\ \emph
  {et~al.}(2025{\natexlab{d}})\citenamefont {Abac} \emph
  {et~al.}}]{LIGOScientific:2025rsn}%
  \BibitemOpen
  \bibfield  {author} {\bibinfo {author} {\bibfnamefont {A.~G.}\ \bibnamefont
  {Abac}} \emph {et~al.} (\bibinfo {collaboration} {LIGO Scientific, VIRGO,
  KAGRA}),\ }\bibfield  {title} {\bibinfo {title} {{GW231123: A Binary Black
  Hole Merger with Total Mass of 190{\textendash}265 solar masses}},\ }\href
  {https://doi.org/10.3847/2041-8213/ae0c9c} {\bibfield  {journal} {\bibinfo
  {journal} {Astrophys. J. Lett.}\ }\textbf {\bibinfo {volume} {993}},\
  \bibinfo {pages} {L25} (\bibinfo {year} {2025}{\natexlab{d}})},\ \Eprint
  {https://arxiv.org/abs/2507.08219} {arXiv:2507.08219 [astro-ph.HE]}
  \BibitemShut {NoStop}%
\bibitem [{LIG(2025)}]{LIGOScientific:2025cwb}%
  \BibitemOpen
  \bibfield  {title} {\bibinfo {title} {{GWTC-4.0: Searches for
  Gravitational-Wave Lensing Signatures}},\ }\href@noop {} {\  (\bibinfo {year}
  {2025})},\ \Eprint {https://arxiv.org/abs/2512.16347} {arXiv:2512.16347
  [gr-qc]} \BibitemShut {NoStop}%
\bibitem [{\citenamefont {Chan}\ \emph {et~al.}(2025)\citenamefont {Chan},
  \citenamefont {Ezquiaga}, \citenamefont {Lo}, \citenamefont {Bowman},
  \citenamefont {Maga{\~n}a~Zertuche},\ and\ \citenamefont
  {Vujeva}}]{Chan:2025kyu}%
  \BibitemOpen
  \bibfield  {author} {\bibinfo {author} {\bibfnamefont {J.~C.~L.}\
  \bibnamefont {Chan}}, \bibinfo {author} {\bibfnamefont {J.~M.}\ \bibnamefont
  {Ezquiaga}}, \bibinfo {author} {\bibfnamefont {R.~K.~L.}\ \bibnamefont {Lo}},
  \bibinfo {author} {\bibfnamefont {J.}~\bibnamefont {Bowman}}, \bibinfo
  {author} {\bibfnamefont {L.}~\bibnamefont {Maga{\~n}a~Zertuche}},\ and\
  \bibinfo {author} {\bibfnamefont {L.}~\bibnamefont {Vujeva}},\ }\bibfield
  {title} {\bibinfo {title} {{Discovering gravitational waveform distortions
  from lensing: a deep dive into GW231123}},\ }\href@noop {} {\  (\bibinfo
  {year} {2025})},\ \Eprint {https://arxiv.org/abs/2512.16916}
  {arXiv:2512.16916 [astro-ph.CO]} \BibitemShut {NoStop}%
\bibitem [{\citenamefont {Goyal}\ \emph {et~al.}(2025)\citenamefont {Goyal},
  \citenamefont {Villarrubia-Rojo},\ and\ \citenamefont
  {Zumalacarregui}}]{Goyal:2025eqo}%
  \BibitemOpen
  \bibfield  {author} {\bibinfo {author} {\bibfnamefont {S.}~\bibnamefont
  {Goyal}}, \bibinfo {author} {\bibfnamefont {H.}~\bibnamefont
  {Villarrubia-Rojo}},\ and\ \bibinfo {author} {\bibfnamefont {M.}~\bibnamefont
  {Zumalacarregui}},\ }\bibfield  {title} {\bibinfo {title} {{Across the
  Universe: GW231123 as a magnified and diffracted black hole merger}},\
  }\href@noop {} {\  (\bibinfo {year} {2025})},\ \Eprint
  {https://arxiv.org/abs/2512.17631} {arXiv:2512.17631 [astro-ph.GA]}
  \BibitemShut {NoStop}%
\bibitem [{\citenamefont {Chakraborty}\ and\ \citenamefont
  {Mukherjee}(2025)}]{Chakraborty:2025pxt}%
  \BibitemOpen
  \bibfield  {author} {\bibinfo {author} {\bibfnamefont {A.}~\bibnamefont
  {Chakraborty}}\ and\ \bibinfo {author} {\bibfnamefont {S.}~\bibnamefont
  {Mukherjee}},\ }\bibfield  {title} {\bibinfo {title} {{The First
  Model-Independent Upper Bound on Micro-lensing Signature of the Highest Mass
  Binary Black Hole Event GW231123}},\ }\href@noop {} {\  (\bibinfo {year}
  {2025})},\ \Eprint {https://arxiv.org/abs/2512.19077} {arXiv:2512.19077
  [gr-qc]} \BibitemShut {NoStop}%
\bibitem [{\citenamefont {Jung}\ and\ \citenamefont
  {Shin}(2019)}]{Jung:2017flg}%
  \BibitemOpen
  \bibfield  {author} {\bibinfo {author} {\bibfnamefont {S.}~\bibnamefont
  {Jung}}\ and\ \bibinfo {author} {\bibfnamefont {C.~S.}\ \bibnamefont
  {Shin}},\ }\bibfield  {title} {\bibinfo {title} {{Gravitational-Wave Fringes
  at LIGO: Detecting Compact Dark Matter by Gravitational Lensing}},\ }\href
  {https://doi.org/10.1103/PhysRevLett.122.041103} {\bibfield  {journal}
  {\bibinfo  {journal} {Phys. Rev. Lett.}\ }\textbf {\bibinfo {volume} {122}},\
  \bibinfo {pages} {041103} (\bibinfo {year} {2019})},\ \Eprint
  {https://arxiv.org/abs/1712.01396} {arXiv:1712.01396 [astro-ph.CO]}
  \BibitemShut {NoStop}%
\bibitem [{\citenamefont {Diego}(2020)}]{Diego:2019rzc}%
  \BibitemOpen
  \bibfield  {author} {\bibinfo {author} {\bibfnamefont {J.~M.}\ \bibnamefont
  {Diego}},\ }\bibfield  {title} {\bibinfo {title} {{Constraining the abundance
  of primordial black holes with gravitational lensing of gravitational waves
  at LIGO frequencies}},\ }\href {https://doi.org/10.1103/PhysRevD.101.123512}
  {\bibfield  {journal} {\bibinfo  {journal} {Phys. Rev. D}\ }\textbf {\bibinfo
  {volume} {101}},\ \bibinfo {pages} {123512} (\bibinfo {year} {2020})},\
  \Eprint {https://arxiv.org/abs/1911.05736} {arXiv:1911.05736 [astro-ph.CO]}
  \BibitemShut {NoStop}%
\bibitem [{\citenamefont {Liao}\ \emph {et~al.}(2020)\citenamefont {Liao},
  \citenamefont {Tian},\ and\ \citenamefont {Ding}}]{Liao:2020hnx}%
  \BibitemOpen
  \bibfield  {author} {\bibinfo {author} {\bibfnamefont {K.}~\bibnamefont
  {Liao}}, \bibinfo {author} {\bibfnamefont {S.}~\bibnamefont {Tian}},\ and\
  \bibinfo {author} {\bibfnamefont {X.}~\bibnamefont {Ding}},\ }\bibfield
  {title} {\bibinfo {title} {{Probing compact dark matter with gravitational
  wave fringes detected by the Einstein Telescope}},\ }\href
  {https://doi.org/10.1093/mnras/staa1388} {\bibfield  {journal} {\bibinfo
  {journal} {Mon. Not. Roy. Astron. Soc.}\ }\textbf {\bibinfo {volume} {495}},\
  \bibinfo {pages} {2002} (\bibinfo {year} {2020})},\ \Eprint
  {https://arxiv.org/abs/2001.07891} {arXiv:2001.07891 [astro-ph.CO]}
  \BibitemShut {NoStop}%
\bibitem [{\citenamefont {Urrutia}\ and\ \citenamefont
  {Vaskonen}(2021)}]{Urrutia:2021qak}%
  \BibitemOpen
  \bibfield  {author} {\bibinfo {author} {\bibfnamefont {J.}~\bibnamefont
  {Urrutia}}\ and\ \bibinfo {author} {\bibfnamefont {V.}~\bibnamefont
  {Vaskonen}},\ }\bibfield  {title} {\bibinfo {title} {{Lensing of
  gravitational waves as a probe of compact dark matter}},\ }\href
  {https://doi.org/10.1093/mnras/stab3118} {\bibfield  {journal} {\bibinfo
  {journal} {Mon. Not. Roy. Astron. Soc.}\ }\textbf {\bibinfo {volume} {509}},\
  \bibinfo {pages} {1358} (\bibinfo {year} {2021})},\ \Eprint
  {https://arxiv.org/abs/2109.03213} {arXiv:2109.03213 [astro-ph.CO]}
  \BibitemShut {NoStop}%
\bibitem [{\citenamefont {Basak}\ \emph {et~al.}(2022)\citenamefont {Basak},
  \citenamefont {Ganguly}, \citenamefont {Haris}, \citenamefont {Kapadia},
  \citenamefont {Mehta},\ and\ \citenamefont {Ajith}}]{Basak:2021ten}%
  \BibitemOpen
  \bibfield  {author} {\bibinfo {author} {\bibfnamefont {S.}~\bibnamefont
  {Basak}}, \bibinfo {author} {\bibfnamefont {A.}~\bibnamefont {Ganguly}},
  \bibinfo {author} {\bibfnamefont {K.}~\bibnamefont {Haris}}, \bibinfo
  {author} {\bibfnamefont {S.}~\bibnamefont {Kapadia}}, \bibinfo {author}
  {\bibfnamefont {A.~K.}\ \bibnamefont {Mehta}},\ and\ \bibinfo {author}
  {\bibfnamefont {P.}~\bibnamefont {Ajith}},\ }\bibfield  {title} {\bibinfo
  {title} {{Constraints on Compact Dark Matter from Gravitational Wave
  Microlensing}},\ }\href {https://doi.org/10.3847/2041-8213/ac4dfa} {\bibfield
   {journal} {\bibinfo  {journal} {Astrophys. J.}\ }\textbf {\bibinfo {volume}
  {926}},\ \bibinfo {pages} {L28} (\bibinfo {year} {2022})},\ \Eprint
  {https://arxiv.org/abs/2109.06456} {arXiv:2109.06456 [gr-qc]} \BibitemShut
  {NoStop}%
\bibitem [{\citenamefont {Zumalac{\'a}rregui}(2024)}]{Zumalacarregui:2024ocb}%
  \BibitemOpen
  \bibfield  {author} {\bibinfo {author} {\bibfnamefont {M.}~\bibnamefont
  {Zumalac{\'a}rregui}},\ }\bibfield  {title} {\bibinfo {title} {{Lens
  Stochastic Diffraction: A Signature of Compact Structures in
  Gravitational-Wave Data}},\ }\href@noop {} {\  (\bibinfo {year} {2024})},\
  \Eprint {https://arxiv.org/abs/2404.17405} {arXiv:2404.17405 [gr-qc]}
  \BibitemShut {NoStop}%
\bibitem [{\citenamefont {Kim}\ \emph {et~al.}(2025)\citenamefont {Kim},
  \citenamefont {Gil~Choi},\ and\ \citenamefont {Jung}}]{Kim:2025njb}%
  \BibitemOpen
  \bibfield  {author} {\bibinfo {author} {\bibfnamefont {S.}~\bibnamefont
  {Kim}}, \bibinfo {author} {\bibfnamefont {H.}~\bibnamefont {Gil~Choi}},\ and\
  \bibinfo {author} {\bibfnamefont {S.}~\bibnamefont {Jung}},\ }\bibfield
  {title} {\bibinfo {title} {{Probing small-scale power spectrum with
  gravitational-wave diffractive lensing}},\ }\href
  {https://doi.org/10.1007/JHEP07(2025)006} {\bibfield  {journal} {\bibinfo
  {journal} {JHEP}\ }\textbf {\bibinfo {volume} {07}},\ \bibinfo {pages}
  {006}},\ \Eprint {https://arxiv.org/abs/2501.14904} {arXiv:2501.14904
  [hep-ph]} \BibitemShut {NoStop}%
\end{thebibliography}%

\end{document}